\documentclass[%
 superscriptaddress,
preprintnumbers,
nofootinbib,
 amsmath,amssymb,
 aps,
longbibliography,
notitlepage
]{revtex4-1}

\usepackage{graphicx}
\usepackage{hyperref}
\usepackage{color}
\usepackage{braket}
\usepackage{leftindex}
\usepackage{bm}
\usepackage[normalem]{ulem} 
\usepackage[dvipsnames]{xcolor}
\usepackage{comment}
\usepackage{array}
\usepackage{relsize}
\usepackage{amsmath}

\usepackage{makecell}
\usepackage{tikz}
\usetikzlibrary{shapes}
\pgfdeclarelayer{bg}
\pgfsetlayers{bg,main}
\usetikzlibrary{decorations.pathmorphing}
\usetikzlibrary{arrows.meta}
\tikzset{snake it/.style={decorate, decoration=snake}}
\usetikzlibrary{calc}
\usepackage{tikz-3dplot} 
\tdplotsetmaincoords{70}{120}  
\tdplotsetrotatedcoords{0}{0}{0} 
\usepackage{tikzit}

\tikzstyle{pointoperator}=[fill=blue, draw=none, shape=circle, minimum size=.1 cm, inner sep=0 pt]
\tikzstyle{pointred}=[fill=red, draw=none, shape=circle, minimum size=.1 cm, inner sep=0 pt]
\tikzstyle{pointblack}=[fill=black, draw=none, shape=circle, minimum size=.1 cm, inner sep=0 pt]

\tikzstyle{dashedline}=[dashed, -, thick]
\tikzstyle{blueline}=[-, draw=blue, thick]
\tikzstyle{redline}=[-, draw=red, thick]
\tikzstyle{greenline}=[-, draw=green, thick]
\tikzstyle{arrowline}=[<-, thick]
\tikzstyle{bluearrowline}=[<-, draw=blue, thick]
\tikzstyle{normalline}=[-, thick]
\tikzstyle{greenfillline}=[-, thick, fill=green, fill opacity=.5]
\tikzstyle{grayfillline}=[-, fill opacity=.5, fill={rgb,255: red,128; green,128; blue,128}, thick]
\tikzstyle{redfillline}=[-, fill opacity=.5, fill=red, thick]
\tikzstyle{bluefillline}=[-, thick, fill=blue, fill opacity=.5]
\tikzstyle{bluedashedline}=[-, draw=blue, thick, dashed]
\tikzstyle{reddashedline}=[-, draw=red, thick, dashed]
\tikzstyle{grayfilldashedline}=[dashed, -, fill opacity=.5, fill={rgb,255: red,128; green,128; blue,128}, thick]

\newcommand{\hexagon}{\mathord{\raisebox{0.6pt}{\tikz{\node[draw,scale=.65,regular polygon, regular polygon sides=6](){};}}}}
\begin{document}

\title{Non-invertible defects in generalized Ising models via strange correlator}

\author{Aswin Parayil Mana \& Yaman Sanghavi}

\email{aswin.parayilmana@stonybrook.edu \& yaman.sanghavi@stonybrook.edu}

\affiliation{C.N. Yang Institute for Theoretical Physics \& Department of Physics and Astronomy, Stony Brook University, Stony Brook, NY 11794, USA}

\begin{abstract}
Defects associated with non-invertible symmetries have attracted significant attention in recent years. Among them, Kramers–Wannier (KW) duality defects have been investigated in both classical statistical systems and quantum Hamiltonian models. Aasen et al. analyzed duality defects in the 2D Ising model and in statistical models built from fusion categories, while Koide et al. later constructed a duality defect in 4D lattice gauge theory. In this work, we extend these developments by providing a systematic construction of KW duality defects/KW defects for a broad class of models formulated within the chain complex framework. Our construction employs the strange correlator—an overlap between a topologically ordered state and a product state—to realize these KW defects. 
\end{abstract}

\maketitle

\def\thefootnote{\arabic{footnote}}
\setcounter{footnote}{0}

\tableofcontents
\section{Introduction}
Generalized symmetries and their associated defects have recently emerged as central themes across condensed matter physics, high-energy theory, and mathematics (see \cite{Gaiotto:2014kfa,McGreevy:2022oyu,Cordova:2022ruw,Brennan:2023mmt,Bhardwaj:2023kri,Schafer-Nameki:2023jdn,Luo:2023ive,Shao:2023gho,Carqueville:2023jhb}). Such symmetries can impose powerful constraints on the structure of phases of matter and the nature of their transitions. They may, for example, restrict the space of renormalization group (RG) flows~\cite{Chang:2018iay,Apte:2022xtu} or enforce Lieb–Schultz–Mattis–type obstructions in lattice systems~\cite{Seiberg:2024gek,Pace:2024acq}. Within this broad landscape, a particularly intriguing class is that of non-invertible symmetries. While their systematic study in lattice models is relatively recent, one of the earliest and most celebrated realizations is the Kramers–Wannier (KW) duality of the Ising model~\cite{Kramers:1941kn,Kramers:1941zz,Aasen:2016dop,Grimm:1992yr,Oshikawa:1996dj,Ho:2014vla,Li:2023ani,Hauru:2015abi,Seiberg:2023cdc}. Extensions of KW duality to higher-dimensional systems have also been developed~\cite{Moradi:2023dan,Cao:2023doz,ParayilMana:2024txy,Ebisu:2024lie,Kaidi:2021xfk}. More generally, non-invertible symmetries arising from categorical structures and their associated defects have been extensively investigated in both field-theoretic and lattice settings~\cite{Aasen:2020jwb,Thorngren:2019iar,Thorngren:2021yso,Inamura:2021szw,Inamura:2023qzl,Cui:2024cav,Bhardwaj:2022yxj,Lootens:2021tet,Delcamp:2023kew,Bhardwaj:2024wlr,Bhardwaj:2024xcx,Decoppet:2023bay}. Recent works have further demonstrated their realization and physical consequences in a wide range of lattice models~\cite{Lootens:2021tet,Lootens:2022avn,Eck:2023gic,Seiberg:2023cdc,Fechisin:2023odt,Inamura:2021szw,Inamura:2024jke,Pace:2024acq,Seifnashri:2024dsd,Li:2024fhy,Lu:2025rwd,Jia:2024bng,Cao:2024qjj,Lu:2024ytl,Li:2024gwx,Jia:2024zdp,Jia:2024wnu,Meng:2024nxx,Gorantla:2024ocs,Pace:2024tgk,Chatterjee:2024ych,Pace:2024oys,Bhardwaj:2024kvy,Bhardwaj:2024wlr,Cao:2025qhg,Cao:2025qnc,Seifnashri:2025fgd,ParayilMana:2025nxw,Lu:2025yru}.

The KW duality~\cite{Kramers:1941kn,Kramers:1941zz} relates the partition function of the two-dimensional classical Ising model at high temperature (disordered phase) to that of its dual lattice at low temperature (ordered phase). At the critical temperature, the two partition functions coincide, and the duality is realized as an exact symmetry. An analogous structure appears in the $1+1$D quantum Ising chain, where the self-dual point exhibits KW duality as a symmetry of the Hamiltonian. Operator formulations of this symmetry in the quantum chain have been constructed in recent works~\cite{Seiberg:2023cdc,Chen:2023qst}.

Given any symmetry of a quantum Hamiltonian, it is natural to ask about its corresponding defects. Symmetry defects modify boundary conditions, even on closed manifolds, and thereby serve as direct probes of universal data. For instance, in the transverse-field Ising chain at the self-dual point, the duality defect allows one to extract conformal data of the associated two-dimensional Ising CFT~\cite{Hauru:2015abi}. In general, a KW defect can be realized by gauging the symmetries associated with KW duality in only part of the system. More broadly, defects associated with non-invertible symmetries have been explored in a wide variety of settings~\cite{Frohlich:2006ch,Frohlich:2004ef,Frohlich:2009gb,Chang:2018iay,Lin:2019hks,Choi:2021kmx,Kaidi:2021xfk,Kaidi:2022cpf,Choi:2022zal,Koide:2021zxj,Bhardwaj:2024xcx}.

A systematic framework for constructing topological defects in the 2D Ising model and related classical lattice systems with fusion-category symmetries was developed in~\cite{Aasen:2016dop,Aasen:2020jwb}. In this formalism, defects are characterized by “defect commutation relations,” analogues of the Yang–Baxter equations, together with additional constraints that encode the quantum dimensions of the defect lines. This approach successfully recovers the KW defect in the 2D Ising model. More recently, such methods have been extended to four-dimensional lattice gauge theories~\cite{Koide:2021zxj}\footnote{For example, the four-dimensional $\mathbb{Z}_2$ lattice gauge theory with Hamiltonian $H_{\text{LG}}=-J\sum_{\rm p}\prod_{\rm e\in\partial p}S_{\rm e}$ (where $\rm p$ runs over plaquettes and $\rm e$ over their bounding edges) is KW dual to the same theory at coupling $J^*=-\frac{1}{2}\log\tanh(J)$.}. Nevertheless, a general and flexible framework for constructing duality defects in arbitrary dimensions remains elusive.

In this work, we propose an alternative construction of KW-type defects based on the strange correlator. Originally introduced as a diagnostic for symmetry-protected topological (SPT) phases~\cite{You:2013cpa}, the strange correlator has since been extended to capture topological order in two spatial dimensions~\cite{Vanhove:2018wlb}. More recently, it has been used to uncover KW dualities among Wegner models and generalized Ising models defined via chain complexes~\cite{Okuda:2024jzh,Okuda:2024azp}. In that framework, the duality arises by gauging the global $\mathbb{Z}_2$ symmetry of the model at coupling $J$, which produces the dual model at $J^*=-\tfrac{1}{2}\log\tanh J$. Concretely, one constructs a cluster state from the chain complex of a CSS code, whose overlap with a particular product state yields the partition function of a generalized Ising model. A different cluster state defined on the same chain complex gives, upon overlap with another product state, the partition function of the dual model. The duality can then be interpreted as arising from a relation between the two topologically ordered states obtained from the two cluster states.

Here we extend this perspective to defects. Instead of gauging the full system, we gauge only a subregion, thereby generating a KW defect on the interface. Operationally, we construct two cluster states from the chain complex: one identical to that used in~\cite{Okuda:2024jzh,Okuda:2024azp}, and another obtained by stitching together the two cluster states of those works along appropriate regions. Then we take overlap of these cluster states with some appropriate product states, as part of the strange correlator framework, and relate them. The former gives us the partition function of the corresponding generalized Ising model and the latter, i.e., stitched cluster state gives us the partition function of the corresponding generalized Ising model with a subregion being gauged by $\mathbb{Z}_2$ symmetry. We can then read off the KW defect from it. This procedure is highly flexible and applies to a broad family of chain-complex–derived Ising models, including the square- and triangular-lattice 2D Ising models, the 3D anisotropic plaquette Ising model, and the 4D Ising gauge theory, among others\footnote{See Section~\ref{sec:KWdefectgenIsingmodel} for the general chain-complex formulation.}.\\
In this paper we:
\begin{itemize}
    \item Introduce a strange-correlator construction of KW defects.
    \item Explicitly realize these defects in 2D and  3D lattice models.
    \item Analyze their fusion rules and associated condensation defects.
    \item Derive the corresponding Hamiltonian realizations.
    \item Generalize the construction to generalized Ising models constructed from a chain-complex in arbitrary dimensions.
\end{itemize}
The paper is organized as follows. In Section~\ref{sec:Dualitydefect2DIsingmodel}, we revisit the 2D Ising model and obtain its duality defect using quantum-information–theoretic tools, analyzing both fusion rules and Hamiltonian realizations. Section~\ref{sec:DualitydefectAPImodel} applies the method to the anisotropic plaquette Ising model. Section~\ref{sec:Duslitydefect3DIG3DImodel} constructs an interface between the 3D Ising gauge theory and the 3D Ising model, interpretable as a KW defect, and studies the fusion of two such interfaces, including the emergence of condensation defects. Section~\ref{sec:KWdefectgenIsingmodel} generalizes the framework to chain-complex–based Ising models in arbitrary dimensions. Technical details, including computations for the 2D Ising case and normalization conventions for topologically ordered states, are provided in Appendices~\ref{app:2DIsingSC} and~\ref{app:topologicallyorderedstatenorm}, while Appendix~\ref{app:triangularlatticeIsing} works out the KW defect between the triangular- and hexagonal-lattice Ising models.
\section{Duality defect in 2D Ising model}\label{sec:Dualitydefect2DIsingmodel}
In this section, we will construct duality defects in 2D Ising model using the concept of strange correlators. Later, we will derive its fusion rules and the corresponding quantum Hamiltonian. 

The two-dimensional (2D) Ising model is known to undergo a phase transition between an ordered phase at low temperatures and a disordered phase at high temperatures~\cite{peierls1936ising,onsager1944crystal}. A notable feature of this model is the Kramers-Wannier duality, which relates the partition function at low temperature to that at high temperature~\cite{Kramers:1941kn,Kramers:1941zz}. In this work, we revisit the derivation of Kramers-Wannier duality using an alternative approach based on strange correlators~\cite{Okuda:2024azp,Okuda:2024jzh}.  This method not only provides a new perspective on the duality in two dimensions but also offers a natural framework for generalization to higher-dimensional systems, as well as various other lattice structures such as triangular lattices. 
\subsection{$2$D Ising model and Kramers-Wannier duality}
\subsubsection{Hamiltonian}
We consider the two-dimensional (2D) Ising model defined on a square lattice with periodic boundary conditions, where classical spin variables $s_{\rm v}=\pm 1$ reside on each vertex $\rm v$. Let $\Delta_{\rm v}$, $\Delta_{\rm e}$ be the set of all the vertices and edges respectively. The pair of vertices at the boundary of an edge $\rm e$ is defined as $\partial \rm e$. The 2D Ising Hamiltonian is given by
\begin{align}\label{eq:IsingH}
    \mathcal{H}_{\text{Ising}}(\{s_{\rm v}\})=-J\sum_{\rm e\in\Delta_{\rm e}}\prod_{\rm v\in\partial e}s_{\rm v}\, ,
\end{align}
where $J$ is the coupling between the spins. The partition function is given by
\begin{align}\label{eq:IsingZ}
    Z_{\text{Ising}}(J)=\sum_{\{s_{\rm v}\}}e^{-\mathcal{H}_{\text{Ising}}(\{s_{\rm v}\})}\, ,
\end{align}
where we have kept the temperature $T=1$ without loss of generality. 
\subsubsection{$\mathbb{Z}_2$ symmetry defect}
The model has a $\mathbb{Z}_2$ symmetry $s_{\rm v} \to -s_{\rm v}$ $\forall$ ${\rm v}$ that leaves the Hamiltonian \eqref{eq:IsingH} invariant.
Given a non-trivial proper subset $S$ of spins $s_{\rm v}$, if we flip them $s_{\rm v} \to -s_{\rm v}$, the partition function \eqref{eq:IsingZ} will remain invariant since all the spins $\{s_{\rm v}\}$ are being summed. However, the Hamiltonian changes in the following way
\begin{align}\label{eq:IsingZ2Defect}
    H_{\eta(z_1)} = -J\sum_{\rm e} (-1)^{\#({\rm e} \cap z_1)} \prod_{\rm v\in\partial e}s_{\rm v}\,,
\end{align}
where $\eta(z_1)$ denotes the $\mathbb{Z}_2$ defect over the closed loop $z_1$ in the dual lattice which bounds $S$ and ${\#({\rm e} \cap z_1)}$ denotes the intersection number between the link $\rm e$ and $z_1$. That is, the coupling $J$ is flipped to $-J$ for all the links $\rm e$ that intersect $z_1$ an odd number of times. Although the loop $z_1$ we created is homologically trivial, we can take \eqref{eq:IsingZ2Defect} as the definition of a $\mathbb{Z}_2$ defect for any cycle $z_1$ in $H_1(M,\mathbb{Z}_2)$. 

\begin{figure}
    \centering 
    \input{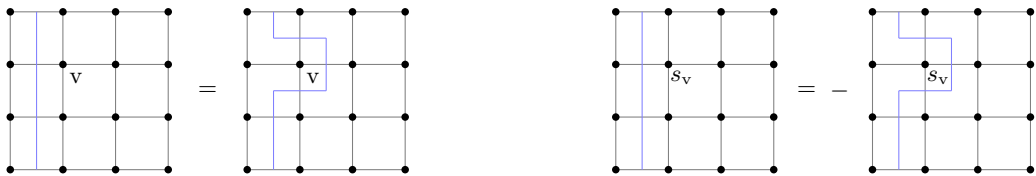}
    \caption{Left: $\mathbb{Z}_2$ defect (blue) being moved by transforming $s_{\rm v} \to -s_{\rm v}$. Right: Moving the $\mathbb{Z}_2$ defect across $v$ in presence of $s_{\rm v}$ inside the correlation function gives a $-1$ factor.}
    \label{fig:IsingZ2TopMove}
\end{figure}

This defect is topological in the way that we can move the defect from the cycle $z_1$ to a cycle $z'_1$ in the same homology class without changing the partition function. It can be seen by transforming the spins $s_{\rm v} \to -s_{\rm v}$ for the vertices ${\rm v}$ that belong to the region bounded by $z_1$ and $z'_1$. This transformation does not change the partition function, since all spins $\{s_{\rm v}\}$ are being summed. For example, if we take a spin $s_{\rm v}$ where ${\rm v}$ is adjacent to $z$ and transform $s_{\rm v} \to - s_{\rm v}$, then we essentially move the defect by one dual plaquette as shown in the left Figure~\ref{fig:IsingZ2TopMove}. Moreover, if there was a spin $s_{\rm v}$ within a correlation function and if we transform $s_{\rm v} \to -s_{\rm v}$, we will move the defect and get a $-1$ factor. In other words, inside a correlation function, sweeping a $\mathbb{Z}_2$ defect across $s_{\rm v}$ produces a factor of $-1$ as shown in the right Figure~\ref{fig:IsingZ2TopMove}.

\subsubsection{Kramers-Wannier Duality}
Kramers-Wannier can be reframed as a strong-weak duality for a constant temperature. Specifically, the weak coupling $J \ll 1$ expansion of the Ising model is related to the strong coupling $J^* \gg 1$ expansion of the original model via the duality relation $J^*=-\frac{1}{2}\log \tanh J$. In the following, we review~\cite{Okuda:2024jzh,Okuda:2024azp} a rederivation of this duality using tools from quantum information theory, providing a formulation that will be instrumental in extending the duality to higher-dimensional systems.

Let us reconsider a square lattice where qubits are positioned on both vertices and edges. We again label the vertices as $\rm v$, edges as $\rm e$, and plaquettes as $\rm p$. The collection of vertices is specified by $\Delta_{\rm v}$, the collection of edges by $\Delta_{\rm e}$, and the collection of plaquettes by $\Delta_{\rm p}$. Let's denote a generic qubit on an edge $\rm e$ as $\ket{\psi}_{\rm e}$ and on a vertex $\rm v$ as $\ket{\psi}_{\rm v}$. Initially, all qubits are set in the $\ket{+}$ state, which is the +1 eigenstate of the Pauli-$X$ operator. Then, a $CZ$ gate is applied between adjacent edges and vertices. This configuration represents the cluster state on this lattice. In detail,
\begin{align}\label{eq:2dIsingClusterState}
    |\Psi_{\text{cluster}}\rangle=\prod_{\rm v\in\Delta_v}\prod_{\rm e\in\partial^*v}CZ_{\rm v,e}\ket{+}^{\rm\Delta_v}\ket{+}^{\rm\Delta_e}\, .
\end{align}
where $\ket{+}^{\Delta_{\rm e}} = \otimes_{\rm e} \ket{+}_{\rm e}$ and $\ket{+}^{\Delta_{\rm v}} = \otimes_{\rm v} \ket{+}_{\rm v}$ and the notation $\rm \partial^*v$ denotes the set of all edges $\rm e$ for which $\rm v$ is one of the boundary vertex. The state above is the ground state of the cluster Hamiltonian: \begin{align}\label{eq:2Dclusterham}
    \rm H_{cluster}=-\sum_v \raisebox{-18pt}
{\begin{tikzpicture}
\draw[-,black!30,line width=1.0] (0.0,0.0) -- (0.5,0.0) -- (1.0,0.0);
\draw[-,black!30,line width=1.0] (0.5,-0.5) -- (0.5,0.0) -- (0.5,0.5);
\node at (0.0,0.0) {$Z$};
\node at (1.0,0.0) {$Z$};
\node at (0.5,0.0) {$X$};
\node at (0.5,0.5) {$Z$};
\node at (0.5,-0.5) {$Z$};
\end{tikzpicture}} 
-\sum_{e} 
\raisebox{-5pt}
{\begin{tikzpicture}
\draw[-,black!30,line width=1.0] (0.0,0.0) -- (0.5,0.0) -- (1.0,0.0);
\node at (0.0,0.0) {$Z$};
\node at (1.0,0.0) {$Z$};
\node at (0.5,0.0) {$X$};
\end{tikzpicture} }\,. 
\end{align} The cluster state is the common $+1$ eigenstate of all Hamiltonian terms~\eqref{eq:2Dclusterham}. It is observed that having an overlap of $|\Psi_{\text{cluster}}\rangle$ with the product state $\ket{+}^{\Delta_{\rm v}}$ results in the symmetric ground state of the toric code\footnote{The resulting toric code ground state is symmetric under the 1-form symmetry generated by product of $X$ operators on a closed loop in the lattice.}. \begin{align}\label{eq:2dIsingTCGS}
    \rm |TC_{GS}\rangle=\langle +|^{\Delta_v}|\Psi_{\text{cluster}}\rangle\, .
\end{align} The toric code Hamiltonian is characterized by \begin{align}\label{eq:toriccodeham}
    \rm H_{TC}=-\sum_v\raisebox{-18pt}
{\begin{tikzpicture}
\draw[-,black!30,line width=1.0] (0.0,0.0) -- (1.0,0.0);
\draw[-,black!30,line width=1.0] (0.5,-0.5) -- (0.5,0.5);
\node at (0.0,0.0) {$Z$};
\node at (1.0,0.0) {$Z$};
\node at (0.5,0.5) {$Z$};
\node at (0.5,-0.5) {$Z$};
\end{tikzpicture} } -\sum_p \raisebox{-18pt}{\begin{tikzpicture}
\draw[-,black!30,line width=1.0] (-0.5,-0.5) -- (0.5,-0.5) -- (0.5,0.5) -- (-0.5,0.5) -- (-0.5,-0.5);
\node at (-0.5,0.05) {$X$};
\node at (0.5,0.05) {$X$};
\node at (0.0,0.53) {$X$};
\node at (0.0,-0.45) {$X$};
\end{tikzpicture} } \,.
\end{align} The logical $X$ operators are defined as the product of $X$ along $z_1$, where $z_1$ constitutes a 1-cycle\footnote{Here we take $z_1$ as the set of edges that form a “cycle" on the lattice as we have not yet introduced the chain complex formalism.} residing in the non-trivial homology class of the square lattice, and is denoted by $X(z_1)$. Furthermore, the logical $Z$ operators are defined by the product of $Z$ along $z_1^*$, where $z_1^*$ is a dual 1-cycle within a non-trivial cohomology class of the square lattice. The logical $X$ and $Z$ operators form the $\mathbb{Z}_2^{(1)}\times\mathbb{Z}_2^{(1)}$ 1-form symmetries of the toric code. The state $\rm |TC_{GS}\rangle$ is stabilized by all the Hamiltonian terms~\eqref{eq:toriccodeham} as well as by the logical operators $X(z_1)$.

Let's denote the eigenstates of $Z$ as $\ket{0}$ and $\ket{1}$ with eigenvalues $+1$ and $-1$ respectively. We now consider a state $(e^{J X_{\rm e}}\ket{0})^{\Delta_{\rm e}}$ and take its overlap with $\ket{\rm TC_{GS}}$. As shown in the Appendix~\ref{app:2DIsingSC}, this overlap yields the partition function of the Ising model at $J$.

\begin{align}
  (\langle 0|e^{J X_{\rm e}})^{\rm \Delta_e}  |\rm TC_{GS}\rangle=\frac{1}{2^{|\Delta_{\rm e}|/2}}\frac{1}{2^{|\Delta_{\rm v}|}}Z_{\text{Ising}}(J)\,.
\end{align}

Now consider the cluster state on the dual lattice of $2$D square lattice, where vertices of the dual lattice correspond one-to-one with the plaquettes of the original lattice. The cluster state on this dual lattice is described by
\begin{align}\label{eq:2dIsingCluster}
    |\Psi^*_{\rm cluster}\rangle=\prod_{\rm p\in\Delta_p}\prod_{\rm e\in\partial p}CZ_{\rm p,e}\ket{+}^{\rm\Delta_p}\ket{+}^{\rm\Delta_e}\, .
\end{align}
The notation $\partial \rm p$ denote the set of all edges on the boundary of the plaquette $\rm p$. This particular state is a ground state of the Hamiltonian
\begin{align}
    \rm H^*_{cluster}=-\sum_{\rm p} \raisebox{-18pt}{\begin{tikzpicture}
\draw[-,black!30,line width=1.0] (-0.5,-0.5) -- (0.5,-0.5) -- (0.5,0.5) -- (-0.5,0.5) -- (-0.5,-0.5);
\node at (-0.5,0.05) {$Z$};
\node at (0.5,0.05) {$Z$};
\node at (0.0,0.5) {$Z$};
\node at (0.0,-0.5) {$Z$};
\node at (0.0,0.0) {$X$};
\end{tikzpicture} }-\sum_{\rm e}\raisebox{-28pt}{\begin{tikzpicture}
\draw[-,black!30,line width=1.0] (1,0) -- (1,1) -- (0,1) --(0,-1)--(1,-1)--(1,0);
\draw[-,black!30,line width=1.0]  (0,0) --(1,0) ;
\node at (0.5,0.5) {$Z$};
\node at (0.5,-0.5) {$Z$};
\node at (0.5,0) {$X$};
\node at (1,0){};
\node at (1,1){};
\node at (0,1){};
\node at (0,-1){};
\node at (1,-1){};
\end{tikzpicture} }\,.
\end{align} 
An overlap with $\ket{+}^{\Delta_{\rm p}}$ results in the ground state of the toric code on the dual lattice
\begin{align}
    |\rm TC^*_{GS}\rangle=\bra{+}^{\Delta_{\rm p}}|\Psi^*_{\rm cluster}\rangle\, .
\end{align} 
Nevertheless, the Hamiltonian can be expressed on the original lattice as follows
\begin{align}\label{eq:dualtoriccodeham}
    \rm H^*_{TC}=-\sum_v\raisebox{-18pt}
{\begin{tikzpicture}
\draw[-,black!30,line width=1.0] (0.0,0.0) -- (1.0,0.0);
\draw[-,black!30,line width=1.0] (0.5,-0.5) -- (0.5,0.5);
\node at (0.0,0.0) {$X$};
\node at (1.0,0.0) {$X$};
\node at (0.5,0.5) {$X$};
\node at (0.5,-0.5) {$X$};
\end{tikzpicture} } -\sum_p \raisebox{-18pt}{\begin{tikzpicture}
\draw[-,black!30,line width=1.0] (-0.5,-0.5) -- (0.5,-0.5) -- (0.5,0.5) -- (-0.5,0.5) -- (-0.5,-0.5);
\node at (-0.5,0.05) {$Z$};
\node at (0.5,0.05) {$Z$};
\node at (0.0,0.53) {$Z$};
\node at (0.0,-0.45) {$Z$};
\end{tikzpicture} }\,.
\end{align} 
The logical $X$ operator is  $X(z_1^*)$, given by product of $X$ operators along the edges in the dual 1-cycle $z_1^*$ residing in the nontrivial cohomological class. Similarly, the logical $Z$ operator is defined by $Z(z_1)$, with $z_1$ representing a 1-cycle situated in the nontrivial homological class. The logical operator that stabilizes $\ket{\Psi_{\text{cluster}}^*}$ is $X(z_1^*)$.

The Hamiltonian $\rm H_{TC}$ in~\eqref{eq:toriccodeham} and $\rm H_{TC}^{*}$ in \eqref{eq:dualtoriccodeham} are related by a Hadamard operator $\mathsf{H}\equiv \bigotimes_{\rm e\in \Delta_e}H_e$. However, their ground states $|\rm TC_{GS}\rangle$ and $|\rm TC_{GS}^*\rangle$ are not just related by the Hadamard operator $\mathsf{H}$ as they are stabilized by different logical operators. To relate them, we can use the logical operators $Z(z_1)$ that stabilize $\mathsf{H}|\mathrm{TC}_{\text{GS}}\rangle$ as
\begin{align}\label{}
   \frac{1}{\mathcal{N}_{\text{TC}}}\mathsf{H}|\mathrm{TC}_{\text{GS}}\rangle=\frac{1}{\mathcal{N}^*_{\text{TC}}} \frac{1}{H_1(M,\mathbb{Z}_2)}\sum_{z_1\in H_1(M,\mathbb{Z}_2)}Z(z_1)|\rm TC^*_{GS}\rangle\,.
   \label{eq:TCGSandTCGS*relation}
\end{align}
where $\mathcal{N}_{\text{TC}}$ and $\mathcal{N}^*_{\text{TC}}$ are normalization factors, shown below, that ensure that both sides of the equation above are normalized to 1. Their values are
\begin{align}
  \mathcal{N}_{\text{TC}} = \sqrt{ \braket{\rm TC_{GS} | TC_{GS}}}=\frac{\sqrt{2}}{2^{\frac{|\Delta_{\rm v}|}{2}}} \quad \quad \quad \quad   \mathcal{N}^*_{\text{TC}} = \sqrt{\bra{\rm TC_{GS}^*} \frac{1}{H_1(M,\mathbb{Z}_2)}\sum_{z_1 \in H_1(M, \mathbb{Z}_2)} Z(z_1)\ket{\rm TC_{GS}^*} }=\frac{1}{\sqrt{H_1(M,\mathbb{Z}_2)}}\frac{\sqrt{2}}{2^{\frac{|\Delta_{\rm p}|}{2}}}\,.
\end{align}

Now we note the fact that $\bra{0}_{\rm e}e^{J X_{\rm e}} H_e=(\sinh (2J))^{1/2}\bra{0}_{\rm e}e^{J^*X_{\rm e}}$ with $e^{-2J}=\tanh J^*$. Taking an overlap of \eqref{eq:TCGSandTCGS*relation} with  $ (\bra{0}_{\rm e}e^{J X_{\rm e}} H_e)^{\bigotimes \Delta_{\rm e}}$ relation, as explained in Appendix~\ref{app:2DIsingSC} we obtain,
\begin{align}\label{eq:IsingGauge}
    Z_{\text{Ising}}(J)=\frac{2^{\frac{|\Delta_{\rm v}|}{2}}(\sinh(2J))^{|\Delta_{\rm e}|/2}}{2^{\frac{|\Delta_{\rm p}|}{2}}\sqrt{|H_1(M,\mathbb{Z}_2)|}}\sum_{z_1\in H_1(M,\mathbb{Z}_2)}Z_{\text{Ising}}(J^*,z_1)\, ,
\end{align}
where
\begin{align}
    Z_{\text{Ising}}(J^*,z_1)=\sum_{\{s_{\rm p}=\pm 1\}}e^{J^*\sum_{\langle \rm p,p'\rangle}(-1)^{\# (z_1\cap \langle \rm p,p'\rangle)}s_{\rm p}s_{\rm p'}}\, .
    \label{eq:twistedIsingpartitionfn}
\end{align}

\begin{figure}
    \centering 
    \input{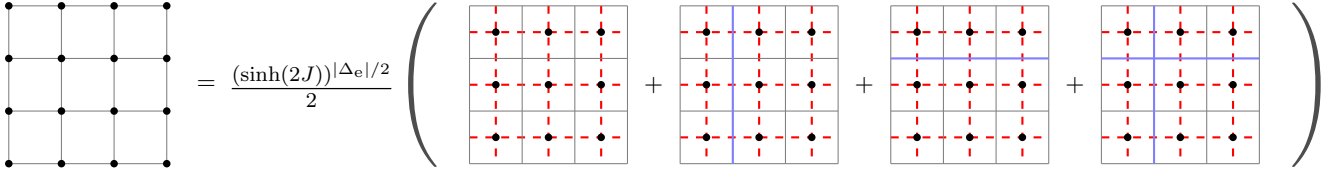}
    \caption{Kramers-Wannier Duality on a periodic square lattice relating Ising Model with coupling $J$ on the left to the Ising Model with coupling $J^*$ on the dual lattice (red) with its $\mathbb{Z}_2$ symmetry (denoted as blue) being gauged, on the right.}
    \label{fig:IsingGauge}
\end{figure}

We note that the above partition function is in the dual lattice and that can be thought of as interactions between plaquettes connected by dual edges of the original lattice. $\#$ denotes the intersection number between the 1-cycle and the dual edge connecting the plaquettes. The sum over $z_1$ in \eqref{eq:IsingGauge} is summing $\mathbb{Z}_2$ symmetry defects over all the cycles in $H_1(M,\mathbb{Z}_2)$ essentially gauging the symmetry. So Kramers-Wannier Duality here relates Ising model at coupling $J$ to $\mathbb{Z}_2$ gauged Ising model at coupling $J^*$. See Figure~\ref{fig:IsingGauge} for instance.

\subsection{Duality defect via strange correlator}\label{sec:dualitydefectIsing}
In this section, we systematically construct the duality defect through the application of a strange correlator. The duality defect arises by executing the gauging of the global symmetry inherent in the Ising model over a specified subset $\Delta_{\rm e}^{\rm A}$ of all edges $\Delta_{\rm e}$. At the boundary between region $\Delta_{\rm e}^{\rm A}$ and region $(\Delta_{\rm e}^{\rm A})^{\rm c}$, the presence of the duality defect is observed. Within region $\Delta_{\rm e}^{\rm A}$, we analyze the system under the framework of the cluster state $|\Psi^*_{\rm cluster}\rangle$, while in region $(\Delta_{\rm e}^{\rm A})^{\rm c}$, the cluster state $|\Psi_{\rm cluster}\rangle$ is considered. The task then involves the integration of these two distinct cluster states using some additional unitary. By taking an overlap of this cluster state with the state $\ket{+}$ across the plaquettes in $\Delta_{\rm e}^{\rm A}$ and vertices in $(\Delta_{\rm e}^{\rm A})^{\rm c}$, we obtain the ground state of the Hamiltonian. This ground state is characterized by stabilizers equivalent to those found in the toric code in $(\Delta_{\rm e}^{\rm A})^{\rm c}$, alongside stabilizers derived from the Hadamard conjugated toric code configuration in $\Delta_{\rm e}^{\rm A}$. Furthermore, the interface features additional stabilizers that provide a smooth transition between the stabilizers in $\Delta_{\rm e}^{\rm A}$ and $\Delta_{\rm e}^{\rm A^c}$. Consequently, regions $\Delta_{\rm e}^{\rm A}$ and the associated cluster states are strategically chosen to ensure that, following the overlap, the stabilizers locally replicate those in $\mathsf{H}^{\rm A}|\rm TC_{GS}\rangle$, with $\mathsf{H}^{\rm A}=\bigotimes_{\rm e\in A}H_e$ ($H_e$ representing the Hadamard operator on edge $e$).

\begin{figure}
    \centering
    \input{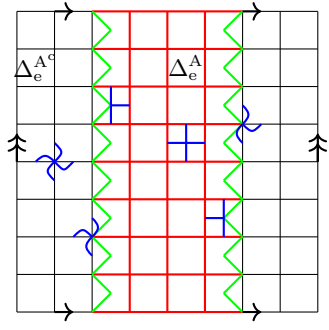}
    \caption{The figure illustrate the cluster state \eqref{eq:2Dclusterstateinterface}. The black edges denote the collection $\Delta_{\rm e}^{\rm A^c}$ and the red edges denote the collection $\Delta_{\rm e}^{\rm A}$. The blue edges denote the cluster entangler between vertices and edges or plaquette centers and edges. The green edges denote the cluster entangler between vertices and the plaquette centers. Samples of entanglement patterns on the black edges and red edges away from the interface and near the interface region are shown in the figure. The whole square lattice is on a torus as indicated by the arrows on the boundary edges.}
    \label{fig:clusterstateAAcomp}
\end{figure}

Now let us construct the cluster state explicitly. We define the region $\Delta_e^{\rm A}\subset \Delta_e$ by choosing a nonempty proper subset of $\Delta_{\rm e}$. Then $(\Delta_{\rm e}^{\rm A})^{\rm c}=\Delta_e^{\rm A^c}\subset \Delta_{\rm e}$ is defined as the complement region to $\Delta_{\rm e}^{\rm A}$. Let us define the following sets:
\begin{subequations}
\begin{align}
    \Delta_{\rm p}^{\rm A}&=\{\rm p|\partial p\cap \Delta_{\rm e}^{\rm A}\neq \emptyset\}\subset \Delta_{\rm p}\, ,\\
    \quad \Delta_{\rm v}^{\rm A^c}&=\{\rm v|\partial^*v\cap \Delta_{\rm e}^{\rm A^c}\neq \emptyset \}\,,\\
    \Delta_{\rm p}^{\rm A|A^c}&=\{\rm p\in\Delta_{\rm p}^{A}|\partial p\neq \partial p\cap \Delta_{\rm e}^{\rm A}\}\,,\\
    \quad \Delta_{\rm v}^{\rm A|A^c}&=\{\rm v\in \Delta_v^{A^c}|\partial^*v\neq \partial^*v\cap \Delta_{\rm e}^{\rm A^c}\}\, .
\end{align}
\end{subequations}
Consider the cluster state that is similar to $\ket{\Psi^*_{\text{cluster}}}$ in the region $\Delta_e^{\rm A}$ and $\ket{\Psi_{\text{cluster}}}$ in region $\Delta_e^{\rm A^c}$ and interpolate between the two using additional $CZ$ operators. This interpolated cluster state is 
\begin{widetext}
\begin{align}\label{eq:2Dclusterstateinterface}
    |\Psi_{\rm cluster}^{\rm A|A^c}\rangle=\prod_{\rm v\in\Delta_{\rm v}^{\rm A^c}}\prod_{\rm e\in \partial^*v\cap \Delta_e^{\rm A^c}}CZ_{\rm v,e}\prod_{\rm p\in\Delta_p^A}\prod_{\rm e\in\partial p\cap \Delta_e^{\rm A}}CZ_{\rm p,e}\prod_{\rm p\in\Delta_p^{\rm A|A^c}}\prod_{\rm v\in\partial(\partial p\cap \Delta_e^{\rm A})}CZ_{\rm p,v}\ket{+}^{\Delta_{\rm e}}\ket{+}^{\Delta_{\rm p}^{\rm A}}\ket{+}^{\Delta_{\rm v}^{\rm A^c}}\, .
\end{align}
The above state is stabilized by
\begin{subequations}
\begin{align} \label{eq:2DCSStabilizer}
    X_{\rm e}Z(\partial^* \mathrm{e})\quad \mathrm{e}\in \Delta_e^{\rm A}\, &, \quad X_eZ(\partial \mathrm{e})\quad \mathrm{e}\in \Delta_e^{\rm A^c}\\
    X_{\rm p}Z(\partial\mathrm{p})\quad \mathrm{p}\in\Delta_{\rm p}^{\rm A}\setminus \Delta_{\rm p}^{\rm A|A^c}\, &,\quad X_{\rm v}Z(\partial^*\mathrm{v})\quad \mathrm{v}\in\Delta_{\rm v}^{\rm A^c}\setminus \Delta_{\rm v}^{\rm A|A^c}\\
    X_{\rm p}Z(\partial\mathrm{p}\cap\Delta_e^{\rm A})Z(\partial(\partial\mathrm{p}\cap\Delta_e^{\rm A}))\quad \mathrm{p}\in\Delta_{\rm p}^{\rm A|A^c}\, &, \quad X_{\rm v}Z(\partial^*\mathrm{v}\cap\Delta_e^{\rm A^c})Z(\partial^*(\partial^*\mathrm{v}\cap\Delta_e^{\rm A^c}))\quad \mathrm{v}\in \Delta_{\rm v}^{\rm A|A^c}\, .
\end{align}
\end{subequations}
\end{widetext}
Consider, as an illustrative example, a square lattice placed on a torus. The torus is partitioned into two cylindrical sections, denoted as $\Delta_e^{\rm A}$ and $\Delta_e^{\rm A^c}$. The regions $\Delta_e^{\rm A}$ and $\Delta_e^{\rm A^c}$ are characterized by the specific edges they encompass within the torus. Subsequently, for this particular selection of region $\Delta_e^{\rm A}$, one can examine the cluster state~\eqref{eq:2Dclusterstateinterface}. An illustration of this configuration is provided in Figure~\ref{fig:clusterstateAAcomp}. 
Specifically, we show the stabilizers here following
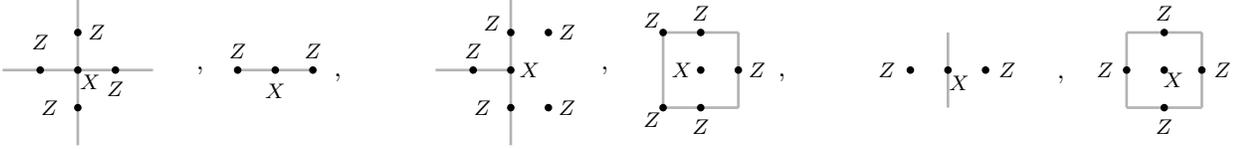
\begin{tikzpicture}
	\begin{pgfonlayer}{nodelayer}
		\node [style=none] (0) at (-3.25, 2) {};
		\node [style=none] (1) at (-5.25, 0) {};
		\node [style=none] (2) at (-3.25, -2) {};
		\node [style=none] (3) at (-1.25, 0) {};
		\node [style=none] (4) at (-3.25, 0) {};
		\node [style=pointoperator, color = black] (5) at (-3.25, 0) {};
		\node [style=pointoperator, color = black] (6) at (-3.25, 1) {};
		\node [style=pointoperator, color = black] (7) at (-2.25, 0) {};
		\node [style=pointoperator, color = black] (8) at (-3.25, -1) {};
		\node [style=pointoperator, color = black] (9) at (-4.25, 0) {};
		\node [style=none] (10) at (-2.75, 1) { \footnotesize $Z$};
		\node [style=none] (11) at (-2.25, -0.5) { \footnotesize $Z$};
		\node [style=none] (12) at (-4, -1) { \footnotesize $Z$};
		\node [style=none] (13) at (-4.25, 0.75) { \footnotesize $Z$};
		\node [style=none] (15) at (-3, -0.25) {};
		\node [style=none] (16) at (-2.925, -0.325) { \footnotesize $X$};
		\node [style=none] (17) at (0, 0) {,};
		\node [style=none] (18) at (1, 0) {};
		\node [style=none] (19) at (3, 0) {};
		\node [style=pointoperator, color = black] (20) at (2, 0) {};
		\node [style=pointoperator, color = black] (21) at (3, 0) {};
		\node [style=pointoperator, color = black] (22) at (1, 0) {};
		\node [style=none] (23) at (1, 0.5) { \footnotesize $Z$};
		\node [style=none] (24) at (3, 0.5) { \footnotesize $Z$};
		\node [style=none] (25) at (2, -0.55) { \footnotesize $X$};
	\end{pgfonlayer}
	\begin{pgfonlayer}{edgelayer}
		\draw [black!30,line width=1.0] (0.center) to (4.center);
		\draw [black!30,line width=1.0] (4.center) to (3.center);
		\draw [black!30,line width=1.0] (4.center) to (2.center);
		\draw [black!30,line width=1.0] (4.center) to (1.center);
		\draw [black!30,line width=1.0] (18.center) to (19.center);
	\end{pgfonlayer}
\end{tikzpicture}, $\quad\quad$ 
\begin{tikzpicture}
	\begin{pgfonlayer}{nodelayer}
		\node [style=none] (0) at (-2.5, 2) {};
		\node [style=none] (1) at (-2.5, -2) {};
		\node [style=none] (2) at (-4.5, 0) {};
		\node [style=none] (3) at (-2.5, 0) {};
		\node [style=none] (4) at (-1.5, 1) {};
		\node [style=none] (5) at (-1.5, -1) {};
		\node [style=pointoperator, color = black] (6) at (-1.5, 1) {};
		\node [style=pointoperator, color = black] (7) at (-1.5, -1) {};
		\node [style=pointoperator, color = black] (8) at (-2.5, 1) {};
		\node [style=pointoperator, color = black] (9) at (-2.5, -1) {};
		\node [style=pointoperator, color = black] (10) at (-3.5, 0) {};
		\node [style=pointoperator, color = black] (11) at (-2.5, 0) {};
		\node [style=none] (12) at (-2, 0) { \footnotesize $X$};
		\node [style=none] (13) at (-3.5, 0.5) { \footnotesize $Z$};
		\node [style=none] (14) at (-3, 1.25) { \footnotesize $Z$};
		\node [style=none] (15) at (-1, 1) { \footnotesize $Z$};
		\node [style=none] (16) at (-1, -1) { \footnotesize $Z$};
		\node [style=none] (17) at (-3.25, -1) { \footnotesize $Z$};
		\node [style=none] (18) at (0, 0) {,};
		\node [style=none] (19) at (1.55, 1) {};
		\node [style=none] (20) at (1.55, -1) {};
		\node [style=none] (21) at (3.55, 1) {};
		\node [style=none] (22) at (3.55, -1) {};
		\node [style=none] (23) at (2.55, 1) {};
		\node [style=none] (24) at (2.55, -1) {};
		\node [style=none] (25) at (3.55, 0) {};
		\node [style=none] (26) at (2.55, 0) {};
		\node [style=pointoperator, color = black] (27) at (1.55, 1) {};
		\node [style=pointoperator, color = black] (28) at (1.55, -1) {};
		\node [style=pointoperator, color = black] (29) at (2.55, 0) {};
		\node [style=pointoperator, color = black] (30) at (2.55, 1) {};
		\node [style=pointoperator, color = black] (32) at (3.55, 0) {};
		\node [style=pointoperator, color = black] (34) at (2.55, -1) {};
		\node [style=none] (35) at (2.05, 0) { \footnotesize $X$};
		\node [style=none] (36) at (1.25, 1.325) { \footnotesize $Z$};
		\node [style=none] (37) at (2.55, 1.5) { \footnotesize $Z$};
		\node [style=none] (38) at (4.05, 0) { \footnotesize $Z$};
		\node [style=none] (39) at (2.55, -1.5) { \footnotesize $Z$};
		\node [style=none] (40) at (1.25, -1.325) { \footnotesize $Z$};
	\end{pgfonlayer}
	\begin{pgfonlayer}{edgelayer}
		\draw [black!30,line width=1.0] (0.center) to (3.center);
		\draw [black!30,line width=1.0] (2.center) to (3.center);
		\draw [black!30,line width=1.0] (3.center) to (1.center);
		\draw [black!30,line width=1.0] (19.center) to (20.center);
		\draw [black!30,line width=1.0] (20.center) to (22.center);
		\draw [black!30,line width=1.0] (22.center) to (21.center);
		\draw [black!30,line width=1.0] (21.center) to (19.center);
	\end{pgfonlayer}
\end{tikzpicture}, $\quad\quad$ 
\begin{tikzpicture}
	\begin{pgfonlayer}{nodelayer}
		\node [style=none] (0) at (-3, 1) {};
		\node [style=none] (1) at (-3, -1) {};
		\node [style=none] (2) at (-2, 0) {};
		\node [style=none] (3) at (-4, 0) {};
		\node [style=pointoperator, color = black] (8) at (-4, 0) {};
		\node [style=pointoperator, color = black] (9) at (-2, 0) {};
		\node [style=pointoperator, color = black] (10) at (-3, 0) {};
		\node [style=none] (11) at (-4.625, 0) { \footnotesize $Z$};
		\node [style=none] (12) at (-1.425, 0) { \footnotesize $Z$};
		\node [style=none] (13) at (-2.7, -0.35) { \footnotesize $X$};
		\node [style=none] (14) at (1.75, 1) {};
		\node [style=none] (15) at (3.75, 1) {};
		\node [style=none] (16) at (1.75, -1) {};
		\node [style=none] (17) at (3.75, -1) {};
		\node [style=none] (18) at (2.75, 1) {};
		\node [style=none] (19) at (3.75, 0) {};
		\node [style=none] (20) at (2.75, -1) {};
		\node [style=none] (21) at (1.75, 0) {};
		\node [style=pointoperator, color = black] (22) at (2.75, 1) {};
		\node [style=pointoperator, color = black] (23) at (3.75, 0) {};
		\node [style=pointoperator, color = black] (24) at (2.75, -1) {};
		\node [style=pointoperator, color = black] (25) at (1.75, 0) {};
		\node [style=pointoperator, color = black] (26) at (2.75, 0) {};
		\node [style=none] (27) at (3, -0.25) { \footnotesize $X$};
		\node [style=none] (28) at (2.75, 1.5) { \footnotesize $Z$};
		\node [style=none] (29) at (1.175, 0) { \footnotesize $Z$};
		\node [style=none] (30) at (4.3, 0) { \footnotesize $Z$};
		\node [style=none] (31) at (2.75, -1.5) { \footnotesize $Z$};
		\node [style=none] (32) at (0, -0.25) {,};
	\end{pgfonlayer}
	\begin{pgfonlayer}{edgelayer}
		\draw [black!30,line width=1.0] (0.center) to (1.center);
		\draw [black!30,line width=1.0] (14.center) to (15.center);
		\draw [black!30,line width=1.0] (15.center) to (17.center);
		\draw [black!30,line width=1.0] (17.center) to (16.center);
		\draw [black!30,line width=1.0] (16.center) to (14.center);
	\end{pgfonlayer}
\end{tikzpicture}$\quad\quad\quad\quad$ 
where the first two terms are stabilizers in the bulk of $\Delta_{\rm v}^{A^c}$ and $\Delta_{\rm e}^{\rm A^c}$, the middle two terms are at the interface $\Delta_{\rm v}^{\rm A|A^c}$ and $\Delta_{\rm p}^{\rm A|A^c}$, and the last two terms are in the bulk of $\Delta_{\rm e}^{\rm A}$ and $\Delta_{\rm p}^{\rm A}$. The plaquettes spins are shown as points at their respective centers.

Next, we compute the overlap with the regions $\ket{+}^{\Delta_{\rm p}^{\rm A}}$ and $\ket{+}^{\Delta_{\rm v}^{\rm A^c}}$, resulting in \begin{align}
    |\rm TC^{*A}_{GS}\rangle= 
    \langle +|^{\Delta_{\rm p}^{\rm A}}\langle +|^{\Delta_{\rm v}^{\rm A^c}}|\Psi_{\rm cluster}^{\rm A|A^c}\rangle\, .
\end{align}
The aforementioned state possesses an identical set of local stabilizers as $\mathsf{H}^{\rm A}|\rm TC_{GS}\rangle$\footnote{$|\rm TC^{*A}_{GS}\rangle$ is a ground state of a Hamiltonian that implements $e\leftrightarrow m$ domain wall between toric codes, i.e., a domain wall that convert electric $e$ excitation to magnetic $m$ excitation. Such domain walls were obtained using SPT-sewing procedure in \cite{Li:2024crt}. The additional $CZ$ operators in \eqref{eq:2Dclusterstateinterface} that connect between the cluster states on the original and dual lattices is similar to the SPT-sewing procedure.}. However, the distinction arises from being stabilized by distinct logical operators. For instance, $\mathsf{H}^{\rm A}|\rm TC_{GS}\rangle$ is stabilized by $\mathsf{H}^{\rm A}X(z_1)\mathsf{H}^{\rm A}$ for a non-trivial 1-cycle $z_1$. Thus, we can represent this as \begin{align}
    &\frac{1}{\mathcal{N}_{\text{TC}}}\mathsf{H}^{\rm A}|\mathrm {TC}_{\rm GS}\rangle=
    \frac{1}{\mathcal{N}_{\text{TC}}^{*\rm A}}\frac{1}{ H_1(M,\mathbb{Z}_2)}\sum_{z_1\in H_1(M,\mathbb{Z}_2)}X(z_1\cap \Delta_e^{\rm A^c})Z(z_1\cap\Delta_e^{\rm A})|\rm TC^{*A}_{GS}\rangle\, .
\end{align}
Both states are normalized to 1 and $\mathcal{N}_{\text{TC}}^{*\rm A}$ is given by
\begin{align}
    \mathcal{N}_{\text{TC}}^{*\rm A}=\sqrt{\langle  \mathrm{TC}^{*\rm A}_{\rm GS}|\frac{1}{ H_1(M,\mathbb{Z}_2)}\sum_{z_1\in H_1(M,\mathbb{Z}_2)}X(z_1\cap \Delta_{\rm e}^{\rm A^c})Z(z_1\cap\Delta_{\rm e}^{\rm A})|\rm TC^{*A}_{GS}\rangle}\,.
\end{align}
Let us take $M=T^2$ to be torus and $\Delta_{\rm e}^{\rm A}$ to be set of edges in a cylinder that wound around one of the cycles of the torus as shown in Figure~\ref{fig:clusterstateAAcomp}. In this case, $\Delta_e^{\rm A}$ and $\Delta_e^{\rm A^c}$ partition the torus into two disconnected regions. For this particular choice of region, $\mathcal{N}_{\text{TC}}^{*\rm A}=\frac{1}{2^{\frac{|\Delta_{\rm p}^{\rm A}|}{2}+\frac{|\Delta_{\rm v}^{\rm A^c}|}{2}}}$ (see Appendix~\ref{app:topologicallyorderedstatenorm} for the norm $\mathcal{N}_{\text{TC}}^{*\rm A}$ for general choices of region $\Delta_{\rm e}^{\rm A}$). Subsequently, we examine the overlap with the product state $(\bra{0}e^{J X})^{\Delta_e^{\rm A^c}}(\bra{0}e^{J^* X})^{\Delta_e^{\rm A}}$, yielding \begin{widetext}
\begin{align}\label{eq:AgaugedIsingmodel}
\begin{split}
    &(\sinh (2J^*))^{|\Delta_e^{\rm A}|/2}Z_{\text{Ising}}(J)=\\
    &\qquad\frac{2^{-\frac{|\Delta_{\rm p}^{\rm A}|}{2}}2^{-\frac{|\Delta_{\rm v}^{\rm A^c}|}{2}}}{2^{-\frac{|\Delta_{\rm v}|}{2}}}\frac{1}{4}\sum_{z_1\in H_1(M,\mathbb{Z}_2)}\sum_{\bigl\{\substack{s_{\rm v}=\pm 1\, ,\\
    (-1)^{h_{\rm v}}=s_{\rm v}}\bigl\}\, ,\bigl\{\substack{s_{\rm p}=\pm 1\, ,\\
    (-1)^{h_{\rm p}}=s_{\rm p}}\bigl\}}e^{J\sum_{\langle \rm v,v'\rangle\in\Delta_e^{\rm A^c}}s_{\rm v}s_{\rm v'}}e^{J^*\sum_{\langle \rm p,p'\rangle\in \Delta_e^{\rm A}}(-1)^{\# (z_1\cap \langle \rm p,p'\rangle)}s_{\rm p}s_{\rm p'}}\\
    &\qquad\qquad \times\prod_{\substack{
    \rm p\in \Delta_p^{A|A^c}\, , \\
  \rm v\in\partial(\partial p\cap \Delta_e^{\rm A})}}(-1)^{h_{\rm v}h_{\rm p}}\times (-1)^{\sum_{\mathrm{e}\in \partial (z_1\cap \Delta_e^{\rm A^c}) }h_{\rm e}}\,  .
\end{split}
\end{align}
\end{widetext} 
\begin{figure}
    \centering
    \input{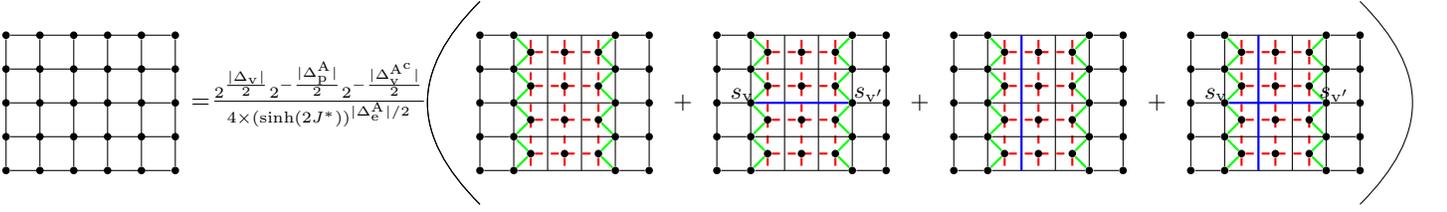}
    \caption{A relation on a periodic square lattice (torus) between the Ising model with coupling $J$ on the left and Ising model with both couplings $J$ (for interaction on black edges) and $J^*$ (for interaction on red edges) separated by an interface (green edges) on the right. The $\mathbb{Z}_2$ symmetry is being gauged in the region with coupling $J^*$ (between the two green interface zig-zag lines). }
    \label{fig:2DIsing_partial_gauging}
\end{figure}
The final equation in~\eqref{eq:AgaugedIsingmodel} can be understood as establishing a relationship between the partition function of the Ising model and that of the Ising model with a duality defect. The sign factors $(-1)^{h_{\rm v}h_{\rm p}}$ are in accordance with the solution to defect commutation relations in~\cite{Aasen:2016dop}. The last sign factor $(-1)^{\sum_{\mathrm{e}\in \partial (z_1\cap \Delta_e^{\rm A^c}) }h_{\rm e}}$ arises from the junction between duality defect and $\mathbb{Z}_2$ symmetry defect. See figure~\ref{fig:2DIsing_partial_gauging} for pictorial illustration of \eqref{eq:AgaugedIsingmodel}. From \eqref{eq:AgaugedIsingmodel}, we can read off the expression for the partition function with two duality defects inserted 
\begin{align}
    Z_{\text{Ising}}^{\mathcal{D}}\equiv\frac{2^{-\frac{|\Delta_{\rm p}^{\rm A}|}{2}}2^{-\frac{|\Delta_{\rm v}^{\rm A^c}|}{2}}}{2^{-\frac{|\Delta_{\rm v}|}{2}}(\sinh (2J^*))^{\frac{|\Delta_e^{\rm A}|}{2}}}\sum_{\bigl\{\substack{s_{\rm v}=\pm 1\, ,\\
    (-1)^{h_{\rm v}}=s_{\rm v}}\bigl\}\, ,\bigl\{\substack{s_{\rm p}=\pm 1\, ,\\
    (-1)^{h_{\rm p}}=s_{\rm p}}\bigl\}}e^{J\sum_{\langle \rm v,v'\rangle\in\Delta_e^{\rm A^c}}s_{\rm v}s_{\rm v'}}e^{J^*\sum_{\langle \rm p,p'\rangle\in \Delta_e^{\rm A}}s_{\rm p}s_{\rm p'}} \times\prod_{\substack{
    \rm p\in \Delta_p^{A|A^c}\, , \\
  \rm v\in\partial(\partial p\cap \Delta_e^{\rm A})}}(-1)^{h_{\rm v}h_{\rm p}}\,  .
  \label{eq:ZIsingD^2}
\end{align}
The same partition function can be interpreted as the partition function in the presence of duality defect for a general choice of $\Delta_{\rm e}^{\rm A}$. We note the following features in the duality defect partition function
\begin{itemize}
   \item The gauged region has dual couplings $J^*$ and interactions between plaquettes.
    \item The ungauged region retains the original couplings $J$ and the usual Ising interaction between vertices
    \item At the interface, there are extra sign factors (the product over $(-1)^{h_{\rm v}h_{\rm p}}$).
\end{itemize}
\subsection{Moving and fusing defects}
We denote the elements in $\Delta_{\rm p}^{\rm A}$ by lowercase Greek letters and those in $\Delta_{\rm v}^{\rm A^c}$ by lowercase Roman letters. We define the following diagrammatic representation of terms in the partition function with duality defect
\begin{align}
    \begin{tikzpicture}
	\begin{pgfonlayer}{nodelayer}
		\node [style=none] (0) at (-3, 3) {};
		\node [style=none] (1) at (0, 3) {};
		\node [style=pointblack] (2) at (-3, 3) {};
		\node [style=pointblack] (3) at (0, 3) {};
		\node [style=none] (4) at (0.5, 3) {$b$};
		\node [style=none] (5) at (-3.5, 3) {$a$};
		\node [style=none] (6) at (1.25, 3) {$=$};
		\node [style=none] (7) at (2.75, 3) {$e^{J s_as_b}\,,$};
		\node [style=none] (8) at (-3, 1) {};
		\node [style=none] (9) at (0, 1) {};
		\node [style=pointblack] (10) at (0, 1) {};
		\node [style=pointblack] (11) at (-3, 1) {};
		\node [style=none] (12) at (-3.5, 1) {$\alpha$};
		\node [style=none] (13) at (0.5, 1) {$\beta$};
		\node [style=none] (14) at (1.25, 1) {$=$};
		\node [style=none] (15) at (3, 1) {$e^{J^*s_{\alpha}s_{\beta}}\,,$};
		\node [style=pointblack] (16) at (-3, -1) {};
		\node [style=pointblack] (17) at (0, -1) {};
		\node [style=none] (18) at (-3.5, -1) {$a$};
		\node [style=none] (19) at (0.5, -1) {$\alpha$};
		\node [style=none] (20) at (1.25, -1) {$=$};
		\node [style=none] (21) at (3.25, -1) {$(-1)^{h_{a}h_{\alpha}}\,.$};
	\end{pgfonlayer}
	\begin{pgfonlayer}{edgelayer}
		\draw (0.center) to (1.center);
		\draw [style=greenline] (16) to (17);
		\draw [style=reddashedline] (11) to (10);
	\end{pgfonlayer}
\end{tikzpicture}
\end{align}
The solid black line is an edge in the original lattice, dashed red line is an edge on the dual lattice, and the green line is an edge connecting the vertices in the original and dual lattices. We use this diagrammatic representation to write down the relations that move the duality defect.
\subsubsection{Movement of duality defect}
In this section, we examine the movement of duality defects. We observe certain identities. Consider an interface between regions $\Delta_{\rm e}^{\rm A}$ and $\Delta_{\rm e}^{\rm A^c}$.  Furthermore, consider an identity concerning the deformation of region $\Delta_e^{\rm A}$
\begin{align}
\begin{tikzpicture}
	\begin{pgfonlayer}{nodelayer}
		\node [style=pointblack] (0) at (6.75, 2) {};
		\node [style=pointblack] (1) at (6.75, -2) {};
		\node [style=pointblack] (2) at (4.75, 0) {};
		\node [style=none] (3) at (2, -0.25) {$\frac{1}{\sqrt{2\sinh(2J^*)}}$};
		\node [style=none] (4) at (8, 0) {$=$};
		\node [style=none] (5) at (9, 0) {$\frac{1}{2}$};
		\node [style=none] (6) at (10.25, 0) {\Large$\sum$};
		\node [style=pointblack] (7) at (11.5, 0) {};
		\node [style=pointblack] (8) at (13.25, 2) {};
		\node [style=pointblack] (9) at (13.25, -2) {};
		\node [style=pointblack] (10) at (15.5, 0) {};
		\node [style=none] (11) at (6.75, 2.5) {$\alpha$};
		\node [style=none] (12) at (6.75, -2.5) {$\beta$};
		\node [style=none] (13) at (4.5, 0) {$b$};
		\node [style=none] (14) at (13.25, 2.5) {$\alpha$};
		\node [style=none] (15) at (16.5, 0) {$a$};
		\node [style=none] (16) at (13.25, -2.5) {$\beta$};
		\node [style=none] (17) at (11.25, 0) {$b$};
		\node [style=none] (18) at (10.25, -0.75) {\tiny $ h_a=0$};
		\node [style=none] (19) at (10.25, 0.75) {\tiny $1$};
	\end{pgfonlayer}
	\begin{pgfonlayer}{edgelayer}
		\draw [style=greenline] (2) to (0);
		\draw [style=greenline] (2) to (1);
		\draw [style=reddashedline] (0) to (1);
		\draw [style=greenline] (8) to (10);
		\draw [style=greenline] (10) to (9);
		\draw (7) to (10);
	\end{pgfonlayer}
\end{tikzpicture}\,.
\label{eq:Isingdefectmovement}
\end{align}
 On the L.H.S., the red dashed line indicate an Ising interaction on the dual lattice, the gauged Ising model. On the R.H.S., the black line indicate an Ising interaction present on the original lattice. The green line can be considered as the link where $CZ$ is incorporated into the original cluster state $|\Psi_{\rm cluster}^{\rm A|A^c}\rangle$.  Transitioning from the L.H.S. to R.H.S. of \eqref{eq:Isingdefectmovement}, we note $|\Delta_e^{\rm A}|\rightarrow |\Delta_e^{\rm A}|-1$ and $|\Delta_{\rm v}^{\rm A^c}|\rightarrow |\Delta_{\rm v}^{\rm A^c}|+1$ where $|S|$ denotes the cardinality of the set $S$. We can express \eqref{eq:Isingdefectmovement} equivalently in terms of equations as \begin{align}
    &(-1)^{h_{b}(h_{\alpha}+h_{\beta})}\frac{e^{J^*s_{\alpha}s_{\beta}}}{(2\sinh(2J^*))^{\frac{1}{2}}}
     =\frac{1}{2}\sum_{h_a=0,1}(-1)^{h_a(h_{\alpha}+h_{\beta})}e^{Js_a s_b}\, ,
\end{align}
where $h_b$, $h_{\alpha}$, and $h_{\beta}$ assume values of either 0 or 1. This equation can be further simplified to obtain
\begin{subequations}
   \begin{align}
    \frac{e^{J^*}}{(2\sinh (2J^*))^{\frac{1}{2}}}=\cosh J\qquad \alpha=\beta\\
    \frac{e^{-J^*}}{(2\sinh (2J^*))^{\frac{1}{2}}}=\sinh J\qquad \alpha\neq\beta
\end{align} 
\end{subequations}
which in turn give $J^*=-\frac{1}{2}\log \tanh(J)$. This is exactly the relation that relate the couplings between Ising model and  gauged Ising model. 
We write down another such identity
\begin{align}
    \begin{tikzpicture}
	\begin{pgfonlayer}{nodelayer}
		\node [style=pointblack] (0) at (8, 2) {};
		\node [style=pointblack] (1) at (8, -2) {};
		\node [style=pointblack] (2) at (6, 0) {};
		\node [style=none] (3) at (2, -0.25) {$\frac{\sqrt{2}}{\sqrt{\sinh(2J^*)}}$};
		\node [style=none] (4) at (11.25, 0) {$=$};
		\node [style=pointblack] (8) at (12.5, 2) {};
		\node [style=pointblack] (9) at (12.5, -2) {};
		\node [style=pointblack] (10) at (14.5, 0) {};
		\node [style=none] (11) at (8, 2.5) {$a$};
		\node [style=none] (12) at (8, -2.5) {$b$};
		\node [style=none] (13) at (5.5, 0) {$\alpha$};
		\node [style=none] (14) at (12.5, 2.75) {$a$};
		\node [style=none] (15) at (15, 0) {$\beta$};
		\node [style=none] (16) at (12.5, -2.5) {$b$};
		\node [style=pointblack] (20) at (10, 0) {};
		\node [style=none] (21) at (12, 0) {2};
		\node [style=none] (22) at (4.5, 0) {\Large $\sum$};
		\node [style=none] (23) at (4.5, -0.75) {\tiny $h_{\alpha}=0$};
		\node [style=none] (24) at (4.5, 0.75) {\tiny $1$};
		\node [style=none] (25) at (10.5, 0) {$\beta$};
	\end{pgfonlayer}
	\begin{pgfonlayer}{edgelayer}
		\draw [style=greenline] (2) to (0);
		\draw [style=greenline] (2) to (1);
		\draw [style=greenline] (8) to (10);
		\draw [style=greenline] (10) to (9);
		\draw [style=reddashedline] (2) to (20);
		\draw (8) to (9);
	\end{pgfonlayer}
\end{tikzpicture}\,.
    \label{eq:Isingdefectmovement2}
\end{align}
 Transitioning from the L.H.S. to R.H.S. of \eqref{eq:Isingdefectmovement2}, we note $|\Delta_e^{\rm A}|\rightarrow |\Delta_e^{\rm A}|-1$ and $|\Delta_{\rm p}^{\rm A}|\rightarrow |\Delta_{\rm p}^{\rm A}|-1$ where $|S|$ denotes the cardinality of the set $S$. We can express \eqref{eq:Isingdefectmovement2} equivalently in terms of equations as \begin{align}
    &\sum_{h_{\alpha}=0,1}(-1)^{h_{\alpha}(h_{b}+h_{a})}\frac{\sqrt{2}e^{J^*s_{\alpha}s_{\beta}}}{(\sinh(2J^*))^{\frac{1}{2}}}=2(-1)^{h_{\beta}(h_{a}+h_{b})}e^{Js_as_b}\, ,
\end{align}
where $h_b$, $h_{\alpha}$, and $h_{\beta}$ assume values of either 0 or 1. This equation again give the relation $J^*=-\frac{1}{2}\log \tanh(J)$.
\subsubsection{Fusion of duality defects}
Now let us consider two duality defects close to each other and apply the identity \eqref{eq:Isingdefectmovement} everywhere to move the defect by one step to the right
\begin{align}
    \begin{tikzpicture}
	\begin{pgfonlayer}{nodelayer}
		\node [style=pointblack] (0) at (2.5, 0) {};
		\node [style=pointblack] (1) at (2.5, 4) {};
		\node [style=pointblack] (2) at (2.5, -4) {};
		\node [style=pointblack] (3) at (0.5, 2) {};
		\node [style=pointblack] (4) at (0.5, -2) {};
		\node [style=pointblack] (5) at (4.5, 2) {};
		\node [style=pointblack] (6) at (4.5, -2) {};
		\node [style=none] (7) at (5.5, 0) {$=$};
		\node [style=none] (8) at (6.75, 0) {\Large\Large $\sum$};
		\node [style=none] (9) at (6.75, -1) {\tiny $\{h_{b_i}\}$};
		\node [style=pointblack] (10) at (9, 2) {};
		\node [style=pointblack] (11) at (11, 0) {};
		\node [style=pointblack] (12) at (11, 4) {};
		\node [style=pointblack] (13) at (13.5, 2) {};
		\node [style=pointblack] (14) at (9, -2) {};
		\node [style=pointblack] (15) at (13.5, -2) {};
		\node [style=pointblack] (16) at (11, -4) {};
		\node [style=pointblack] (17) at (12.5, 2) {};
		\node [style=pointblack] (18) at (12.5, -2) {};
		\node [style=none] (19) at (3, 4) {$\alpha_3$};
		\node [style=none] (20) at (3, 0) {$\alpha_2$};
		\node [style=none] (21) at (3, -4) {$\alpha_1$};
		\node [style=none] (22) at (0, -2) {$a_1$};
		\node [style=none] (23) at (0, 2) {$a_2$};
		\node [style=none] (24) at (5, -2) {$c_1$};
		\node [style=none] (25) at (5, 2) {$c_2$};
		\node [style=none] (26) at (11.5, 4) {$\alpha_3$};
		\node [style=none] (27) at (11.5, 0) {$\alpha_2$};
		\node [style=none] (28) at (11.5, -4) {$\alpha_1$};
		\node [style=none] (29) at (8.5, 2) {$a_2$};
		\node [style=none] (30) at (8.5, -2) {$a_1$};
		\node [style=none] (31) at (14, 2) {$c_2$};
		\node [style=none] (32) at (14, -2) {$c_1$};
		\node [style=none] (33) at (13, 2) {$b_2$};
		\node [style=none] (34) at (13, -2) {$b_1$};
		\node [style=none] (35) at (-2, 0) {$\frac{1}{(2\sinh(2J^*))^{\frac{L_y}{2}}}$};
		\node [style=none] (36) at (8, 0) {$\frac{1}{2^{L_y}}$};
		\node [style=none] (37) at (11, 5) {};
		\node [style=none] (38) at (11, 7) {};
		\node [style=none] (39) at (11, -5) {};
		\node [style=none] (40) at (11, -7) {};
		\node [style=none] (41) at (2.5, -5) {};
		\node [style=none] (42) at (2.5, -7) {};
		\node [style=none] (43) at (2.5, 5) {};
		\node [style=none] (44) at (2.5, 7) {};
	\end{pgfonlayer}
	\begin{pgfonlayer}{edgelayer}
		\draw [style=greenline] (1) to (3);
		\draw [style=greenline] (3) to (0);
		\draw [style=greenline] (0) to (5);
		\draw [style=greenline] (1) to (5);
		\draw [style=greenline] (4) to (0);
		\draw [style=greenline] (4) to (2);
		\draw [style=greenline] (2) to (6);
		\draw [style=greenline] (0) to (6);
		\draw [style=reddashedline] (1) to (0);
		\draw [style=reddashedline] (0) to (2);
		\draw [style=greenline] (12) to (17);
		\draw [style=greenline] (17) to (11);
		\draw [style=greenline] (11) to (18);
		\draw [style=greenline] (18) to (16);
		\draw [style=greenline] (12) to (13);
		\draw [style=greenline] (13) to (11);
		\draw [style=greenline] (11) to (15);
		\draw [style=greenline] (15) to (16);
		\draw (10) to (17);
		\draw (14) to (18);
		\draw (10) to (14);
		\draw (13) to (15);
		\draw (3) to (4);
		\draw (5) to (6);
		\draw [style=dotted] (37.center) to (38.center);
		\draw [style=dotted] (43.center) to (44.center);
		\draw [style=dotted] (41.center) to (42.center);
		\draw [style=dotted] (39.center) to (40.center);
	\end{pgfonlayer}
\end{tikzpicture}\,.
\end{align}
Observe that on the right-hand side, the Ising interaction is absent on the dual lattice. Nonetheless, degrees of freedom $\{h_{\alpha_i}\}$ located at the vertices of the dual lattice are subject to summation \begin{align}
    \sum_{\{h_{\alpha_i}\}}(-1)^{\sum_{h_{\alpha_i}}h_{\alpha_i}(h_{b_i}+h_{c_i}+h_{b_{i-1}}+h_{c_{i-1}})}\,.\nonumber
\end{align} 
For the aforementioned sum to be non-zero, it must satisfy condition $h_{b_i}+h_{c_i}+h_{b_{i-1}}+h_{c_{i-1}}=0$ modulo 2 $\forall\, i$. This requirement is fulfilled when $h_{b_i}=h_{c_i}$ $\forall\, i$ or $h_{b_i}=h_{c_i}+1$ $\forall\, i$ holds true. Subsequently, the summation over $\{h_{b_i}\}$ introduces both an identity defect and a spin-flip defect. In other words
\begin{align}
    \begin{tikzpicture}
	\begin{pgfonlayer}{nodelayer}
		\node [style=pointblack] (0) at (5.25, 2) {};
		\node [style=pointblack] (1) at (7.25, 0) {};
		\node [style=pointblack] (2) at (7.25, 4) {};
		\node [style=pointblack] (3) at (9.75, 2) {};
		\node [style=pointblack] (4) at (5.25, -2) {};
		\node [style=pointblack] (5) at (9.75, -2) {};
		\node [style=pointblack] (6) at (7.25, -4) {};
		\node [style=pointblack] (7) at (8.75, 2) {};
		\node [style=pointblack] (8) at (8.75, -2) {};
		\node [style=none] (9) at (7.75, 4) {$\alpha_3$};
		\node [style=none] (10) at (7.75, 0) {$\alpha_2$};
		\node [style=none] (11) at (7.75, -4) {$\alpha_1$};
		\node [style=none] (12) at (4.75, 2) {$a_2$};
		\node [style=none] (13) at (4.75, -2) {$a_1$};
		\node [style=none] (14) at (10.25, 2) {$c_2$};
		\node [style=none] (15) at (10.25, -2) {$c_1$};
		\node [style=none] (16) at (9.25, 2) {$b_2$};
		\node [style=none] (17) at (9.25, -2) {$b_1$};
		\node [style=none] (18) at (2, 0) {\Large $\sum$};
		\node [style=none] (19) at (2, -1.25) {\tiny $\{h_{b_i}\},\{h_{\alpha_i}\}$};
		\node [style=none] (20) at (11, 0) {$=$};
		\node [style=pointblack] (21) at (13, 2) {};
		\node [style=pointblack] (22) at (16.5, 2) {};
		\node [style=pointblack] (23) at (13, -2) {};
		\node [style=pointblack] (24) at (16.5, -2) {};
		\node [style=none] (25) at (18.5, 0) {\Large $+$};
		\node [style=pointblack] (26) at (20, 2) {};
		\node [style=pointblack] (27) at (23.5, 2) {};
		\node [style=pointblack] (28) at (20, -2) {};
		\node [style=pointblack] (29) at (23.5, -2) {};
		\node [style=none] (30) at (13, 2.5) {$a_2$};
		\node [style=none] (31) at (16.5, 2.5) {$c_2$};
		\node [style=none] (32) at (13, -2.5) {$a_1$};
		\node [style=none] (33) at (16.5, -2.5) {$c_1$};
		\node [style=none] (34) at (20, 2.5) {$a_2$};
		\node [style=none] (35) at (20, -2.5) {$a_1$};
		\node [style=none] (36) at (23.5, 2.5) {$c'_2$};
		\node [style=none] (37) at (23.5, -2.5) {$c'_1$};
		\node [style=pointblack] (38) at (17.5, 2) {};
		\node [style=pointblack] (39) at (17.5, -2) {};
		\node [style=none] (40) at (17.5, 2.5) {$c_2$};
		\node [style=none] (41) at (17.5, -2.5) {$c_1$};
		\node [style=pointblack] (42) at (24.5, 2) {};
		\node [style=pointblack] (43) at (24.5, -2) {};
		\node [style=none] (44) at (24.5, 2.5) {$c_2$};
		\node [style=none] (45) at (24.5, -2.5) {$c_1$};
		\node [style=none] (46) at (0, 0) {$\frac{1}{2^{L_y}}$};
	\end{pgfonlayer}
	\begin{pgfonlayer}{edgelayer}
		\draw [style=greenline] (2) to (7);
		\draw [style=greenline] (7) to (1);
		\draw [style=greenline] (1) to (8);
		\draw [style=greenline] (8) to (6);
		\draw [style=greenline] (2) to (3);
		\draw [style=greenline] (3) to (1);
		\draw [style=greenline] (1) to (5);
		\draw [style=greenline] (5) to (6);
		\draw (0) to (7);
		\draw (4) to (8);
		\draw (0) to (4);
		\draw (3) to (5);
		\draw (21) to (23);
		\draw (21) to (22);
		\draw (23) to (24);
		\draw (26) to (28);
		\draw (26) to (27);
		\draw (28) to (29);
		\draw (38) to (39);
		\draw (42) to (43);
	\end{pgfonlayer}
\end{tikzpicture}
    \, ,\qquad \text{ with }h_{c_i'}=h_{c_i}+1\, .
\end{align} 
If we represent the duality defect as $\mathcal{D}$, the $\mathbb{Z}_2$ symmetry defect as $D_{\eta}$, and the absence of defect insertion as $\rm I$, the resultant fusion of duality defects corresponds to \begin{align}
    \mathcal{D}\times\mathcal{D}=\mathrm{I}+D_{\eta}\,.
    \label{eq:D^2fusion}
\end{align} This equality can be thought of as being embedded within the partition function.

When the duality defect is at the boundary of a simply connected region $\Delta_{\rm e}^{\rm A}$ consisting of a set of edges, the configuration can be reduced to one without any defect, up to an overall multiplicative factor in front of the Ising partition function. To determine this factor explicitly, we start from the defect partition function \eqref{eq:ZIsingD^2} with $\Delta_{\rm e}^{\rm A}$ containing only a single edge. In this case, we have $|\Delta_{\rm p}^{\rm A}| = 2$ and $|\Delta_{\rm v}^{\rm A^c}| = |\Delta_{\rm v}|$. Using the local relation
\begin{align}
    \begin{tikzpicture}
	\begin{pgfonlayer}{nodelayer}
		\node [style=none] (0) at (0, 2) {};
		\node [style=none] (1) at (0, -2) {};
		\node [style=none] (2) at (-2, 0) {};
		\node [style=none] (3) at (2, 0) {};
		\node [style=pointblack] (4) at (0, 2) {};
		\node [style=pointblack] (5) at (0, -2) {};
		\node [style=pointblack] (6) at (2, 0) {};
		\node [style=pointblack] (7) at (-2, 0) {};
		\node [style=none] (8) at (3.5, 0) {$=$};
		\node [style=none] (9) at (-7.75, 0) {$\sum\limits_{h{\alpha},h_{\beta}=0}^1$};
		\node [style=none] (10) at (-2.5, 0) {$\alpha$};
		\node [style=none] (11) at (2.5, 0) {$\beta$};
		\node [style=none] (12) at (0, 2.5) {$a$};
		\node [style=none] (13) at (0, -2.5) {$b$};
		\node [style=none] (14) at (5.5, 2) {};
		\node [style=none] (15) at (5.5, -2) {};
		\node [style=pointblack] (16) at (5.5, 2) {};
		\node [style=pointblack] (17) at (5.5, -2) {};
		\node [style=none] (18) at (5.5, 2.5) {$a$};
		\node [style=none] (19) at (5.5, -2.5) {$b$};
		\node [style=none] (20) at (-4.75, 0) {$\frac{1}{\sqrt{2\sinh(2J^*)}}$};
		\node [style=none] (21) at (4.5, 0) {$2$};
	\end{pgfonlayer}
	\begin{pgfonlayer}{edgelayer}
		\draw [style=greenline] (2.center) to (0.center);
		\draw [style=greenline] (2.center) to (1.center);
		\draw [style=greenline] (0.center) to (3.center);
		\draw [style=greenline] (3.center) to (1.center);
		\draw [style=reddashedline] (2.center) to (3.center);
		\draw (14.center) to (15.center);
	\end{pgfonlayer}
\end{tikzpicture}\,,
\end{align}
we find that a contractible duality defect loop contributes a multiplicative factor of $\sqrt{2}$ to the Ising partition function, 
\begin{align}
    \raisebox{-0.3cm}{\begin{tikzpicture}
	\begin{pgfonlayer}{nodelayer}
		\node [style=none] (8) at (1.75, 1) {};
		\node [style=none] (9) at (1.75, 0.25) {};
		\node [style=none] (10) at (2.75, 0.5) {$\mathcal{D}$};
		\node [style=none] (12) at (4, 0.5) {$=$};
		\node [style=none] (22) at (6.5, 0.5) {$\sqrt{2}\,Z_{\text{Ising}}(J)$};
		\node [style=none] (23) at (0, 0.5) {$Z_{\text{Ising}}($};
		\node [style=none] (24) at (3.25, 0.5) {$)$};
	\end{pgfonlayer}
	\begin{pgfonlayer}{edgelayer}
		\draw [style=greenline, bend right=90, looseness=1.75] (8.center) to (9.center);
		\draw [style=greenline, bend left=90, looseness=1.75] (8.center) to (9.center);
	\end{pgfonlayer}
\end{tikzpicture}}\,,
\end{align}
where the L.H.S. is the partition function \eqref{eq:ZIsingD^2} of the Ising model with a contractible duality defect inserted. 

In other words, the duality defect $\mathcal{D}$ has quantum dimension $\sqrt{2}$. This reproduces, directly on the lattice, the expected quantum dimension of the duality line in the continuum Ising conformal field theory. Our lattice derivation, obtained within the strange correlator framework, parallels the result of Ref.~\cite{Aasen:2016dop}. 

\subsection{Quantum Hamiltonian from partition function}
\subsubsection{Quantum Hamiltonian from partition function with duality defect}
Consider the partition function of the two-dimensional Ising model with a duality defect inserted. To derive the corresponding quantum Hamiltonian, we take the direction normal to the duality defect to be infinite, so that a single vertical duality defect is valid. Let the vertices be denoted by coordinates $\mathrm{v} = (x, y)$ and the plaquettes by $\mathrm{p} = (x+\frac{1}{2}, y+\frac{1}{2})$, where $x$ and $y$ are positive integers. We assume the presence of a single defect oriented along the vertical ($y$) direction located at $x = i_0$, and take the $x$ direction to be infinite. We introduce anisotropic couplings $J_x$ and $J_y$ along the $x$ and $y$ directions, respectively. Partition function with a single duality defect in this notation is 
\begin{widetext}\begin{align}
    Z_{\text{Ising}}^{\mathcal{D}}(J_x,J_y)=\sum_{\bigl\{\substack{s_{\rm v}=\pm 1\, ,\\
    (-1)^{h_{\rm v}}=s_{\rm v}}\bigl\}\, ,\bigl\{\substack{s_{\rm p}=\pm 1\, ,\\
    (-1)^{h_{\rm p}}=s_{\rm p}}\bigl\}}&\exp \left[J_x\sum_{x<i_0}\sum_{y}s_{(x,y)}s_{(x+1,y)}+J_y\sum_{x\leq i_0}\sum_{y}s_{(x,y)}s_{(x,y+1)}\right.\nonumber\\
    &\left.+J_x^*\sum_{x\geq i_0}\sum_{y}s_{(x+\frac{1}{2},y+\frac{1}{2})}s_{(x+\frac{3}{2},y+\frac{1}{2})}+ J_y^*\sum_{x\geq i_0}\sum_{y}s_{(x+\frac{1}{2},y+\frac{1}{2})}s_{(x+\frac{1}{2},y+\frac{3}{2})}\right]\nonumber\\
    &(-1)^{\sum_y\left(h_{(i_0,y)}h_{(i_0+\frac{1}{2},y+\frac{1}{2})}+h_{(i_0+\frac{1}{2},y+\frac{1}{2})}h_{(i_0,y+1)}\right)}\,.
\end{align}
To construct the transfer matrix, we express the Boltzmann weight in the partition function as a product of $L_y$ rows, taking $L_y \to \infty$. As shown below, \begin{subequations}
    \begin{align}
    Z_{\text{Ising}}^{\mathcal{D}}(J_x,J_y)=\sum_{\bigl\{\substack{s_{\rm v}=\pm 1\, ,\\
    (-1)^{h_{\rm v}}=s_{\rm v}}\bigl\}\, ,\bigl\{\substack{s_{\rm p}=\pm 1\, ,\\
    (-1)^{h_{\rm p}}=s_{\rm p}}\bigl\}}&\prod_y\left\{\exp \left[J_x\sum_{x<i_0}s_{(x,y)}s_{(x+1,y)}-J_y/2\sum_{x\leq i_0}(s_{(x,y+1)}-s_{(x,y)})^2\right.\right.\nonumber\\
    &\left.\left.+J_x^*\sum_{x\geq i_0}s_{(x+\frac{1}{2},y+\frac{1}{2})}s_{(x+\frac{3}{2},y+\frac{1}{2})} -J_y^*/2\sum_{x\geq i_0}(s_{(x+\frac{1}{2},y+\frac{3}{2})}-s_{(x+\frac{1}{2},y+\frac{1}{2})})^2\right]\right.\nonumber\\
    &\left.\qquad\times(-1)^{\left(h_{(i_0,y)}h_{(i_0+\frac{1}{2},y+\frac{1}{2})}+h_{(i_0+\frac{1}{2},y+\frac{1}{2})}h_{(i_0,y+1)}\right)}\right\}\\
    \equiv \text{Tr}\left(\hat{\rm T}^{L_y}\right)\, ,
\end{align}
\end{subequations}
where an overall normalization constant has been absorbed.

We now compute the matrix elements of the transfer matrix $\hat{\rm T}$. For a fixed $y$, consider a specific spin configuration $\{s_{(x,y)}\} \cup \{s_{(x+\frac{1}{2},y+\frac{1}{2})}\}$. We define a “no-flip” configuration when spins at rows $y$ and $y+1$ are identical, i.e., $s_{(x,y)}=s_{(x,y+1)}$ and $s_{(x+\frac{1}{2},y+\frac{1}{2})}=s_{(x+\frac{1}{2},y+\frac{3}{2})}$. A “single-flip” configuration occurs when the two rows differ by exactly one spin. The quantum Hamiltonian is obtained in the limit $e^{-2J_y}\longrightarrow 0$ and $J_x\longrightarrow 0$.
\begin{subequations}
\begin{align}
    \hat{\rm T}\rvert_{0\text{-flips}}&= \exp\left[J_x\sum_{x<i_0}s_{(x,y)}s_{(x+1,y)}+J_x^*\sum_{x\geq i_0}s_{(x+\frac{1}{2},y+\frac{1}{2})}s_{(x+\frac{3}{2},y+\frac{1}{2})}\right]\, ,\\
    \hat{\rm T}\rvert_{1\text{-flip}}&=\begin{cases}
        \exp[-2J_y]*\exp\left[J_x\sum_{x<i_0}s_{(x,y)}s_{(x+1,y)}+J_x^*\sum_{x\geq i_0}s_{(x+\frac{1}{2},y+\frac{1}{2})}s_{(x+\frac{3}{2},y+\frac{1}{2})}\right]\quad \text{flip at } (x,y)\text{ for } x<i_0\, ,\\
        \exp[-2J_y^*]*\exp\left[J_x\sum_{x<i_0}s_{(x,y)}s_{(x+1,y)}+J_x^*\sum_{x\geq i_0}s_{(x+\frac{1}{2},y+\frac{1}{2})}s_{(x+\frac{3}{2},y+\frac{1}{2})}\right]\quad \text{flip at }(x+\frac{1}{2},y+\frac{1}{2})\text{ for } x\geq i_0\, ,\\
        \exp[-2J_y]*\exp\left[J_x\sum_{x<i_0}s_{(x,y)}s_{(x+1,y)}+J_x^*\sum_{x\geq i_0}s_{(x+\frac{1}{2},y+\frac{1}{2})}s_{(x+\frac{3}{2},y+\frac{1}{2})}\right](-1)^{h_{(i_0+1,y)}}\quad \text{flip at } (x,y)\text{ for } x=i_0\, .
    \end{cases}
\end{align}
\label{eq:transfermatrix}
\end{subequations} \end{widetext}
In the $\tau$-continuum limit (see Ref.~\cite{Kogut:1979wt} for a detailed derivation of the transverse field Ising model from the classical 2D Ising model), we have 
\begin{align}
    \hat{\rm T}=e^{-\tau\hat{\rm H}}\approx 1-\tau \hat{\rm H}\, .
\end{align}
Hence, 
\begin{align}
    &\hat{\rm T}\rvert_{0\text{-flips}}\approx 1-\tau \hat{\rm H}\rvert_{0\text{-flips}}\, ,\quad \hat{\rm T}\rvert_{1\text{-flip}}\approx -\tau \hat{\rm H}\rvert_{1\text{-flip}}\,,
    \qquad \hat{\rm T}\rvert_{n\text{-flip}}\approx -\tau \hat{\rm H}\rvert_{n\text{-flip}}\, .
\end{align}
Comparing with~\eqref{eq:transfermatrix}, we obtain the relations
\begin{align}
   e^{-2J_y}\sim \tau\,, \quad J_x\sim \tau\,,\quad e^{-2J_y^*}\sim\tau\,,\quad J_x^*\sim\tau \, .
\end{align}
This implies that, in the $\tau$-continuum limit, the matrix elements that remain nonzero correspond to $0$- and $1$-flip configurations, since $\hat{\rm H}\rvert_{n\text{-flip}} \sim \tau^{n-1}$. We fix the normalization by setting $\tau = e^{-2J_y}$ and define $J_x = \lambda e^{-2J_y}$. Using $J_y^* = -\frac{1}{2}\log\tanh J_x$ and $J_x^* = -\frac{1}{2}\log\tanh J_y$, and the approximation $\tanh J \approx J$ for small $J$, we find
\begin{align}
    e^{-2J_y}=\tau\,,\quad J_x=\lambda\tau\,,\quad J_x^*=\tau\,,\quad e^{-2J_y^*}=\lambda \tau\, .
\end{align}
We then relabel the plaquette coordinates $(x+\frac{1}{2},y+\frac{1}{2})$ for $x \geq i_0$ as $(x+1,y)$ for fixed $y$. With this redefinition, the $1+1$D quantum Hamiltonian with a duality defect on the link $\langle i_0, i_0+1 \rangle$ takes the form
\begin{align}
    \hat{\rm H}^{(i_0,i_0+1)}_{\text{TFI};\mathcal{D}}&=-\lambda\sum_{i<i_0}Z_{i}Z_{i+1}-\sum_{i>i_0}Z_{i}Z_{i+1}-\sum_{i<i_0}X_i
    -\lambda\sum_{i>i_0}X_i-X_{i_0}Z_{i_0+1}\, ,
\end{align}
where $X_i$ and $Z_i$ denote Pauli operators at site $i$. Notably, the duality defect located on the link $\langle i_0, i_0+1\rangle$ can be shifted to $\langle i_0+1, i_0+2\rangle$ by a local unitary transformation $H_{i_0+1} CZ_{i_0,i_0+1}$ where $H_{i_0+1}$ is the Hadamard operator at site $i_0+1$.
\subsubsection{Quantum Hamiltonian with condensation defect}
Now let us consider the fusion of two duality defects in the quantum transverse-field Ising (TFI) Hamiltonian. Suppose the duality defects are located at $\langle i_0,i_0+1\rangle$ and $\langle i_0+2,i_0+3\rangle$. The corresponding Hamiltonian is
\begin{align}
    \hat{\rm H}^{(i_0,i_0+1);(i_0+2,i_0+3)}_{\text{TFI};\mathcal{D}}&=-\lambda\sum_{i<i_0}Z_iZ_{i+1}-\sum_{i<i_0}X_i-X_{i_0}Z_{i_0+1}-\lambda X_{i_0+1}-\lambda X_{i_0+2}-Z_{i_0+1}Z_{i_0+2}-Z_{i_0+2}X_{i_0+3}\nonumber\\
    &\hspace{2cm}-\lambda\sum_{i>i_0+2}Z_{i}Z_{i+1}-\sum_{i>i_0+3}X_i\,.
\end{align}
We now apply the following sequence of local unitaries to move and fuse the defects:\\ $CX_{i_0+2,i_0+3}H_{i_0+2}CZ_{i_0+1,i_0+2}H_{i_0+1}CZ_{i_0,i_0+1}$. This sequence transforms the Hamiltonian into
\begin{align}
    \hat{\rm H}_{\text{TFI};\mathcal{D}^2}=-\lambda\sum_{i<i_0+2}Z_iZ_{i+1}-\sum_{i<i_0+3}X_i-\lambda\sum_{i>i_0+3}Z_iZ_{i+1}-\sum_{i>i_0+3}X_i-\lambda Z_{i_0+2}Z_{i_0+3}Z_{i_0+4}\,.
    \label{eq:TFID^2}
\end{align}
which can be interpreted as the sum of two TFI Hamiltonians: one without any defect (without the site at $i_0+3$) and another with a spin-flip defect between sites $\langle i_0+2,i_0+4\rangle$. More precisely,
\begin{align}
    \hat{\rm H}_{\text{TFI};\mathcal{D}^2}=\begin{pmatrix}
        \hat{\rm H}_{\text{TFI}} & \\
        & \hat{\rm H}_{\text{TFI};\eta}^{(i_0+2,i_0+4)}
    \end{pmatrix}\,,
\end{align}
where $\hat{\rm H}{\text{TFI}}$ corresponds to the sector with eigenvalue $Z_{i_0+3}=+1$, and $\hat{\rm H}{\text{TFI};\eta}^{(i_0+2,i_0+4)}$ corresponds to $Z_{i_0+3}=-1$. Explicitly, the two Hamiltonians are
\begin{subequations}
\begin{align}
   \hat{\rm H}_{\text{TFI}}&= -\lambda\sum_{i<i_0+2}Z_iZ_{i+1}-\sum_{i<i_0+3}X_i-\lambda\sum_{i>i_0+3}Z_iZ_{i+1}-\sum_{i>i_0+3}X_i-\lambda Z_{i_0+2}Z_{i_0+4}\,,\\
   \hat{\rm H}_{\text{TFI};\eta}^{(i_0+2,i_0+4)}&=-\lambda\sum_{i<i_0+2}Z_iZ_{i+1}-\sum_{i<i_0+3}X_i-\lambda\sum_{i>i_0+3}Z_iZ_{i+1}-\sum_{i>i_0+3}X_i+\lambda Z_{i_0+2}Z_{i_0+4}\,.
\end{align}
\end{subequations}
We note that the Hamiltonian \eqref{eq:TFID^2} can be equivalently obtained from finding the transfer matrix of a sum of statical models with identity and spin flip defect as in the R.H.S. of \eqref{eq:D^2fusion}. Therefore, the Hamiltonian \eqref{eq:TFID^2} realizes the fusion rule~\eqref{eq:D^2fusion}
since it decomposes into sectors corresponding to the identity and spin-flip defect.
\section{Duality defect in 3D anisotropic plaquette Ising model}\label{sec:DualitydefectAPImodel}
This section focuses on the anisotropic Ising model in three dimensions, which exhibits a Kramers–Wannier duality relating its high- and low-temperature phases. We restrict our discussion to the case of a three-torus geometry. For a treatment applicable to general three-manifolds $M_3$, we refer the reader to Section~\ref{sec:KWdefectgenIsingmodel}, where a more general formalism based on chain complexes is developed.

Before introducing the specific details of the model, we first establish some relevant notation. We consider a three-dimensional cubic lattice with a single cube as the unit cell. The cubes are denoted by ${\rm c}$, and the set of all cubes is written as $\Delta_{\rm c}$. In addition to the cubes, we distinguish the following types of cells:
\begin{itemize}
    \item Plaquettes lying in the $xy$-plane, denoted by ${\rm p}_{xy}$, with the set of all such plaquettes written as $\Delta_{\rm p_{xy}}$;
    \item Edges oriented along the $z$-direction, denoted by ${\rm e}_z$, with the set $\Delta_{\rm e_z}$;
    \item Vertices, denoted by ${\rm v}$, forming the set $\Delta_{\rm v}$.
\end{itemize}
The boundary operation is denoted by $\delta$. The boundaries of the various cell types are given as follows:
\begin{itemize}
    \item The boundary of a cube consists of the two $xy$-plaquettes and the four $z$-oriented edges.
    \item The boundary of an $xy$-plaquette contains its four corner vertices.
    \item The boundary of a $z$-oriented edge is its two endpoint vertices.
\end{itemize}
The corresponding dual boundary operator $\delta^*$ is defined as follows:
\begin{itemize}
    \item $\delta^*{\rm v}$ consists of the four $xy$-plaquettes and the two $z$-oriented edges which share $\rm v$.
    \item $\delta^*{\rm e_z}$ consists of the four cubes which share $\rm e_z$.
    \item $\delta^*{\rm p}_{{xy}}$ consists of the two cubes which share ${\rm p}_{xy}$.
\end{itemize}
These conventions and definitions will be employed throughout the construction and analysis of the anisotropic three-dimensional Ising model.
\subsection{Plaquette Ising model and Kramers-Wannier duality}
\subsubsection{Anisotropic plaquette Ising model and Kramers-Wannier duality}
\begin{figure}[h!]
    \centering
    \begin{tikzpicture}
\draw[-,black!30,line width=1.0]
(0,0)--(1.5,0)--(2,0.5)--(0.5,0.5)--(0,0);
\draw[-,black!30,line width=1.0]
(6,-0.5)--(6,1);
    \node at (0,0) {$s_1$};
    \node at (1.5,0) {$s_2$};
    \node at (0.5,0.5) {$s_3$};
    \node at (2,0.5) {$s_4$};
    \node at (6,-0.5) {$s_1$};
     \node at (6,1) {$s_2$};
\end{tikzpicture}
    \caption{Two interactions in the anisotropic plaquette Ising model. Left diagram represent the four-body plaquette interaction in the $xy$ plane while the right diagram represent the two body Ising interaction in the $z$ direction.}
    \label{fig:anisotropic}
\end{figure}
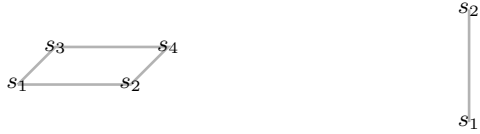

We now define the anisotropic Ising model, which is characterized by two distinct types of Ising-like interactions. The first is a four-body plaquette interaction acting on the $xy$-planes, while the second is a two-body interaction acting along the $z$-direction (see Figure~\ref{fig:anisotropic} for an illustration).
The classical energy function of the anisotropic plaquette Ising (API) model is given by
\begin{align}\label{eq:APIHamiltonian}
    \mathcal{H}(\{s_i\})=-J\sum_{\rm p_{xy}\in\Delta_{\rm p_{xy}}}\prod_{\rm v\in \delta p_{xy}}s_{\rm v}-K\sum_{\rm e_z\in \Delta_{\rm e_z}}\prod_{\rm v\in\delta e_z}s_{\rm v}\, ,
\end{align}
where $J$ and $K$ denote the coupling constants corresponding to the plaquette and edge interactions, respectively.
The associated partition function is
\begin{align}
    Z_{\text{API}}=\sum_{\{s_i\}} \exp(-\mathcal{H}(\{s_i\}))\, ,
    \label{eq:APIpartition}
\end{align}
where the sum runs over all spin configurations ${s_i=\pm1}$ on the vertices of the cubic lattice.
\subsubsection{$\mathbb{Z}_2$ symmetry}
The Hamiltonian in~\eqref{eq:APIHamiltonian} possesses a $\mathbb{Z}_2$ subsystem symmetry, which can be identified as follows.
Let $\Delta^{xz}{\rm v}$, $\Delta^{yz}{\rm v}$, and $\Delta^{xy}{\rm v}$ denote the sets of all spins residing in the $xz$-, $yz$-, and $xy$-planes, respectively. If we perform the transformation $s_{\rm v}\longrightarrow -s_{\rm v}$, $\forall\, \mathrm{v}\in \Delta_{\rm v}^{xz}$, the Hamiltonian remains invariant. Similarly, flipping all the spins in a $yz$-plane leaves the Hamiltonian unchanged. However, performing the same transformation on spins in the $xy$-plane does not leave the Hamiltonian invariant. Therefore, the model exhibits a $\mathbb{Z}_2$ subsystem symmetry acting independently within each vertical ($xz$ and $yz$) plane, but not within horizontal ($xy$) planes.

\begin{figure}
    \centering
    \input{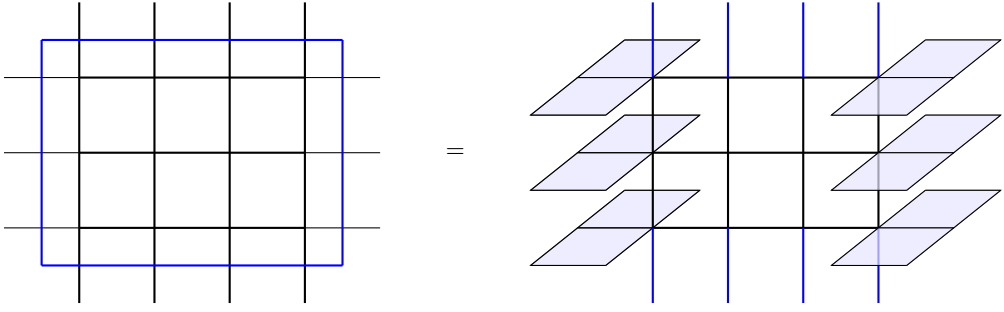}
    \caption{A $\mathbb{Z}_2$ subsystem symmetry defect in 3D Anisotropic Ising Model (shown on the left in Blue) lives on the vertical plane, i.e., $xz$ (or $yz$) plane. When it intersects the vertical links, i.e., $z$-directed (shown on the right in blue), it flips the link interaction and when it intersects horizontal link, i.e., $x$-directed (or $y$-directed), it flips the two plaquette interactions for the $xy$ plaquettes (shown on the right in blue) that contain that horizontal link.}
    \label{fig:APIZ2Defect}
\end{figure}

To describe the corresponding $\mathbb{Z}_2$ symmetry defect, let us focus on one of the vertical planes, say the $xz$-plane. We choose a subset $S^{xz} \subset \Delta^{xz}_{\rm v}$ and perform the transformation $s_{\rm v}\longrightarrow -s_{\rm v}$, $\forall\, \mathrm{v}\in S^{xz}$. This operation introduces a $\mathbb{Z}_2$ defect along the interface separating $S^{xz}$ from the rest of the $xz$-plane. More precisely, consider the dual lattice to the square lattice in the $xz$-plane. The links of this dual lattice that bound the region $S^{xz}$ form a closed 1-cycle $C$. The $\mathbb{Z}_2$ defect, denoted $\eta(C)$, can thus be regarded as living on this cycle $C$.
\begin{itemize}
    \item When $C$ intersects a $z$-directed link $\mathrm{e}_z=\langle {\rm v,v'} \rangle$ of the original lattice, the corresponding interaction term is flipped:
    \begin{align*}
        K s_{\rm v}s_{\rm v'}\longrightarrow -K s_{\rm v}s_{\rm v'}\,.
    \end{align*}
    \item When $C$ intersects an $x$-directed link ${\rm e}_x$ of the original lattice, it flips the signs of the two adjacent $xy$-plaquette terms sharing ${\rm e}_x$:
    \begin{align*}
        J\prod_{\rm v\in\delta p_{xy}}s_{\rm v}\longrightarrow -J\prod_{\rm v\in\delta p_{xy}}s_{\rm v}\,,
    \end{align*}
    for ${\rm p}_{xy}$ and ${\rm p}'_{xy}$ that share the edge ${\rm e}_x$.
\end{itemize}
The resulting defect Hamiltonian takes the form
\begin{align}\label{eq:APIZ2DefectHam}
    \mathcal{H}_{\eta(C)}=J\sum_{{\rm p}_{xy}\in\Delta_{{\rm p}_{xy}}} \left(\prod_{{{\rm e} \in \partial {\rm p}_{xy}}}(-1)^{\#(C,e)} \right) \prod_{\rm v\in \delta p_{xy}}s_{\rm v}+K\sum_{\rm e_z\in \Delta_{\rm e_z}}(-1)^{\# (C,{\rm e}_z)} \prod_{\rm v\in\delta e_z}s_{\rm v}\, ,
\end{align}
where $\partial {\rm p}_{xy}$ denotes the set of links forming the plaquette ${\rm p}_{xy}$ (see Figure~\ref{fig:APIZ2Defect}), and $\#(C,{\rm e})$ is $1$ if $C$ intersects link ${\rm e}$ and $0$ otherwise. Although we have defined the defect on a homologically trivial cycle $C$, equation~\eqref{eq:APIZ2DefectHam} serves as the general definition of the $\mathbb{Z}_2$ defect for any planar cycle $C$ lying in an $xz$ (or $yz$) plane. Importantly, the defect $\eta(C)$ is restricted to move within the plane on which it resides. To move the defect from a cycle $C$ to another cycle $C'$ on the same plane, one must flip all spins $s_{\rm v} \to -s_{\rm v}$ for vertices ${\rm v}$ contained in the region bounded by $C$ and $C'$.

\subsubsection{Kramers-Wannier duality}
This model exhibits a Kramers–Wannier self–duality, which exchanges the strong-coupling and weak-coupling descriptions, i.e., $(J,K) \longleftrightarrow (J^*,K^*)$ where $J=-\frac{1}{2}\log\tanh J^*$ and $K=-\frac{1}{2}\log\tanh K^*$. This duality has been studied previously in the literature \cite{xu2004strong,2005NuPhB.716..487X}. Below we state the duality explicitly on a three–torus with $L_x$, $L_y$ and $L_z$ vertices along the $x$, $y$ and $z$ directions respectively, after introducing a few convenient notations.

Denote by $l_{xy}$ the straight line along the $z$–direction located at the coordinate $(x,y)$. Likewise, denote by $st_x$ and $st_y$ the $x$–directed and $y$–directed strips in the $xy$–plane at fixed $z$ (we may take $z=0$ for definiteness). Define the collections of such lines and strips by
\begin{subequations}
\begin{align}
    &\mathrm{L}=\bigg\{l_{xy}=\bigcup_{z\in\mathbb{Z}_{L_z}} \{\mathrm{e}_{(x,y,z+\frac{1}{2})}\}|(x,y)\in(\cup_{i}\{(i,1)\})\cup(\cup_j\{(1,j)\})\,,i\in\mathbb{Z}_{L_x}\,,j\in\mathbb{Z}_{L_y}\bigg\}\, ,\\
    & \mathrm{S}=\bigg\{st_x=\bigcup_{y\in \mathbb{Z}_{L_y} }\{\mathrm{p}_{(x+\frac{1}{2},y+\frac{1}{2},z)}\}|x\in\mathbb{Z}_{L_x}\bigg\}\cup\bigg\{st_y=\bigcup_{x\in \mathbb{Z}_{L_x} }\{\mathrm{p}_{(x+\frac{1}{2},y+\frac{1}{2},z)}\}|y\in\mathbb{Z}_{L_y}\bigg\}\, .
\end{align}
\end{subequations}
Using these, introduce the set $H_q$ of 1–cycles generated by the chosen family of vertical lines and planar strips:
\begin{align}
H_q=\bigg\{\mathcal{S}\cup\mathfrak{L}|\mathfrak{L}\subset \mathrm{L},\mathcal{S}\subset\mathrm{S}\bigg\}\,
\end{align}
the set of all subsets of $\rm L\cup S$. There are in total $2^{2(L_x+L_y-1)}$ subsets of $\rm L\cup S$.
The Kramers–Wannier duality then relates the original anisotropic plaquette Ising partition function $Z_{\mathrm{API}}(J,K)$ to a sum of dual partition functions on the dual lattice. Explicitly,
\begin{widetext}
\begin{align}
    Z_{\text{API}}(J,K)=\frac{2^{\frac{|\Delta_{\rm v}|}{2}}(\sinh 2J)^{|\Delta_{\rm p_{xy}}|/2}(\sinh 2K)^{|\Delta_{\rm e_{z}}|/2}}{2^{\frac{|\Delta_{\rm c}|}{2}}\sqrt{|H_q|}}\sum_{\bm z\in H_q}Z_{\text{API}}(J^*,K^*,\bm z)\,,
\end{align}
where the dual partition function $Z_{\text{API}}(J^*,K^*,\bm z)$ on the dual lattice is
\begin{align}
    Z_{\text{API}}(J^*,K^*,\bm z)=\sum_{\{s_{\rm c}=\pm 1\}}\exp\left(J^*\sum_{\rm p_{xy}\in\Delta_{\rm p_{xy}}}(-1)^{\#(\bm z\cap \rm p_{xy})}\prod_{\substack{\rm c\\
    \rm p_{xy}\in \delta c}}s_{\rm c}+K^*\sum_{\rm e_z\in\Delta_{\rm e_z}}(-1)^{\#(\bm z\cap \rm e_z)}\prod_{\substack{\rm c\\
    \rm e_z\in \delta c}}s_{\rm c}\right)\, .
\end{align}
\end{widetext}
In these expressions $\#(\bm z\cap\cdot)$ denotes the intersection number\footnote{$\#(\bm z\cap\cdot)=\begin{cases}
    0\quad\text{ if }\quad\bm z\cap \cdot= \emptyset\\
    1\quad\text{ otherwise } 
\end{cases}$ } between the 1–cycle $\bm z\in H_q$ and the dual edge (or plaquette) in question, and $|\cdot|$ denotes cardinality. Note that $Z_{\text{API}}(J^*,K^*,\bm z)$ lives on the dual cubic lattice; the $s_{\rm c}$ variables are spins on cube centers (equivalently sites of the dual lattice), and the dual couplings $J^*,K^*$ are related to $J,K$ by the usual Kramers–Wannier relations. The prefactor accounts for the change of variables and normalization between the original and dual descriptions, while the sum over $\bm z\in H_q$ implements the sum over symmetry defects  appropriate for the three–torus.

This formulation makes the self–duality manifest: the original API partition function is equivalent to a (normalized) sum over dual API partition functions with possible symmetry defect insertion determined by the 1–cycles in $H_q$.
\subsection{Duality defect via strange correlator}
We now construct the Kramers–Wannier duality defect for the anisotropic plaquette Ising (API) model by following the same strategy used in the two–dimensional Ising case. We assume periodic boundary condition on all three directions and hence the lattice can be taken to embedded in a 3-torus. Partition the set of plaquettes and $z$–directed edges as
\begin{align}
    \Delta_{\rm p_{xy}\cup e_z}=\Delta_{\rm p_{xy}\cup e_z}^{\rm A}\cup \Delta_{\rm p_{xy}\cup e_z}^{\rm A^c}\,,
\end{align}
so that $\Delta_{\rm p_{xy}\cup e_z}^{\rm A}$ and $\Delta_{\rm p_{xy}\cup e_z}^{\rm A^c}$ are separated by an interface on which the duality defect will be supported. On the region $\Delta_{\rm p_{xy}\cup e_z}^{\rm A}$ we place the gauged API model, while on the complementary region $\Delta_{\rm p_{xy}\cup e_z}^{\rm A^c}$ we keep the original API degrees of freedom. To implement this construction, we first introduce the following auxiliary sets of cells that probe the neighbourhood of the interface:
\begin{subequations}
\begin{align}
    &\Delta_{\rm c}^{\rm A}=\{\rm c| \delta c\cap \Delta_{\rm p_{xy}\cup e_z}^{\rm A}\neq \emptyset\}\subset \Delta_{\rm c}\, ,\\
    &\Delta_{\rm v}^{\rm A^c}=\{\rm v|\delta^*v\cap \Delta_{\rm p_{xy}\cup e_z}^{\rm A^c}\neq \emptyset \}\,,\\
    &\Delta_{\rm c}^{\rm A|A^c}=\{\rm c\in\Delta_{\rm c}^{A}|\delta c\neq \delta c\cap \Delta_{\rm p_{xy}\cup e_z}^{\rm A}\}\, ,\\
    &\Delta_{\rm v}^{\rm A|A^c}=\{\rm v\in \Delta_v^{A^c}|\delta^*v\neq \delta^*v\cap \Delta_{\rm p_{xy}\cup e_z}^{\rm A^c}\}\, .
\end{align}
\end{subequations}
The cluster entangler that prepares the gauged cluster region and couples it to the ungauged region is given by a product of controlled–$Z$ gates acting between the indicated cell degrees of freedom. Explicitly, \begin{widetext}
\begin{align}
    \mathcal{U}_{CZ}&=\prod_{\rm v\in\Delta_{\rm v}^{\rm A^c}}\left(\prod_{\rm e_z\in\delta^*v\cap \Delta_{\rm p_{xy}\cup e_z}^{\rm A^c}}CZ_{\rm v,e_z}\prod_{\rm p_{xy}\in\delta^*v\cap \Delta_{\rm p_{xy}\cup e_z}^{\rm A^c}}CZ_{\rm v,p_{xy}}\right)\prod_{\rm c\in\Delta_{\rm c}^{\rm A}}\left(\prod_{\rm e_z\in \delta c\cap \Delta_{\rm p_{xy}\cup e_z}^{\rm A}}CZ_{\rm c, e_z}\prod_{\rm p_{xy}\in\delta c\cap \Delta_{\rm p_{xy}\cup e_z}^{\rm A}}CZ_{\rm c, p_{xy}}\right)\nonumber\\
    &\hspace{2cm}\times\prod_{\rm c\in\Delta_{\rm c}^{\rm A|A^c}}\prod_{\rm v\in\delta(\delta c\cap \Delta_{\rm p_{xy}\cup e_z}^{\rm A})}CZ_{\rm c,v}\,.
\end{align}
\end{widetext}
Using $\mathcal{U}_{CZ}$, we define the joint cluster state on the combined cell sets,
\begin{align}
    |\Psi_{\text{cluster}}^{\text{3D};\rm A|A^c}\rangle=\mathcal{U}_{CZ}\ket{+}^{\Delta_{\rm p_{xy}}}\ket{+}^{\Delta_{\rm e_{z}}}\ket{+}^{\Delta_{\rm c}^{\rm A}}\ket{+}^{\Delta_{\rm v}^{\rm A^c}}\, .
    \label{eq:3DclusterAAc}
\end{align}
The stabilizer generators of this cluster state decompose according to whether the supporting cell lies entirely in the gauged region, entirely in the ungauged region, or intersects the interface. Concretely, the stabilizers are
\begin{subequations}
\begin{align}
    &X_{\sigma}Z(\delta^*\sigma)\quad \sigma\in \Delta_{\rm p_{xy}\cup e_z}^{\rm A}\, , \\ &X_{\sigma}Z(\delta \sigma)\quad \sigma\in \Delta_{\rm p_{xy}\cup e_z}^{\rm A^c}\, ,\\
    &X_{\rm c}Z(\delta \mathrm{c})\quad \mathrm{c}\in\Delta_{\rm c}^{\rm A}\setminus \Delta_{\rm c}^{\rm A|A^c}\, , \\
    & X_{\rm v}Z(\delta^* \mathrm{v})\quad \mathrm{v}\in\Delta_{\rm v}^{\rm A^c}\setminus \Delta_{\rm v}^{\rm A|A^c}\, ,\\
    &X_{\rm c}Z(\delta \mathrm{c}\cap \Delta_{\rm p_{xy}\cup e_z}^{\rm A})Z(\delta(\delta \mathrm{c}\cap \Delta_{\rm p_{xy}\cup e_z}^{\rm A}))\quad \mathrm{c}\in\Delta_{\rm c}^{\rm A|A^c}\, , \\
    &X_{\rm v}Z(\delta^* \mathrm{v}\cap \Delta_{\rm p_{xy}\cup e_z}^{\rm A^c})Z(\delta^*(\delta^* \mathrm{v}\cap \Delta_{\rm p_{xy}\cup e_z}^{\rm A^c}))\quad \mathrm{v}\in\Delta_{\rm v}^{\rm A|A^c}\, .
\end{align}
\end{subequations}
We reconsider a region $\Delta_{\rm p_{xy}\cup e_z}^{\rm A}$ as a connected area that we take to be $T^2\times I$, a 2-torus times an interval, to facilitate the identification of a duality defect. Without loss of generality, we take the interval $I$ in the $x$-direction. We do not include the $\rm e_z$ edges in $\Delta_{\rm p_{xy}\cup e_z}^{\rm A}$ at both interfaces. Let us take the number of plaquettes in the $x$ direction in $\Delta_{\rm p_{xy}}^{\rm A}$ to be $l_x$. The stabilizers are explicitly shown as following. \newline In the bulk of $\Delta_{{\rm p}_{xy \cup {\rm e}_{\rm z}}}^{\rm A^c}$ and $\Delta_{\rm v}^{\rm A^c}$, we have
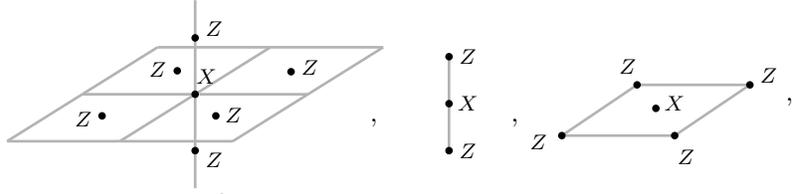
\begin{tikzpicture}
	\begin{pgfonlayer}{nodelayer}
		\node [style=none] (0) at (-4.25, 0) {};
		\node [style=none] (1) at (-2.25, 1.25) {};
		\node [style=none] (2) at (-6.25, -1.25) {};
		\node [style=none] (3) at (-5.25, 1.25) {};
		\node [style=none] (4) at (-7.25, 0) {};
		\node [style=none] (5) at (-9.25, -1.25) {};
		\node [style=none] (6) at (-4.25, 2.5) {};
		\node [style=none] (7) at (-4.25, -2.5) {};
		\node [style=pointoperator, color = black] (10) at (-4.25, 1.5) {};
		\node [style=pointoperator, color = black] (11) at (-4.25, -1.5) {};
		\node [style=pointoperator, color = black] (16) at (-4.25, 0) {};
		\node [style=none] (17) at (0.75, 1.25) {};
		\node [style=none] (18) at (-1.25, 0) {};
		\node [style=none] (19) at (-3.25, -1.25) {};
		\node [style=pointoperator, color = black] (24) at (-4.725, 0.625) {};
		\node [style=pointoperator, color = black] (25) at (-6.725, -0.575) {};
		\node [style=pointoperator, color = black] (26) at (-3.7, -0.575) {};
		\node [style=pointoperator, color = black] (27) at (-1.7, 0.6) {};
		\node [style=none] (28) at (-3.75, 1.75) { \footnotesize $Z$};
		\node [style=none] (29) at (-3.75, -1.75) { \footnotesize $Z$};
		\node [style=none] (30) at (-3.225, -0.55) { \footnotesize $Z$};
		\node [style=none] (31) at (-1.2, 0.725) { \footnotesize $Z$};
		\node [style=none] (32) at (-5.25, 0.65) { \footnotesize $Z$};
		\node [style=none] (33) at (-7.225, -0.675) { \footnotesize $Z$};
		\node [style=none] (34) at (-3.95, 0.475) { \footnotesize $X$};
		\node [style=none] (35) at (7.5, 0.25) {};
		\node [style=none] (36) at (10.5, 0.25) {};
		\node [style=none] (37) at (5.5, -1.1) {};
		\node [style=none] (38) at (8.5, -1.1) {};
		\node [style=pointoperator, color = black] (39) at (7.5, 0.25) {};
		\node [style=pointoperator, color = black] (40) at (10.5, 0.25) {};
		\node [style=pointoperator, color = black] (41) at (5.5, -1.1) {};
		\node [style=pointoperator, color = black] (42) at (8.5, -1.1) {};
		\node [style=pointoperator, color = black] (43) at (8, -0.375) {};
		\node [style=none] (44) at (8.5, -0.25) { \footnotesize $X$};
		\node [style=none] (45) at (7.25, 0.75) { \footnotesize $Z$};
		\node [style=none] (46) at (11, 0.5) { \footnotesize $Z$};
		\node [style=none] (47) at (8.8, -1.675) { \footnotesize $Z$};
		\node [style=none] (48) at (4.875, -1.275) { \footnotesize $Z$};
		\node [style=none] (49) at (0.5, -0.75) {,};
		\node [style=none] (50) at (4.25, -0.75) {,};
		\node [style=none] (51) at (2.5, 1) {};
		\node [style=none] (52) at (2.5, -1.5) {};
		\node [style=none] (53) at (2.5, -0.25) {};
		\node [style=pointoperator, color = black] (54) at (2.5, -0.25) {};
		\node [style=pointoperator, color = black] (55) at (2.5, 1) {};
		\node [style=pointoperator, color = black] (56) at (2.5, -1.5) {};
		\node [style=none] (57) at (3, -0.25) { \footnotesize $X$};
		\node [style=none] (58) at (3, 1) { \footnotesize $Z$};
		\node [style=none] (59) at (3, -1.5) { \footnotesize $Z$};
	\end{pgfonlayer}
	\begin{pgfonlayer}{edgelayer}
		\draw [black!30,line width=1.0] (3.center) to (1.center);
		\draw [black!30,line width=1.0] (4.center) to (3.center);
		\draw [black!30,line width=1.0] (4.center) to (0.center);
		\draw [black!30,line width=1.0] (0.center) to (1.center);
		\draw [black!30,line width=1.0] (5.center) to (4.center);
		\draw [black!30,line width=1.0] (5.center) to (2.center);
		\draw [black!30,line width=1.0] (2.center) to (0.center);
		\draw [black!30,line width=1.0] (6.center) to (0.center);
		\draw [black!30,line width=1.0] (0.center) to (7.center);
		\draw [black!30,line width=1.0] (16) to (18.center);
		\draw [black!30,line width=1.0] (19.center) to (18.center);
		\draw [black!30,line width=1.0] (18.center) to (17.center);
		\draw [black!30,line width=1.0] (1.center) to (17.center);
		\draw [black!30,line width=1.0] (2.center) to (19.center);
		\draw [black!30,line width=1.0] (35.center) to (36.center);
		\draw [black!30,line width=1.0] (36.center) to (38.center);
		\draw [black!30,line width=1.0] (37.center) to (38.center);
		\draw [black!30,line width=1.0] (37.center) to (35.center);
		\draw [black!30,line width=1.0] (51.center) to (52.center);
	\end{pgfonlayer}
\end{tikzpicture}, \newline on the interface $\Delta_{\rm c}^{\rm A|A^c}$ and $\Delta_{\rm v}^{\rm A|A^c}$, we have 
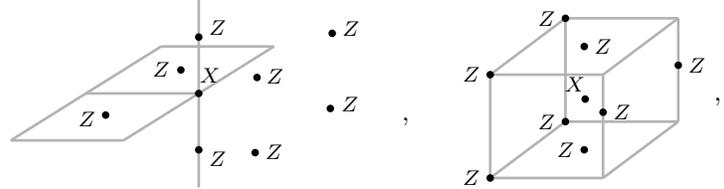
\begin{tikzpicture}
	\begin{pgfonlayer}{nodelayer}
		\node [style=none] (0) at (-6, 0) {};
		\node [style=none] (1) at (-4, 1.25) {};
		\node [style=none] (2) at (-8, -1.25) {};
		\node [style=none] (3) at (-7, 1.25) {};
		\node [style=none] (4) at (-9, 0) {};
		\node [style=none] (5) at (-11, -1.25) {};
		\node [style=none] (6) at (-6, 2.5) {};
		\node [style=none] (7) at (-6, -2.5) {};
		\node [style=pointoperator, color = black] (10) at (-6, 1.5) {};
		\node [style=pointoperator, color = black] (11) at (-6, -1.5) {};
		\node [style=pointoperator, color = black] (16) at (-6, 0) {};
		\node [style=pointoperator, color = black] (24) at (-6.475, 0.625) {};
		\node [style=pointoperator, color = black] (25) at (-8.475, -0.575) {};
		\node [style=pointoperator, color = black] (26) at (-4.45, 0.425) {};
		\node [style=pointoperator, color = black] (27) at (-2.45, 1.6) {};
		\node [style=none] (28) at (-5.5, 1.75) { \footnotesize $Z$};
		\node [style=none] (29) at (-5.5, -1.75) { \footnotesize $Z$};
		\node [style=none] (30) at (-3.975, 0.45) { \footnotesize $Z$};
		\node [style=none] (31) at (-1.95, 1.725) { \footnotesize $Z$};
		\node [style=none] (32) at (-7, 0.65) { \footnotesize $Z$};
		\node [style=none] (33) at (-8.975, -0.675) { \footnotesize $Z$};
		\node [style=none] (34) at (-5.7, 0.475) { \footnotesize $X$};
		\node [style=none] (49) at (-0.5, -0.75) {,};
		\node [style=pointoperator, color = black] (60) at (-4.5, -1.575) {};
		\node [style=pointoperator, color = black] (61) at (-2.5, -0.4) {};
		\node [style=none] (62) at (-4, -1.55) { \footnotesize $Z$};
		\node [style=none] (63) at (-1.975, -0.275) { \footnotesize $Z$};
		\node [style=none] (64) at (3.75, 2) {};
		\node [style=none] (65) at (1.75, 0.5) {};
		\node [style=none] (66) at (6.75, 2) {};
		\node [style=none] (67) at (4.75, 0.5) {};
		\node [style=none] (68) at (3.75, -0.75) {};
		\node [style=none] (69) at (1.75, -2.25) {};
		\node [style=none] (70) at (6.75, -0.75) {};
		\node [style=none] (71) at (4.75, -2.25) {};
		\node [style=none] (72) at (4.275, -0.15) {};
		\node [style=pointoperator, color = black] (73) at (4.275, -0.15) {};
		\node [style=pointoperator, color = black] (74) at (1.75, 0.5) {};
		\node [style=pointoperator, color = black] (75) at (3.75, 2) {};
		\node [style=pointoperator, color = black] (76) at (1.75, -2.25) {};
		\node [style=pointoperator, color = black] (77) at (3.75, -0.75) {};
		\node [style=pointoperator, color = black] (78) at (4.25, 1.25) {};
		\node [style=pointoperator, color = black] (79) at (4.25, -1.5) {};
		\node [style=pointoperator, color = black] (80) at (4.75, -0.5) {};
		\node [style=pointoperator, color = black] (81) at (6.75, 0.75) {};
		\node [style=none] (82) at (4, 0.25) { \footnotesize $X$};
		\node [style=none] (83) at (4.75, 1.25) { \footnotesize $Z$};
		\node [style=none] (84) at (3.75, -1.5) { \footnotesize $Z$};
		\node [style=none] (85) at (7.25, 0.75) { \footnotesize $Z$};
		\node [style=none] (86) at (5.25, -0.5) { \footnotesize $Z$};
		\node [style=none] (87) at (1.25, 0.5) { \footnotesize $Z$};
		\node [style=none] (88) at (1.25, -2.25) { \footnotesize $Z$};
		\node [style=none] (89) at (3.25, -0.75) { \footnotesize $Z$};
		\node [style=none] (90) at (3.25, 2) { \footnotesize $Z$};
	\end{pgfonlayer}
	\begin{pgfonlayer}{edgelayer}
		\draw [black!30,line width=1.0] (3.center) to (1.center);
		\draw [black!30,line width=1.0] (4.center) to (3.center);
		\draw [black!30,line width=1.0] (4.center) to (0.center);
		\draw [black!30,line width=1.0] (0.center) to (1.center);
		\draw [black!30,line width=1.0] (5.center) to (4.center);
		\draw [black!30,line width=1.0] (5.center) to (2.center);
		\draw [black!30,line width=1.0] (2.center) to (0.center);
		\draw [black!30,line width=1.0] (6.center) to (0.center);
		\draw [black!30,line width=1.0] (0.center) to (7.center);
		\draw [black!30,line width=1.0] (65.center) to (67.center);
		\draw [black!30,line width=1.0] (67.center) to (66.center);
		\draw [black!30,line width=1.0] (64.center) to (66.center);
		\draw [black!30,line width=1.0] (65.center) to (64.center);
		\draw [black!30,line width=1.0] (65.center) to (69.center);
		\draw [black!30,line width=1.0] (64.center) to (68.center);
		\draw [black!30,line width=1.0] (67.center) to (71.center);
		\draw [black!30,line width=1.0] (66.center) to (70.center);
		\draw [black!30,line width=1.0] (69.center) to (71.center);
		\draw [black!30,line width=1.0] (71.center) to (70.center);
		\draw [black!30,line width=1.0] (69.center) to (68.center);
		\draw [black!30,line width=1.0] (68.center) to (70.center);
	\end{pgfonlayer}
\end{tikzpicture}, \newline and in the bulk of $\Delta_{{\rm p}_{xy \cup {\rm e}_{\rm z}}}^{\rm A}$ and $\Delta_{\rm c}^{\rm A}$, we have 
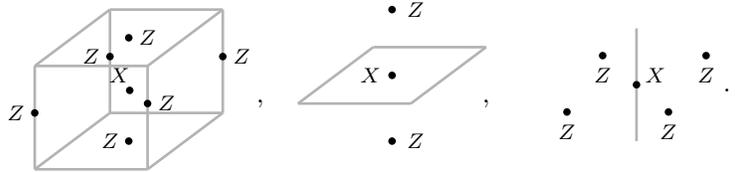
\begin{tikzpicture}
	\begin{pgfonlayer}{nodelayer}
		\node [style=none] (64) at (-5, 2) {};
		\node [style=none] (65) at (-7, 0.5) {};
		\node [style=none] (66) at (-2, 2) {};
		\node [style=none] (67) at (-4, 0.5) {};
		\node [style=none] (68) at (-5, -0.75) {};
		\node [style=none] (69) at (-7, -2.25) {};
		\node [style=none] (70) at (-2, -0.75) {};
		\node [style=none] (71) at (-4, -2.25) {};
		\node [style=none] (72) at (-4.475, -0.15) {};
		\node [style=pointoperator, color = black] (73) at (-4.475, -0.15) {};
		\node [style=pointoperator, color = black] (78) at (-4.5, 1.25) {};
		\node [style=pointoperator, color = black] (79) at (-4.5, -1.5) {};
		\node [style=pointoperator, color = black] (80) at (-4, -0.5) {};
		\node [style=pointoperator, color = black] (81) at (-2, 0.75) {};
		\node [style=none] (82) at (-4.75, 0.25) { \footnotesize $X$};
		\node [style=none] (83) at (-4, 1.25) { \footnotesize $Z$};
		\node [style=none] (84) at (-5, -1.5) { \footnotesize $Z$};
		\node [style=none] (85) at (-1.5, 0.75) { \footnotesize $Z$};
		\node [style=none] (86) at (-3.5, -0.5) { \footnotesize $Z$};
		\node [style=pointoperator, color = black] (87) at (-5, 0.75) {};
		\node [style=pointoperator, color = black] (88) at (-7, -0.75) {};
		\node [style=none] (89) at (-5.5, 0.75) { \footnotesize $Z$};
		\node [style=none] (90) at (-7.5, -0.75) { \footnotesize $Z$};
		\node [style=none] (91) at (2, 1) {};
		\node [style=none] (92) at (0, -0.5) {};
		\node [style=none] (93) at (5, 1) {};
		\node [style=none] (94) at (3, -0.5) {};
		\node [style=pointoperator, color = black] (95) at (2.5, 0.25) {};
		\node [style=pointoperator, color = black] (96) at (2.5, 2) {};
		\node [style=pointoperator, color = black] (97) at (2.5, -1.5) {};
		\node [style=none] (98) at (3.125, 2) { \footnotesize $Z$};
		\node [style=none] (99) at (3.125, -1.5) { \footnotesize $Z$};
		\node [style=none] (100) at (1.925, 0.25) { \footnotesize $X$};
		\node [style=none] (101) at (-1, -0.5) {,};
		\node [style=none] (102) at (5, -0.5) {,};
		\node [style=none] (103) at (9, 1.5) {};
		\node [style=none] (104) at (9, -1.5) {};
		\node [style=none] (105) at (9, 0) {};
		\node [style=pointoperator, color = black] (106) at (9, 0) {};
		\node [style=pointoperator, color = black] (107) at (8.1, 0.775) {};
		\node [style=pointoperator, color = black] (108) at (7.15, -0.725) {};
		\node [style=pointoperator, color = black] (109) at (10.85, 0.775) {};
		\node [style=pointoperator, color = black] (110) at (9.85, -0.725) {};
		\node [style=none] (111) at (7.15, -1.25) { \footnotesize $Z$};
		\node [style=none] (112) at (8.1, 0.25) { \footnotesize $Z$};
		\node [style=none] (113) at (10.85, 0.25) { \footnotesize $Z$};
		\node [style=none] (114) at (9.85, -1.25) { \footnotesize $Z$};
		\node [style=none] (115) at (9.5, 0.25) { \footnotesize $X$};
	\end{pgfonlayer}
	\begin{pgfonlayer}{edgelayer}
		\draw [black!30,line width=1.0] (65.center) to (67.center);
		\draw [black!30,line width=1.0] (67.center) to (66.center);
		\draw [black!30,line width=1.0] (64.center) to (66.center);
		\draw [black!30,line width=1.0] (65.center) to (64.center);
		\draw [black!30,line width=1.0] (65.center) to (69.center);
		\draw [black!30,line width=1.0] (64.center) to (68.center);
		\draw [black!30,line width=1.0] (67.center) to (71.center);
		\draw [black!30,line width=1.0] (66.center) to (70.center);
		\draw [black!30,line width=1.0] (69.center) to (71.center);
		\draw [black!30,line width=1.0] (71.center) to (70.center);
		\draw [black!30,line width=1.0] (69.center) to (68.center);
		\draw [black!30,line width=1.0] (68.center) to (70.center);
		\draw [black!30,line width=1.0] (91.center) to (93.center);
		\draw [black!30,line width=1.0] (93.center) to (94.center);
		\draw [black!30,line width=1.0] (92.center) to (94.center);
		\draw [black!30,line width=1.0] (92.center) to (91.center);
		\draw [black!30,line width=1.0] (103.center) to (104.center);
	\end{pgfonlayer}
\end{tikzpicture}. \newline
The spins on cubes are shown as points at their centers. 
To obtain the gauged–to–ungauged overlap state we project the auxiliary cube and vertex qubits onto $\ket{+}$ on the corresponding regions and form the overlap
\begin{align}
    |\text{ACM}_{\text{GS}}^{*\rm A}\rangle=\bra{+}^{\Delta_{\rm c}^{\rm A}}\bra{+}^{\Delta_{\rm v}^{\rm A^c}}|\Psi_{\text{cluster}}^{\text{3D};\rm A|A^c}\rangle\, .
\end{align}
The state $\big|\text{ACM}_{\text{GS}}^{*\rm A}\big\rangle$ shares the same local stabilizer structure as the gauged ground state $\mathsf{H}^{\rm A}|\text{ACM}{\text{GS}}\rangle$, but differs by global (logical) operator sectors. These sectors are related by the action of the sum of logical operators associated with the subgroup $H_q$. Explicitly,
\begin{align}
     &\frac{1}{\mathcal{N}_{\text{ACM}}}\mathsf{H}^{\rm A}|\text{ACM}_{\text{GS}}\rangle =\frac{1}{\mathcal{N}_{\text{ACM}}^{*\rm A}}\frac{1}{|H_q|}\left(\sum_{\bm z\in H_q}X(\bm z\cap\Delta_{\rm p_{xy}\cup e_z}^{\rm A^c})Z(\bm z\cap\Delta_{\rm p_{xy}\cup e_z}^{\rm A})\right)|\text{ACM}_{\text{GS}}^{*\rm A}\rangle\,.
\end{align}
where $\mathcal{N}_{\text{ACM}}$ and $\mathcal{N}_{\text{ACM}}^{*\rm A}$ are normalization constants chosen so that both states are unit normalized. The normalizations can be written as
\begin{subequations}
    \begin{align}
    &\mathcal{N}_{\text{ACM}}=\sqrt{\langle \text{ACM}_{\text{GS}}|\text{ACM}_{\text{GS}}\rangle}=\sqrt{\frac{2^{L_x+L_y-1}}{2^{|\Delta_{\rm v}|}}}\\
    &\mathcal{N}^{*\rm A}_{\text{ACM}}=\sqrt{\langle \text{ACM}_{\text{GS}}^{*\rm A}|\frac{1}{|H_q|}\left(\sum_{\bm z\in H_q}X(\bm z\cap\Delta_{\rm p_{xy}\cup e_z}^{\rm A^c})Z(\bm z\cap\Delta_{\rm p_{xy}\cup e_z}^{\rm A})\right)|\text{ACM}_{\text{GS}}^{*\rm A}\rangle }=\sqrt{\frac{2^{L_x+L_y-2l_x+1}}{2^{|\Delta_{\rm c}^{\rm A}|+|\Delta_{\rm v}^{\rm A^c}|}}}\,.
\end{align}
\end{subequations}
Next consider the overlap of $\big|\text{ACM}_{\text{GS}}^{*\rm A}\big\rangle$ with product state that produce the classical Boltzmann weights on the ungauged and gauged regions. Concretely, overlap with
\begin{align*}(\bra{0}e^{JX})^{\Delta_{\rm p_{xy}}\cap \Delta_{\rm p_{xy}\cup e_z}^{\rm A^c}}(\bra{0}e^{KX})^{\Delta_{\rm e_z}\cap \Delta_{\rm p_{xy}\cup e_z}^{\rm A^c}}(\bra{0}e^{J^*X})^{\Delta_{\rm p_{xy}}\cap \Delta_{\rm p_{xy}\cup e_z}^{\rm A}}(\bra{0}e^{K^*X})^{\Delta_{\rm e_z}\cap \Delta_{\rm p_{xy}\cup e_z}^{\rm A}}\,
\end{align*} 
yields the following expression (after collecting prefactors and summing over the remaining spins):
\begin{widetext}
\begin{align}
\begin{split}
    Z_{\text{API}}(J,K)&=\frac{2^{-\frac{|\Delta_{\rm c}^{\rm A}|}{2}}2^{-\frac{|\Delta_{\rm v}^{\rm A^c}|}{2}}2^{\frac{|\Delta_{\rm v}|}{2}}2^{l_x-1}}{(\sinh 2J^*)^{|\Delta_{\rm p_{xy}}\cap \Delta_{\rm p_{xy}\cup e_z}^{\rm A}|/2}(\sinh 2K^*)^{|\Delta_{\rm e_z}\cap \Delta_{\rm p_{xy}\cup e_z}^{\rm A}|/2}|H_q|}\sum_{\bm z\in H_q}\sum_{\{s_{\sigma}=\pm 1\}\rvert_{\sigma\in\Delta_{\rm v}^{\rm A^c}}}\sum_{\{s_{\sigma}=\pm 1\}\rvert_{\sigma\in\Delta_{\rm c}^{\rm A}}}\\
    & \exp\left(J \sum_{\sigma'\in \Delta_{\rm p_{xy}}\cap \Delta_{\rm p_{xy}\cup e_z}^{\rm A^c}}s(\delta \sigma')+K\sum_{\sigma''\in\Delta_{\rm e_z}\cap \Delta_{\rm p_{xy}\cup e_z}^{\rm A^c}}s(\delta\sigma'')+J^*\sum_{\sigma'\in\Delta_{\rm p_{xy}}\cap \Delta_{\rm p_{xy}\cup e_z}^{\rm A}}(-1)^{\#(\bm z\cap \sigma')}s(\delta^*\sigma')\right.\\
    &\qquad\left.+K^*\sum_{\sigma''\in\Delta_{\rm e_z}\cap \Delta_{\rm p_{xy}\cup e_z}^{\rm A}}(-1)^{\# (\bm z\cap \sigma'')}s(\delta^*\sigma'')\right)\prod_{\substack{ \rho\in\Delta_{\rm c}^{\rm A|A^c}\\
\tau\in\delta(\delta\rho\cap\rm A)}}(-1)^{\frac{1-s(\tau)}{2}\frac{1-s(\rho)}{2}}\times s(\delta(\bm z\cap \rm A^c))\, .
\label{eq:AgaugedAPI}
\end{split}
\end{align}
\end{widetext}
 In the above formula the shorthand $s(\delta\sigma)=\prod_{\rho\in\delta\sigma}s_{\rho}$ denotes the product of spins on the boundary of the cell $\sigma$, and similarly $s(\delta(\bm z\cap {\rm A^c}))$ denotes the product of spins along the appropriate intersection\footnote{We note that these spins arise at the junction between the duality defect and the $\mathbb{Z}_2$ symmetry defect.}. The last two factors encode the extra sign structure arising from cells that straddle the interface.

 From the overlap \eqref{eq:AgaugedAPI} one can read off the API partition function in the presence of two duality defects (one on each side of the interface). After removing the sum over $\bm z\in H_q$ (i.e., fixing a particular sector) and collecting prefactors, the partition function with two duality defects inserted along the interfaces takes the form
 \begin{align}
     \begin{split}
    Z_{\text{API}}^{\mathcal{D}}(J,K)&=\frac{2^{-\frac{|\Delta_{\rm c}^{\rm A}|}{2}}2^{-\frac{|\Delta_{\rm v}^{\rm A^c}|}{2}}2^{\frac{|\Delta_{\rm v}|}{2}}2^{l_x-1}}{(\sinh 2J^*)^{|\Delta_{\rm p_{xy}}\cap \Delta_{\rm p_{xy}\cup e_z}^{\rm A}|/2}(\sinh 2K^*)^{|\Delta_{\rm e_z}\cap \Delta_{\rm p_{xy}\cup e_z}^{\rm A}|/2}}\sum_{\{s_{\sigma}=\pm 1\}\rvert_{\sigma\in\Delta_{\rm v}^{\rm A^c}}}\sum_{\{s_{\sigma}=\pm 1\}\rvert_{\sigma\in\Delta_{\rm c}^{\rm A}}}\\
    & \exp\left(J \sum_{\sigma'\in \Delta_{\rm p_{xy}}\cap \Delta_{\rm p_{xy}\cup e_z}^{\rm A^c}}s(\delta \sigma')+K\sum_{\sigma''\in\Delta_{\rm e_z}\cap \Delta_{\rm p_{xy}\cup e_z}^{\rm A^c}}s(\delta\sigma'')+J^*\sum_{\sigma'\in\Delta_{\rm p_{xy}}\cap \Delta_{\rm p_{xy}\cup e_z}^{\rm A}}s(\delta^*\sigma')\right.\\
    &\qquad\left.+K^*\sum_{\sigma''\in\Delta_{\rm e_z}\cap \Delta_{\rm p_{xy}\cup e_z}^{\rm A}}s(\delta^*\sigma'')\right)\prod_{\substack{ \rho\in\Delta_{\rm c}^{\rm A|A^c}\\
\tau\in\delta(\delta\rho\cap\rm A)}}(-1)^{\frac{1-s(\tau)}{2}\frac{1-s(\rho)}{2}}\, .
\label{eq:dualitydefectAPI}
\end{split}
 \end{align}
 The formula transparently encodes the following features:
\begin{itemize}
    \item the gauged region contributes dual couplings $J^*,K^*$ and products of dual–site spins $s(\delta^*\sigma)$,
    \item the ungauged region retains the original couplings $J,K$ and the usual boundary products $s(\delta\sigma)$, and
    \item cells straddling the interface produce extra sign factors (the product over $\rho,\tau$).
\end{itemize}
This completes the lattice realization of the duality defect for the anisotropic plaquette Ising model, in direct analogy with the two–dimensional duality defect constructed via strange correlators. 
 \subsection{Moving and fusing defects}
 Analogous to the two-dimensional Ising case, we denote the spins residing in region $\Delta_{\rm c}^{\rm A}$ by lowercase Greek letters, while those in region $\Delta_{\rm v}^{\rm A^c}$ are denoted by lowercase Roman letters. The diagrammatic representation of the terms appearing in the partition function in the presence of the duality defect is defined as follows:
\begin{align}
\begin{tikzpicture}
	\begin{pgfonlayer}{nodelayer}
		\node [style=pointblack] (0) at (4, 1.25) {};
		\node [style=pointblack] (1) at (4, 4.25) {};
		\node [style=none] (2) at (5, 2.75) {$=$};
		\node [style=none] (3) at (4, 0.75) {$a$};
		\node [style=none] (4) at (4, 4.75) {$b$};
		\node [style=none] (5) at (6.75, 3) {$e^{Ks_as_b}\,,$};
		\node [style=none] (6) at (-8, 2) {};
		\node [style=none] (7) at (-3, 2) {};
		\node [style=none] (8) at (-7, 4) {};
		\node [style=none] (9) at (-4, 4) {};
		\node [style=pointblack] (10) at (-8, 2) {};
		\node [style=pointblack] (11) at (-3, 2) {};
		\node [style=pointblack] (12) at (-4, 4) {};
		\node [style=pointblack] (13) at (-7, 4) {};
		\node [style=none] (14) at (-2.5, 2.75) {$=$};
		\node [style=none] (15) at (-0.5, 3) {$e^{Js_as_bs_cs_d}\,,$};
		\node [style=none] (16) at (-8.5, 1.75) {$a$};
		\node [style=none] (17) at (-2.75, 1.75) {$b$};
		\node [style=none] (18) at (-3.75, 4.25) {$c$};
		\node [style=none] (19) at (-7.25, 4.25) {$d$};
		\node [style=none] (28) at (-8.25, -4) {$\alpha$};
		\node [style=none] (29) at (-3, -4) {$\beta$};
		\node [style=none] (30) at (-3.5, -1.25) {$\gamma$};
		\node [style=none] (31) at (-7.5, -1.25) {$\delta$};
		\node [style=none] (32) at (-2.5, -2.5) {$=$};
		\node [style=none] (33) at (-0.25, -2.25) {$e^{K^*s_{\alpha}s_{\beta}s_{\gamma}s_{\delta}}\,,$};
		\node [style=none] (34) at (-8, -3.5) {};
		\node [style=none] (35) at (-3, -3.5) {};
		\node [style=none] (36) at (-4, -1.5) {};
		\node [style=none] (37) at (-7, -1.5) {};
		\node [style=pointblack] (38) at (-8, -3.5) {};
		\node [style=pointblack] (39) at (-3, -3.5) {};
		\node [style=pointblack] (40) at (-4, -1.5) {};
		\node [style=pointblack] (41) at (-7, -1.5) {};
		\node [style=pointblack] (42) at (4, -1.5) {};
		\node [style=pointblack] (43) at (4, -4.5) {};
		\node [style=none] (44) at (5, -2.75) {$=$};
		\node [style=none] (45) at (6.75, -2.5) {$e^{J^*s_{\alpha}s_{\beta}}\,,$};
		\node [style=none] (46) at (4, -5) {$\alpha$};
		\node [style=none] (47) at (4, -1) {$\beta$};
		\node [style=pointblack] (48) at (10, 1) {};
		\node [style=pointblack] (49) at (10, -2) {};
		\node [style=none] (50) at (10, -2.5) {$a$};
		\node [style=none] (51) at (10, 1.5) {$\alpha$};
		\node [style=none] (52) at (11, -0.5) {$=$};
		\node [style=none] (53) at (13, -0.5) {$(-1)^{h_ah_{\alpha}}\,.$};
	\end{pgfonlayer}
	\begin{pgfonlayer}{edgelayer}
		\draw (1) to (0);
		\draw [style=grayfillline] (8.center)
			 to (6.center)
			 to (7.center)
			 to (9.center)
			 to cycle;
		\draw [style=redfillline] (35.center)
			 to (36.center)
			 to (37.center)
			 to (34.center)
			 to cycle;
		\draw [style=reddashedline] (42) to (43);
		\draw [style=greenline] (48) to (49);
	\end{pgfonlayer}
\end{tikzpicture}
\end{align}
In this notation, the solid black line represents a $z$-directed edge of the original lattice, whereas the red dashed line denotes a $z$-directed edge of the dual lattice. The plaquette shaded in grey corresponds to one in the original lattice, while the red-shaded plaquette belongs to the dual lattice. This diagrammatic formalism will be employed to express the relations governing the movement of the duality defect. \subsubsection{Movement of duality defects}
 We examine the fusion of two duality defects by considering an interface between regions $\Delta_{\rm p_{xy}\cup e_z}^{\rm A}$ and $\Delta_{\rm p_{xy}\cup e_z}^{\rm A^c}$.  First, let us address an identity concerning the deformation of region $\Delta_{\rm p_{xy}\cup e_z}^{\rm A}$:
 \begin{align}
     \begin{tikzpicture}
	\begin{pgfonlayer}{nodelayer}
		\node [style=none] (0) at (6, 1) {};
		\node [style=none] (1) at (4, -1) {};
		\node [style=none] (2) at (9, 1) {};
		\node [style=none] (3) at (11, -1) {};
		\node [style=none] (4) at (7.5, 2) {};
		\node [style=pointblack] (5) at (4, -1) {};
		\node [style=pointblack] (6) at (6, 1) {};
		\node [style=pointblack] (7) at (9, 1) {};
		\node [style=pointblack] (8) at (11, -1) {};
		\node [style=pointblack] (9) at (7.5, 2) {};
		\node [style=none] (10) at (12.75, 0) {$=$};
		\node [style=none] (11) at (14, 0) {\Large $\sum$};
		\node [style=none] (12) at (14, -0.75) {\tiny $h_{\beta}=0$};
		\node [style=pointblack] (13) at (19.5, -1) {};
		\node [style=pointblack] (14) at (21.5, 1) {};
		\node [style=pointblack] (15) at (24.5, 1) {};
		\node [style=pointblack] (16) at (26.5, -1) {};
		\node [style=pointblack] (17) at (23, 2) {};
		\node [style=pointblack] (18) at (23, -3) {};
		\node [style=none] (19) at (3.5, -1.25) {$a$};
		\node [style=none] (20) at (11.25, -1.5) {$b$};
		\node [style=none] (21) at (9.5, 1) {$c$};
		\node [style=none] (22) at (5.5, 1) {$d$};
		\node [style=none] (23) at (7.5, 2.5) {$\alpha$};
		\node [style=none] (24) at (23, 2.5) {$\alpha$};
		\node [style=none] (25) at (19.25, -1.25) {$a$};
		\node [style=none] (26) at (26.75, -1.25) {$b$};
		\node [style=none] (27) at (25, 1) {$c$};
		\node [style=none] (28) at (21, 1) {$d$};
		\node [style=none] (29) at (23, -3.5) {$\beta$};
		\node [style=none] (30) at (14, 0.75) {\tiny $1$};
		\node [style=none] (31) at (17, -1) {$\frac{1}{\sqrt{2\sinh(2J^*)}}$};
	\end{pgfonlayer}
	\begin{pgfonlayer}{edgelayer}
		\draw [style=grayfillline] (2.center)
			 to (0.center)
			 to (1.center)
			 to (3.center)
			 to cycle;
		\draw [style=greenline] (1.center) to (4.center);
		\draw [style=greenline] (0.center) to (4.center);
		\draw [style=greenline] (4.center) to (2.center);
		\draw [style=greenline] (4.center) to (3.center);
		\draw [style=grayfillline] (14) to (13);
		\draw [style=grayfillline] (13) to (16);
		\draw [style=grayfillline] (14) to (15);
		\draw [style=grayfillline] (15) to (16);
		\draw [style=reddashedline] (17) to (18);
		\draw [style=greenline] (18) to (13);
		\draw [style=greenline] (18) to (14);
		\draw [style=greenline] (18) to (15);
		\draw [style=greenline] (18) to (16);
	\end{pgfonlayer}
\end{tikzpicture}\,.
     \label{eq:3DAPImovement1}
 \end{align}
In the above equation, shaded grey square on the L.H.S. denote a plaquette Ising interaction while the red line on the R.H.S. denote an Ising interaction on links. The green lines denote the $(-1)^{h_{\rm c}h_{\rm v}}$ type phases or equivalently, it can be thought of as the link where $CZ$ is placed in the original cluster state $|\Psi_{\rm cluster}^{\text{3D};\rm A|A^c}\rangle$. On the L.H.S., we have an interaction on the original lattice, while on the R.H.S. the interaction is on the dual lattice. Going from L.H.S. to R.H.S. we have $|\Delta_{\rm p_{xy}\cup e_z}^{\rm A}|\longrightarrow |\Delta_{\rm p_{xy}\cup e_z}^{\rm A}|+1$ and $|\Delta_{\rm c}^{\rm A}|\longrightarrow|\Delta_{\rm c}^{\rm A}|+1$. In terms of equations, we have
\begin{align}
    &(-1)^{h_{\alpha}(h_a+h_b+h_c+h_d)}e^{Js_as_bs_cs_d}
    =\sum_{h_{\beta}=0,1}(-1)^{h_{\beta}(h_a+h_b+h_c+h_d)}\frac{e^{J^*s_{\alpha}s_{\beta}}}{\sqrt{2}(\sinh 2J^*)^{\frac{1}{2}}}\, .
\end{align}
There are other identities that we need to consider. For $|\Delta_{\rm p_{xy}\cup e_z}^{\rm A}|\longrightarrow |\Delta_{\rm p_{xy}\cup e_z}^{\rm A}|+1$ and $|\Delta_{\rm v}^{\rm A^c}|\longrightarrow |\Delta_{\rm v}^{\rm A^c}|-2$, we have the following identity
\begin{align}
    \begin{tikzpicture}
	\begin{pgfonlayer}{nodelayer}
		\node [style=none] (0) at (6, 1) {};
		\node [style=none] (1) at (4, -1) {};
		\node [style=none] (2) at (9, 1) {};
		\node [style=none] (3) at (11, -1) {};
		\node [style=none] (4) at (7.5, 2) {};
		\node [style=pointblack] (5) at (4, -1) {};
		\node [style=pointblack] (6) at (6, 1) {};
		\node [style=pointblack] (7) at (9, 1) {};
		\node [style=pointblack] (8) at (11, -1) {};
		\node [style=pointblack] (9) at (7.5, 2) {};
		\node [style=none] (10) at (12.75, 0) {$=$};
		\node [style=pointblack] (13) at (17, -1) {};
		\node [style=pointblack] (14) at (19, 1) {};
		\node [style=pointblack] (17) at (20.5, 2) {};
		\node [style=pointblack] (18) at (20.5, -3) {};
		\node [style=none] (19) at (3.5, -1.25) {$a$};
		\node [style=none] (20) at (11.25, -1.5) {$b$};
		\node [style=none] (21) at (9.5, 1) {$c$};
		\node [style=none] (22) at (5.5, 1) {$d$};
		\node [style=none] (23) at (7.5, 2.5) {$\alpha$};
		\node [style=none] (24) at (20.5, 2.5) {$\alpha$};
		\node [style=none] (25) at (16.75, -1.25) {$a$};
		\node [style=none] (28) at (18.5, 1) {$d$};
		\node [style=none] (29) at (20.5, -3.5) {$\beta$};
		\node [style=none] (30) at (2, 0) {\Large $\sum$};
		\node [style=none] (31) at (2, -0.75) {\tiny $h_b,h_c=0$};
		\node [style=none] (32) at (2, 0.75) {\tiny $1$};
		\node [style=pointblack] (33) at (7.5, -3) {};
		\node [style=none] (34) at (22, 1) {};
		\node [style=none] (35) at (24, -1) {};
		\node [style=none] (36) at (7.5, -3.5) {$\beta$};
		\node [style=none] (37) at (14.75, -0.25) {$\frac{2\sqrt{2}}{\sqrt{\sinh(2J^*)}}$};
	\end{pgfonlayer}
	\begin{pgfonlayer}{edgelayer}
		\draw [style=greenline] (33) to (7);
		\draw [style=grayfillline] (2.center)
			 to (0.center)
			 to (1.center)
			 to (3.center)
			 to cycle;
		\draw [style=greenline] (4.center) to (2.center);
		\draw [style=greenline] (4.center) to (3.center);
		\draw [style=grayfillline] (14) to (13);
		\draw [style=greenline] (33) to (8);
		\draw (14) to (34.center);
		\draw (34.center) to (35.center);
		\draw (13) to (35.center);
		\draw [style=greenline] (13) to (17);
		\draw [style=greenline] (14) to (17);
		\draw [style=greenline] (14) to (18);
		\draw [style=greenline] (13) to (18);
		\draw [style=reddashedline] (17) to (18);
	\end{pgfonlayer}
\end{tikzpicture}\,.
    \label{eq:3DAPImovement2}
\end{align}
The above equation is equivalent to
\begin{align}
    &\sum_{h_b,h_c=0,1}(-1)^{(h_b+h_c)(h_{\alpha}+h_{\beta})}e^{Js_{a}s_{b}s_{c}s_{d}}
    =(-1)^{(h_a+h_d)(h_{\alpha}+h_{\beta})}\frac{2\sqrt{2}\times e^{J^*s_{\alpha}s_{\beta}}}{(\sinh 2J^*)^{1/2}}\, .
\end{align}
The preceding two equations can be reduced to derive $J=-\frac{1}{2}\log\tanh J^*$. We shall now examine analogous equations pertaining to vertical edges within the original lattice. Initially, consider the following scenario $|\Delta_{\rm p_{xy}\cup e_z}^{\rm A}|\longrightarrow |\Delta_{\rm p_{xy}\cup e_z}^{\rm A}|+1$ and $|\Delta_{\rm v}^{\rm A^c}|\longrightarrow |\Delta_{\rm v}^{\rm A^c}|-1$:
\begin{align}
    \begin{tikzpicture}
	\begin{pgfonlayer}{nodelayer}
		\node [style=pointblack] (4) at (22, -3) {};
		\node [style=none] (5) at (20.5, 1) {};
		\node [style=none] (6) at (23.5, 1) {};
		\node [style=none] (7) at (25.5, -1) {};
		\node [style=none] (8) at (18.5, -1) {};
		\node [style=pointblack] (9) at (18.5, -1) {};
		\node [style=pointblack] (10) at (25.5, -1) {};
		\node [style=pointblack] (11) at (23.5, 1) {};
		\node [style=pointblack] (12) at (20.5, 1) {};
		\node [style=none] (13) at (18, -1.25) {$\alpha$};
		\node [style=none] (14) at (26, -1.25) {$\beta$};
		\node [style=none] (15) at (24, 1) {$\gamma$};
		\node [style=none] (16) at (20, 1) {$\delta$};
		\node [style=none] (17) at (22, -3.5) {$b$};
		\node [style=none] (18) at (13.5, 0) {$=$};
		\node [style=none] (19) at (2.75, 0) {\Large $\sum$};
		\node [style=none] (20) at (2.75, -0.75) {\tiny $h_a=0$};
		\node [style=pointblack] (21) at (4.75, -1) {};
		\node [style=pointblack] (22) at (6.75, 1) {};
		\node [style=pointblack] (23) at (9.75, 1) {};
		\node [style=pointblack] (24) at (11.75, -1) {};
		\node [style=pointblack] (25) at (8.25, -3) {};
		\node [style=pointblack] (26) at (8.25, 3) {};
		\node [style=none] (27) at (8.25, -3.5) {$b$};
		\node [style=none] (28) at (4.25, -1.25) {$\alpha$};
		\node [style=none] (29) at (12, -1.25) {$\beta$};
		\node [style=none] (30) at (10.25, 1) {$\gamma$};
		\node [style=none] (31) at (6.25, 1.25) {$\delta$};
		\node [style=none] (32) at (8.25, 3.5) {$a$};
		\node [style=none] (33) at (2.75, 0.75) {\tiny $1$};
		\node [style=none] (34) at (15.75, -0.25) {$\frac{\sqrt{2}}{\sqrt{\sinh(2K^*)}}$};
	\end{pgfonlayer}
	\begin{pgfonlayer}{edgelayer}
		\draw [style=greenline] (4) to (11);
		\draw [style=greenline] (4) to (12);
		\draw [style=redfillline] (5.center)
			 to (8.center)
			 to (7.center)
			 to (6.center)
			 to cycle;
		\draw [style=greenline] (9) to (4);
		\draw [style=greenline] (4) to (10);
		\draw (26) to (25);
		\draw [style=greenline] (21) to (26);
		\draw [style=greenline] (22) to (26);
		\draw [style=greenline] (26) to (23);
		\draw [style=greenline] (26) to (24);
	\end{pgfonlayer}
\end{tikzpicture}\,,
\label{eq:3DAPImovement3}
\end{align}
and translate to 
\begin{align}
    &(-1)^{h_b(h_{\alpha}+h_{\beta}+h_{\gamma}+h_{\delta})}\frac{\sqrt{2}\times e^{K^*s_{\alpha}s_{\beta}s_{\gamma}s_{\delta}}}{(\sinh 2K^*)^{1/2}}
  =\sum_{h_a=0}^1(-1)^{h_a(h_{\alpha}+h_{\beta}+h_{\gamma}+h_{\delta})}e^{K s_as_b}\, .
\end{align}
Now consider the scenario $|\Delta_{\rm p_{xy}\cup e_z}^{\rm A}|\longrightarrow |\Delta_{\rm p_{xy}\cup e_z}^{\rm A}|+1$ and $|\Delta_{\rm c}^{\rm A}|\longrightarrow|\Delta_{\rm c}^{\rm A}|+2$:
\begin{align}
    \begin{tikzpicture}
	\begin{pgfonlayer}{nodelayer}
		\node [style=pointblack] (0) at (25.5, -3) {};
		\node [style=none] (1) at (24, 1) {};
		\node [style=none] (2) at (27, 1) {};
		\node [style=none] (3) at (29, -1) {};
		\node [style=none] (4) at (22, -1) {};
		\node [style=pointblack] (5) at (22, -1) {};
		\node [style=pointblack] (6) at (29, -1) {};
		\node [style=pointblack] (7) at (27, 1) {};
		\node [style=pointblack] (8) at (24, 1) {};
		\node [style=none] (9) at (21.5, -1.25) {$\alpha$};
		\node [style=none] (10) at (29.5, -1.25) {$\beta$};
		\node [style=none] (11) at (27.5, 1) {$\gamma$};
		\node [style=none] (12) at (23.5, 1) {$\delta$};
		\node [style=none] (13) at (25.5, -3.5) {$b$};
		\node [style=none] (14) at (13.5, 0) {$=$};
		\node [style=pointblack] (19) at (9.75, 1) {};
		\node [style=pointblack] (20) at (11.75, -1) {};
		\node [style=pointblack] (21) at (8.25, -3) {};
		\node [style=pointblack] (22) at (8.25, 3) {};
		\node [style=none] (23) at (8.25, -3.5) {$b$};
		\node [style=none] (25) at (12, -1.25) {$\beta$};
		\node [style=none] (26) at (10.25, 1) {$\gamma$};
		\node [style=none] (28) at (8.25, 3.5) {$a$};
		\node [style=pointblack] (30) at (25.5, 3) {};
		\node [style=none] (31) at (25.5, 3.5) {$a$};
		\node [style=none] (32) at (15.25, 0) {\Large $\sum$};
		\node [style=none] (33) at (15.25, -0.75) {\tiny $h_{\alpha},h_{\delta}=0$};
		\node [style=none] (34) at (15.25, 0.75) {\tiny $1$};
		\node [style=none] (35) at (19, -0.25) {$\frac{1}{2\sqrt{2\sinh(2K^*)}}$};
	\end{pgfonlayer}
	\begin{pgfonlayer}{edgelayer}
		\draw [style=greenline] (0) to (8);
		\draw [style=redfillline] (1.center)
			 to (4.center)
			 to (3.center)
			 to (2.center)
			 to cycle;
		\draw [style=greenline] (5) to (0);
		\draw (22) to (21);
		\draw [style=greenline] (22) to (19);
		\draw [style=greenline] (22) to (20);
		\draw [style=greenline] (21) to (20);
		\draw [style=greenline] (21) to (19);
		\draw [style=greenline] (5) to (30);
		\draw [style=greenline] (8) to (30);
	\end{pgfonlayer}
\end{tikzpicture}\,,
    \label{fig:3DAPImovement4}
\end{align}
and translate to
\begin{align}
    &(-1)^{(h_a+h_b)(h_{\gamma}+h_{\beta})}e^{Ks_as_b}
    =\sum_{h_{\alpha},h_{\delta}=0,1}(-1)^{(h_a+h_b)(h_{\alpha}+h_{\delta})}\frac{e^{K^*s_{\alpha}s_{\beta}s_{\gamma}s_{\delta}}}{2\sqrt{2}(\sinh 2K^*)^{1/2}}\, .
\end{align}
The last two equations are satisfied when $K=-\frac{1}{2}\log\tanh K^*$. 

Now let us mention some global movement of duality defects. For instance, if we have the duality defect on $xy$-plane, then we can use \eqref{eq:3DAPImovement1} as
\begin{align}
    \begin{tikzpicture}
	\begin{pgfonlayer}{nodelayer}
		\node [style=none] (0) at (-6.25, 2) {};
		\node [style=none] (1) at (-7.25, 1) {};
		\node [style=none] (2) at (-4.25, 2) {};
		\node [style=none] (3) at (-5.25, 1) {};
		\node [style=none] (4) at (-2.25, 2) {};
		\node [style=none] (5) at (-3.25, 1) {};
		\node [style=none] (6) at (-8.25, 0) {};
		\node [style=none] (7) at (-6.25, 0) {};
		\node [style=none] (8) at (-4.25, 0) {};
		\node [style=none] (9) at (-5.75, 3.075) {};
		\node [style=none] (10) at (-6.75, 2.075) {};
		\node [style=none] (11) at (-3.75, 3.075) {};
		\node [style=none] (12) at (-4.75, 2.075) {};
		\node [style=none] (13) at (5, 2) {};
		\node [style=none] (14) at (4, 1) {};
		\node [style=none] (15) at (7, 2) {};
		\node [style=none] (16) at (6, 1) {};
		\node [style=none] (17) at (9, 2) {};
		\node [style=none] (18) at (8, 1) {};
		\node [style=none] (19) at (3, 0) {};
		\node [style=none] (20) at (5, 0) {};
		\node [style=none] (21) at (7, 0) {};
		\node [style=none] (22) at (5.5, -0.175) {};
		\node [style=none] (23) at (4.5, -1.175) {};
		\node [style=none] (24) at (7.5, -0.175) {};
		\node [style=none] (25) at (6.5, -1.175) {};
		\node [style=none] (26) at (-2.1, 1) {$=$};
		\node [style=none] (27) at (5.5, 3.075) {};
		\node [style=none] (28) at (4.5, 2.075) {};
		\node [style=none] (29) at (7.5, 3.075) {};
		\node [style=none] (30) at (6.5, 2.075) {};
		\node [style=pointoperator, color = black] (31) at (-8.25, 0) {};
		\node [style=pointoperator, color = black] (32) at (-6.25, 0) {};
		\node [style=pointoperator, color = black] (33) at (-4.25, 0) {};
		\node [style=pointoperator, color = black] (34) at (-3.25, 1) {};
		\node [style=pointoperator, color = black] (35) at (-2.25, 2) {};
		\node [style=pointoperator, color = black] (36) at (-4.25, 2) {};
		\node [style=pointoperator, color = black] (37) at (-6.25, 2) {};
		\node [style=pointoperator, color = black] (38) at (-7.25, 1) {};
		\node [style=pointoperator, color = black] (39) at (-5.25, 1) {};
		\node [style=pointoperator, color = black] (40) at (-6.75, 2.075) {};
		\node [style=pointoperator, color = black] (41) at (-4.75, 2.075) {};
		\node [style=pointoperator, color = black] (42) at (-3.75, 3.075) {};
		\node [style=pointoperator, color = black] (43) at (-5.75, 3.075) {};
		\node [style=pointoperator, color = black] (44) at (4.5, 2.075) {};
		\node [style=pointoperator, color = black] (45) at (6.5, 2.075) {};
		\node [style=pointoperator, color = black] (46) at (7.5, 3.075) {};
		\node [style=pointoperator, color = black] (47) at (5.5, 3.075) {};
		\node [style=pointoperator, color = black] (48) at (3, 0) {};
		\node [style=pointoperator, color = black] (49) at (5, 0) {};
		\node [style=pointoperator, color = black] (50) at (7, 0) {};
		\node [style=pointoperator, color = black] (51) at (8, 1) {};
		\node [style=pointoperator, color = black] (52) at (9, 2) {};
		\node [style=pointoperator, color = black] (53) at (7, 2) {};
		\node [style=pointoperator, color = black] (54) at (5, 2) {};
		\node [style=pointoperator, color = black] (55) at (4, 1) {};
		\node [style=pointoperator, color = black] (56) at (6, 1) {};
		\node [style=pointoperator, color = black] (57) at (4.5, -1.175) {};
		\node [style=pointoperator, color = black] (58) at (5.5, -0.175) {};
		\node [style=pointoperator, color = black] (59) at (6.5, -1.175) {};
		\node [style=pointoperator, color = black] (60) at (7.5, -0.175) {};
		\node [style=none] (61) at (0.9, 1) {$\frac{1}{(2\sinh(2J^{*}))^{\frac{\Delta_{{\rm p}_{\rm xy}}}{2}}}$};
	\end{pgfonlayer}
	\begin{pgfonlayer}{edgelayer}
		\draw (0.center)
			 to (6.center)
			 to (8.center)
			 to (4.center)
			 to [in=360, out=180] cycle;
		\draw (2.center) to (7.center);
		\draw (1.center) to (5.center);
		\draw [style=normalline, color=green] (10.center) to (1.center);
		\draw [style=normalline, color=green] (10.center) to (3.center);
		\draw [style=normalline, color=green] (10.center) to (7.center);
		\draw [style=normalline, color=green] (10.center) to (6.center);
		\draw [style=normalline, color=green] (9.center) to (0.center);
		\draw [style=normalline, color=green] (9.center) to (1.center);
		\draw [style=normalline, color=green] (9.center) to (3.center);
		\draw [style=normalline, color=green] (9.center) to (2.center);
		\draw [style=normalline, color=green] (11.center) to (2.center);
		\draw [style=normalline, color=green] (11.center) to (4.center);
		\draw [style=normalline, color=green] (11.center) to (5.center);
		\draw [style=normalline, color=green] (11.center) to (3.center);
		\draw [style=normalline, color=green] (12.center) to (3.center);
		\draw [style=normalline, color=green] (12.center) to (5.center);
		\draw [style=normalline, color=green] (12.center) to (8.center);
		\draw [style=normalline, color=green] (12.center) to (7.center);
		\draw (13.center) to (19.center);
		\draw (19.center) to (21.center);
		\draw (21.center) to (17.center);
		\draw [in=360, out=180] (17.center) to (13.center);
		\draw (15.center) to (20.center);
		\draw (14.center) to (18.center);
		\draw [style=normalline, color=green] (23.center) to (14.center);
		\draw [style=normalline, color=green] (23.center) to (16.center);
		\draw [style=normalline, color=green] (23.center) to (20.center);
		\draw [style=normalline, color=green] (23.center) to (19.center);
		\draw [style=normalline, color=green] (22.center) to (13.center);
		\draw [style=normalline, color=green] (22.center) to (14.center);
		\draw [style=normalline, color=green] (22.center) to (16.center);
		\draw [style=normalline, color=green] (22.center) to (15.center);
		\draw [style=normalline, color=green] (24.center) to (15.center);
		\draw [style=normalline, color=green] (24.center) to (17.center);
		\draw [style=normalline, color=green] (24.center) to (18.center);
		\draw [style=normalline, color=green] (24.center) to (16.center);
		\draw [style=normalline, color=green] (25.center) to (16.center);
		\draw [style=normalline, color=green] (25.center) to (18.center);
		\draw [style=normalline, color=green] (25.center) to (21.center);
		\draw [style=normalline, color=green] (25.center) to (20.center);
		\draw [style=grayfillline] (6.center)
			 to (8.center)
			 to (4.center)
			 to (0.center)
			 to cycle;
		\draw [style=dashedline, color = red] (28.center) to (23.center);
		\draw [style=dashedline, color = red] (30.center) to (25.center);
		\draw [style=dashedline, color = red] (29.center) to (24.center);
		\draw [style=dashedline, color = red] (27.center) to (22.center);
	\end{pgfonlayer}
\end{tikzpicture}\,,
\end{align}
which creates an array of new spins which are not interacting among themselves yet. Then we subsequently apply \eqref{eq:3DAPImovement3} on the newly created spins as 
\begin{align}
    \begin{tikzpicture}
	\begin{pgfonlayer}{nodelayer}
		\node [style=none] (13) at (-7.75, 2) {};
		\node [style=none] (14) at (-8.85, 1) {};
		\node [style=none] (15) at (-5.75, 2) {};
		\node [style=none] (16) at (-6.85, 1) {};
		\node [style=none] (17) at (-3.75, 2) {};
		\node [style=none] (18) at (-4.85, 1) {};
		\node [style=none] (19) at (-10, 0) {};
		\node [style=none] (20) at (-8, 0) {};
		\node [style=none] (21) at (-6, 0) {};
		\node [style=none] (22) at (-7.5, 3.075) {};
		\node [style=none] (23) at (-8.5, 2.075) {};
		\node [style=none] (24) at (-5.5, 3.075) {};
		\node [style=none] (25) at (-6.5, 2.075) {};
		\node [style=none] (27) at (-7.5, 3.075) {};
		\node [style=none] (28) at (-8.5, 2.075) {};
		\node [style=none] (29) at (-5.5, 3.075) {};
		\node [style=none] (30) at (-6.5, 2.075) {};
		\node [style=pointoperator, color = black] (48) at (-10, 0) {};
		\node [style=pointoperator, color = black] (49) at (-8, 0) {};
		\node [style=pointoperator, color = black] (50) at (-6, 0) {};
		\node [style=pointoperator, color = black] (51) at (-4.85, 1) {};
		\node [style=pointoperator, color = black] (52) at (-3.75, 2) {};
		\node [style=pointoperator, color = black] (53) at (-5.75, 2) {};
		\node [style=pointoperator, color = black] (54) at (-7.75, 2) {};
		\node [style=pointoperator, color = black] (55) at (-8.85, 1) {};
		\node [style=pointoperator, color = black] (56) at (-6.85, 1) {};
		\node [style=pointoperator, color = black] (57) at (-8.5, 2.075) {};
		\node [style=pointoperator, color = black] (58) at (-7.5, 3.075) {};
		\node [style=pointoperator, color = black] (59) at (-6.5, 2.075) {};
		\node [style=pointoperator, color = black] (60) at (-5.5, 3.075) {};
		\node [style=none] (61) at (-7.75, 4.25) {};
		\node [style=none] (62) at (-8.85, 3.25) {};
		\node [style=none] (63) at (-5.75, 4.25) {};
		\node [style=none] (64) at (-6.85, 3.25) {};
		\node [style=none] (65) at (-3.75, 4.25) {};
		\node [style=none] (66) at (-4.85, 3.25) {};
		\node [style=none] (67) at (-10, 2.25) {};
		\node [style=none] (68) at (-8, 2.25) {};
		\node [style=none] (69) at (-6, 2.25) {};
		\node [style=none] (70) at (-7.5, 0.175) {};
		\node [style=none] (71) at (-8.5, -0.825) {};
		\node [style=none] (72) at (-5.5, 0.175) {};
		\node [style=none] (73) at (-6.5, -0.825) {};
		\node [style=none] (74) at (-7.5, 0.175) {};
		\node [style=none] (75) at (-8.5, -0.825) {};
		\node [style=none] (76) at (-5.5, 0.175) {};
		\node [style=none] (77) at (-6.5, -0.825) {};
		\node [style=pointoperator, color = black] (78) at (-8.5, -0.825) {};
		\node [style=pointoperator, color = black] (79) at (-7.5, 0.175) {};
		\node [style=pointoperator, color = black] (80) at (-6.5, -0.825) {};
		\node [style=pointoperator, color = black] (81) at (-5.5, 0.175) {};
		\node [style=none] (82) at (4.25, 2) {};
		\node [style=none] (83) at (3.15, 1) {};
		\node [style=none] (84) at (6.25, 2) {};
		\node [style=none] (85) at (5.15, 1) {};
		\node [style=none] (86) at (8.25, 2) {};
		\node [style=none] (87) at (7.15, 1) {};
		\node [style=none] (88) at (2, 0) {};
		\node [style=none] (89) at (4, 0) {};
		\node [style=none] (90) at (6, 0) {};
		\node [style=none] (91) at (4.5, -0.175) {};
		\node [style=none] (92) at (3.5, -1.175) {};
		\node [style=none] (93) at (6.5, -0.175) {};
		\node [style=none] (94) at (5.5, -1.175) {};
		\node [style=none] (95) at (4.5, -0.175) {};
		\node [style=none] (96) at (3.5, -1.175) {};
		\node [style=none] (97) at (6.5, -0.175) {};
		\node [style=none] (98) at (5.5, -1.175) {};
		\node [style=pointoperator, color = black] (100) at (4, 0) {};
		\node [style=pointoperator, color = black] (102) at (7.15, 1) {};
		\node [style=pointoperator, color = black] (104) at (6.25, 2) {};
		\node [style=pointoperator, color = black] (106) at (3.15, 1) {};
		\node [style=pointoperator, color = black] (107) at (5.15, 1) {};
		\node [style=pointoperator, color = black] (108) at (3.5, -1.175) {};
		\node [style=pointoperator, color = black] (109) at (4.5, -0.175) {};
		\node [style=pointoperator, color = black] (110) at (5.5, -1.175) {};
		\node [style=pointoperator, color = black] (111) at (6.5, -0.175) {};
		\node [style=none] (112) at (4.25, 4.25) {};
		\node [style=none] (113) at (3.15, 3.25) {};
		\node [style=none] (114) at (6.25, 4.25) {};
		\node [style=none] (115) at (5.15, 3.25) {};
		\node [style=none] (116) at (8.25, 4.25) {};
		\node [style=none] (117) at (7.15, 3.25) {};
		\node [style=none] (118) at (2, 2.25) {};
		\node [style=none] (119) at (4, 2.25) {};
		\node [style=none] (120) at (6, 2.25) {};
		\node [style=none] (133) at (-3.75, 1) {$=$};
		\node [style=none] (134) at (-0.8, 1) {$\left(\frac{2}{\sinh(2K^*)}\right)^{\frac{\Delta_{{\rm p}_{\rm xy}}}{2}}$};
		\node [style=pointoperator, color = black] (135) at (2, 0) {};
		\node [style=pointoperator, color = black] (136) at (6, 0) {};
		\node [style=pointoperator, color = black] (137) at (8.25, 2) {};
		\node [style=pointoperator, color = black] (138) at (4.25, 2) {};
	\end{pgfonlayer}
	\begin{pgfonlayer}{edgelayer}
		\draw (13.center) to (19.center);
		\draw (19.center) to (21.center);
		\draw (21.center) to (17.center);
		\draw [in=360, out=180] (17.center) to (13.center);
		\draw (15.center) to (20.center);
		\draw (14.center) to (18.center);
		\draw [style=blueline, color = green] (23.center) to (14.center);
		\draw [style=blueline, color = green] (23.center) to (16.center);
		\draw [style=blueline, color = green] (23.center) to (20.center);
		\draw [style=blueline, color = green] (23.center) to (19.center);
		\draw [style=blueline, color = green] (22.center) to (13.center);
		\draw [style=blueline, color = green] (22.center) to (14.center);
		\draw [style=blueline, color = green] (22.center) to (16.center);
		\draw [style=blueline, color = green] (22.center) to (15.center);
		\draw [style=blueline, color = green] (24.center) to (15.center);
		\draw [style=blueline, color = green] (24.center) to (17.center);
		\draw [style=blueline, color = green] (24.center) to (18.center);
		\draw [style=blueline, color = green] (24.center) to (16.center);
		\draw [style=blueline, color = green] (25.center) to (16.center);
		\draw [style=blueline, color = green] (25.center) to (18.center);
		\draw [style=blueline, color = green] (25.center) to (21.center);
		\draw [style=blueline, color = green] (25.center) to (20.center);
		\draw [style=dashedline, color = red] (28.center) to (23.center);
		\draw [style=dashedline, color = red] (30.center) to (25.center);
		\draw [style=dashedline, color = red] (29.center) to (24.center);
		\draw [style=dashedline, color = red] (27.center) to (22.center);
		\draw [style=dashedline, color = red] (69.center) to (50);
		\draw [style=dashedline, color = red] (66.center) to (51);
		\draw [style=dashedline, color = red] (65.center) to (52);
		\draw [style=dashedline, color = red] (68.center) to (49);
		\draw [style=dashedline, color = red] (64.center) to (56);
		\draw [style=dashedline, color = red] (63.center) to (53);
		\draw [style=dashedline, color = red] (67.center) to (48);
		\draw [style=dashedline, color = red] (62.center) to (55);
		\draw [style=dashedline, color = red] (61.center) to (54);
		\draw [style=dashedline, color = red] (75.center) to (71.center);
		\draw [style=dashedline, color = red] (77.center) to (73.center);
		\draw [style=dashedline, color = red] (76.center) to (72.center);
		\draw [style=dashedline, color = red] (74.center) to (70.center);
		\draw (57) to (78);
		\draw (59) to (80);
		\draw (60) to (81);
		\draw (58) to (79);
		\draw (82.center) to (88.center);
		\draw (88.center) to (90.center);
		\draw (90.center) to (86.center);
		\draw [in=360, out=180] (86.center) to (82.center);
		\draw (84.center) to (89.center);
		\draw (83.center) to (87.center);
		\draw [style=blueline, color = green] (92.center) to (83.center);
		\draw [style=blueline, color = green] (92.center) to (85.center);
		\draw [style=blueline, color = green] (92.center) to (89.center);
		\draw [style=blueline, color = green] (92.center) to (88.center);
		\draw [style=blueline, color = green] (91.center) to (82.center);
		\draw [style=blueline, color = green] (91.center) to (83.center);
		\draw [style=blueline, color = green] (91.center) to (85.center);
		\draw [style=blueline, color = green] (91.center) to (84.center);
		\draw [style=blueline, color = green] (93.center) to (84.center);
		\draw [style=blueline, color = green] (93.center) to (86.center);
		\draw [style=blueline, color = green] (93.center) to (87.center);
		\draw [style=blueline, color = green] (93.center) to (85.center);
		\draw [style=blueline, color = green] (94.center) to (85.center);
		\draw [style=blueline, color = green] (94.center) to (87.center);
		\draw [style=blueline, color = green] (94.center) to (90.center);
		\draw [style=blueline, color = green] (94.center) to (89.center);
		\draw [style=dashedline, color = red] (96.center) to (92.center);
		\draw [style=dashedline, color = red] (98.center) to (94.center);
		\draw [style=dashedline, color = red] (97.center) to (93.center);
		\draw [style=dashedline, color = red] (95.center) to (91.center);
		\draw [style=dashedline, color = red] (114.center) to (104);
		\draw [style=dashedline, color = red] (113.center) to (106);
		\draw [style=redfillline] (86.center)
			 to (82.center)
			 to (88.center)
			 to (90.center)
			 to cycle;
		\draw [style=dashedline, color = red] (118.center) to (88.center);
		\draw [style=dashedline, color = red] (120.center) to (90.center);
		\draw [style=dashedline, color = red] (116.center) to (86.center);
		\draw [style=dashedline, color = red] (112.center) to (82.center);
		\draw [style=dashedline, color = red] (119.center) to (100);
		\draw [style=dashedline, color = red] (115.center) to (107);
		\draw [style=dashedline, color = red] (117.center) to (102);
	\end{pgfonlayer}
\end{tikzpicture}\,,
\end{align}
which creates an interaction between those newly created spins. The combination of these two steps moves the duality defect on the $xy$-plane by one lattice site along the $z$-direction.

\begin{figure}
    \flushleft
    \input{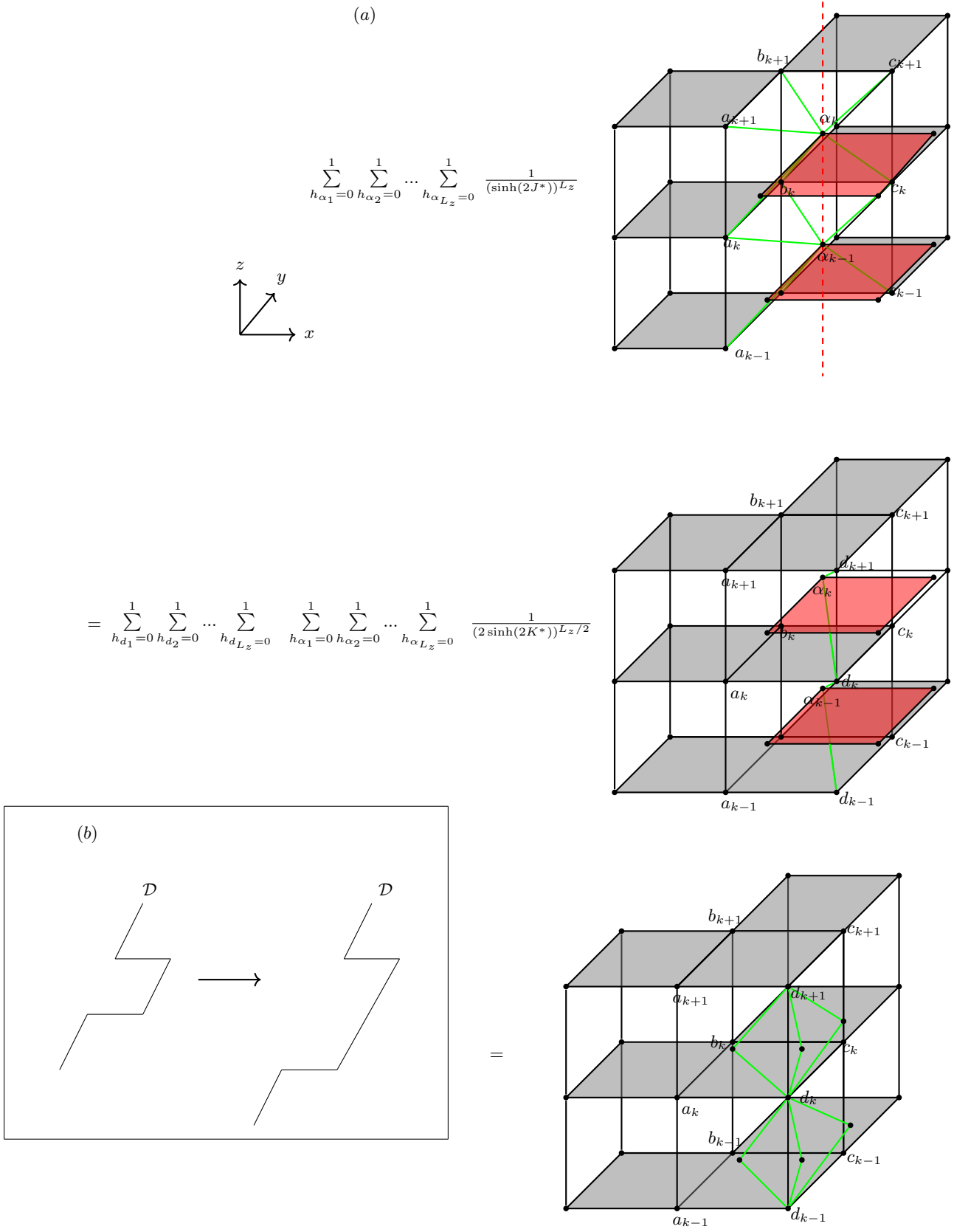}
    \caption{Moving the duality defect along a corner. The corner can be moved in the $y$-direction using the movement relations. $(a)$ Moving the corner using mathematical relations. $(b)$ Moving the duality defect pictorially.}
    \label{fig:3DAPI_duality_defect_movement_corner}
\end{figure}

On the other hand, consider a configuration of duality defect where there is a corner with translation invariance in the $z$-direction as in Figure~\ref{fig:3DAPI_duality_defect_movement_corner}. This corner can be moved along the $y$-direction after applying local movement relations that is a slight modification of \eqref{eq:3DAPImovement2} and \eqref{eq:3DAPImovement3}.
\subsubsection{Duality defect movements that is not allowed}
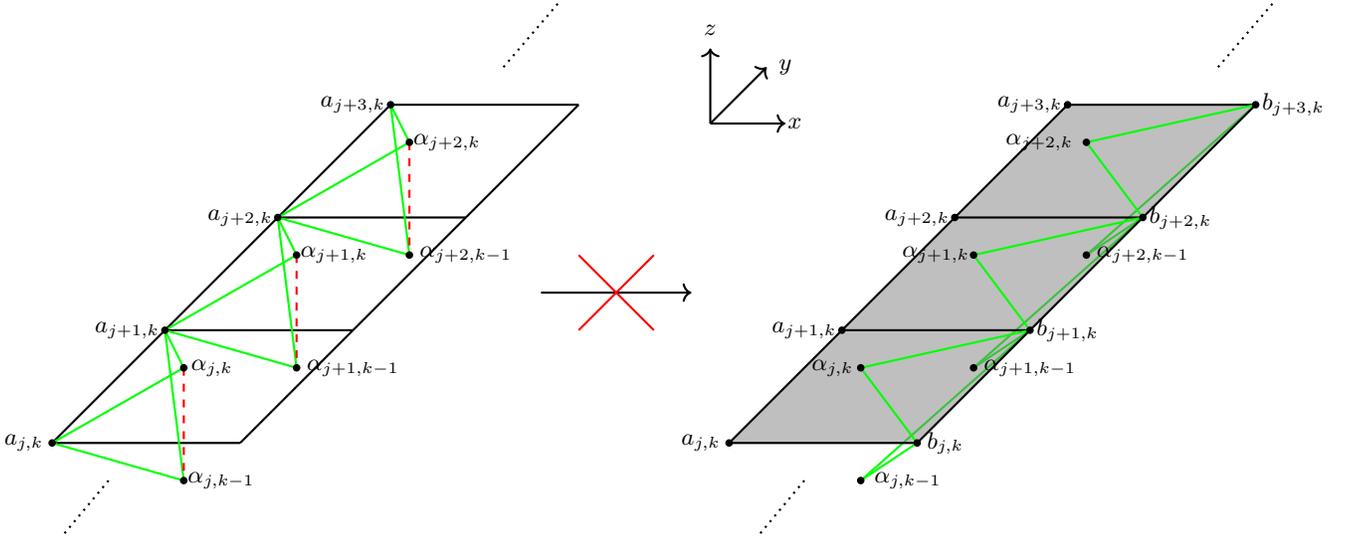
\begin{figure}
    \centering
    \begin{tikzpicture}
	\begin{pgfonlayer}{nodelayer}
		\node [style=none] (40) at (-11, -2) {};
		\node [style=none] (41) at (-8, 1) {};
		\node [style=none] (42) at (-5, 4) {};
		\node [style=none] (43) at (-2, 7) {};
		\node [style=none] (44) at (-6, -2) {};
		\node [style=none] (45) at (-3, 1) {};
		\node [style=none] (46) at (0, 4) {};
		\node [style=none] (47) at (3, 7) {};
		\node [style=none] (48) at (-7.5, 0) {};
		\node [style=none] (49) at (-7.5, -3) {};
		\node [style=none] (50) at (-4.5, 3) {};
		\node [style=none] (51) at (-4.5, 0) {};
		\node [style=none] (52) at (-1.5, 6) {};
		\node [style=none] (53) at (-1.5, 3) {};
		\node [style=pointblack] (54) at (-11, -2) {};
		\node [style=pointblack] (55) at (-8, 1) {};
		\node [style=pointblack] (56) at (-5, 4) {};
		\node [style=pointblack] (57) at (-2, 7) {};
		\node [style=pointblack] (58) at (-1.5, 6) {};
		\node [style=pointblack] (59) at (-4.5, 3) {};
		\node [style=pointblack] (60) at (-7.5, 0) {};
		\node [style=pointblack] (61) at (-7.5, -3) {};
		\node [style=pointblack] (62) at (-4.5, 0) {};
		\node [style=pointblack] (63) at (-1.5, 3) {};
		\node [style=none] (64) at (-6.75, 0) {$\alpha_{j,k}$};
		\node [style=none] (65) at (-3.5, 3) {$\alpha_{j+1,k}$};
		\node [style=none] (66) at (-0.5, 6) {$\alpha_{j+2,k}$};
		\node [style=none] (67) at (-6.5, -3) {$\alpha_{j,k-1}$};
		\node [style=none] (68) at (-3, 0) {$\alpha_{j+1,k-1}$};
		\node [style=none] (69) at (0, 3) {$\alpha_{j+2,k-1}$};
		\node [style=none] (70) at (-11.75, -2) {$a_{j,k}$};
		\node [style=none] (71) at (-9, 1) {$a_{j+1,k}$};
		\node [style=none] (72) at (-6, 4) {$a_{j+2,k}$};
		\node [style=none] (73) at (-3, 7) {$a_{j+3,k}$};
		\node [style=none] (74) at (2, 2) {};
		\node [style=none] (75) at (6, 2) {};
		\node [style=none] (76) at (7, -2) {};
		\node [style=none] (77) at (12, -2) {};
		\node [style=none] (78) at (10, 1) {};
		\node [style=none] (79) at (15, 1) {};
		\node [style=none] (80) at (13, 4) {};
		\node [style=none] (81) at (18, 4) {};
		\node [style=none] (82) at (16, 7) {};
		\node [style=none] (83) at (21, 7) {};
		\node [style=none] (84) at (10.5, 0) {};
		\node [style=none] (85) at (10.5, -3) {};
		\node [style=none] (86) at (13.5, 3) {};
		\node [style=none] (87) at (13.5, 0) {};
		\node [style=none] (88) at (16.5, 6) {};
		\node [style=none] (89) at (16.5, 3) {};
		\node [style=pointblack] (90) at (13.5, 3) {};
		\node [style=pointblack] (91) at (16.5, 6) {};
		\node [style=pointblack] (92) at (16.5, 3) {};
		\node [style=pointblack] (93) at (13.5, 0) {};
		\node [style=pointblack] (94) at (10.5, 0) {};
		\node [style=pointblack] (95) at (10.5, -3) {};
		\node [style=pointblack] (96) at (12, -2) {};
		\node [style=pointblack] (97) at (15, 1) {};
		\node [style=pointblack] (98) at (18, 4) {};
		\node [style=pointblack] (99) at (21, 7) {};
		\node [style=pointblack] (100) at (16, 7) {};
		\node [style=pointblack] (101) at (13, 4) {};
		\node [style=pointblack] (102) at (10, 1) {};
		\node [style=pointblack] (103) at (7, -2) {};
		\node [style=none] (104) at (6.25, -2) {$a_{j,k}$};
		\node [style=none] (105) at (9, 1) {$a_{j+1,k}$};
		\node [style=none] (106) at (12, 4) {$a_{j+2,k}$};
		\node [style=none] (107) at (15, 7) {$a_{j+3,k}$};
		\node [style=none] (108) at (12.75, -2) {$b_{j,k}$};
		\node [style=none] (109) at (16, 1) {$b_{j+1,k}$};
		\node [style=none] (110) at (19, 4) {$b_{j+2,k}$};
		\node [style=none] (111) at (22, 7) {$b_{j+3,k}$};
		\node [style=none] (112) at (9.75, 0) {$\alpha_{j,k}$};
		\node [style=none] (113) at (12.5, 3) {$\alpha_{j+1,k}$};
		\node [style=none] (114) at (15.25, 6) {$\alpha_{j+2,k}$};
		\node [style=none] (115) at (11.75, -3) {$\alpha_{j,k-1}$};
		\node [style=none] (116) at (15, 0) {$\alpha_{j+1,k-1}$};
		\node [style=none] (117) at (18, 3) {$\alpha_{j+2,k-1}$};
		\node [style=none] (118) at (1, 8) {};
		\node [style=none] (119) at (2.5, 9.75) {};
		\node [style=none] (120) at (-9.5, -3) {};
		\node [style=none] (121) at (-10.75, -4.5) {};
		\node [style=none] (122) at (9, -3) {};
		\node [style=none] (123) at (7.75, -4.5) {};
		\node [style=none] (124) at (20, 8) {};
		\node [style=none] (125) at (21.5, 9.75) {};
		\node [style=none] (126) at (3, 3) {};
		\node [style=none] (127) at (5, 1) {};
		\node [style=none] (128) at (3, 1) {};
		\node [style=none] (129) at (5, 3) {};
		\node [style=none] (130) at (6.5, 6.5) {};
		\node [style=none] (131) at (8, 8) {};
		\node [style=none] (132) at (8.5, 6.5) {};
		\node [style=none] (133) at (6.5, 8.5) {};
		\node [style=none] (134) at (8.75, 6.5) {$x$};
		\node [style=none] (135) at (8.5, 8) {$y$};
		\node [style=none] (136) at (6.5, 9) {$z$};
	\end{pgfonlayer}
	\begin{pgfonlayer}{edgelayer}
		\draw [style=greenline] (89.center) to (81.center);
		\draw [style=greenline] (89.center) to (83.center);
		\draw [style=greenline] (87.center) to (79.center);
		\draw [style=greenline] (87.center) to (81.center);
		\draw [style=greenline] (85.center) to (79.center);
		\draw [style=grayfillline] (79.center)
			 to (77.center)
			 to (76.center)
			 to (78.center);
		\draw [style=grayfillline] (40.center) to (44.center);
		\draw [style=grayfillline] (44.center) to (45.center);
		\draw [style=grayfillline] (41.center) to (45.center);
		\draw [style=grayfillline] (40.center) to (41.center);
		\draw [style=grayfillline] (41.center) to (42.center);
		\draw [style=grayfillline] (42.center) to (46.center);
		\draw [style=grayfillline] (45.center) to (46.center);
		\draw [style=grayfillline] (46.center) to (47.center);
		\draw [style=grayfillline] (42.center) to (43.center);
		\draw [style=grayfillline] (43.center) to (47.center);
		\draw [style=greenline] (42.center) to (52.center);
		\draw [style=greenline] (52.center) to (43.center);
		\draw [style=greenline] (42.center) to (53.center);
		\draw [style=greenline] (53.center) to (43.center);
		\draw [style=greenline] (50.center) to (41.center);
		\draw [style=greenline] (42.center) to (50.center);
		\draw [style=greenline] (41.center) to (48.center);
		\draw [style=greenline] (40.center) to (48.center);
		\draw [style=greenline] (51.center) to (41.center);
		\draw [style=greenline] (51.center) to (42.center);
		\draw [style=greenline] (49.center) to (40.center);
		\draw [style=greenline] (49.center) to (41.center);
		\draw [style=reddashedline] (52.center) to (53.center);
		\draw [style=reddashedline] (50.center) to (51.center);
		\draw [style=reddashedline] (48.center) to (49.center);
		\draw [style=arrowline] (75.center) to (74.center);
		\draw [style=grayfillline] (80.center)
			 to (78.center)
			 to (79.center)
			 to (81.center);
		\draw [style=grayfillline] (82.center)
			 to (80.center)
			 to (81.center)
			 to (83.center);
		\draw [style=grayfillline] (82.center) to (83.center);
		\draw [style=greenline] (88.center) to (81.center);
		\draw [style=greenline] (88.center) to (83.center);
		\draw [style=greenline] (86.center) to (79.center);
		\draw [style=greenline] (86.center) to (81.center);
		\draw [style=greenline] (84.center) to (77.center);
		\draw [style=greenline] (84.center) to (79.center);
		\draw [style=greenline] (85.center) to (77.center);
		\draw [thick,style=dotted] (118.center) to (119.center);
		\draw [thick,style=dotted] (124.center) to (125.center);
		\draw [thick, style=dotted] (122.center) to (123.center);
		\draw [thick, style=dotted] (120.center) to (121.center);
		\draw [style=redline] (126.center) to (127.center);
		\draw [style=redline] (128.center) to (129.center);
		\draw [style=arrowline] (132.center) to (130.center);
		\draw [style=arrowline] (131.center) to (130.center);
		\draw [style=arrowline] (133.center) to (130.center);
	\end{pgfonlayer}
\end{tikzpicture}
    \caption{Figure illustrate a configuration of duality defect on the L.H.S. and R.H.S. that are not related by any movement relations. we asume periodic boundary condition along $y$ direction.}
    \label{fig:3DAPI_duality_defect_movement_notallowed}
\end{figure}

In the previous section, we exhibited movement relations for anisotropic plaquette Ising model. Here, we propose a movement that is not allowed. Consider the configuration of defect in Figure~\ref{fig:3DAPI_duality_defect_movement_notallowed}. We assume periodic boundary condition along $y$ direction. The L.H.S. and R.H.S. of the figure depict two duality defect configurations. We show that they are not related by a movement relation.
Going from L.H.S. to R.H.S., we add additional vertices labeled $b_{j,k}$ and convert the vertical edge interaction on the dual lattice to plaquette interaction on the original lattice. The corresponding equation would read as follows
\begin{align}
&C\prod_{j=1}^{L_y}\frac{e^{J^*s_{\alpha_{j,k}}s_{\alpha_{j,k-1}}}}{(\sinh(2J^*))^{\frac{1}{2}}}\prod_{j=1}^{L_y}(-1)^{(h_{\alpha_{j,k}}+h_{\alpha_{j,k-1}})(h_{a_{j,k}}+h_{a_{j+1,k}})}\nonumber\\
&\qquad\overset{?}{=}\sum\limits_{\{h_{b_{j,k}}\}}\prod_{j=1}^{L_y}e^{Js_{a_{j,k}}s_{a_{j+1,k}}s_{b_{j,k}}s_{b_{j+1,k}}}\prod_{j=1}^{L_y}(-1)^{(h_{\alpha_{j,k}}+h_{\alpha_{j,k-1}})(h_{b_{j,k}}+h_{b_{j+1,k}})}\,,
\end{align}
where $C$ is a constant independent of any coupling.
Now we consider the following two cases:
\begin{itemize}
    \item $h_{\alpha_{j,k}}+h_{\alpha_{j,k-1}}=0$ mod $2$ $\forall\, j\in\mathbb{Z}_{L_y}$. This would give the relation
    \begin{align}
        C\frac{e^{L_y J^*}}{(\sinh(2J^*))^{L_y/2}}=2^{L_y}(\cosh(J)^{L_y}+\sinh(J)^{L_y})\,.
        \label{eq:APInotallowed1}
    \end{align}
    Note that the term $\sinh(J)^{L_y}$ appear due to periodic boundary condition in $y$ direction.
    \item $h_{\alpha_{j,k}}+h_{\alpha_{j,k-1}}=1$ mod $2$ $\forall\, j\in\mathbb{Z}_{L_y}$. This would give the relation
    \begin{align}
        C\frac{e^{-L_y J^*}}{(\sinh(2J^*))^{L_y/2}}=2^{L_y}(\cosh(J)^{L_y}+\sinh(J)^{L_y})\,.
        \label{eq:APInotallowed2}
    \end{align}
\end{itemize}
Dividing \eqref{eq:APInotallowed1} by \eqref{eq:APInotallowed2} results in $e^{2L_yJ^*}=1$ and $J^*=0$. However, we assume a nonzero coupling and the above relation contradicts that. Hence, we can not move the defect in a way given in \ref{fig:3DAPI_duality_defect_movement_notallowed}.
\subsubsection{Fusion of two duality defects}
\begin{figure}
    \centering
    \begin{tikzpicture}
	\begin{pgfonlayer}{nodelayer}
		\node [style=pointblack] (53) at (-0.75, 3) {};
		\node [style=pointblack] (54) at (-3, 3) {};
		\node [style=none] (2) at (-7.75, 1) {};
		\node [style=none] (3) at (-4, 1) {};
		\node [style=none] (10) at (0.25, 1) {};
		\node [style=none] (11) at (3.75, 1) {};
		\node [style=none] (12) at (1.75, 3) {};
		\node [style=none] (13) at (-5.25, 3) {};
		\node [style=none] (14) at (-3, 3) {};
		\node [style=none] (15) at (-0.75, 3) {};
		\node [style=none] (16) at (-5.25, 3) {};
		\node [style=none] (17) at (-2.25, 3) {};
		\node [style=none] (18) at (1, 3) {};
		\node [style=none] (21) at (-6.25, 2.25) {};
		\node [style=none] (22) at (-2.25, 2.25) {};
		\node [style=none] (23) at (1.5, 2.25) {};
		\node [style=none] (25) at (-4.25, 3.75) {};
		\node [style=none] (26) at (-2.25, 3.75) {};
		\node [style=none] (27) at (0.25, 3.75) {};
		\node [style=none] (0) at (-7, 4) {};
		\node [style=none] (4) at (-5, 5.25) {};
		\node [style=none] (1) at (-3.75, 4) {};
		\node [style=none] (5) at (-2.75, 5.25) {};
		\node [style=none] (7) at (3.5, 4) {};
		\node [style=none] (9) at (2, 5.25) {};
		\node [style=none] (6) at (0.5, 4) {};
		\node [style=none] (8) at (-0.25, 5.25) {};
		\node [style=none] (28) at (10, 4) {};
		\node [style=none] (29) at (8, 2) {};
		\node [style=none] (30) at (12, 4) {};
		\node [style=none] (31) at (14, 2) {};
		\node [style=none] (32) at (11, 4.5) {};
		\node [style=none] (33) at (11, 0.75) {};
		\node [style=pointblack] (34) at (-5, 5.25) {};
		\node [style=pointblack] (35) at (-2.75, 5.25) {};
		\node [style=pointblack] (36) at (-0.25, 5.25) {};
		\node [style=pointblack] (37) at (2, 5.25) {};
		\node [style=pointblack] (38) at (3.5, 4) {};
		\node [style=pointblack] (39) at (-7, 4) {};
		\node [style=pointblack] (40) at (-3.75, 4) {};
		\node [style=pointblack] (41) at (0.5, 4) {};
		\node [style=pointblack] (42) at (0.25, 1) {};
		\node [style=pointblack] (43) at (3.75, 1) {};
		\node [style=pointblack] (44) at (1.75, 3) {};
		\node [style=pointblack] (45) at (-7.75, 1) {};
		\node [style=pointblack] (46) at (-6.25, 2.25) {};
		\node [style=pointblack] (47) at (-5.25, 3) {};
		\node [style=pointblack] (48) at (-4.25, 3.75) {};
		\node [style=pointblack] (49) at (-2.25, 3.75) {};
		\node [style=pointblack] (50) at (0.25, 3.75) {};
		\node [style=pointblack] (51) at (1.5, 2.25) {};
		\node [style=pointblack] (52) at (-2.25, 2.25) {};
		\node [style=pointblack] (55) at (-2.25, 3) {};
		\node [style=pointblack] (56) at (1, 3) {};
		\node [style=pointblack] (57) at (10, 4) {};
		\node [style=pointblack] (58) at (11, 4.5) {};
		\node [style=pointblack] (59) at (12, 4) {};
		\node [style=pointblack] (60) at (14, 2) {};
		\node [style=pointblack] (61) at (11, 0.75) {};
		\node [style=pointblack] (62) at (8, 2) {};
		\node [style=none] (63) at (7.5, 1.75) {$\alpha$};
		\node [style=none] (64) at (14.5, 1.75) {$\beta$};
		\node [style=none] (65) at (12.5, 4.25) {$\gamma$};
		\node [style=none] (66) at (9.5, 4.25) {$\delta$};
		\node [style=none] (67) at (11, 5) {$a$};
		\node [style=none] (68) at (11, 0.25) {$b$};
		\node [style=none] (69) at (4.5, -1) {};
		\node [style=none] (70) at (4.5, -3) {};
		\node [style=none] (71) at (6.5, -3) {};
		\node [style=none] (72) at (5.75, -1.5) {};
		\node [style=none] (73) at (7, -3) {$x$};
		\node [style=none] (74) at (6, -1.25) {$y$};
		\node [style=none] (75) at (4.5, -0.5) {$z$};
		\node [style=none] (76) at (-7, 6) {$(a)$};
		\node [style=none] (77) at (7, 6) {$(b)$};
	\end{pgfonlayer}
	\begin{pgfonlayer}{edgelayer}
		\draw [style=greenline] (33.center) to (30.center);
		\draw [style=greenline] (28.center) to (33.center);
		\draw [style=greenline] (16.center) to (14.center);
		\draw [style=greenline] (18.center) to (15.center);
		\draw [style=greenline] (17.center) to (15.center);
		\draw [style=greenline] (14.center) to (17.center);
		\draw [style=greenline] (16.center) to (14.center);
		\draw [style=greenline] (16.center) to (13.center);
		\draw [style=grayfillline] (9.center) to (7.center);
		\draw [style=grayfillline] (11.center) to (12.center);
		\draw [style=grayfillline] (11.center)
			 to (10.center)
			 to (15.center)
			 to (12.center);
		\draw [style=grayfillline] (10.center)
			 to (3.center)
			 to (14.center)
			 to (15.center);
		\draw [style=grayfillline] (3.center)
			 to (2.center)
			 to (13.center)
			 to (14.center);
		\draw [style=redfillline] (22.center)
			 to (21.center)
			 to (16.center)
			 to (17.center);
		\draw [style=redfillline] (18.center)
			 to (17.center)
			 to (22.center)
			 to (23.center);
		\draw [style=redfillline] (18.center) to (23.center);
		\draw [style=redfillline] (25.center) to (16.center);
		\draw [style=redfillline] (16.center)
			 to (17.center)
			 to (26.center)
			 to (25.center);
		\draw [style=redfillline] (26.center)
			 to (27.center)
			 to (18.center)
			 to (17.center);
		\draw [style=greenline] (16.center) to (1.center);
		\draw [style=greenline] (16.center) to (0.center);
		\draw [style=greenline] (16.center) to (4.center);
		\draw [style=greenline] (16.center) to (5.center);
		\draw [style=greenline] (17.center) to (6.center);
		\draw [style=greenline] (17.center) to (1.center);
		\draw [style=greenline] (17.center) to (5.center);
		\draw [style=greenline] (17.center) to (8.center);
		\draw [style=greenline] (18.center) to (6.center);
		\draw [style=greenline] (18.center) to (7.center);
		\draw [style=greenline] (18.center) to (9.center);
		\draw [style=greenline] (18.center) to (8.center);
		\draw [style=grayfillline] (1.center)
			 to (0.center)
			 to (4.center)
			 to (5.center);
		\draw [style=grayfillline] (6.center)
			 to (1.center)
			 to (5.center)
			 to (8.center);
		\draw [style=grayfillline] (9.center)
			 to (8.center)
			 to (6.center)
			 to (7.center);
		\draw [style=greenline] (18.center) to (11.center);
		\draw [style=greenline] (18.center) to (12.center);
		\draw [style=greenline] (16.center) to (2.center);
		\draw [style=greenline] (13.center) to (16.center);
		\draw [style=greenline] (17.center) to (10.center);
		\draw [style=greenline] (18.center) to (10.center);
		\draw [style=greenline] (17.center) to (3.center);
		\draw [style=greenline] (16.center) to (3.center);
		\draw [style=redfillline] (30.center)
			 to (31.center)
			 to (29.center)
			 to (28.center);
		\draw [style=redfillline] (28.center) to (30.center);
		\draw [style=greenline] (29.center) to (32.center);
		\draw [style=greenline] (28.center) to (32.center);
		\draw [style=greenline] (32.center) to (30.center);
		\draw [style=greenline] (32.center) to (31.center);
		\draw [style=greenline] (33.center) to (31.center);
		\draw [style=greenline] (29.center) to (33.center);
		\draw [style=arrowline] (71.center) to (70.center);
		\draw [style=arrowline] (72.center) to (70.center);
		\draw [style=arrowline] (69.center) to (70.center);
	\end{pgfonlayer}
\end{tikzpicture}
    \caption{The figure illustrate two duality defects in the anisotropic plaquette Ising model placed close to each other. $(a)$ is the configuration of two duality defects. The red plaquette indicate a dual Ising plaquette interaction. Everywhere else, we have anisotropic plaquette Ising model on the original lattice. $(b)$ is a figure where we have a single plaquette interaction connecting the cube centers.}
    \label{fig:3DAPIdefect_configuration}
\end{figure}
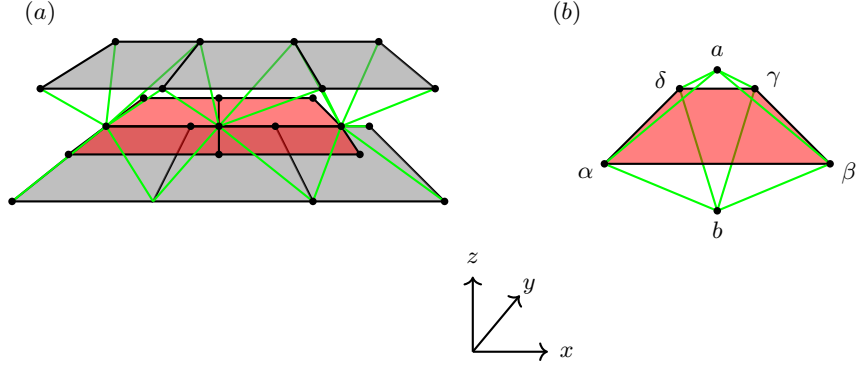

\begin{figure}
    \centering
    \input{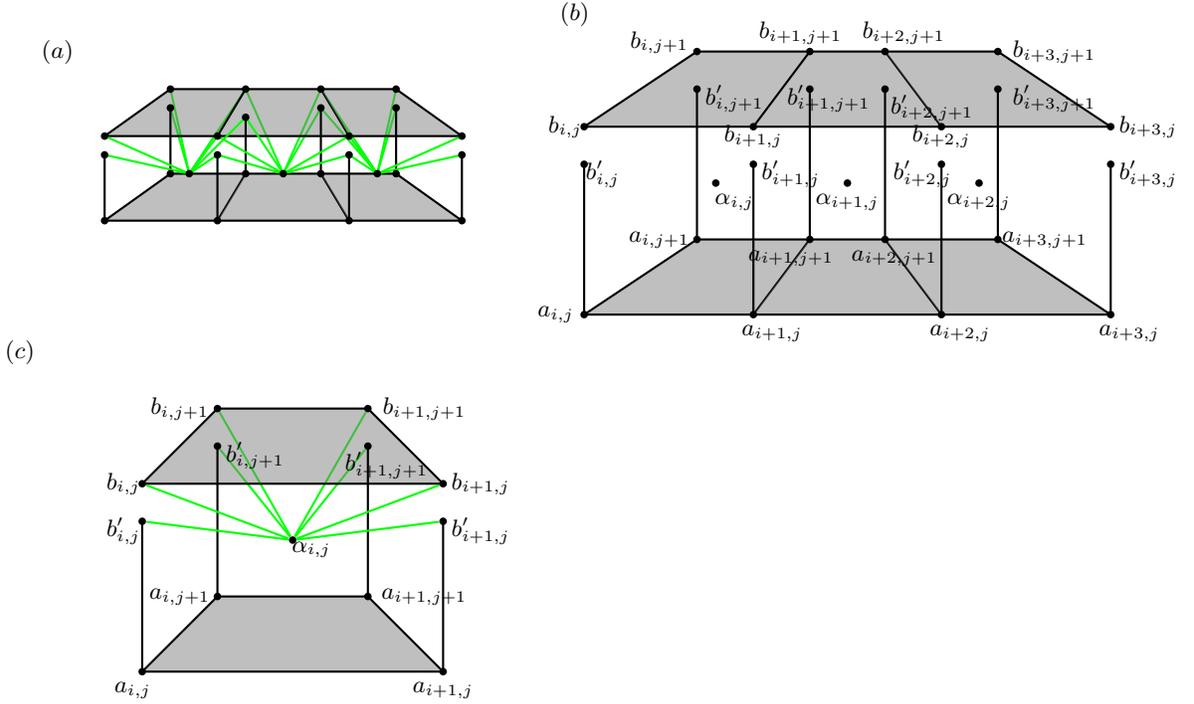}
\caption{The figure illustrate the configuration of spins after applying the movement relations to the configuration in \ref{fig:3DAPIdefect_configuration}. $(a)$ is the full configuration while $(c)$ is for a single cube. $(b)$ define the labels of vertices and cube centres.}
    \label{fig:3DAPIdefect_fusion}
\end{figure}

We now examine the scenario in which two duality defects are positioned in close proximity. A representative configuration of this setup is shown in Figure~\ref{fig:3DAPIdefect_configuration}. Note that the two defects should lie on $xy$ plane\footnote{If the two parallel defects lie in $yz$ or $xz$ plane, we cannot move them close to each other to fuse them. Certain movements of duality defects are not allowed as we show in Figure~\ref{fig:3DAPI_duality_defect_movement_notallowed}.}. 
This configuration corresponds to the situation just before the two duality defects fuse. By applying the movement relation~\eqref{eq:3DAPImovement3}, we can move the defect and fuse them, yielding the configuration depicted in Figure~\ref{fig:3DAPIdefect_fusion}. 
After moving the defect to the top, there are no plaquette interactions between vertices $\{\alpha_{ij}\}$. Instead the new interaction is a vertical edge interaction. However, there are d.o.f $\{h_{\alpha_{ij}}\}$ on the vertices of the dual lattice that are summed over
\begin{widetext}
    \begin{align*}
    \sum_{\{h_{\alpha_{i,j}}\}}(-1)^{\sum_{h_{\alpha_{i,j}}}h_{\alpha_{i,j}}(h_{b_{i,j}}+h_{b_{i+1,j}}+h_{b_{i,j+1}}+h_{b_{i+1,j+1}}+h_{b'_{i,j}}+h_{b'_{i+1,j}}+h_{b'_{i,j+1}}+h_{b'_{i+1,j+1}})}\,.
\end{align*}
\end{widetext}
This sum is non-zero only when $h_{b_{i,j}}+h_{b_{i+1,j}}+h_{b_{i,j+1}}+h_{b_{i+1,j+1}}+h_{b'_{i,j}}+h_{b'_{i+1,j}}+h_{b'_{i,j+1}}+h_{b'_{i+1,j+1}}$ is zero mod 2. This can be satisfied in multiple possible ways. Let us denote $\mathcal{X}_{i,j}$ as the relative spin-flip operation: $\mathcal{X}_{i,j}h_{b_{i,j}}=h_{b_{i,j}}+1$ and do not flip any other spin. Let us assume that the fusion is happening in the $xy$ plane and at $z=k_0$. Then we define the operation $\mathcal{\eta}_j^x=\prod_{i}\mathcal{X}_{ij}$ and $\mathcal{\eta}_i^y=\prod_{j}\mathcal{X}_{ij}$ as the relative spin flip operation along a line in the $x$ direction and in the $y$ direction, respectively. Let $\mathfrak{l}_x\subset\{1,...,L_x\}$ and $\mathfrak{l}_{y}\subset\{1,...,L_y\}$. Then the possible solutions are $h'_{b_{i,j}}=\prod\limits_{m\in\mathfrak{l}_x}\eta^y_m\prod\limits_{n\in\mathfrak{l}_y}\eta^x_{n}h_{b_{i,j}}$ and the solutions are indexed by the sets $\mathfrak{l}_x$ and $\mathfrak{l}_y$. 

Now suppose that we denote the duality defect by $\mathcal{D}$ and the spin flip defect along a line in the two directions by $D_{\mathcal{\eta}_j^x}$ and $D_{\mathcal{\eta}_i^y}$\footnote{The partition function with spin flip defects are obtained by changing the couplings $J$ to $-J$ and $K$ to $-K$ in the plaquette terms or edge terms with centers at $z\geq k_0$ in the anisotropic plaquette Ising model that share the vertex along which the line flip operations $\eta^x_j$ or $\eta^y_i$ are defined.}, then, we have
\begin{align}
    \mathcal{D}\times\mathcal{D}=\prod_{j}(1+D_{\mathcal{\eta}_j^x})\prod_i(1+D_{\mathcal{\eta}_i^y})\, .
    \label{eq:APID^2}
\end{align}
Equation~\eqref{eq:APID^2} should be interpreted as a statement about defect insertions in the partition function: the products correspond to summing over partition functions with the appropriate combinations of defects inserted.
\subsection{Quantum Hamiltonian from partition function}
\subsubsection{Quantum Hamiltonian from partition function with duality defect}
We now consider the partition function of the three-dimensional anisotropic plaquette Ising model with a single duality defect inserted. The vertices of the cubic lattice are labeled by $\mathrm{v}=(x,y,z)$, and the cube centers by $\mathrm{c}=(x+\tfrac{1}{2},y+\tfrac{1}{2},z+\tfrac{1}{2})$. A duality defect is placed along the $yz$ plane at $x=i_0$.

To analyze this system, it is convenient to rewrite the Boltzmann weights as a product over $xy$-plane layers, thereby expressing the partition function in the form
\begin{align}
Z_{\text{API}}^{\mathcal{D}}(J,K) = \mathrm{Tr}\left(\hat{\rm T}^{L_z}\right),
\end{align}
where $\hat{\rm T}$ denotes the transfer matrix acting between adjacent layers in the $z$ direction.

In the $\tau$-continuum limit (see \cite{Kogut:1979wt}), the transfer matrix takes the standard form
\begin{align}
\hat{\rm T} = e^{-\tau \hat{\rm H}_{\text{PI},\mathcal{D}}} \approx 1 - \tau \hat{\rm H}_{\text{PI},\mathcal{D}} \,.
\end{align}
We then separate the matrix elements according to the number of spin flips between successive layers:
\begin{align}
&\hat{\rm T}\rvert_{0\text{-flip}} \approx 1 - \tau \hat{\rm H}_{\text{PI},\mathcal{D}}\rvert_{0\text{-flip}},\quad
\hat{\rm T}\rvert_{1\text{-flip}} \approx -\tau \hat{\rm H}_{\text{PI},\mathcal{D}}\rvert_{1\text{-flip}},\quad
\hat{\rm T}\rvert_{n\text{-flip}} \approx -\tau \hat{\rm H}_{\text{PI},\mathcal{D}}\rvert_{n\text{-flip}}, .
\end{align}
Following the same reasoning as in~\eqref{eq:transfermatrix}, we identify the relevant scaling relations
\begin{align}
e^{-2K}\sim \tau\,, \quad J\sim \tau\,,\quad K^*\sim\tau\,,\quad e^{-2J^*}\sim\tau\,,
\end{align}
which imply that only the $0$- and $1$-flip matrix elements contribute in the $\tau$-continuum limit. Adopting the scaling $J = \lambda e^{-2K}$ and choosing $\tau = e^{-2K}$ to fix the overall constant, together with
\begin{align}
K^*=-\frac{1}{2}\log\tanh K, \quad J^*=-\frac{1}{2}\log\tanh J, \quad \tanh J \approx J,
\end{align}
we obtain
\begin{align}
e^{-2K} = \tau,\quad J = \lambda\tau,\quad e^{-2J^*} = \lambda\tau,\quad K^* = \tau\,.
\end{align}

To facilitate the construction of the Hamiltonian, we identify the cube centers in each $xy$ plane, labeled by $(x+\tfrac{1}{2},y+\tfrac{1}{2},z+\tfrac{1}{2})$, as effective vertices $(x+\tfrac{1}{2},y+\tfrac{1}{2},z)$. Under this identification, the resulting $(2+1)$D quantum Hamiltonian takes the form
\begin{widetext}
\begin{align}
\hat{\rm H}_{\text{PI},\mathcal{D}}&=-\lambda\sum_{i<i_0,j}Z_{i,j}Z_{i+1,j}Z_{i,j+1}Z_{i+1,j+1}
-\sum_{i\geq i_0,j}Z_{i+\frac{1}{2},j+\frac{1}{2}}Z_{i+\frac{3}{2},j+\frac{1}{2}}Z_{i+\frac{1}{2},j+\frac{3}{2}}Z_{i+\frac{3}{2},j+\frac{3}{2}}\nonumber\\
&\qquad-\sum_{i<i_0,j}X_{i,j}
-\lambda\sum_{i\geq i_0,j}X_{i+\frac{1}{2},j+\frac{1}{2}}
-\sum_{j}X_{i_0,j}Z_{i_0+\frac{1}{2},j+\frac{1}{2}}Z_{i_0+\frac{1}{2},j-\frac{1}{2}}\,,
\end{align}
\end{widetext}
where $X_{i,j}$ and $Z_{i,j}$ are Pauli operators acting on the effective sites $(i,j)$.
\section{Kramers-Wannier defect between 3D Ising model and 3D Ising gauge theory}\label{sec:Duslitydefect3DIG3DImodel}
In this section, we examine another example in three dimensions. It is well known that the three-dimensional Ising model and the three-dimensional Ising gauge theory\footnote{The 3D Ising gauge theory is also known as the 3D $\mathbb{Z}_2$ gauge theory.} are related by a Kramers–Wannier duality \cite{carroll1976lattice,Wegner}. As in the two-dimensional case, this correspondence represents a strong-coupling/weak-coupling duality. Our goal here is to provide an explicit construction of the Kramers–Wannier defect in this setting.

We consider a three-dimensional cubic lattice, where each cube, plaquette, edge, and vertex is denoted respectively by $\mathrm{c}$, $\mathrm{p}$, $\mathrm{e}$, and $\mathrm{v}$. The sets of all cubes, plaquettes, edges, and vertices are denoted by $\Delta_{\mathrm{c}}$, $\Delta_{\mathrm{p}}$, $\Delta_{\mathrm{e}}$, and $\Delta_{\mathrm{v}}$, respectively. We use $\partial$ to denote the standard boundary operator acting on these cells; that is, $\partial$ maps cubes to their bounding plaquettes, plaquettes to their bounding edges, and edges to their endpoint vertices.
\subsection{3D Ising model, 3D Ising gauge theory and Kramers-Wannier duality}
\subsubsection{3D Ising model and 3D Ising gauge theory}
The Hamiltonian of the classical three-dimensional Ising model consists of pairwise interactions along the links of the lattice, represented by products of spins residing on the vertices at the endpoints of each link:
\begin{align}
    \mathcal{H}_{\text{3D-I}}(\{s_{\rm v}\})=-J\sum_{\rm e\in\Delta_{\rm e}}\prod_{\rm v\in\partial e}s_{\rm v}\, .
\end{align}
The corresponding partition function is then given by
\begin{align}
    Z_{\text{3D-I}}(J)=\sum_{\{s_{\rm v}\}}e^{-\mathcal{H}_{\text{3D-I}}(\{s_{\rm v}\})}\, .
\end{align}
In contrast, the Hamiltonian of the three-dimensional Ising gauge theory involves plaquette interactions, where the energy is determined by the product of edge spins around the boundary of each plaquette:
\begin{align}
    \mathcal{H}_{\text{3D-IG}}(\{s_{\rm e}\})=-J\sum_{\rm p\in\Delta_{\rm p}}\prod_{\rm e\in\partial p}s_{\rm e}\, .
\end{align}
Accordingly, the partition function of the 3D Ising gauge theory reads
\begin{align}\label{eq:3DIGpartitionfn}
     Z_{\text{3D-IG}}(J)=\sum_{\{s_{\rm e}\}}e^{-\mathcal{H}_{\text{3D-IG}}(\{s_{\rm e}\})}\, .
\end{align}

\subsubsection{$\mathbb{Z}_2$ Symmetries}
\begin{figure}
    \centering \begin{tikzpicture}
	\begin{pgfonlayer}{nodelayer}
            \node [style=none] (24) at (-5.5, 0) {};
            \node [style=none] (25) at (-4, 0) {};
		\node [style=none] (26) at (-4, 0) {};
		\node [style=none] (27) at (-3.5, 0) {};
		\node [style=none] (28) at (-3.5, 2) {}; 
		\node [style=none] (29) at (-3.5, -2) {};
		\node [style=none] (30) at (-2.275, 1.25) {};
		\node [style=none] (31) at (-4.925, -1.5) {};
		\node [style=none] (32) at (-1.5, 0) {};
		\node [style=none] (38) at (-3.725, 4.075) {};
		\node [style=none] (39) at (-6.725, 1.075) {};
		\node [style=none] (40) at (-6.725, -3.925) {};
		\node [style=none] (42) at (-3.725, -0.775) {};
		\node [style=none] (43) at (2.725, 0) {};
		\node [style=none] (44) at (4.725, 0) {};
            \node [style=none] (44') at (5.0, -0.3) {$v$};
            \node [style=none] (44'') at (-3.3, -0.3) {$v$};
		\node [style=none] (45) at (4.725, 2) {};
		\node [style=none] (46) at (4.725, -2) {};
		\node [style=none] (47) at (5.925, 1.25) {};
		\node [style=none] (48) at (3.3, -1.5) {};
		\node [style=none] (49) at (6.725, 0) {};
		\node [style=none] (50) at (2.3, 0.325) {};
		\node [style=none] (51) at (4.05, 2.075) {};
		\node [style=none] (52) at (2.3, -1.85) {};
		\node [style=none] (53) at (4.05, -0.175) {};
		\node [style=none] (54) at (4.5, 4.075) {};
		\node [style=none] (55) at (1.5, 1.075) {};
		\node [style=none] (56) at (1.5, -3.925) {};
		\node [style=none] (57) at (4.5, -0.775) {};
		\node [style=none] (58) at (4.225, 0.325) {};
		\node [style=none] (59) at (5.975, 2.075) {};
		\node [style=none] (60) at (4.225, -1.85) {};
		\node [style=none] (61) at (5.975, -0.175) {};
		\node [style=none] (62) at (0, 0) {$=$};

            \node [style=pointoperator, color=black] (63) at (-3.5, 2) {};
		\node [style=pointoperator, color=black] (64) at (-2.25, 1.25) {};
		\node [style=pointoperator, color=black] (65) at (-1.5, 0) {};
		\node [style=pointoperator, color=black] (66) at (-3.5, -2) {};
		\node [style=pointoperator, color=black] (67) at (-4.925, -1.5) {};
		\node [style=pointoperator, color=black] (69) at (-3.5, 0) {};
            \node [style=pointoperator, color=black] (68) at (-5.5, 0) {};

		\node [style=pointoperator, color=black] (70) at (4.725, 0) {};
		\node [style=pointoperator, color=black] (71) at (4.725, 2) {};
		\node [style=pointoperator, color=black] (72) at (2.725, 0) {};
		\node [style=pointoperator, color=black] (73) at (6.725, 0) {};
		\node [style=pointoperator, color=black] (74) at (4.725, -2) {};
		\node [style=pointoperator, color=black] (75) at (3.3, -1.5) {};
		\node [style=pointoperator, color=black] (76) at (5.925, 1.25) {};

            \node [style=none] (77) at (4.725, 1.25) {};
		\node [style=none] (78) at (4.725, -1.3) {};
		\node [style=none] (79) at (3.675, -1.1) {};
		\node [style=none] (80) at (5.5, 0) {};
	\end{pgfonlayer}
	\begin{pgfonlayer}{edgelayer}

            \draw (24.center) to (25.center);
            \draw[fill=blue!10, opacity=0.70] (39.center) -- (38.center) -- (42.center) -- (40.center) -- cycle;

		\draw (28.center) to (27.center);
		\draw (27.center) to (29.center);
		\draw (27.center) to (30.center);
		\draw (27.center) to (31.center);
		\draw (27.center) to (32.center);
		\draw (26.center) to (27.center);
		\draw (40.center) to (39.center);
		\draw (40.center) to (42.center);
		\draw (38.center) to (42.center);
		\draw (39.center) to (38.center);
        
            \draw (44.center) to (77.center);
            \draw (44.center) to (78.center);
            \draw (44.center) to (79.center);
            \draw (44.center) to (80.center);
		\draw (44.center) to (47.center);
		\draw (43.center) to (44.center);
        
            \draw[fill=white, opacity=0.70] (50.center) -- (51.center) -- (53.center) -- (52.center) -- cycle;
            
            \draw[fill=blue!10, opacity=0.70, even odd rule] (55.center) -- (56.center) -- (57.center) -- (54.center) -- cycle
            (50.center) -- (51.center) -- (53.center) -- (52.center) -- cycle;
            
            \draw[color =blue!10, fill=blue!10, opacity=0.70] (50.center) -- (51.center) -- (59.center) -- (58.center) -- cycle;
            
            \draw[color =blue!10, fill=blue!10, opacity=0.70] (59.center) -- (61.center) -- (60.center) -- (58.center) -- cycle;

            \draw[color =blue!10, fill=blue!10, opacity=0.70] (50.center) -- (58.center) -- (60.center) -- (52.center) -- cycle;

		\draw (56.center) to (55.center);
		\draw (56.center) to (57.center);
		\draw [in=90, out=-90] (54.center) to (57.center);
		\draw (55.center) to (54.center);
		\draw (58.center) to (59.center);
		\draw (59.center) to (61.center);
		\draw (61.center) to (60.center);
		\draw (60.center) to (58.center);
		\draw (51.center) to (59.center);
		\draw (50.center) to (58.center);
		\draw (52.center) to (60.center);
		\draw (53.center) to (61.center);
        
            \draw (77.center) to (45.center);
            \draw (78.center) to (46.center);
            \draw (79.center) to (48.center);
            \draw (80.center) to (49.center);

	\end{pgfonlayer}
\end{tikzpicture}
    \caption{Movement of the $\mathbb{Z}_2$ 0-form symmetry defect from a 2-cycle $z_{2} \in H_2(M,\mathbb{Z}_2)$ to another homologous 2-cycle $z'_{2}$ in 3D Ising Model done by transforming $s_{\rm v} \to -s_{\rm v}$ for $\rm v$ that lie in the region $S'$ such that $\partial S' = z_{2} - z'_{2}$ }
    \label{fig:3DIsingZ2TopMove}
\end{figure}
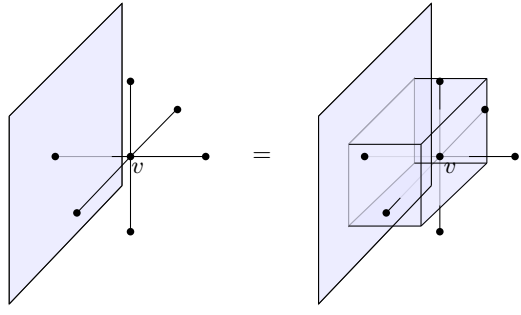

Let's discuss the invertible symmetries for 3D Ising model. If we transform all the spins $s_{\rm v} \to -s_{\rm v}$, then the Hamiltonian remains invariant. However, if we take a subset $S$ of the spins $s_{\rm v}$ and transform $s_{\rm v} \to -s_{\rm v}$ for $s_{\rm v} \in S$, then the Hamiltonian changes in the following way
\begin{align}\label{eq:3DIsingZ2Defect}
    H_{\eta(z_{2})} = -J\sum_{\rm e\in\Delta_{\rm e}}(-1)^{\# ({\rm e} \cap z_{2})} \prod_{\rm v\in\partial e}s_{\rm v}
\end{align}
where $z_{2}$ is the 2-cycle made up of plaquettes on the dual lattice bounding the vertices $S$ and $\# (z_{2} \cap {\rm e})$ denotes the intersection number of $z_{2}$ and the link ${\rm e}$. The subscript $\eta(z_{2})$ denotes the defect $\mathbb{Z}_2$ thus created. Although the defect we created is homologically trivial, we can take \eqref{eq:3DIsingZ2Defect} as the definition for the $\mathbb{Z}_2$ defect on any 2-cycle $z_{2}$ in $H_2(M,\mathbb{Z}_2)$. In the language of higher-form symmetries, it is a 0-form symmetry defect. 

This defect is topological, i.e., we can move the defect from the 2-cycle $z_{2}$ to a 2-cycle $z'_{2}$ in the same homology class without changing the partition function. It can be seen by transforming the spins $s_{\rm v} \to -s_{\rm v}$ for the vertices $\rm v$ that belong in $S'$ where $S'$ is the region bounded by $z_2 - z_2'$. This transformation doesn't change the partition function, since all spins $\{s_{\rm v}\}$ are being summed. For example, if we take a spin $s_{\rm v}$ where $\rm v$ is adjacent to $z_{2}$ (at least one link in connecting $\rm v$ intersects $z_{2}$) and transform $s_{\rm v} \to - s_{\rm v}$, then we essentially move the defect by one dual cube as shown in the left Figure~\ref{fig:3DIsingZ2TopMove}. Moreover, if there was a spin $s_{\rm v}$ inside a correlation function and if we transform $s_{\rm v} \to -s_{\rm v}$, we will move the defect and get a $-1$ factor. In other words, inside a correlation function, sweeping a $\mathbb{Z}_2$ defect across $s_{\rm v}$ makes it act on the spin and produce a factor of $-1$.

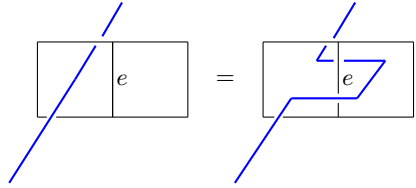
\begin{figure}
    \centering 
    \begin{tikzpicture}
	\begin{pgfonlayer}{nodelayer}
		\node [style=none] (0) at (-2, 0) {};
		\node [style=none] (1) at (0, 0) {};
		\node [style=none] (2) at (-2, -2) {};
		\node [style=none] (3) at (0, -2) {};
		\node [style=none] (4) at (0, 0) {};
		\node [style=none] (5) at (2, 0) {};
		\node [style=none] (6) at (0, -2) {};
		\node [style=none] (7) at (2, -2) {};
		\node [style=none] (8) at (-1, -1) {};
		\node [style=none] (9) at (-2.75, -3.75) {};
		\node [style=none] (10) at (-0.45, -0.1) {};
		\node [style=none] (11) at (-1, -1) {};
		\node [style=none] (12) at (0.25, 1.05) {};
		\node [style=none] (13) at (-0.275, 0.15) {};
		\node [style=none] (14) at (4, 0) {};
		\node [style=none] (15) at (6, 0) {};
		\node [style=none] (16) at (4, -2) {};
		\node [style=none] (17) at (6, -2) {};
		\node [style=none] (18) at (6, 0) {};
		\node [style=none] (19) at (8, 0) {};
		\node [style=none] (20) at (6, -2) {};
		\node [style=none] (21) at (8, -2) {};
		\node [style=none] (22) at (4.75, -1.5) {};
		\node [style=none] (23) at (3.25, -3.75) {};
		\node [style=none] (26) at (6.45, 1.05) {};
		\node [style=none] (28) at (6.5, -1.5) {};
		\node [style=none] (29) at (7.25, -0.5) {};
		\node [style=none] (30) at (5.425, -0.5) {};
		\node [style=none] (31) at (5.875, -0.5) {};
		\node [style=none] (32) at (6.15, -0.5) {};
		\node [style=none] (33) at (5.675, -0.1) {};
		\node [style=none] (34) at (5.875, 0.1) {};
		\node [style=none] (35) at (6, -1.375) {};
		\node [style=none] (36) at (6, -1.625) {};
		\node [style=none] (37) at (-1.75, -2) {};
		\node [style=none] (38) at (-1.5, -2) {};
		\node [style=none] (39) at (4, -2) {};
		\node [style=none] (40) at (6, -2) {};
		\node [style=none] (41) at (6, -2) {};
		\node [style=none] (42) at (4.3, -2) {};
		\node [style=none] (43) at (4.525, -2) {};
		\node [style=none] (44) at (3, -1) {$=$};
		\node [style=none] (46) at (0.25, -1) {$e$};
		\node [style=none] (47) at (6.25, -1) {$e$};
	\end{pgfonlayer}
	\begin{pgfonlayer}{edgelayer}
		\draw (0.center) to (2.center);
		\draw (0.center) to (1.center);
		\draw (1.center) to (3.center);
		\draw (4.center) to (6.center);
		\draw (6.center) to (7.center);
		\draw (4.center) to (5.center);
		\draw (5.center) to (7.center);
		\draw [style=blueline] (9.center) to (11.center);
		\draw [style=blueline] (11.center) to (10.center);
		\draw [style=blueline] (13.center) to (12.center);
		\draw (14.center) to (16.center);
		\draw (14.center) to (15.center);
		\draw (20.center) to (21.center);
		\draw (18.center) to (19.center);
		\draw (19.center) to (21.center);
		\draw [style=blueline] (23.center) to (22.center);
		\draw [style=blueline] (22.center) to (28.center);
		\draw [style=blueline] (28.center) to (29.center);
		\draw [style=blueline] (30.center) to (31.center);
		\draw [style=blueline] (32.center) to (29.center);
		\draw [style=blueline] (30.center) to (33.center);
		\draw [style=blueline] (34.center) to (26.center);
		\draw (18.center) to (35.center);
		\draw (36.center) to (20.center);
		\draw (2.center) to (37.center);
		\draw (38.center) to (6.center);
		\draw (39.center) to (42.center);
		\draw (43.center) to (41.center);
	\end{pgfonlayer}
\end{tikzpicture}
    \caption{Movement of the $\mathbb{Z}_2$ 1-form symmetry defect from a 1-cycle $z_1 \in H_1(M,\mathbb{Z}_2)$ to another homologous 1-cycle $z_1'$ in 3D $\mathbb{Z}_2$ gauge theory done by transforming $s_{\rm e} \to -s_{\rm e}$ for all $\rm e$ such that $\#(B \cap {\rm e})=1$ where $\partial B = z_1-z_1'$.}
    \label{fig:3DZ2Gauge1Form}
\end{figure}
Let us discuss the invertible symmetry of the 3D Ising gauge theory. Consider the $x$-$y$ plane on the dual lattice and consider the set $\Delta_{e_z}$ of all $z$-directed links. If we transform $s_{\rm e} \to -s_{\rm e}$ $\forall$ ${\rm e} \in \Delta_{e_z}$, then the Hamiltonian remains invariant. Similarly, we can do this with all the $x$ directed links in the $yz$ plane and $y$ directed links in the $xz$ plane. If we take a subset $B$ of spins all $z$ directed links and transform them then the Hamiltonian becomes
\begin{align}\label{eq:3DZ2GaugeDefect}
    H_{\eta^{(1)}(\rm z_1)} = - J \sum_{\rm p\in\Delta_{\rm p}} (-1)^{\# (\mathrm{p} \cap z_1)} \prod_{\rm e\in\partial p}s_{\rm e}
\end{align}
where $\eta^{(1)}(z_1)$ denotes the $\mathbb{Z}_2$ symmetry defect on a 1-cycle $z_1$ made up of links on the dual lattice and $\# (\rm p \cap z_1)$ denotes the intersection number between plaquette $\rm p$ and $z_1$. Even though, by the above construction, the 1-cycle $z_1$ is planar and homologically trivial, \eqref{eq:3DZ2GaugeDefect} can be taken as a definition of a $\mathbb{Z}_2$ defect in for any generic 1-cycle $z_1 \in H_1(M,\mathbb{Z}_2)$. In the language of higher form symmetry, this symmetry is a one-form symmetry.

Being a symmetry defect, it is topological in nature. That is, we can move the defect from a 1-cycle $z_1$ to another homologous 1-cycle $z_1'$, by transforming all the spins $s_{\rm e} \to -s_{\rm e}$ on links $\rm e$ that intersect the surface $B'$ (made up of dual plaquettes), i.e., $\#( {\rm e} \cap B') = 1$ such that $\partial B' = z_1 - z_1'$. See Figure~\ref{fig:3DZ2Gauge1Form}. If there was a Wilson line $\prod_{\gamma_1} s_e$ in the correlation function, where $\gamma_1$ is a 1-cycle made up of links on the original lattice and if the intersection number $\# (B\cap\gamma_1)$ is $1$, then we will get a $-1$ factor upon doing this transformation. Essentially, when moving the $\mathbb{Z}_2$ one-form symmetry across a Wilson line, it acts on it by producing a $-1$ factor. 
\subsubsection{Kramers-Wannier duality}
We can employ the Raussendorf–Briegel–Harrington (RBH) cluster state \cite{raussendorf2005long} to explicitly realize the duality between the three-dimensional Ising model and the three-dimensional Ising gauge theory. This construction is a straightforward generalization of the approach used to derive the Kramers–Wannier duality in the two-dimensional statistical Ising model. For a detailed description of the method, we refer the reader to \cite{Okuda:2024azp}. 

On a three-torus, the duality takes the form
\begin{align}
    Z_{\text{3D-IG}}(J)=\frac{2^{\frac{|\Delta_{\rm e}|}{2}}(\sinh 2J)^{|\Delta_{\rm p}|/2}2^{\frac{|\Delta_{\rm v}|}{2}}}{2^{\frac{|\Delta_{\rm c}|}{2}}\times 4}\sum_{z_2\in H_2(M_3,\mathbb{Z}_2)}Z_{\text{3D-I}}(J^*,z_2)\, ,
\end{align}
where
\begin{align}
    Z_{\text{3D-I}}(J^*,z_2)=\sum_{\{s_{\rm c}=\pm 1\}}e^{J^*\sum_{\langle\rm c,c'\rangle}(-1)^{\# (z_2\cap\langle\rm c,c'\rangle)}s_{\rm c}s_{\rm c'}}\, .
\end{align}
Here, $\langle\rm c,c'\rangle$ denotes a pair of neighboring cube centers connected by a dual edge, and the sign factor $(-1)^{\# (z_2\cap\langle\rm c,c'\rangle)}$ indicates that the Ising coupling associated with a dual edge is flipped whenever the edge intersects the 2-cycle $z_2$. This expression captures the Kramers–Wannier duality between the three-dimensional Ising gauge theory and the three-dimensional Ising model.
\subsection{Duality defect via strange correlator}
To construct the duality defect, we employ the same methodology previously applied to the two-dimensional Ising model and the three-dimensional anisotropic plaquette Ising model. Specifically, we divide the system into two distinct regions, denoted $\Delta_{\rm p}^{\rm A}$ and $\Delta_{\rm p}^{\rm A^c}$. In region $\Delta_{\rm p}^{\rm A^c}$, we consider the Ising gauge theory, while in region $\Delta_{\rm p}^{\rm A}$, we consider the Ising model obtained by gauging the Ising gauge theory. The region $\Delta_{\rm p}^{\rm A}$ is a proper subset of $\Delta_{\rm p}$, that is, $\Delta_{\rm p}^{\rm A}\subset \Delta_{\rm p}$, and its complement is $\Delta_{\rm p}^{\rm A^c}=\Delta_{\rm p}\setminus \Delta_{\rm p}^{\rm A}$.

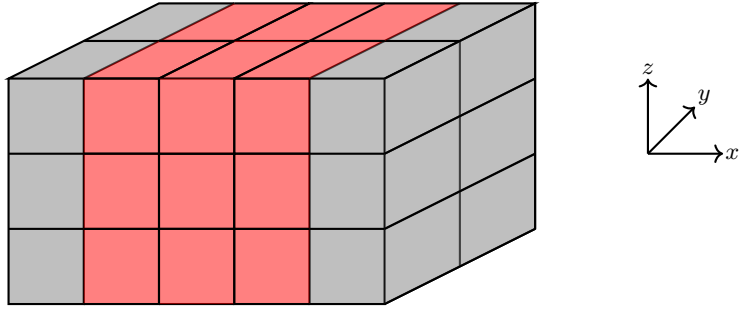
\begin{figure}
    \centering
    \begin{tikzpicture}
	\begin{pgfonlayer}{nodelayer}
		\node [style=none] (0) at (3, 7) {};
		\node [style=none] (1) at (5, 8) {};
		\node [style=none] (2) at (7, 9) {};
		\node [style=none] (3) at (7, 8) {};
		\node [style=none] (4) at (9, 8) {};
		\node [style=none] (5) at (9, 9) {};
		\node [style=none] (6) at (11, 9) {};
		\node [style=none] (7) at (5, 7) {};
		\node [style=none] (8) at (7, 7) {};
		\node [style=none] (9) at (3, 5) {};
		\node [style=none] (10) at (3, 3) {};
		\node [style=none] (11) at (5, 5) {};
		\node [style=none] (12) at (5, 3) {};
		\node [style=none] (13) at (7, 5) {};
		\node [style=none] (14) at (7, 3) {};
		\node [style=none] (19) at (9, 7) {};
		\node [style=none] (20) at (11, 8) {};
		\node [style=none] (21) at (13, 9) {};
		\node [style=none] (22) at (9, 5) {};
		\node [style=none] (23) at (11, 6) {};
		\node [style=none] (24) at (13, 7) {};
		\node [style=none] (25) at (13, 5) {};
		\node [style=none] (26) at (11, 4) {};
		\node [style=none] (27) at (9, 3) {};
		\node [style=none] (28) at (1, 7) {};
		\node [style=none] (29) at (3, 8) {};
		\node [style=none] (30) at (5, 9) {};
		\node [style=none] (31) at (1, 5) {};
		\node [style=none] (32) at (1, 3) {};
		\node [style=none] (33) at (1, 1) {};
		\node [style=none] (34) at (3, 1) {};
		\node [style=none] (35) at (5, 1) {};
		\node [style=none] (36) at (7, 1) {};
		\node [style=none] (37) at (9, 1) {};
		\node [style=none] (38) at (11, 2) {};
		\node [style=none] (39) at (13, 3) {};
		\node [style=none] (40) at (-1, 7) {};
		\node [style=none] (41) at (1, 8) {};
		\node [style=none] (42) at (3, 9) {};
		\node [style=none] (43) at (-1, 5) {};
		\node [style=none] (44) at (-1, 3) {};
		\node [style=none] (45) at (-1, 1) {};
		\node [style=none] (46) at (16, 5) {};
		\node [style=none] (47) at (16, 7) {};
		\node [style=none] (48) at (18, 5) {};
		\node [style=none] (49) at (17.25, 6.25) {};
		\node [style=none] (50) at (18.25, 5) {$x$};
		\node [style=none] (51) at (17.5, 6.5) {$y$};
		\node [style=none] (52) at (16, 7.25) {$z$};
	\end{pgfonlayer}
	\begin{pgfonlayer}{edgelayer}
		\draw [style=grayfillline] (8.center) to (6.center);
		\draw [style=grayfillline] (8.center)
			 to (14.center)
			 to (36.center)
			 to (37.center)
			 to (27.center)
			 to (19.center);
		\draw [style=grayfillline] (14.center) to (27.center);
		\draw [style=grayfillline] (27.center) to (25.center);
		\draw [style=grayfillline] (8.center)
			 to (19.center)
			 to (21.center)
			 to (6.center);
		\draw [style=grayfillline] (4.center) to (20.center);
		\draw [style=grayfillline] (20.center) to (26.center);
		\draw [style=grayfillline] (21.center) to (25.center);
		\draw [style=grayfillline] (13.center) to (22.center);
		\draw [style=grayfillline] (22.center) to (24.center);
		\draw [style=grayfillline] (28.center) to (30.center);
		\draw [style=grayfillline] (40.center)
			 to (45.center)
			 to (33.center)
			 to (32.center)
			 to (28.center);
		\draw [style=grayfillline] (37.center) to (39.center);
		\draw [style=grayfillline] (25.center) to (39.center);
		\draw [style=grayfillline] (28.center)
			 to (40.center)
			 to (42.center)
			 to (30.center);
		\draw [style=grayfillline] (41.center) to (29.center);
		\draw [style=grayfillline] (44.center) to (32.center);
		\draw [style=grayfillline] (43.center) to (31.center);
		\draw [style=grayfillline] (26.center) to (38.center);
		\draw [style=redfillline] (28.center)
			 to (0.center)
			 to (1.center)
			 to (29.center);
		\draw [style=redfillline] (5.center)
			 to (2.center)
			 to (1.center)
			 to (3.center);
		\draw [style=redfillline] (29.center)
			 to (1.center)
			 to (2.center)
			 to (30.center);
		\draw [style=redfillline] (4.center)
			 to (3.center)
			 to (5.center)
			 to (6.center);
		\draw [style=redfillline] (1.center)
			 to (0.center)
			 to (7.center)
			 to (3.center);
		\draw [style=redfillline] (31.center)
			 to (9.center)
			 to (0.center)
			 to (28.center);
		\draw [style=redfillline] (8.center)
			 to (7.center)
			 to (11.center)
			 to (13.center);
		\draw [style=redfillline] (10.center)
			 to (9.center)
			 to (11.center)
			 to (12.center);
		\draw [style=redfillline] (10.center) to (12.center);
		\draw [style=redfillline] (32.center)
			 to (10.center)
			 to (34.center)
			 to (33.center);
		\draw [style=redfillline] (14.center)
			 to (12.center)
			 to (35.center)
			 to (36.center);
		\draw [style=redfillline] (34.center) to (35.center);
		\draw [style=redfillline] (29.center) to (1.center);
		\draw [style=redfillline] (4.center)
			 to (3.center)
			 to (7.center)
			 to (8.center);
		\draw [style=redfillline] (0.center)
			 to (9.center)
			 to (11.center)
			 to (7.center);
		\draw [style=redfillline] (14.center)
			 to (12.center)
			 to (11.center)
			 to (13.center);
		\draw [style=redfillline] (31.center)
			 to (9.center)
			 to (10.center)
			 to (32.center);
		\draw [style=redfillline] (34.center)
			 to (10.center)
			 to (12.center)
			 to (35.center);
		\draw [style=grayfillline] (19.center)
			 to (21.center)
			 to (39.center)
			 to (37.center);
		\draw [style=grayfillline] (22.center) to (24.center);
		\draw [style=grayfillline] (27.center) to (25.center);
		\draw [style=arrowline] (47.center) to (46.center);
		\draw [style=arrowline] (48.center) to (46.center);
		\draw [style=arrowline] (49.center) to (46.center);
	\end{pgfonlayer}
\end{tikzpicture}
    \caption{An example configuration of the plaquette set $\Delta_{\rm p}^{\rm A}$. The red plaquettes belong to $\Delta_{\rm p}^{\rm A}$, while the grey plaquettes are in $\Delta_{\rm p}^{\rm A^c}$. Periodic boundary conditions are imposed in all directions so that $M_3=T^3$. The red plaquettes form the region $T^2\times I$. We note that at the interface along the $yz$ plane, we take the plaquettes to be grey. They can be taken to be red and give another choice of the configuration set $\Delta_{\rm p}^{\rm A}$.}
    \label{fig:Delta_pA_configuration}
\end{figure}

We define the associated sets as follows:
\begin{subequations}
\begin{align}
    \Delta_{\rm c}^{\rm A}&=\{\rm c| \partial c\cap \Delta_{\rm p}^{\rm A}\neq \emptyset\}\subset \Delta_{\rm c}\, ,\\
    \Delta_{\rm e}^{\rm A^c}&=\{\rm e|\partial^*e\cap \Delta_{\rm p}^{\rm A^c}\neq \emptyset \}\,,\\
    \Delta_{\rm c}^{\rm A|A^c}&=\{\rm c\in\Delta_{\rm c}^{A}|\partial c\neq \partial c\cap \Delta_{\rm p}^{\rm A}\}\, ,\\
    \Delta_{\rm e}^{\rm A|A^c}&=\{\rm e\in \Delta_e^{A^c}|\partial^*e\neq \partial^*e\cap \Delta_{\rm p}^{\rm A^c}\}\, ,\\
    \Delta_{\rm v}^{\rm A^c}&=\{\mathrm{v}\in\Delta_{\rm v}|\partial ^*\mathrm{v}\cap \Delta_{\rm e}^{\rm A^c}\neq \emptyset\}
\end{align}
\end{subequations}
Next, we define the cluster entangler as
\begin{align}
    &\mathcal{U}_{CZ}=\prod_{\rm e\in\Delta_{\rm e}^{\rm A^c}}\left(\prod_{\rm p\in\partial^*e\cap \Delta_{\rm p}^{\rm A^c}}CZ_{\rm e,p}\right)\prod_{\rm c\in\Delta_{\rm c}^{\rm A}}\left(\prod_{\rm p\in \partial c\cap \Delta_{\rm p}^{\rm A}}CZ_{\rm c, p}\right)\nonumber\\
    &\hspace{3cm}\times\prod_{\rm c\in\Delta_{\rm c}^{\rm A|A^c}}\prod_{\rm e\in\partial(\partial c\cap \Delta_{\rm p}^{\rm A})}CZ_{\rm c,e}\,.
\end{align}
The corresponding cluster state is then given by
\begin{align}
    |\Psi_{\text{RBH}}^{\text{3D};\rm A|A^c}\rangle=\mathcal{U}_{CZ}\ket{+}^{\Delta_{\rm p}}\ket{+}^{\Delta_{\rm c}^{\rm A}}\ket{+}^{\Delta_{\rm e}^{\rm A^c}}\, .
    \label{eq:3DRBHAAc}
\end{align}
The state above is stabilized by the following operators:
\begin{subequations}
\begin{align} \label{eq:3DCSStabilizer}
    &X_{\rm p}Z(\partial^*\rm p)\quad \mathrm{p}\in \Delta_{\rm p}^{\rm A}\, ,\\
    &X_{\rm p}Z(\partial \rm p)\quad \mathrm{p}\in \Delta_{\rm p}^{\rm A^c}\, ,\\
    &X_{\rm c}Z(\partial \mathrm{c})\quad \mathrm{c}\in\Delta_{\rm c}^{\rm A}\setminus \Delta_{\rm c}^{\rm A|A^c}\, , \\
    &X_{\rm e}Z(\partial^* \mathrm{e})\quad \mathrm{e}\in\Delta_{\rm e}^{\rm A^c}\setminus \Delta_{\rm e}^{\rm A|A^c}\, ,\\
    &X_{\rm c}Z(\partial \mathrm{c}\cap \Delta_{\rm p}^{\rm A})Z(\partial(\partial \mathrm{c}\cap \Delta_{\rm p}^{\rm A}))\quad \mathrm{c}\in\Delta_{\rm c}^{\rm A|A^c}\, , \\
    &X_{\rm e}Z(\partial^* \mathrm{e}\cap \Delta_{\rm p}^{\rm A^c})Z(\partial^*(\partial^* \mathrm{e}\cap \Delta_{\rm p}^{\rm A^c}))\quad \mathrm{e}\in\Delta_{\rm e}^{\rm A|A^c}\, .
\end{align}
\end{subequations}
As before, we take $\Delta_{\rm p}^{\rm A}$ to be a connected region so that the interface between $\Delta_{\rm p}^{\rm A}$ and $\Delta_{\rm p}^{\rm A^c}$ supports the duality defect.
In particular, for the case $M_3=T^3$, we choose $\Delta_{\rm p}^{\rm A}$ to be the set of plaquettes lying on $T^2\times I$, i.e., a two-torus times an interval, as illustrated in Figure~\ref{fig:Delta_pA_configuration}. 
We now display the local stabilizers for the cluster state, following equation~\eqref{eq:3DCSStabilizer}, as
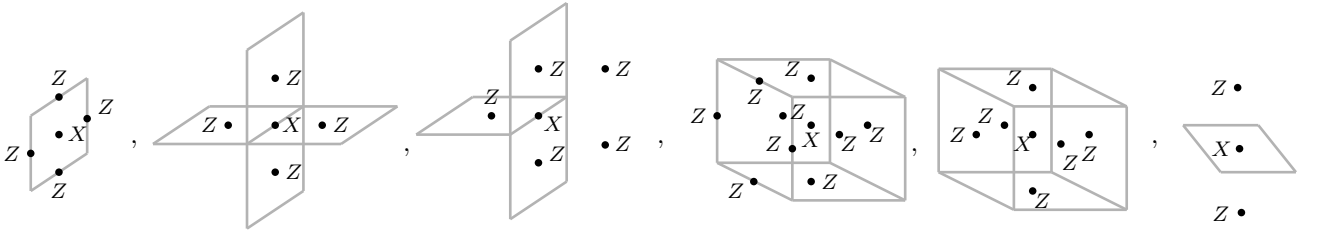
\begin{align*}
    \begin{tikzpicture}
	\begin{pgfonlayer}{nodelayer}
		\node [style=none] (0) at (2.25, 0) {};
		\node [style=none] (1) at (3.75, 1) {};
		\node [style=none] (2) at (3.75, 3.5) {};
		\node [style=none] (3) at (2.25, 2.5) {};
		\node [style=none] (4) at (6.25, 1) {};
		\node [style=none] (5) at (4.75, 0) {};
		\node [style=none] (6) at (1.25, 1) {};
		\node [style=none] (7) at (-0.25, 0) {};
		\node [style=none] (8) at (3.75, -1.25) {};
		\node [style=none] (9) at (2.25, -2.25) {};
		\node [style=none] (10) at (3, 0.5) {};
		\node [style=none] (11) at (3, 1.75) {};
		\node [style=none] (12) at (1.75, 0.5) {};
		\node [style=none] (13) at (3, -0.75) {};
		\node [style=none] (14) at (4.25, 0.5) {};
		\node [style=pointoperator, color = black] (15) at (3, 0.5) {};
		\node [style=pointoperator, color = black] (16) at (3, 1.75) {};
		\node [style=pointoperator, color = black] (17) at (1.75, 0.5) {};
		\node [style=pointoperator, color = black] (18) at (3, -0.75) {};
		\node [style=pointoperator, color = black] (19) at (4.25, 0.5) {};
		\node [style=none] (97) at (-3.5, 0.75) {};
		\node [style=none] (98) at (-2, 1.75) {};
		\node [style=none] (99) at (-3.5, -1.25) {};
		\node [style=none] (100) at (-2, -0.25) {};
		\node [style=none] (101) at (-2.75, 1.25) {};
		\node [style=none] (102) at (-3.5, -0.25) {};
		\node [style=none] (103) at (-2.75, -0.75) {};
		\node [style=none] (104) at (-2, 0.675) {};
		\node [style=pointoperator, color = black] (105) at (-2.75, 1.25) {};
		\node [style=pointoperator, color = black] (106) at (-3.5, -0.25) {};
		\node [style=pointoperator, color = black] (107) at (-2.75, -0.75) {};
		\node [style=pointoperator, color = black] (108) at (-2, 0.675) {};
		\node [style=pointoperator, color = black] (109) at (-2.75, 0.25) {};
		\node [style=none] (117) at (-2.75, 1.75) { \footnotesize $Z$};
		\node [style=none] (118) at (-1.5, 1) { \footnotesize $Z$};
		\node [style=none] (119) at (-2.75, -1.25) { \footnotesize $Z$};
		\node [style=none] (120) at (-4, -0.25) { \footnotesize $Z$};
		\node [style=none] (121) at (3.5, 1.75) { \footnotesize $Z$};
		\node [style=none] (122) at (3.5, -0.75) { \footnotesize $Z$};
		\node [style=none] (123) at (4.75, 0.5) { \footnotesize $Z$};
		\node [style=none] (124) at (1.25, 0.5) { \footnotesize $Z$};
		\node [style=none] (125) at (3.45, 0.5) { \footnotesize $X$};
		\node [style=none] (126) at (-2.25, 0.275) { \footnotesize $X$};
		\node [style=none] (127) at (-0.75, 0) {,};
	\end{pgfonlayer}
	\begin{pgfonlayer}{edgelayer}
		\draw [black!30,line width=1.0] (0.center) to (1.center);
		\draw [black!30,line width=1.0] (2.center) to (1.center);
		\draw [black!30,line width=1.0] (3.center) to (2.center);
		\draw [black!30,line width=1.0] (3.center) to (0.center);
		\draw [black!30,line width=1.0] (1.center) to (4.center);
		\draw [black!30,line width=1.0] (0.center) to (5.center);
		\draw [black!30,line width=1.0] (5.center) to (4.center);
		\draw [black!30,line width=1.0] (7.center) to (6.center);
		\draw [black!30,line width=1.0] (7.center) to (0.center);
		\draw [black!30,line width=1.0] (6.center) to (1.center);
		\draw [black!30,line width=1.0] (9.center) to (8.center);
		\draw [black!30,line width=1.0] (0.center) to (9.center);
		\draw [black!30,line width=1.0] (1.center) to (8.center);
		\draw [black!30,line width=1.0] (97.center) to (99.center);
		\draw [black!30,line width=1.0] (98.center) to (100.center);
		\draw [black!30,line width=1.0] (99.center) to (100.center);
		\draw [black!30,line width=1.0] (97.center) to (98.center);
	\end{pgfonlayer}
\end{tikzpicture}\,, \begin{tikzpicture}
	\begin{pgfonlayer}{nodelayer}
		\node [style=none] (20) at (-4, 0.25) {};
		\node [style=none] (21) at (-2.5, 1.25) {};
		\node [style=none] (22) at (-2.5, 3.75) {};
		\node [style=none] (23) at (-4, 2.75) {};
		\node [style=none] (26) at (-5, 1.25) {};
		\node [style=none] (27) at (-6.5, 0.25) {};
		\node [style=none] (28) at (-2.5, -1) {};
		\node [style=none] (29) at (-4, -2) {};
		\node [style=none] (30) at (-3.25, 0.75) {};
		\node [style=none] (31) at (-3.25, 2) {};
		\node [style=none] (32) at (-4.5, 0.75) {};
		\node [style=none] (33) at (-3.25, -0.5) {};
		\node [style=none] (34) at (-1.475, 2) {};
		\node [style=pointoperator, color = black] (35) at (-3.25, 0.75) {};
		\node [style=pointoperator, color = black] (36) at (-3.25, 2) {};
		\node [style=pointoperator, color = black] (37) at (-4.5, 0.75) {};
		\node [style=pointoperator, color = black] (38) at (-3.25, -0.5) {};
		\node [style=pointoperator, color = black] (39) at (-1.475, 2) {};
		\node [style=none] (40) at (-1.475, -0.025) {};
		\node [style=pointoperator, color = black] (41) at (-1.475, -0.025) {};
		\node [style=none] (42) at (1.5, 2.25) {};
		\node [style=none] (43) at (3.5, 1.25) {};
		\node [style=none] (44) at (1.5, -0.5) {};
		\node [style=none] (45) at (3.5, -1.5) {};
		\node [style=none] (46) at (4.5, 2.25) {};
		\node [style=none] (47) at (6.5, 1.25) {};
		\node [style=none] (48) at (4.5, -0.5) {};
		\node [style=none] (49) at (6.5, -1.5) {};
		\node [style=none] (50) at (3.25, 0.75) {};
		\node [style=none] (51) at (5.5, 0.5) {};
		\node [style=none] (52) at (4.75, 0.25) {};
		\node [style=none] (53) at (4, 1.75) {};
		\node [style=none] (54) at (4, -1) {};
		\node [style=none] (55) at (2.625, 1.675) {};
		\node [style=none] (56) at (1.5, 0.75) {};
		\node [style=none] (57) at (3.5, -0.125) {};
		\node [style=none] (58) at (2.475, -1) {};
		\node [style=pointoperator, color = black] (59) at (2.625, 1.675) {};
		\node [style=pointoperator, color = black] (60) at (1.5, 0.75) {};
		\node [style=pointoperator, color = black] (61) at (3.5, -0.125) {};
		\node [style=pointoperator, color = black] (62) at (2.475, -1) {};
		\node [style=pointoperator, color = black] (63) at (4, 1.75) {};
		\node [style=pointoperator, color = black] (64) at (5.5, 0.5) {};
		\node [style=pointoperator, color = black] (65) at (4.75, 0.25) {};
		\node [style=pointoperator, color = black] (66) at (3.25, 0.75) {};
		\node [style=pointoperator, color = black] (67) at (4, -1) {};
		\node [style=pointoperator, color = black] (68) at (4, 0.5) {};
		\node [style=none] (127) at (-4.5, 1.25) { \footnotesize $Z$};
		\node [style=none] (128) at (-2.75, 2) { \footnotesize $Z$};
		\node [style=none] (129) at (-2.75, -0.25) { \footnotesize $Z$};
		\node [style=none] (130) at (-1, 2) { \footnotesize $Z$};
		\node [style=none] (131) at (-1, 0) { \footnotesize $Z$};
		\node [style=none] (132) at (2.5, 1.25) { \footnotesize $Z$};
		\node [style=none] (133) at (1, 0.75) { \footnotesize $Z$};
		\node [style=none] (134) at (3, 0) { \footnotesize $Z$};
		\node [style=none] (135) at (2, -1.25) { \footnotesize $Z$};
		\node [style=none] (136) at (3.5, 2) { \footnotesize $Z$};
		\node [style=none] (137) at (5.75, 0.25) { \footnotesize $Z$};
		\node [style=none] (138) at (4.5, -1) { \footnotesize $Z$};
		\node [style=none] (139) at (3.65, 0.85) { \footnotesize $Z$};
		\node [style=none] (140) at (5, 0) { \footnotesize $Z$};
		\node [style=none] (141) at (4, 0.125) { \footnotesize $X$};
		\node [style=none] (152) at (-2.85, 0.55) { \footnotesize $X$};
		\node [style=none] (153) at (0, 0) {,};
	\end{pgfonlayer}
	\begin{pgfonlayer}{edgelayer}
		\draw [black!30,line width=1.0] (20.center) to (21.center);
		\draw [black!30,line width=1.0] (22.center) to (21.center);
		\draw [black!30,line width=1.0] (23.center) to (22.center);
		\draw [black!30,line width=1.0] (23.center) to (20.center);
		\draw [black!30,line width=1.0] (27.center) to (26.center);
		\draw [black!30,line width=1.0] (27.center) to (20.center);
		\draw [black!30,line width=1.0] (26.center) to (21.center);
		\draw [black!30,line width=1.0] (29.center) to (28.center);
		\draw [black!30,line width=1.0] (20.center) to (29.center);
		\draw [black!30,line width=1.0] (21.center) to (28.center);
		\draw [black!30,line width=1.0] (42.center) to (43.center);
		\draw [black!30,line width=1.0] (43.center) to (45.center);
		\draw [black!30,line width=1.0] (45.center) to (44.center);
		\draw [black!30,line width=1.0] (42.center) to (44.center);
		\draw [black!30,line width=1.0] (46.center) to (47.center);
		\draw [black!30,line width=1.0] (47.center) to (49.center);
		\draw [black!30,line width=1.0] (49.center) to (48.center);
		\draw [black!30,line width=1.0] (46.center) to (48.center);
		\draw [black!30,line width=1.0] (42.center) to (46.center);
		\draw [black!30,line width=1.0] (43.center) to (47.center);
		\draw [black!30,line width=1.0] (45.center) to (49.center);
		\draw [black!30,line width=1.0] (44.center) to (48.center);
	\end{pgfonlayer}
\end{tikzpicture}\,, \;\; \begin{tikzpicture}
	\begin{pgfonlayer}{nodelayer}
		\node [style=none] (69) at (-5.75, 2) {};
		\node [style=none] (70) at (-3.75, 1) {};
		\node [style=none] (71) at (-5.75, -0.75) {};
		\node [style=none] (72) at (-3.75, -1.75) {};
		\node [style=none] (73) at (-2.75, 2) {};
		\node [style=none] (74) at (-0.75, 1) {};
		\node [style=none] (75) at (-2.75, -0.75) {};
		\node [style=none] (76) at (-0.75, -1.75) {};
		\node [style=none] (77) at (-4, 0.5) {};
		\node [style=none] (78) at (-1.75, 0.25) {};
		\node [style=none] (79) at (-2.5, 0) {};
		\node [style=none] (80) at (-3.25, 1.5) {};
		\node [style=none] (81) at (-3.25, -1.25) {};
		\node [style=pointoperator, color = black] (90) at (-3.25, 1.5) {};
		\node [style=pointoperator, color = black] (91) at (-1.75, 0.25) {};
		\node [style=pointoperator, color = black] (92) at (-2.5, 0) {};
		\node [style=pointoperator, color = black] (93) at (-4, 0.5) {};
		\node [style=pointoperator, color = black] (94) at (-3.25, -1.25) {};
		\node [style=pointoperator, color = black] (95) at (-3.25, 0.25) {};
		\node [style=pointoperator, color = black] (96) at (-4.75, 0.25) {};
		\node [style=none] (110) at (0.75, 0.5) {};
		\node [style=none] (111) at (1.75, -0.75) {};
		\node [style=none] (112) at (2.75, 0.5) {};
		\node [style=none] (113) at (3.75, -0.75) {};
		\node [style=pointoperator, color = black] (114) at (2.25, -0.125) {};
		\node [style=pointoperator, color = black] (115) at (2.2, 1.5) {};
		\node [style=pointoperator, color = black] (116) at (2.3, -1.825) {};
		\node [style=none] (142) at (-5.25, 0.25) { \footnotesize $Z$};
		\node [style=none] (143) at (-3.75, 1.75) { \footnotesize $Z$};
		\node [style=none] (144) at (-1.75, -0.25) { \footnotesize $Z$};
		\node [style=none] (145) at (-3, -1.5) { \footnotesize $Z$};
		\node [style=none] (146) at (-4.5, 0.75) { \footnotesize $Z$};
		\node [style=none] (147) at (-2.25, -0.5) { \footnotesize $Z$};
		\node [style=none] (148) at (-3.5, 0) { \footnotesize $X$};
		\node [style=none] (149) at (1.775, -0.125) { \footnotesize $X$};
		\node [style=none] (150) at (1.7, 1.5) { \footnotesize $Z$};
		\node [style=none] (151) at (1.75, -1.85) { \footnotesize $Z$};
		\node [style=none] (152) at (0, 0) {,};
	\end{pgfonlayer}
	\begin{pgfonlayer}{edgelayer}
		\draw [black!30,line width=1.0] (69.center) to (70.center);
		\draw [black!30,line width=1.0] (70.center) to (72.center);
		\draw [black!30,line width=1.0] (72.center) to (71.center);
		\draw [black!30,line width=1.0] (69.center) to (71.center);
		\draw [black!30,line width=1.0] (73.center) to (74.center);
		\draw [black!30,line width=1.0] (74.center) to (76.center);
		\draw [black!30,line width=1.0] (76.center) to (75.center);
		\draw [black!30,line width=1.0] (73.center) to (75.center);
		\draw [black!30,line width=1.0] (69.center) to (73.center);
		\draw [black!30,line width=1.0] (70.center) to (74.center);
		\draw [black!30,line width=1.0] (72.center) to (76.center);
		\draw [black!30,line width=1.0] (71.center) to (75.center);
		\draw [black!30,line width=1.0] (110.center) to (112.center);
		\draw [black!30,line width=1.0] (110.center) to (111.center);
		\draw [black!30,line width=1.0] (112.center) to (113.center);
		\draw [black!30,line width=1.0] (111.center) to (113.center);
	\end{pgfonlayer}
\end{tikzpicture}
\end{align*}
where the first two diagrams correspond to the bulk of $\Delta_{\rm p}^{\rm A^c}$ and $\Delta_{\rm e}^{\rm A^c}$, the middle two to the interface regions $\Delta_{\rm e}^{\rm A|A^c}$ and $\Delta_{\rm c}^{\rm A|A^c}$, and the last two to the bulk of $\Delta_{\rm c}^{\rm A}$ and $\Delta_{\rm p}^{\rm A}$. Cube centers are represented by points in these diagrams.

Next, we evaluate the overlap between the cluster state \eqref{eq:3DRBHAAc} and the product state $\ket{+}^{\Delta_{\rm c}^{\rm A}}\ket{+}^{\Delta_{\rm e}^{\rm A^c}}$, obtaining
\begin{align}
    |\text{3D-TC}_{\text{GS}}^{*\rm A}\rangle=\bra{+}^{\Delta_{\rm c}^{\rm A}}\bra{+}^{\Delta_{\rm e}^{\rm A^c}}|\Psi_{\text{RBH}}^{\text{3D};\rm A|A^c}\rangle\, .
\end{align}
This state has the same set of local stabilizers as $\mathsf{H}^{\rm A}|\text{3D-TC}{\text{GS}}\rangle$, but differs in the logical operators by which it is stabilized:
   \begin{align}
     \frac{1}{\mathcal{N}_{\text{3D-TC}}}\mathsf{H}^{\rm A}|\text{3D-TC}_{\text{GS}}\rangle=\frac{1}{\mathcal{N}_{\text{3D-TC}}^{*\rm A}}\frac{1}{|H_2(M_3,\mathbb{Z}_2)|}\left(\sum_{ z_2\in H_2(M_3,\mathbb{Z}_2)}X( z_2\cap\Delta_{\rm p}^{\rm A^c})Z( z_2\cap\Delta_{\rm p}^{\rm A})\right)|\text{3D-TC}_{\text{GS}}^{*\rm A}\rangle\,.
\end{align} 
Here $\mathcal{N}_{\text{3D-TC}}$ and $\mathcal{N}_{\text{3D-TC}}^{\rm A}$ are normalization factors ensuring unit norm, explicitly given by
\begin{subequations}
\begin{align}
    &\mathcal{N}_{\text{3D-TC}}=\sqrt{\langle\text{3D-TC}_{\text{GS}}|\text{3D-TC}_{\text{GS}}\rangle}=\frac{2^{\frac{|\Delta_{\rm v}|}{2}}\sqrt{H_2(M_3,\mathbb{Z}_2)}}{\sqrt{2}\times2^{\frac{|\Delta_{\rm e}|}{2}}}=\frac{2\times 2^{\frac{|\Delta_{\rm v}|}{2}}}{2^{\frac{|\Delta_{\rm e}|}{2}}}\\
    &\mathcal{N}_{\text{3D-TC}}^{*\rm A}=\sqrt{\langle\text{3D-TC}^{*\rm A}_{\text{GS}}|\frac{1}{|H_2(M_3,\mathbb{Z}_2)|}\left(\sum_{ z_2\in H_2(M_3,\mathbb{Z}_2)}X( z_2\cap\Delta_{\rm p}^{\rm A^c})Z( z_2\cap\Delta_{\rm p}^{\rm A})\right)|\text{3D-TC}^{*\rm A}_{\text{GS}}\rangle}=\sqrt{\frac{2\times 2^{|\Delta_{\rm v}^{\rm A^c}|}}{2^{|\Delta_{\rm c}^{\rm A}|}2^{|\Delta_{\rm e}^{\rm A^c}|}}}\,.
\end{align}
\end{subequations}

We now take the overlap with $(\bra{0}e^{JX_{\rm p}})^{\Delta_{\rm p}^{\rm A^c}}(\bra{0}e^{J^*X_{\rm p}})^{\Delta_{\rm p}^{\rm A}}$, yielding 
\begin{widetext}
\begin{align}
\begin{split}
    Z_{\text{3D-IG}}(J)&=\frac{2^{-\frac{|\Delta_{\rm c}^{\rm A}|}{2}}2^{-\frac{|\Delta_{\rm e}^{\rm A^c}|}{2}}2^{\frac{|\Delta_{\rm e}|}{2}}2^{\frac{|\Delta_{\rm v}|-|\Delta_{\rm v}^{\rm A^c}|}{2}}\sqrt{2}}{(\sinh 2J^*)^{|\Delta_{\rm p}^{\rm A}|/2}|H_2(M_3,\mathbb{Z}_2)|}\sum_{z_2\in H_2(M_3,\mathbb{Z}_2)}\sum_{\{s_{\rm e}=\pm 1\}\rvert_{\mathrm{e}\in\Delta_{\rm e}^{\rm A^c}}}\sum_{\{s_{\rm c}=\pm 1\}\rvert_{\mathrm{c}\in\Delta_{\rm c}^{\rm A}}}\\
    & \exp\left(J \sum_{\mathrm{p}\in  \Delta_{\rm p}^{\rm A^c}}s(\partial \mathrm{p})+J^*\sum_{\mathrm{p}\in \Delta_{\rm p}^{\rm A}}(-1)^{\#( z_2\cap \mathrm{p})}s(\partial^*\mathrm{p})\right)\prod_{\substack{ \mathrm{c}\in\Delta_{\rm c}^{\rm A|A^c}\\
\mathrm{v}\in\partial(\partial\mathrm{p}\cap\Delta_{\rm p}^{\rm A})}}(-1)^{h_{\rm v}h_{\rm c}}\times s( \partial( z_2\cap \Delta_{\rm p}^{\rm A^c}))\, .
\end{split}
\label{eq:API_duality_defect}
\end{align}
\end{widetext}
The notation $s(\partial\mathrm{p})$ denotes the product $\prod_{\mathrm{e}\in\partial\mathrm{p}}s_{e}$ and similarly $s(\partial( z_2\cap \rm A^c))$ also represents a product of spins $\prod_{\mathrm{v}\in \partial( z_2\cap \rm A^c})s_{\rm v}$. From \eqref{eq:API_duality_defect}, we infer the partition function with two duality defects inserted  
\begin{align}
\begin{split}
    Z_{\text{3D-IG}}^{\mathcal{D}}(J)&=\frac{2^{-\frac{|\Delta_{\rm c}^{\rm A}|}{2}}2^{-\frac{|\Delta_{\rm e}^{\rm A^c}|}{2}}2^{\frac{|\Delta_{\rm e}|}{2}}2^{\frac{|\Delta_{\rm v}|-|\Delta_{\rm v}^{\rm A^c}|}{2}}\sqrt{2}}{(\sinh 2J^*)^{|\Delta_{\rm p}^{\rm A}|/2}}\sum_{\{s_{\rm e}=\pm 1\}\rvert_{\mathrm{e}\in\Delta_{\rm e}^{\rm A^c}}}\sum_{\{s_{\rm c}=\pm 1\}\rvert_{\mathrm{c}\in\Delta_{\rm c}^{\rm A}}}\\
    &\exp\left(J \sum_{\mathrm{p}\in  \Delta_{\rm p}^{\rm A^c}}s(\partial \mathrm{p})+J^*\sum_{\mathrm{p}\in \Delta_{\rm p}^{\rm A}}s(\partial^*\mathrm{p})\right)\prod_{\substack{ \mathrm{c}\in\Delta_{\rm c}^{\rm A|A^c}\\
\mathrm{v}\in\partial(\partial\mathrm{p}\cap\Delta_{\rm p}^{\rm A})}}(-1)^{h_{\rm v}h_{\rm c}}\, .
\end{split}
\end{align}
\subsection{Moving and fusing defects}
Analogous to the case of the two-dimensional Ising model, we denote the spins in region $\Delta_{\rm c}^{\rm A}$ by lowercase Greek letters and those in region $\Delta_{\rm e}^{\rm A^c}$ by lowercase Roman letters. We introduce the following diagrammatic representation to depict the terms appearing in the partition function in the presence of a duality defect:
\begin{align}
    \begin{tikzpicture}
	\begin{pgfonlayer}{nodelayer}
		\node [style=none] (0) at (-3, 2) {};
		\node [style=none] (1) at (-1, 3) {};
		\node [style=none] (2) at (-3, 0) {};
		\node [style=none] (3) at (-1, 1) {};
		\node [style=none] (4) at (0.5, 1.5) {$=$};
		\node [style=none] (5) at (2.75, 1.75) {$e^{Js_as_bs_cs_d}\,,$};
		\node [style=none] (6) at (-2, 3) {$a$};
		\node [style=none] (7) at (-0.5, 2) {$b$};
		\node [style=none] (8) at (-2, 0) {$c$};
		\node [style=none] (9) at (-3.5, 1) {$d$};
		\node [style=none] (10) at (-3.5, -2) {};
		\node [style=none] (11) at (-0.5, -2) {};
		\node [style=pointblack] (12) at (-3.5, -2) {};
		\node [style=pointblack] (13) at (-0.5, -2) {};
		\node [style=none] (14) at (0.75, -2) {$=$};
		\node [style=none] (15) at (2.5, -1.75) {$e^{J^*s_{\alpha}s_{\beta}}\,,$};
		\node [style=none] (16) at (-4.25, -2) {$\alpha$};
		\node [style=none] (17) at (0, -2) {$\beta$};
		\node [style=none] (18) at (-3.5, -5.5) {};
		\node [style=none] (19) at (-0.5, -4.5) {};
		\node [style=none] (20) at (-2.5, -6) {};
		\node [style=none] (21) at (-1.5, -5.25) {};
		\node [style=pointblack] (22) at (-3.5, -5.5) {};
		\node [style=pointblack] (23) at (-1.5, -5.25) {};
		\node [style=none] (24) at (0.75, -5) {$=$};
		\node [style=none] (25) at (2.75, -5) {$(-1)^{h_{\alpha}h_a}\,.$};
		\node [style=none] (26) at (-4, -5.75) {$\alpha$};
		\node [style=none] (27) at (-1, -5.25) {$a$};
		\node [style=pointblack] (28) at (-2, 2.5) {};
		\node [style=pointblack] (29) at (-3, 1) {};
		\node [style=pointblack] (30) at (-1, 2) {};
		\node [style=pointblack] (31) at (-2, 0.5) {};
	\end{pgfonlayer}
	\begin{pgfonlayer}{edgelayer}
		\draw [style=grayfillline] (3.center)
			 to (2.center)
			 to (0.center)
			 to (1.center)
			 to cycle;
		\draw [style=reddashedline] (10.center) to (11.center);
		\draw (20.center) to (19.center);
		\draw [style=greenline] (18.center) to (21.center);
	\end{pgfonlayer}
\end{tikzpicture}
\end{align}
In this diagram, the plaquette shaded in grey belongs to the original lattice, while the red dashed line represents an edge in the dual lattice. The green line connects an edge in the original lattice to a vertex in the dual lattice. We employ this diagrammatic representation to formulate the local relations that describe the movement of the duality defect.
\subsubsection{Movement of duality defect}
 We now examine the fusion of two duality defects by considering an interface between the regions $\Delta_{\rm p}^{\rm A}$ and $\Delta_{\rm p}^{\rm A^c}$. Let us begin with an identity obtained by deforming the region $\Delta_{\rm p}^{\rm A}$:
\begin{align}
\begin{tikzpicture}
	\begin{pgfonlayer}{nodelayer}
		\node [style=none] (0) at (-2.25, 2.25) {};
		\node [style=none] (1) at (-4.25, 0.25) {};
		\node [style=none] (2) at (-4.25, -2.75) {};
		\node [style=none] (3) at (-2.25, -0.75) {};
		\node [style=none] (4) at (-3.5, 1.7) {$a$};
		\node [style=none] (5) at (-1.8, 0.5) {$b$};
		\node [style=none] (6) at (-2.85, -1.825) {$c$};
		\node [style=none] (7) at (-4.7, -1.475) {$d$};
		\node [style=pointblack] (8) at (-6, -0.25) {};
            \node [style=none] (A) at (-6.5, -0.25) {$\alpha$};
            \node [style=none] (A) at (-8.5, -0.25) {$\frac{\sqrt{2}}{\sqrt{ \sinh(2 J)}}$};

		\node [style=none] (9) at (-3.25, 1.25) {};
		\node [style=none] (10) at (-2.25, 0.5) {};
		\node [style=none] (11) at (-3, -1.5) {};
		\node [style=none] (12) at (-4.25, -1.25) {};
		\node [style=none] (25) at (-1, -0.3) {$=$};
		\node [style=none] (26) at (0, -0.2) {\Large $\sum$};
		\node [style=none] (27) at (0, -1.15) {$h_{\beta}$};
		\node [style=none] (28) at (5, 2.25) {};
		\node [style=none] (29) at (3, 0.25) {};
		\node [style=none] (30) at (3, -2.75) {};
		\node [style=none] (31) at (5, -0.75) {};
		\node [style=none] (32) at (3.75, 1.7) {$a$};
		\node [style=none] (33) at (4.7, 0.3) {$b$};
		\node [style=none] (34) at (4.4, -1.825) {$c$};
		\node [style=none] (35) at (2.55, -1.475) {$d$};
		\node [style=pointblack] (36) at (1.5, -0.25) {};
            \node [style=none] (A2) at (1.1, -0.25) {$\alpha$};

		\node [style=pointblack] (37) at (4, 1.25) {};
		\node [style=pointblack] (38) at (5, 0.5) {};
		\node [style=pointblack] (39) at (4.25, -1.5) {};
		\node [style=pointblack] (40) at (3, -1.25) {};
		\node [style=pointblack] (41) at (7, -0.25) {};
        \node [style=none] (B) at (7.5, -0.25) {$\beta$};
        
		\node [style=pointblack] (4) at (-3.3, 1.2) {};
        \node [style=pointblack] (4) at (-3.0, -1.5) {};
        \node [style=pointblack] (4) at (-4.25, -1.3) {};
        \node [style=pointblack] (4) at (-2.25, 0.5) {};

	\end{pgfonlayer}
	\begin{pgfonlayer}{edgelayer}
		\draw (0.center) to (1.center);
		\draw (1.center) to (2.center);
		\draw (2.center) to (3.center);
		\draw (0.center) to (3.center);
		\draw [style=normalline, color=green] (8.center) to (9.center);
		\draw [style=normalline, color=green] (8.center) to (10.center);
		\draw [style=normalline, color=green] (8.center) to (11.center);
		\draw [style=normalline, color=green] (8.center) to (12.center);
		\draw (28.center) to (29.center);
		\draw (29.center) to (30.center);
		\draw (30.center) to (31.center);
		\draw (28.center) to (31.center);
		\draw [style=normalline, color=green] (37.center) to (41.center);
		\draw [style=normalline, color=green] (38.center) to (41.center);
		\draw [style=normalline, color=green] (39.center) to (41.center);
		\draw [style=normalline, color=green] (40.center) to (41.center);
		\draw [style=dashedline, color = red] (36.center) to (41.center);
	\end{pgfonlayer}

    \filldraw[fill=gray!30, opacity = 0.6, draw=black] (0.center) -- (1.center) -- (2.center) -- (3.center) -- cycle;
\end{tikzpicture}\,.
\label{eq:3Dmovementid1}
\end{align}
On the left-hand side (L.H.S.) of the above equation, the shaded grey region represents the plaquette Ising interaction among the links in the three-dimensional Ising gauge theory, while the red dashed line on the right-hand side (R.H.S.) denotes the dual Ising interaction. The green lines indicate the interaction terms of the form $(-1)^{h_{\rm c}h_{\rm e}}$ between cube centers and edges, which arise from the $CZ$ gates in~\eqref{eq:3DRBHAAc} coupling these degrees of freedom.

Transitioning from the L.H.S. to the R.H.S. corresponds to the change $|\Delta_{\rm p}^{\rm A}|\longrightarrow |\Delta_{\rm p}^{\rm A}|+1$. Mathematically, this transformation is expressed as
\begin{align}
(-1)^{h_{\alpha}\left(h_{a} + h_{b} + h_{c} + h_{d}\right)} \frac{e^{Js_{a}s_{b}s_{c}s_{d}}\sqrt{2}}{\sqrt{ \sinh(2 J)}} = \sum_{h_{\beta}=0,1} (-1)^{h_{\beta}\left(h_{a} + h_{b} + h_{c} + h_{d}\right)} e^{J^*s_{\alpha}s_{\beta}}\,.
\end{align}
Similarly, we can consider the complementary identity corresponding to the deformation
$|\Delta_{\rm p}^{\rm A^c}|\longrightarrow|\Delta_{\rm p}^{\rm A^c}|+1$:
\begin{align}
\begin{tikzpicture}
	\begin{pgfonlayer}{nodelayer}
		\node [style=none] (0) at (-3.75, 0.75) {};
		\node [style=none] (1) at (-6.25, -1.25) {};
		\node [style=none] (2) at (-3.25, -1.25) {};
		\node [style=none] (3) at (-0.75, 0.75) {};
		\node [style=pointblack] (4) at (-3.25, 1.75) {};
		\node [style=pointblack] (5) at (-3.25, -2.25) {};
		\node [style=none] (6) at (-4.85, -0.125) {};
		\node [style=none] (7) at (3.9, 0.75) {};
		\node [style=none] (8) at (1.5, -1.25) {};
		\node [style=none] (9) at (4.5, -1.25) {};
		\node [style=none] (10) at (6.9, 0.75) {};
		\node [style=pointblack] (11) at (4.375, 1.75) {};
		\node [style=pointblack] (12) at (4.375, -2.25) {};
		\node [style=none] (14) at (0.25, 0) {$=$};
		\node [style=none] (15) at (5.4, 0.75) {};
		\node [style=none] (16) at (5.75, -0.25) {};
		\node [style=none] (17) at (3.225, -1.25) {};
		\node [style=none] (18) at (-7, 0) {\Large $\sum$};
            \node [style=none] () at (-9.5, 0) { $\frac{1}{\sqrt{2 \sinh(2 J)}}$};
		\node [style=none] (19) at (-7, -0.95) {$s_{a}$};
		\node [style=none] (20) at (-2.25, 1.25) {$b$};
		\node [style=none] (21) at (-1.575, -0.5) {$c$};
		\node [style=none] (22) at (-4.75, -1.75) {$d$};
		\node [style=none] (23) at (-5.6, 0) {$a$};
		\node [style=none] (24) at (5.5, 1.25) {$b$};
		\node [style=none] (25) at (6.175, -0.5) {$c$};
		\node [style=none] (26) at (3, -1.75) {$d$};
		\node [style=none] (27) at (-3.25, 2.2) {$\alpha$};
		\node [style=none] (28) at (-3.25, -2.75) {$\beta$};
		\node [style=none] (29) at (4.375, 2.2) {$\alpha$};
		\node [style=none] (30) at (4.375, -2.75) {$\beta$};
	\end{pgfonlayer}
	\begin{pgfonlayer}{edgelayer}
		\draw (0.center) to (1.center);
		\draw (1.center) to (2.center);
		\draw (2.center) to (3.center);
		\draw (3.center) to (0.center);
		\draw [style=normalline, color = green] (4) to (6.center);
		\draw [style=normalline, color = green] (6.center) to (5);
		\draw (8.center) to (9.center);
		\draw (9.center) to (10.center);
		\draw (10.center) to (7.center);
		\draw [style=normalline, color = green] (11) to (15.center);
		\draw [style=normalline, color = green] (15.center) to (12);
		\draw [style=normalline, color = green] (16.center) to (12);
		\draw [style=normalline, color = green] (11) to (16.center);
		\draw [style=normalline, color = green] (11) to (17.center);
		\draw [style=normalline, color = green] (17.center) to (12);
		\draw [style=dashedline, color = red] (11) to (12);
	\end{pgfonlayer}
    \filldraw[fill=gray!30, opacity = 0.6, draw=black] (0.center) -- (1.center) -- (2.center) -- (3.center) -- cycle;

        \fill [black] (6.center) circle (3pt);
        \fill [black] ($0.5*(1.center) + 0.5*(2.center)$) circle (3pt);
        \fill [black] ($0.5*(2.center) + 0.5*(3.center)$) circle (3pt);
        \fill [black] ($0.5*(3.center) + 0.5*(0.center)$) circle (3pt);
        \fill [black] (15.center) circle (3pt);
        \fill [black] (16.center) circle (3pt);
        \fill [black] (17.center) circle (3pt);

\end{tikzpicture}\,.
\label{eq:3Dmovementid2}
\end{align}
The corresponding algebraic relation reads
\begin{align}
\sum_{h_{a}=0,1}(-1)^{h_{a}\left(h_{\alpha} + h_{\beta}\right)} \frac{e^{Js_{a}s_{b}s_{c}s_{d}}}{\sqrt{2\sinh(2 J)}} = (-1)^{(h_{\alpha}+h_{\beta})\left(h_{b} + h_{c} + h_{d}\right)} e^{J^*s_{\alpha}s_{\beta}}\,.
\end{align}
Both relations yield the standard Kramers–Wannier mapping between the couplings, $J^*=-\frac{1}{2}\ln \tanh(J)$.

Equations~\eqref{eq:3Dmovementid1} and \eqref{eq:3Dmovementid2} provide the movement relations for the duality defect. For instance, starting from a planar defect on $yz$-plane, we can first apply~\eqref{eq:3Dmovementid1}:
\begin{align}
\begin{tikzpicture}
	\begin{pgfonlayer}{nodelayer}
		\node [style=none] (1) at (-2, 2.25) {};
		\node [style=none] (2) at (-3, 1.25) {};
		\node [style=none] (3) at (-4, 0.25) {};
		\node [style=none] (5) at (-2, 0.75) {};
		\node [style=none] (6) at (-3, -0.25) {};
		\node [style=none] (7) at (-4, -1.25) {};
		\node [style=none] (9) at (-2, -0.75) {};
		\node [style=none] (10) at (-3, -1.75) {};
		\node [style=none] (11) at (-4, -2.75) {};
		\node [style=none] (17) at (-3.5, 1.25) {};
		\node [style=none] (18) at (-4.5, 0.25) {};
		\node [style=none] (20) at (-3.5, -0.25) {};
		\node [style=none] (21) at (-4.5, -1.25) {};
		\node [style=pointoperator, color=black] (25) at (-3.5, 1.25) {};
		\node [style=pointoperator, color=black] (27) at (-4.5, 0.25) {};
		\node [style=pointoperator, color=black] (28) at (-4.5, -1.25) {};
		\node [style=pointoperator, color=black] (32) at (-3.5, -0.25) {};
		\node [style=none] (36) at (-2.5, 1.75) {};
		\node [style=none] (37) at (-3.5, 0.75) {};
		\node [style=none] (39) at (-2.5, 0.25) {};
		\node [style=none] (40) at (-3.5, -0.75) {};
		\node [style=none] (42) at (-2.5, -1.25) {};
		\node [style=none] (43) at (-3.5, -2.25) {};
		\node [style=none] (50) at (-2, 1.5) {};
		\node [style=none] (51) at (-2, 0) {};
		\node [style=none] (53) at (-3, 0.5) {};
		\node [style=none] (54) at (-3, -1) {};
		\node [style=none] (56) at (-4, -0.5) {};
		\node [style=none] (57) at (-4, -2) {};
		\node [style=none] (60) at (7, 2.25) {};
		\node [style=none] (61) at (6, 1.25) {};
		\node [style=none] (62) at (5, 0.25) {};
		\node [style=none] (64) at (7, 0.75) {};
		\node [style=none] (65) at (6, -0.25) {};
		\node [style=none] (66) at (5, -1.25) {};
		\node [style=none] (68) at (7, -0.75) {};
		\node [style=none] (69) at (6, -1.75) {};
		\node [style=none] (70) at (5, -2.75) {};
		\node [style=none] (94) at (6.5, 1.75) {};
		\node [style=none] (95) at (5.5, 0.75) {};
		\node [style=none] (97) at (6.5, 0.25) {};
		\node [style=none] (98) at (5.5, -0.75) {};
		\node [style=none] (100) at (6.5, -1.25) {};
		\node [style=none] (101) at (5.5, -2.25) {};
		\node [style=none] (108) at (7, 1.5) {};
		\node [style=none] (109) at (7, 0) {};
		\node [style=none] (111) at (6, 0.5) {};
		\node [style=none] (112) at (6, -1) {};
		\node [style=none] (114) at (5, -0.5) {};
		\node [style=none] (115) at (5, -2) {};
		\node [style=pointoperator, color=black] (117) at (5.5, 1.325) {};
		\node [style=pointoperator, color=black] (119) at (4.5, 0.325) {};
		\node [style=pointoperator, color=black] (120) at (4.5, -1.175) {};
		\node [style=pointoperator, color=black] (124) at (5.5, -0.175) {};
		\node [style=pointoperator, color=black] (126) at (7.75, 0.5) {};
		\node [style=pointoperator, color=black] (128) at (6.5, -0.425) {};
		\node [style=pointoperator, color=black] (129) at (6.5, -2.025) {};
		\node [style=pointoperator, color=black] (133) at (7.5, -0.925) {};
		\node [style=none] (135) at (-0.75, 0) {$=$};
		\node [style=none] (137) at (-3.5, 2.75) {};
		\node [style=none] (138) at (-4.5, 1.75) {};
		\node [style=none] (140) at (5.5, 2.725) {};
		\node [style=none] (141) at (4.5, 1.725) {};
		\node [style=none] (145) at (6.5, -2.025) {};
		\node [style=none] (148) at (-5.5, -0.75) {};
		\node [style=none] (149) at (-5.5, -2.25) {};
		\node [style=none] (154) at (3.45, -0.75) {};
		\node [style=none] (155) at (3.5, -2.25) {};
		\node [style=none] (159) at (7.75, 0.5) {};
		\node [style=none] (160) at (-4.5, -2.75) {};
		\node [style=none] (161) at (-3.5, -2) {};
		\node [style=none] (162) at (-2.5, 2.25) {};
		\node [style=none] (163) at (-2.5, 0.75) {};
		\node [style=none] (164) at (4.5, -2.5) {};
		\node [style=none] (165) at (5.5, -1.5) {};
		\node [style=none] (166) at (1.8, 0) {$\left(\frac{\sinh(2J)}{2}\right)^{N_{\rm yz}/2}$};
		\node [style=none] (167) at (6.425, 2.25) {};
		\node [style=none] (168) at (6.475, 0.8) {};

        \node [style=none] (169) at (-4, 0.25) {};
		\node [style=none] (170) at (-2, 2.25) {};
		\node [style=none] (171) at (-2, -0.75) {};
		\node [style=none] (172) at (-4, -2.75) {};

        \node [style=pointoperator, color=black] (A) at (-3.5, 0.75) {};
		\node [style=pointoperator, color=black] (B) at (-2.5, 1.75) {};
		\node [style=pointoperator, color=black] (C) at (-3.5, -0.75) {};
		\node [style=pointoperator, color=black] (D) at (-4, -0.5) {};
		\node [style=pointoperator, color=black] (173) at (-2.5, 0.25) {};
		\node [style=pointoperator, color=black] (174) at (-2, 1.5) {};
		\node [style=pointoperator, color=black] (175) at (-2, 0) {};
		\node [style=pointoperator, color=black] (176) at (-2.5, -1.25) {};
		\node [style=pointoperator, color=black] (177) at (-3.5, -2.25) {};
		\node [style=pointoperator, color=black] (178) at (-4, -2) {};
		\node [style=pointoperator, color=black] (179) at (-3, 0.5) {};
		\node [style=pointoperator, color=black] (180) at (-3, -1) {};
		\node [style=pointoperator, color=black] (181) at (5.5, 0.75) {};
		\node [style=pointoperator, color=black] (182) at (6.5, 1.75) {};
		\node [style=pointoperator, color=black] (183) at (5.5, -0.75) {};
		\node [style=pointoperator, color=black] (184) at (6.5, 0.25) {};
		\node [style=pointoperator, color=black] (185) at (5.5, -2.25) {};
		\node [style=pointoperator, color=black] (186) at (6.5, -1.25) {};
		\node [style=pointoperator, color=black] (187) at (5, -2) {};
		\node [style=pointoperator, color=black] (188) at (6, -1) {};
		\node [style=pointoperator, color=black] (189) at (7, 0) {};
		\node [style=pointoperator, color=black] (190) at (5, -0.5) {};
		\node [style=pointoperator, color=black] (191) at (6, 0.5) {};
		\node [style=pointoperator, color=black] (192) at (7, 1.5) {};
	\end{pgfonlayer}
	\begin{pgfonlayer}{edgelayer}
		\draw [style=normalline, color=green] (27) to (37.center);
		\draw [style=normalline, color=green] (27) to (56.center);
		\draw [style=normalline, color=green] (27) to (53.center);
		\draw [style=normalline, color=green] (27) to (40.center);
		\draw [style=normalline, color=green] (25) to (36.center);
		\draw [style=normalline, color=green] (25) to (53.center);
		\draw [style=normalline, color=green] (25) to (39.center);
		\draw [style=normalline, color=green] (25) to (50.center);
		\draw [style=normalline, color=green] (28) to (40.center);
		\draw [style=normalline, color=green] (28) to (57.center);
		\draw [style=normalline, color=green] (28) to (43.center);
		\draw [style=normalline, color=green] (28) to (54.center);
		\draw [style=normalline, color=green] (32) to (54.center);
		\draw [style=normalline, color=green] (32) to (51.center);
		\draw [style=normalline, color=green] (32) to (39.center);
		\draw [style=normalline, color=green] (32) to (42.center);
		\draw [style=dashedline, color=red] (119) to (128);
		\draw [style=dashedline, color=red] (117) to (126);
		\draw [style=dashedline, color=red] (120) to (129);
		\draw [style=dashedline, color=red] (124) to (133);
		\draw [style=normalline, color=green] (95.center) to (128);
		\draw [style=normalline, color=green] (114.center) to (128);
		\draw [style=normalline, color=green] (111.center) to (128);
		\draw [style=normalline, color=green] (98.center) to (128);
		\draw [style=normalline, color=green] (111.center) to (126);
		\draw [style=normalline, color=green] (94.center) to (126);
		\draw [style=normalline, color=green] (108.center) to (126);
		\draw [style=normalline, color=green] (97.center) to (126);
		\draw [style=normalline, color=green] (115.center) to (129);
		\draw [style=normalline, color=green] (98.center) to (129);
		\draw [style=normalline, color=green] (112.center) to (129);
		\draw [style=normalline, color=green] (101.center) to (129);
		\draw [style=normalline, color=green] (100.center) to (133);
		\draw [style=normalline, color=green] (112.center) to (133);
		\draw [style=normalline, color=green] (97.center) to (133);
		\draw [style=normalline, color=green] (109.center) to (133);
		\draw [style=dashedline, color=red] (137.center) to (25);
		\draw [style=dashedline, color=red] (138.center) to (27);
		\draw [style=dashedline, color=red] (148.center) to (27);
		\draw [style=dashedline, color=red] (149.center) to (28);
		\draw [style=dashedline, color=red] (154.center) to (119);
		\draw [style=dashedline, color=red] (155.center) to (120);
		\draw [style=dashedline, color=red] (141.center) to (119);
		\draw [style=dashedline, color=red] (140.center) to (117);
		\draw (3.center) to (11.center);
		\draw (2.center) to (10.center);
		\draw (1.center) to (9.center);
		\draw (3.center) to (1.center);
		\draw (11.center) to (9.center);
		\draw (7.center) to (5.center);
		\draw [style=dashedline, color=red] (27) to (25);
		\draw [style=dashedline, color=red] (27) to (28);
		\draw [style=dashedline, color=red] (28) to (32);
		\draw [style=dashedline, color=red] (25) to (32);
		\draw [style=dashedline, color=red] (28) to (160.center);
		\draw [style=dashedline, color=red] (32) to (161.center);
		\draw [style=dashedline, color=red] (32) to (163.center);
		\draw [style=dashedline, color=red] (25) to (162.center);
		\draw [style=dashedline, color=red] (119) to (117);
		\draw [style=dashedline, color=red] (119) to (120);
		\draw [style=dashedline, color=red] (120) to (124);
		\draw [style=dashedline, color=red] (117) to (124);
		\draw [style=dashedline, color=red] (120) to (164.center);
		\draw [style=dashedline, color=red] (124) to (165.center);
		\draw (62.center) to (60.center);
		\draw (62.center) to (70.center);
		\draw (70.center) to (68.center);
		\draw (60.center) to (68.center);
		\draw (66.center) to (64.center);
		\draw (61.center) to (69.center);
		\draw [style=dashedline, color=red] (124) to (168.center);
		\draw [style=dashedline, color=red] (117) to (167.center);
	\end{pgfonlayer}
        \filldraw[fill=gray!30, opacity = 0.6, draw=black] (169.center) -- (170.center) -- (171.center) -- (172.center) -- cycle;
\end{tikzpicture}
\end{align}
where $N_{\rm yz}$ denotes the number of vertices on $yz$-plane. We see new spins being created by the movement equation, which are not interacting among themselves yet. Starting from these newly created spins, we apply~\eqref{eq:3Dmovementid2} and get
\begin{align}
\input{3dDualityMovement3.tikz}
\end{align}
where in the first equality, the movement equation \eqref{eq:3Dmovementid2} makes the spins interacting, and in the second equality, we sum over the newly rendered free link variables and get $2^{N_{\rm yz}}$. This sequence of transformations effectively shifts the defect by one lattice unit, demonstrating that the duality defect is topological, i.e., its precise position does not affect the partition function.

\subsubsection{Fusion of duality defects}
\begin{figure}
    \centering\begin{tikzpicture}
	\begin{pgfonlayer}{nodelayer}
		\node [style=none] (21) at (2, 4) {};
		\node [style=none] (22) at (4, 5) {};
		\node [style=none] (23) at (6, 6) {};
		\node [style=none] (24) at (2, 2) {};
		\node [style=none] (25) at (2, 0) {};
		\node [style=none] (26) at (4, 1) {};
		\node [style=none] (27) at (6, 2) {};
		\node [style=none] (28) at (4, 3) {};
		\node [style=none] (29) at (6, 4) {};
		\node [style=none] (30) at (3.5, 4) {};
		\node [style=none] (31) at (5.5, 5) {};
		\node [style=none] (32) at (7.5, 6) {};
		\node [style=none] (33) at (3.5, 2) {};
		\node [style=none] (34) at (3.5, 0) {};
		\node [style=none] (35) at (5.5, 1) {};
		\node [style=none] (36) at (7.5, 2) {};
		\node [style=none] (37) at (5.5, 3) {};
		\node [style=none] (38) at (7.5, 4) {};
		\node [style=none] (39) at (3.75, 3.5) {};
		\node [style=none] (40) at (5.75, 4.5) {};
		\node [style=none] (41) at (3.75, 1.5) {};
		\node [style=none] (42) at (5.75, 2.5) {};
		\node [style=none] (43) at (3, 4.5) {};
		\node [style=none] (44) at (5, 5.5) {};
		\node [style=none] (45) at (3, 2.5) {};
		\node [style=none] (46) at (6.5, 3.5) {};
		\node [style=none] (47) at (4.5, 2.5) {};
		\node [style=none] (48) at (4.5, 0.5) {};
		\node [style=none] (49) at (3, 0.5) {};
		\node [style=none] (50) at (6.5, 1.5) {};
		\node [style=none] (51) at (5, 1.5) {};
		\node [style=none] (52) at (6.5, 5.5) {};
		\node [style=none] (53) at (4.5, 4.5) {};
		\node [style=none] (54) at (5, 3.5) {};
		\node [style=none] (55) at (3.75, 5.5) {};
		\node [style=none] (56) at (5.75, 6.75) {};
		\node [style=none] (57) at (5.75, 0.75) {};
		\node [style=none] (58) at (3.75, -0.5) {};
		\node [style=none] (59) at (2.25, 2.75) {};
		\node [style=none] (60) at (7, 5) {};
		\node [style=none] (61) at (2.25, 0.75) {};
		\node [style=none] (62) at (7, 3) {};
		\node [style=none] (63) at (14.25, 3.75) {};
		\node [style=none] (64) at (13, 3) {};
		\node [style=none] (65) at (16, 3) {};
		\node [style=none] (66) at (17.25, 3.75) {};
		\node [style=none] (67) at (15.25, 4.75) {};
		\node [style=none] (68) at (15.25, 2.25) {};
		\node [style=none] (69) at (13.75, 3.5) {};
		\node [style=none] (70) at (16.75, 3.5) {};
		\node [style=none] (71) at (13, 3.25) {$d'$};
		\node [style=none] (72) at (15.25, 5) {$\alpha$};
		\node [style=none] (73) at (15.25, 1.75) {$\alpha'$};
		\node [style=none] (74) at (17, 3.25) {$d$};
		\node [style=none] (75) at (1, 7) {$(a)$};
		\node [style=none] (76) at (12, 7) {$(b)$};
		\node [style=none] (95) at (10, -2) {};
		\node [style=none] (96) at (10, -0.5) {};
		\node [style=none] (97) at (11.5, -2) {};
		\node [style=none] (98) at (11.25, -0.75) {};
		\node [style=none] (99) at (11.75, -2) {$x$};
		\node [style=none] (100) at (11.5, -0.75) {$y$};
		\node [style=none] (101) at (10, -0.25) {$z$};
		\node [style=none] (102) at (-0.5, 7.5) {};
		\node [style=pointblack] (103) at (3.75, 3.5) {};
		\node [style=pointblack] (104) at (5.75, 4.5) {};
		\node [style=pointblack] (105) at (3.75, 1.5) {};
		\node [style=pointblack] (106) at (5.75, 2.5) {};
		\node [style=pointblack] (107) at (15.25, 4.75) {};
		\node [style=pointblack] (108) at (15.25, 2.25) {};
		\node [style=pointblack] (109) at (3, 4.5) {};
		\node [style=pointblack] (110) at (4.5, 4.5) {};
		\node [style=pointblack] (111) at (3, 2.5) {};
		\node [style=pointblack] (112) at (4.5, 2.5) {};
		\node [style=pointblack] (113) at (3, 0.5) {};
		\node [style=pointblack] (114) at (4.5, 0.5) {};
		\node [style=pointblack] (115) at (6.5, 1.5) {};
		\node [style=pointblack] (116) at (5, 1.5) {};
		\node [style=pointblack] (117) at (6.5, 3.5) {};
		\node [style=pointblack] (118) at (5, 3.5) {};
		\node [style=pointblack] (119) at (5, 5.5) {};
		\node [style=pointblack] (120) at (6.5, 5.5) {};
		\node [style=pointblack] (121) at (13.75, 3.5) {};
		\node [style=pointblack] (122) at (16.75, 3.5) {};
		\node [style=pointblack] (123) at (2, 3) {};
		\node [style=pointblack] (124) at (2, 1) {};
		\node [style=pointblack] (125) at (3.5, 1) {};
		\node [style=pointblack] (126) at (5.5, 2) {};
		\node [style=pointblack] (127) at (7.5, 2.75) {};
		\node [style=pointblack] (128) at (7.5, 5) {};
		\node [style=pointblack] (129) at (5.5, 4) {};
		\node [style=pointblack] (131) at (4, 4) {};
		\node [style=pointblack] (132) at (6, 5) {};
		\node [style=pointblack] (133) at (3.5, 3) {};
		\node [style=pointblack] (134) at (6, 3) {};
		\node [style=pointblack] (135) at (4, 2) {};
	\end{pgfonlayer}
	\begin{pgfonlayer}{edgelayer}
		\draw (21.center) to (23.center);
		\draw (23.center) to (27.center);
		\draw (21.center) to (25.center);
		\draw (25.center) to (27.center);
		\draw (24.center) to (29.center);
		\draw (22.center) to (26.center);
		\draw (30.center) to (32.center);
		\draw (32.center) to (36.center);
		\draw (30.center) to (34.center);
		\draw (34.center) to (36.center);
		\draw (33.center) to (38.center);
		\draw (31.center) to (35.center);
		\draw (24.center) to (33.center);
		\draw (28.center) to (37.center);
		\draw (29.center) to (38.center);
		\draw (21.center) to (30.center);
		\draw (22.center) to (31.center);
		\draw (23.center) to (32.center);
		\draw (25.center) to (34.center);
		\draw (26.center) to (35.center);
		\draw (27.center) to (36.center);
		\draw [style=greenline] (39.center) to (43.center);
		\draw [style=greenline] (39.center) to (47.center);
		\draw [style=greenline] (39.center) to (45.center);
		\draw [style=greenline] (41.center) to (45.center);
		\draw [style=greenline] (41.center) to (47.center);
		\draw [style=greenline] (41.center) to (48.center);
		\draw [style=greenline] (41.center) to (49.center);
		\draw [style=greenline] (40.center) to (44.center);
		\draw [style=greenline] (40.center) to (46.center);
		\draw [style=greenline] (42.center) to (46.center);
		\draw [style=greenline] (42.center) to (50.center);
		\draw [style=greenline] (42.center) to (51.center);
		\draw [style=greenline] (40.center) to (54.center);
		\draw [style=greenline] (40.center) to (52.center);
		\draw [style=greenline] (39.center) to (53.center);
		\draw [style=greenline] (54.center) to (42.center);
		\draw [style=reddashedline] (39.center) to (41.center);
		\draw [style=reddashedline] (40.center) to (42.center);
		\draw [style=reddashedline] (39.center) to (40.center);
		\draw [style=reddashedline] (41.center) to (42.center);
		\draw [style=reddashedline] (39.center) to (55.center);
		\draw [style=reddashedline] (40.center) to (56.center);
		\draw [style=reddashedline] (42.center) to (57.center);
		\draw [style=reddashedline] (41.center) to (58.center);
		\draw [style=reddashedline] (59.center) to (39.center);
		\draw [style=reddashedline] (40.center) to (60.center);
		\draw [style=reddashedline] (42.center) to (62.center);
		\draw [style=reddashedline] (61.center) to (41.center);
		\draw (63.center) to (66.center);
		\draw (66.center) to (65.center);
		\draw (64.center) to (65.center);
		\draw (64.center) to (63.center);
		\draw [style=greenline] (69.center) to (67.center);
		\draw [style=greenline] (67.center) to (70.center);
		\draw [style=greenline] (68.center) to (70.center);
		\draw [style=greenline] (69.center) to (68.center);
		\draw [style=reddashedline] (67.center) to (68.center);
		\draw [style=arrowline] (96.center) to (95.center);
		\draw [style=arrowline] (97.center) to (95.center);
		\draw [style=arrowline] (98.center) to (95.center);
	\end{pgfonlayer}
\end{tikzpicture}
    \caption{Configuration of two duality defects placed in close proximity. (a) shows two parallel duality defects, with the red dashed lines indicating dual Ising interactions localized on a plane, while elsewhere the system hosts Ising gauge theory interactions. (b) depicts the corresponding local Ising interaction.}\label{fig:3Ddefect_configuration}
\end{figure}
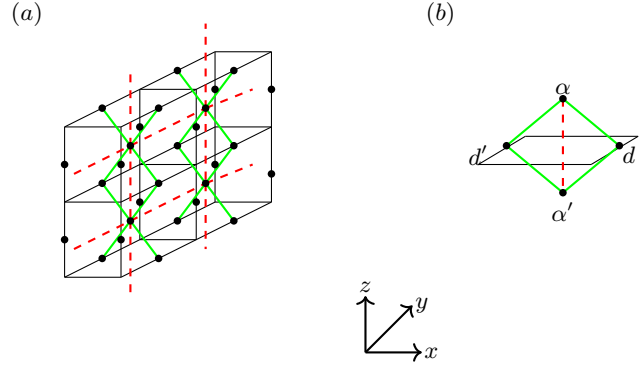

\begin{figure}
    \centering
    \input{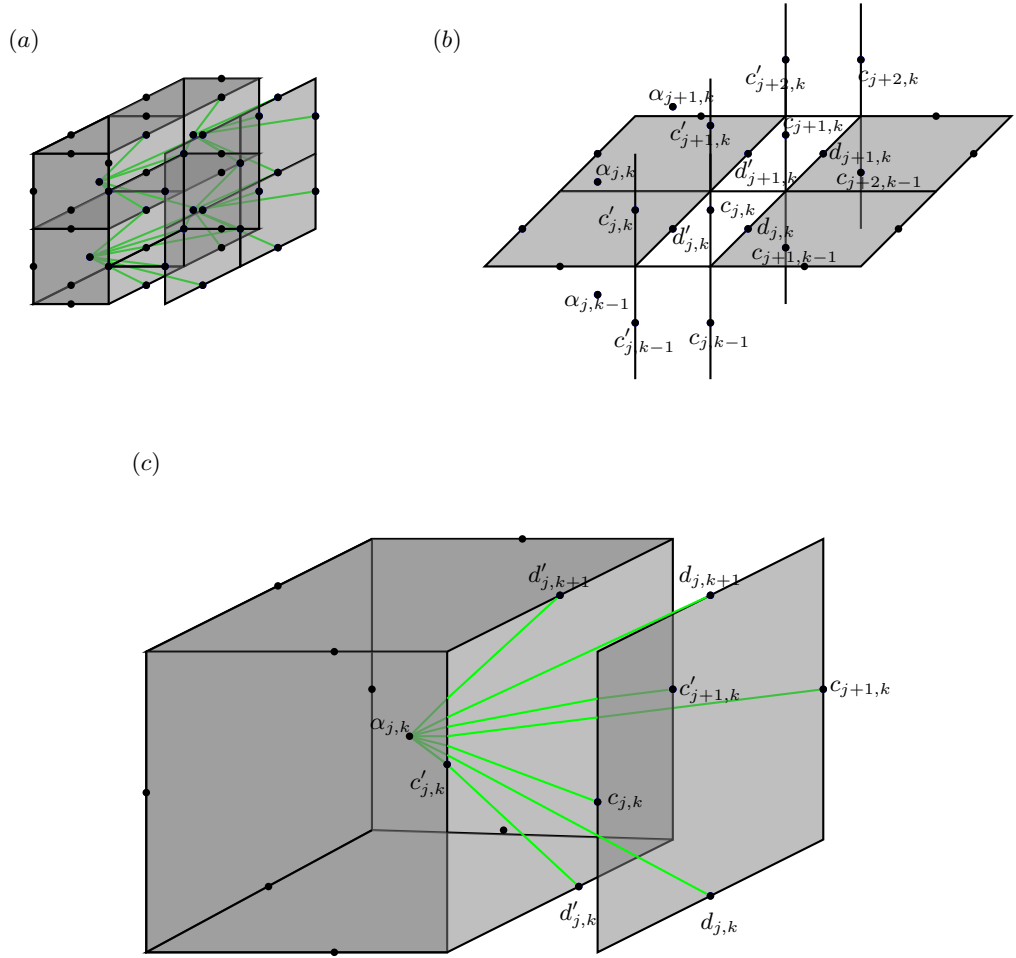}
    \caption{Configuration of spins after applying the movement relations to the setup in Figure~\ref{fig:3Ddefect_configuration}. (a) shows the full configuration, (b) defines the labels for edges and cube centers, and (c) zooms into a single cube.}
    \label{fig:3Ddefect_fusion}
\end{figure}

\begin{figure}
    \centering
    \input{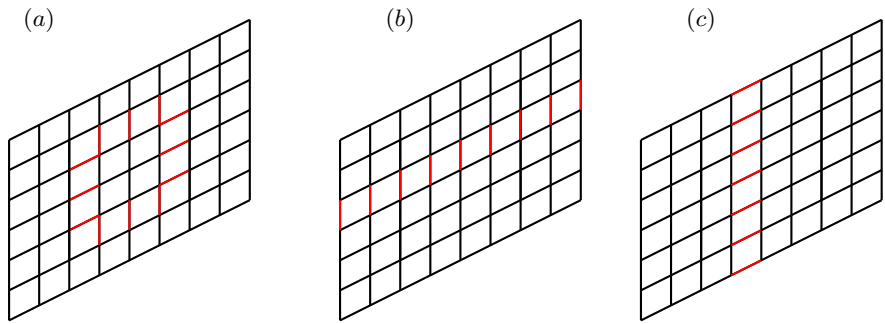}
    \caption{Closed-loop configurations where spins get a relative flip, for instance $c'_{j,k} = c_{j,k} + 1$ where $c'_{j,k},c_{j,k}$ are from fig \ref{fig:3Ddefect_fusion}, as a result of fusing two duality defects. (a) corresponds to a contractible loop, while (b) and (c) represent non-contractible loops of the $\mathbb{Z}_2$ one-form symmetry that wind around the $x$ and $y$ directions, respectively.}
    \label{fig:3Dcondensation_defect_loop_configuration}
\end{figure}

Now let us examine the situation where two duality defects are placed close to each other. Consider the configuration of two defects located parallel to each other along the $xz$ or $yz$ direction, as illustrated in Figure~\ref{fig:3Ddefect_configuration}.
We now apply the relation \eqref{eq:3Dmovementid2} on all dual Ising interactions (indicated by the red dashed lines) in order to fuse the two defects. This procedure yields the configuration shown in Figure~\ref{fig:3Ddefect_fusion}, and leads to the following sum:
\begin{align}
    \sum_{\{\alpha_{j,k}\}}(-1)^{\sum_{j,k}h_{\alpha_{j,k}}\left(h_{d_{j,k}}+h_{d'_{j,k}}+h_{d_{j,k+1}}+h_{d'_{j,k+1}}+h_{c_{j,k}}+h_{c'_{j,k}}+h_{c_{j+1,k}}+h_{c'_{j+1,k}}\right)}\,.
    \label{eq:3DIG_constraint_sum}
\end{align}
The above sum is non-zero only when $h_{d_{j,k}}+h_{d'_{j,k}}+h_{d_{j,k+1}}+h_{d'_{j,k+1}}+h_{c_{j,k}}+h_{c'_{j,k}}+h_{c_{j+1,k}}+h_{c'_{j+1,k}}=0$ mod $2$. This can be satisfied when $h_{d_{j,k}}=h_{d'_{j,k}}+1$ mod $2$ and $h_{c_{j,k}}=h_{c'_{j,k}}+1$ mod $2$, i.e., a relative spin flip, on edges that form a closed loop on the dual plane as shown in for example in Figure~\ref{fig:3Dcondensation_defect_loop_configuration}.

Let us assume the fusion occurs on the $yz$ plane. Without loss of generality, we fix the $x$-coordinate to $x = i_0$. The vertices on this plane have coordinates $(i_0, j, k)$, with $j \in \mathbb{Z}_{L_y}$ and $k \in \mathbb{Z}_{L_z}$. The edges on the $yz$ plane incident on vertex $\mathrm{v} = (i_0, j, k)$ are denoted by $\tilde{\partial}\mathrm{v}$, and explicitly given by $(i_0, j \pm \frac{1}{2}, k)$ and $(i_0, j, k \pm \frac{1}{2})$.

Denote by $\chi_{\mathrm{e}}$ the relative spin flip operation on edge $\mathrm{e}$, i.e., $\chi_{\rm e}h_{d_{i,j}}=h_{d_{i,j}}+1$ or $\chi_{\rm e}h_{c_{i,j}}=h_{c_{i,j}}+1$ depending on $\rm e$ is a $y$-directed edge or a $z$-directed edge. The relative spin flip operation on the loop $C_{\mathrm{v}} \equiv \tilde{\partial}\mathrm{v}$ ($\rm v$ is a vertex on the $yz$ plane where the fusion happens) is then given by $ \prod\limits_{\rm e\in \tilde{\partial}\rm v}\chi_{\rm e}$. Let $C_y$ and $C_z$ denote non-contractible loops on the dual plane winding around the $y$ and $z$ directions, respectively. Let $\Delta_{\rm v}^{x=i_0}$ be the set of vertices on the $yz$ plane at $x=i_0$ and  $\mathrm{V}\subset\Delta_{\rm v}^{x=i_0}$ be subset of it. The solutions to the constraint coming from the sum~\eqref{eq:3DIG_constraint_sum} can be expressed in terms of the relative spin flip operation, the set $\rm V$ and the numbers $c_y,c_z=0,1$.
\begin{align}
    h'_{d_{i,j}}=\prod_{\substack{\rm v\in V\\
    \mathrm{V}\in\Delta_{\rm v}^{x=i_0}}}\prod_{\rm e\in\tilde{\partial}v}\chi_{\rm e}\left(\prod_{\mathrm{e}\in C_y}\chi_{\rm e}\right)^{c_y}\left(\prod_{\mathrm{e}\in C_z}\chi_{\rm e}\right)^{c_z}h_{d_{i,j}}\,,\nonumber\\
    h'_{c_{i,j}}=\prod_{\substack{\rm v\in V\\
    \mathrm{V}\in\Delta_{\rm v}^{x=i_0}}}\prod_{\rm e\in\tilde{\partial}v}\chi_{\rm e}\left(\prod_{\mathrm{e}\in C_y}\chi_{\rm e}\right)^{c_y}\left(\prod_{\mathrm{e}\in C_z}\chi_{\rm e}\right)^{c_z}h_{c_{i,j}}\,.
\end{align}
It once we impose this and express everything in variables $h_{d_{i,j}}$, this essentially creates one form $\mathbb{Z}_2$ symmetry defects along the cycles $C_y$ and $C_z$.
Let $\mathcal{D}$ denote the duality defect, and $D_{C_{\mathrm{v}}}$, $D_{C_y}$, and $D_{C_z}$ denote the spin-flip defects along the loops $C_{\mathrm{v}}$, $C_y$, and $C_z$\footnote{The partition function with spin-flip defects along a loop is obtained by changing the coupling $J$ to $-J$ for plaquettes with centers lying at $x > i_0$ that share an edge with the loop.}. Then we obtain the fusion rule\footnote{We note that fusing $\mathcal{D}$ in the opposite way such that we get the 3D Ising interaction in the bulk  produces the fusion rule $\mathcal{D}\times \mathcal{D}\propto \mathrm{I}+D_{\eta}$ where $D_{\eta}$ is the defect of the 0-form symmetry in the 3D Ising model. $D_{\eta}$ is obtained by flipping the coupling of the 3D Ising along the $x$-directed edges on the $yz$ plane where the fusion takes place. We thank Arkya Chatterjee for pointing this out.}
\begin{align}
    \mathcal{D}\times \mathcal{D}=\frac{\sqrt{2}}{2^{L_yL_z}}\prod_{\rm v}(1+D_{\rm v})\times (1+D_{C_y})(1+D_{C_z})\,.
\end{align}

\subsection{Quantum Hamiltonian from partition function}
\subsubsection{Quantum Hamiltonian for $\mathbb{Z}_2$ gauge theory}
In order to understand the quantum Hamiltonian with a duality interface between $\mathbb{Z}_2$ gauge theory and Ising Model, it is fruitful to first review quantum Hamiltonian of $\mathbb{Z}_2$ gauge theory without any defects inserted. Let's start from the classical statistical Hamiltonian. 
\begin{align}
    \mathcal{H}_{\text{3D-IG}}(\{s_{\rm e}\})=-J\sum_{\rm p\in\Delta_{\rm p}}\prod_{\rm e\in\partial p}s_{\rm e}\, .
\end{align}
This Hamiltonian is invariant under a gauge transformation $g_{\rm v}$, which can be defined at each vertex $\rm v$ as $s_{\rm e} \to -s_{\rm e}$ $\forall$ ${\rm e} \in \partial^*{\rm v}$. It is clear that applying $g_{\rm v}$ forms a $\mathbb{Z}_2$ group under composition. The group $G$ of all gauge transformations that we can perform on the lattice is a direct product of $\mathbb{Z}_2$ for all vertices, i.e., $\mathbb{Z}_2^{\Delta_{\rm v}} \equiv \bigotimes_{\rm v} \mathbb{Z}_2$. The set of all operators that we can insert into the correlation functions can be decomposed into a one-dimensional representation of this group $G$. Without loss of generality, we will consider operators that transform in this one-dimensional representation of $G$. In this representation, a non-trivial transformation of an operator is $\mathcal{O} \to -\mathcal{O}$. Consider a correlation function $\langle \mathcal{O} \rangle$ where $\mathcal{O}$ could be a product of local and non-local operators. We now show that if $\mathcal{O}$ transforms non-trivially under the gauge transformations, then the correlation function $\langle \mathcal{O} \rangle$ vanishes. We start with the definition of $\langle \mathcal{O} \rangle$ 
\begin{align}
\langle \mathcal{O}\rangle = \frac{\sum_{\{s_e\}} \mathcal{O} e^{-H} }{\sum_{\{s_e\}} e^{-H} }\,.
\end{align}
To perform the gauge transformation, we essentially transform the variables $s_{\rm e} \to - s_{\rm e}$, $\mathrm{e} \in \partial^*{\rm v}$ on desired vertices $\rm v$'s. The correlation function should remain the same under this, since it is just a change of variable of spins $s_{\rm e}$ which are summed. Moreover since the Hamiltonian is also invariant under gauge transformation, we essentially get $\langle \mathcal{O} \rangle = -\langle \mathcal{O} \rangle$, i.e., $\langle \mathcal{O} \rangle = 0$. This means that we can restrict ourselves to operators that are gauge invariant, without loss of generality.

In order to define a Hilbert space, let's choose $z$ direction as the temporal direction and $x$-$y$ plane as the spatial plane. In what follows, we will assume periodic boundary conditions in all directions. Let's denote the coordinates of each vertex as $(x,y,z)$ where $x \in \{1,2,...L_x\}$, $y \in \{1,2,...L_y\}$ and $z \in \{1,2,...L_z\}$ where $1$ unit represents unit lattice spacing. To denote a link, we use the coordinates of its midpoint, for example, $\mathrm{e}=(x+\frac{1}{2},y,z)$, $\mathrm{e}=(x,y+\frac{1}{2},z)$, or $\mathrm{e}=(x,y,z+\frac{1}{2})$. For each vertex with coordinates $(x,y,0)$ in the $x$-$y$ plane, we will refer the set of $z$ directed links with coordinates $(x,y,z +\frac{1}{2})$ as the $z$-cycle at $v=(x,y)$. Using the gauge freedom, we can set $s_{(x,y,z +\frac{1}{2})}=1$ for all $z \in \{1,2,...L_z-1\}$ but not for $z = L_z$. The last link at $(x,y,L_z + \frac{1}{2})$ can't be set to $1$ because the product $\prod_{z} s_{(x,y,z +\frac{1}{2})}$ of all links in the $z$-cycle at any vertex $v=(x,y)$, is gauge invariant and represents the $\mathbb{Z}_2$ holonomy of the gauge field on that cycle. The choice $s_{(x,y,z +\frac{1}{2})}=1$ for all $z \in \{1,2,...L_z-1\}$ is called the temporal gauge choice. 

With the temporal gauge choice, the last $z$-directed link $(x,y,L_z + \frac{1}{2})$ can take two values $\{-1,1\}$ depending on the holonomy. We will see that presence of this last $z$-directed link will essentially impose Gauss law constraint on our Hilbert space. After choosing temporal gauge, we are still left with residual gauge transformations. For a fixed $x,y$, we can apply the gauge transformation $g_{\rm v}$ simultaneously at each vertex ${\rm v} = (x,y,z)$ for all $z \in \{1,2,...L_z\}$ and it preserves the temporal gauge. We will see that this residual gauge symmetry implies that the Gauss law will commute with quantum Hamiltonian.

We will work in the temporal gauge in what follows. Let $\{s_{\rm e}\}_z$ and $\{s_{\rm e}\}_{z+1}$ denote the spin configurations on a spatial plane at $z$ and $z+1$ respectively. We can define a state $\ket{\{s_{\rm e}\}_z}$ and define corresponding Pauli operators $Z_e,X_e$ where $\ket{\{s_{\rm e}\}_z}$ is an eigenstate of $Z_e$ with eigenvalue $s_{\rm e}$. Let's denote the corresponding Hilbert space which is the linear span of states $\ket{\{s_{\rm e}\}_z}$ as $\bar{\mathcal{H}}$. We define the transfer matrix as
\begin{align}\label{eq:TMatrixZ2GT}
&\bra{\{s_{\rm e}\}_{z+1}}\hat{\mathrm{T}}\ket{\{s_{\rm e}\}_z} = \exp\left(\frac{1}{2} J\sum_{x,y} \prod_{{\rm e} \in \partial\mathrm{p}_{\left(x,y,z\right)}}s_{\rm e} + \frac{1}{2}J\sum_{x,y}\prod_{{\rm e} \in \partial\mathrm{p}_{(x,y,z+1)}}s_{\rm e}\right.\nonumber\\
&\left.\hspace{5cm}+ J\sum_{x,y}s_{(x,y+\frac{1}{2},z)}s_{(x,y+\frac{1}{2},z+1)} + J\sum_{x,y}s_{(x+\frac{1}{2},y,z)}s_{(x+\frac{1}{2},y,z+1)} \right)\,,
\end{align}
where $\mathrm{p}_{(x,y,z)}$ denotes the plaquette with the center $\left(x+\frac{1}{2} , y+\frac{1}{2}, z\right)$. Consider the $z$-directed link $s_{(x,y,L_z+\frac{1}{2})}$ for all $x,y$, where in the temporal gauge each of them represents the holonomies along their respective $z$-cycles. Let's denote the partition function as $Z_{V}$ where $V$ denotes the set of vertices $v = (x,y)$ for which their corresponding $z$-cycle has non-trivial holonomy. 
\begin{figure}
    \centering \begin{tikzpicture}
	\begin{pgfonlayer}{nodelayer}
		\node [style=none] (0) at (0, 2) {};
		\node [style=none] (1) at (-1.475, 1) {};
		\node [style=none] (2) at (-1.475, -1.5) {};
		\node [style=none] (3) at (0, -0.5) {};
		\node [style=none] (4) at (2, 2) {};
		\node [style=none] (5) at (2, -0.5) {};
		\node [style=none] (6) at (1.525, 3) {};
		\node [style=none] (7) at (1.525, 0.5) {};
		\node [style=none] (8) at (-2, 2) {};
		\node [style=none] (9) at (-2, -0.5) {};
		\node [style=none] (10) at (0, 2) {};

            \node [style=none] (I) at (-1.475, 1) {};
		\node [style=none] (J) at (-1.475, -1.5) {};
		\node [style=none] (B) at (0, -0.5) {};
		\node [style=none] (C) at (2, 2) {};
		\node [style=none] (D) at (2, -0.5) {};
		\node [style=none] (G) at (1.525, 3) {};
		\node [style=none] (H) at (1.525, 0.5) {};
		\node [style=none] (E) at (-2, 2) {};
		\node [style=none] (F) at (-2, -0.5) {};
		\node [style=none] (A) at (0, 2) {};
            \filldraw[fill=gray!30, opacity = 0.6] (A.center) -- (C.center) -- (D.center) -- (B.center) -- cycle;
            \filldraw[fill=gray!30, opacity = 0.6] (A.center) -- (G.center) -- (H.center) -- (B.center) -- cycle;
            \filldraw[fill=gray!30, opacity = 0.6] (A.center) -- (E.center) -- (F.center) -- (B.center) -- cycle;       
            \filldraw[fill=gray!30, opacity = 0.6] (A.center) -- (I.center) -- (J.center) -- (B.center) -- cycle;    
        
		\node [style=none] (11) at (0.25, 0.75) {$e$};
	\end{pgfonlayer}
	\begin{pgfonlayer}{edgelayer}
		\draw (1.center) to (0.center);
		\draw (2.center) to (3.center);
		\draw (0.center) to (3.center);
		\draw (0.center) to (4.center);
		\draw (3.center) to (5.center);
		\draw (0.center) to (6.center);
		\draw (7.center) to (3.center);
		\draw (8.center) to (0.center);
		\draw (9.center) to (3.center);
	\end{pgfonlayer}

\end{tikzpicture}
    \caption{Link $\mathrm{e}=\left(x,y,L_z+\frac{1}{2}\right)$ here shows the last link $z$-directed link in the temporal gauge which is equal to the holonomy on that $z$-cycle. If the holonomy is $-1$, then each of the four plaquette terms for $\mathrm{p} \in \partial^*\rm e$ essentially get a relative $-1$ sign compared to the case when the holonomy is trivial.}
    \label{fig:HolonomyGauss}
\end{figure}
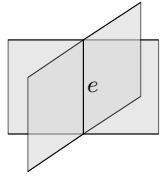
If we consider the case where all of them are $1$, i.e., $V = \phi$, then our partition function can be written as 
$Z_{\phi} = {\rm Tr}_{\bar{\mathcal{H}}}\left(\hat{\rm T}^{L_z}\right)$. Now consider the case $s_{(x,y,L_z+\frac{1}{2})} =-1$. Having this holonomy, essentially gives us a relative $-1$ sign, to each of the four plaquette terms for plaquettes $\mathrm{p} \in \partial^*\rm e$ for $\mathrm{e} = \left(x,y,L_z + \frac{1}{2}\right)$, compared to when the holonomy is trivial. We can mimic this relative sign in the partition function by inserting a Gauss Law operator $G_v = \prod_{e \in \partial^*v} X_e$ for $v = (x,y)$ in \eqref{eq:TMatrixZ2GT}, i.e., $Z_{(v)} = {\rm Tr}_{\bar{\mathcal{H}}}\left(G_{v}\hat{\rm T}^{L_z}\right)$. If we consider all the configurations of $s_{(x,y,L_z+\frac{1}{2})}$ for all $x,y$, i.e., considering the sum over the partition function for all possible holonomy combinations along $z$-cycles, we will get
\begin{align}
Z = \sum_{V}Z_{V} = {\rm Tr}_{\bar{\mathcal{H}}}\left(\prod_v (1+G_v) \hat{\rm T}^{L_z}\right)\,.
\end{align}
Note that the presence of operators $1+G_v$ imply that any state $\ket{\psi}$ which obeys $G_v \ket{\psi} = -\ket{\psi}$ will not contribute to the trace. We can restrict to the states which obey the Gauss Law condition \footnote{If we insert a Wilson loop inside the correlation function along the $z$-cycle passing through a vertex $v'=(x,y)$, then we will get $Z = \sum_{V}Z_{V} = {\rm Tr}_{\bar{\mathcal{H}}}\left((1-G_{v'})\prod_{v\neq v'} (1+G_v)  \hat{\rm T}^{L_z}\right)$. This will imply that our Hilbert space should be restricted to $G_{v'} \ket{\psi} = -\ket{\psi}$ and $G_v \ket{\psi} =  \ket{\psi}$ for $v \neq v'$.}
\begin{align}\label{eq:GaussCondition}
    G_v \ket{\psi} = \ket{\psi} \; \forall \; v\,.
\end{align}
Let's denote the Hilbert space made up of linear span of states obeying \eqref{eq:GaussCondition} as $\mathcal{H}$. We can write the partition function as
\begin{align}
    Z = 2^{L_x L_y}{\rm Tr}_{\mathcal{H}}\left(\hat{\rm T}^{L_z}\right).
\end{align}

Let's now analyze the consequences of the residual gauge transformations in the temporal gauge. For a point $v = (x,y)$ in the $x$-$y$ plane, consider the set $V_{ v} = \cup_z (x,y,z)$ of all points on the $z$-axis passing through $v$. The residual gauge transformation denoted by $\tilde{g}_v$ can be defined for each $v=(x,y)$ as
\begin{align}
\tilde{g}_{v=(x,y)}=\prod_{z}g_{\mathrm{v}=(x,y,z)}\,.
\end{align} 
It acts by flipping all the links $\mathrm{e} \in \cup_{{\rm v} \in V_{v}} \partial^*{\rm v}$. In terms of the states $\ket{\psi}$, this residual gauge acts as $G_{v}\ket{\psi}$. Let's look at the effect of applying $\tilde{g}_v$ in \eqref{eq:TMatrixZ2GT}. It can be seen that the right hand side of \eqref{eq:TMatrixZ2GT} is invariant under $\tilde{g}_v$. So we can write
\begin{align}
    \bra{\{s_{\rm e}\}_{z+1}}G_v^{\dagger}\hat{\mathrm{T}}G_v\ket{\{s_{\rm e}\}_z} = \bra{\{s_{\rm e}\}_{z+1}}\hat{\mathrm{T}}\ket{\{s_{\rm e}\}_z}\,,
\end{align}
and $G_v^{\dagger} = G_v^{-1} = G_v$. Since $\bra{\{s_{\rm e}\}_{z+1}}$ and $\ket{\{s_{\rm e}\}_{z}}$ are arbitrary, we get
\begin{align}
    G_v \hat{\rm T} = \hat{\rm T} G_v \;\forall \;v\,.
\end{align}
So, the transfer matrix $\hat{\rm T}$ commutes with the Gauss Law. Let's obtain the quantum Hamiltonian that we obtain by assuming the $L_z \to \infty$ and the spacing $\tau$ between two consecutive spins in $z$-direction goes to $0$. Again, here we work in the temporal gauge. First, we express the partition function as
\begin{align}
    Z_{IG} \propto \text{Tr}(\hat{\rm T}^{L_z}) 
\end{align}
and expand the transfer matrix
\begin{align}
    {\rm \hat{T}} = e^{-\tau \hat{\rm H}_{\rm IG}} \approx 1 - \tau \hat{\rm H}_{\rm IG}\,.
\end{align}
Doing the similar procedure that we did in 2D Ising, we get the quantum Hamiltonian as
\begin{align}
    \hat{\rm H}_{\rm IG} = -\lambda \sum_{\rm p}\prod_{{\rm e} \in \partial {\rm p}} Z_{\rm e} - \sum_{\rm e} X_{\rm e} =     \begin{tikzpicture}
	\begin{pgfonlayer}{nodelayer}
		\node [style=none] (0) at (-0.5, 0.65) {};
		\node [style=none] (1) at (-2, -0.75) {};
		\node [style=none] (2) at (1.75, 0.65) {};
		\node [style=none] (3) at (0.25, -0.75) {};
		\node [style=pointoperator, color = black] (4) at (0.525, 0.65) {};
		\node [style=pointoperator, color = black] (5) at (-1.15, 0.05) {};
		\node [style=pointoperator, color = black] (6) at (-0.725, -0.75) {};
		\node [style=pointoperator, color = black] (7) at (0.825, -0.2) {};
		\node [style=none] (8) at (-4.75, 0) {$-\lambda$};
		\node [style=none] (9) at (2.5, 0) {$-$};
		\node [style=none] (10) at (5, 0) {};
		\node [style=none] (11) at (7, 0) {};
		\node [style=none] (12) at (-3.25, 0) {$\sum_{\rm p}$};
		\node [style=none] (13) at (0, 0) {};
		\node [style=none] (14) at (6, 0) {};
		\node [style=pointoperator, color = black] (15) at (6, 0) {};
		\node [style=none] (16) at (6, -0.6) {$X_{\rm e}$};
		\node [style=none] (17) at (3.9, 0) {$\sum_{\rm e}$};
		\node [style=none] (18) at (0.5, 1.25) {$Z$};
		\node [style=none] (19) at (1.45, -0.2) {$Z$};
		\node [style=none] (20) at (-0.75, -1.25) {$Z$};
		\node [style=none] (21) at (-2, 0) {$Z$};
        \node [style=none] (21) at (0, 0) {$\rm p$};
	\end{pgfonlayer}
	\begin{pgfonlayer}{edgelayer}
		\draw [black!30,line width=1.0] (0.center) to (2.center);
		\draw [black!30,line width=1.0] (2.center) to (3.center);
		\draw [black!30,line width=1.0] (1.center) to (3.center);
		\draw [black!30,line width=1.0] (1.center) to (0.center);
		\draw [black!30,line width=1.0] (10.center) to (11.center);
	\end{pgfonlayer}
\end{tikzpicture}
\end{align}
where $J_{xy}=\lambda \tau$ and $e^{-2J_z}=\tau$. It is clear that the Hamiltonian commutes with gauss law $G_v$. 

\subsubsection{Quantum Hamiltonian from partition function with duality defect}
We consider the partition function of the three-dimensional Ising gauge theory and the three-dimensional Ising model, which are distinguished by the presence of a duality defect. The link degrees of freedom on which the Ising gauge theory is defined are denoted by
$\mathrm{e}=(x+\frac{1}{2},y,z)$, $\mathrm{e}=(x,y+\frac{1}{2},z)$, or $\mathrm{e}=(x,y,z+\frac{1}{2})$,
while the vertices of the dual lattice, where the Ising model lives, are indicated by
$\mathrm{v}=(x+\frac{1}{2},y+\frac{1}{2},z+\frac{1}{2})$.
We take the duality defect to lie within the $yz$ plane, specifically between $x=i_0$ and $x=i_0+\frac{1}{2}$.

The partition function of the system with a single defect on an infinite lattice is then given by
\begin{align}
    Z_{\text{IG}}^{\mathcal{D}}(J)&=\sum_{\bigl\{\substack{s_{\rm e}=\pm 1\, ,\\
    (-1)^{h_{\rm e}}=s_{\rm e}}\bigl\}\, ,\bigl\{\substack{s_{\rm c}=\pm 1\, ,\\
    (-1)^{h_{\rm c}}=s_{\rm c}}\bigl\}}\exp\left[J\sum_{x\leq i_0,y,z}\left(s_{(x,y+\frac{1}{2},z)}s_{(x,y+1,z+\frac{1}{2})}s_{(x,y,z+\frac{1}{2})}s_{(x,y+\frac{1}{2},z+1)}\right.\right.\nonumber\\
    &\left.\left.\hspace{1cm}+s_{(x-\frac{1}{2},y,z)}s_{(x-1,y,z+\frac{1}{2})}s_{(x,y,z+\frac{1}{2})}s_{(x-\frac{1}{2},y,z+1)}+s_{(x-\frac{1}{2},y,z)}s_{(x-1,y+\frac{1}{2},z)}s_{(x,y+\frac{1}{2},z)}s_{(x-\frac{1}{2},y+1,z)}\right)\right.\nonumber\\
    &\left.\hspace{1cm}+J^*\sum_{x\geq i_0,y,z}s_{(x+\frac{1}{2},y+\frac{1}{2},z+\frac{1}{2})}s_{(x+\frac{1}{2},y+\frac{1}{2},z+\frac{3}{2})}+s_{(x+\frac{1}{2},y+\frac{1}{2},z+\frac{1}{2})}s_{(x+\frac{1}{2},y+\frac{3}{2},z+\frac{1}{2})}+s_{(x+\frac{1}{2},y+\frac{1}{2},z+\frac{1}{2})}s_{(x+\frac{3}{2},y+\frac{1}{2},z+\frac{1}{2})}\right]\nonumber\\
    &\hspace{2cm}\times (-1)^{h_{(x+\frac{1}{2},y+\frac{1}{2},z+\frac{1}{2})}\left(h_{(x,y+\frac{1}{2},z)}+h_{(x,y,z+\frac{1}{2})}+h_{(x,y+1,z+\frac{1}{2})}+h_{(x,y+\frac{1}{2},z+1)}\right)}\,.
\end{align}
The 3D Ising gauge theory away from defect possesses a local $\mathbb{Z}_2$ gauge symmetry generated by flipping all six spins on the links adjacent to a given vertex, denoted by $g_{\rm v}$. This gauge redundancy allows one to impose gauge-fixing conditions. For instance, in the temporal gauge, all spins on links oriented along the $z$-direction can be set to $+1$, except for the final links at $(x,y,L_z-\frac{1}{2})$ that encode the $\mathbb{Z}_2$ holonomy along the $z$-direction. This holonomy enforces Gauss’s law, as in the defect-free Ising gauge theory discussed previously. Near the defect localized at $x=i_0$, the situation is slightly modified: for vertices $\mathrm{v}=(i_0,y,z)$ on the defect plane, only five links are incident. Nonetheless, flipping all five spins at such a vertex still leaves the partition function invariant. Hence, the local symmetry persists, and one can again fix all $z$-oriented links to $+1$ except for those at $(i_0,y,L_z-\frac{1}{2})$. However, summing over the holonomy now introduces a modified Gauss’s law term in the trace over the transfer matrix. If $s_{(x,y,L_z-\frac{1}{2})}=1$ for all links with $x\leq i_0$, the partition function is simply
$\mathrm{Tr}\left(\hat{\mathrm{T}}^{L_z}\right)$.
If instead $s_{(i_0,y,L_z-\frac{1}{2})}=-1$ for a single link (while all others are fixed to $+1$), the resulting partition function is
$\mathrm{Tr}\left(\tilde{G}_{(i_0,y)}\hat{\mathrm{T}}^{L_z}\right)$,
where $\tilde{G}_{(i_0,y)} = X_{(i_0,y+\frac{1}{2})} X_{(i_0-\frac{1}{2},y)} X_{(i_0,y-\frac{1}{2})}$ is the modified Gauss’s law operator near the defect. After imposing the temporal gauge and setting all holonomy degrees of freedom to $+1$, the partition function simplifies to
\begin{align}
    Z_{\text{IG}}^{\mathcal{D}}(J)&=\sum_{\bigl\{\substack{s_{\rm e}=\pm 1\, ,\\
    (-1)^{h_{\rm e}}=s_{\rm e}}\bigl\}\, ,\bigl\{\substack{s_{\rm c}=\pm 1\, ,\\
    (-1)^{h_{\rm c}}=s_{\rm c}}\bigl\}}\exp\left[J\sum_{x\leq i_0,y,z}\left(s_{(x,y+\frac{1}{2},z)}s_{(x,y+\frac{1}{2},z+1)}+s_{(x-\frac{1}{2},y,z)}s_{(x-\frac{1}{2},y,z+1)}\right.\right.\nonumber\\
    &\left.\left.\hspace{2cm}+s_{(x-\frac{1}{2},y,z)}s_{(x-1,y+\frac{1}{2},z)}s_{(x,y+\frac{1}{2},z)}s_{(x-\frac{1}{2},y+1,z)}\right)+J^*\sum_{x\geq i_0,y,z}\left(s_{(x+\frac{1}{2},y+\frac{1}{2},z+\frac{1}{2})}s_{(x+\frac{1}{2},y+\frac{1}{2},z+\frac{3}{2})}\right.\right.\nonumber\\
    &\left.\left.+s_{(x+\frac{1}{2},y+\frac{1}{2},z+\frac{1}{2})}s_{(x+\frac{1}{2},y+\frac{3}{2},z+\frac{1}{2})}
    +s_{(x+\frac{1}{2},y+\frac{1}{2},z+\frac{1}{2})}s_{(x+\frac{3}{2},y+\frac{1}{2},z+\frac{1}{2})}\right) \right]\times (-1)^{h_{(x+\frac{1}{2},y+\frac{1}{2},z+\frac{1}{2})}\left(h_{(x,y+\frac{1}{2},z)}+h_{(x,y+\frac{1}{2},z+1)}\right)}\,.
\end{align}
We observe that the partition function remains invariant under the transformation
\begin{align}
\tilde{g}_{v=(x,y)} = \prod_z g_{\mathrm{v}=(x,y,z)},
\end{align}
which represents the residual gauge symmetry. This ensures that the Gauss’s law operator commutes with both the transfer matrix and the quantum Hamiltonian as in the defect free Ising gauge theory.

To obtain the corresponding quantum model, we construct the transfer matrix using the same strategy as before. Introducing anisotropic couplings $J_z$, $J_{xy}$, $J_z^*$, and $J_{xy}^*$, we take the $\tau$-continuum limit by defining
\begin{align}
Z_{\text{IG}}^{\mathcal{D}}(J) \propto \mathrm{Tr}\left(\hat{\mathrm{T}}^{L_z}\right),
\end{align}
and expanding
\begin{align}
\hat{\mathrm{T}} = e^{-\tau \hat{\mathrm{H}}_{\text{IG-TFI}}} \approx 1 - \tau \hat{\mathrm{H}}_{\text{IG-TFI}}.
\end{align}
The couplings are related as
\begin{align}
e^{-2J_z} = \tau,\quad J_{xy} = \lambda\tau,\quad e^{-2J_z^*} = \lambda\tau,\quad J_{xy}^* = \tau.
\end{align}
The resulting $(2+1)$D quantum Hamiltonian reads
\begin{align}
    \hat{\rm H}_{\text{IG-TFI}}=-\lambda\sum_{\substack{v=(x,y),\\
    x\leq i_0}}\raisebox{-10pt}{\begin{tikzpicture}
    \draw[-,black!30,line width=1.0]
(0,0)--(1,0)--(1.5,0.5)--(0.5,0.5)--(0,0);
        \node at (0.5,0) {$Z$};
        \node at (1.25,0.25) {$Z$};
        \node at (1,0.5) {$Z$};
        \node at (0.25,0.25) {$Z$};
        \node at (1,0) {$v$};
    \end{tikzpicture}}-\sum_{\substack{v=(x,y),\\
    x\leq i_0}}\left(\raisebox{-5pt}{\begin{tikzpicture}
    \draw[-,black!30,line width=1.0]
(0,0)--(1,0);
        \node at (0.5,0) {$X$};
        \node at (1,0) {$v$};
    \end{tikzpicture}}+\raisebox{-8pt}{\begin{tikzpicture}
    \draw[-,black!30,line width=1.0]
(0,0)--(0,1);
        \node at (0,0.5) {$X$};
        \node at (0,1) {$v$};
    \end{tikzpicture}}\right)-\sum_{\substack{v=(x+\frac{1}{2},y+\frac{1}{2}),\\
    x\geq i_0}}\raisebox{-15pt}{\begin{tikzpicture}
    \draw[dashed,black!30,line width=1.0]
(0,0)--(0,1);
        \node at (0,0) {$Z_{v}$};
        \node at (0,1) {$Z$};
    \end{tikzpicture}}-\lambda\sum_{\substack{v=(x+\frac{1}{2},y+\frac{1}{2}),\\
    x\geq i_0}} \raisebox{-5pt}{\begin{tikzpicture}
        \node at (0,0) {$X_{v}$};
    \end{tikzpicture}}-\sum_{v=(i_0,y)}\raisebox{-15pt}{\begin{tikzpicture}
    \draw[-,black!30,line width=1.0]
(0,0)--(0,1);
        \node at (0,0.5) {$X$};
        \node at (0,0) {$v$};
         \filldraw[black] (0.4,0.5) circle (1pt);
        \node at (0.5,0.5) {$Z$};
    \end{tikzpicture}}\,,
\end{align}
with the Gauss's law
\begin{align}
\raisebox{-10pt}{
    \begin{tikzpicture}
    \draw[-,black!30,line width=1.0]
(0,0)--(1,0)--(2,0);
\draw[-,black!30,line width=1.0](1.5,0.5)--(1,0)--(0.5,-0.5);
        \node at (0.5,0) {$X$};
        \node at (1.25,0.25) {$X$};
        \node at (1.5,0) {$X$};
        \node at (0.75,-0.25) {$X$};
        \node at (1,0) {$v$};
    \end{tikzpicture}}=1\,\qquad \text{ for } x<i_0 \,,\text{ and }\qquad\raisebox{-10pt}{\begin{tikzpicture}
    \draw[-,black!30,line width=1.0]
(0,0)--(1,0);
\draw[-,black!30,line width=1.0](1.5,0.5)--(1,0)--(0.5,-0.5);
        \node at (0.5,0) {$X$};
        \node at (1.25,0.25) {$X$};
        \node at (0.75,-0.25) {$X$};
        \filldraw[black] (1.75,0.25) circle (1pt);
        \filldraw[black] (1.25,-0.25) circle (1pt);
        \node at (1,0) {$v$};
    \end{tikzpicture}}=1\qquad \text{ for } x=i_0 \,.
\end{align}
\subsubsection{Quantum Hamiltonian with condensation defect}
Here, we examine the quantum Hamiltonian with condensation defect. We start with the classical statistical model with two duality defects just before the fusion as in Figure~\ref{fig:3Ddefect_fusion}. Using the local symmetry, we fix the temporal links($z$-directed links) to $+1$ except at $z=L_z-\frac{1}{2}$. Now, the sum over the spins $s_{\alpha_{i,j}}$ located at the cube centers is
\begin{align}
    \sum_{\{s_{\alpha_{j,k}}\}}(-1)^{\sum\limits_{k\neq L_z, j}h_{\alpha_{j,k}}\left(h_{d_{j,k}}+h_{d'_{j,k}}+h_{d_{j,k+1}}+h_{d'_{j,k+1}}\right)}(-1)^{h_{\alpha_{j,L_z}}\left(h_{d_{j,L_z}}+h_{d'_{j,L_z}}+h_{d_{j,1}}+h_{d'_{j,1}}+h_{c_{j,L_z}}+h_{c'_{j,L_z}}+h_{c_{j+1,L_z}}+h_{c'_{j+1,L_z}}\right)}\,.
    \label{eq:sumoveralpha}
\end{align}
This sum is non-zero when
\begin{enumerate}
    \item $h_{d'_{j,k}}=h_{d_{j,k}}$ $\forall\, k$ or $h_{d'_{j,k}}=h_{d_{j,k}}+1$ $\forall\, k$\,. This correspond to spin flip along edges of the form $\mathrm{e}=(i_0,j+\frac{1}{2},k)$ $\forall\, k$ and some particular set of $j$ values. For a fixed $j$ and varying values of $k$, this is a line ( in the $z$-direction ) of edges on the plane where the fusion takes place. Let us define one such line as a set of edges: $C^z_{y=j}=\{(i_0,j+\frac{1}{2},k)|k\in\{1,...,L_z\}\}$.
    \item $h_{c'_{j,L_z}}=h_{c_{j,L_z}}$ $\forall\, j$ or $h_{c'_{j,L_z}}=h_{c_{j,L_z}}+1$ $\forall\, j$\,. This correspond to spin flip along edges of the form $\mathrm{e}=(i_0,j,L_z+\frac{1}{2})$ $\forall\, j$. This again is a line ( in the $y$-direction ) of edges on the plane where the fusion takes place. Let us define this line as a set of edges: $C^y_{z=L_z}=\{(i_0,j,L_z+\frac{1}{2})|j\in\{1,...,L_y\}$.
\end{enumerate}
Performing the sum in \eqref{eq:sumoveralpha} results in the fusion of duality defects. If we denote the duality defect by $\mathcal{D}$ and the spin flip defects along the line of edges $C^z_{y=j}$ and $C^y_{z=L_z}$ by $D_{C^z_{y=j}}$ and $D_{C^y_{z=L_z}}$, we have
\begin{align}
    \mathcal{D}^2=\frac{\sqrt{2}}{2^{L_yL_z}}\prod_{j=1}^{L_y}(1+D_{C^z_{y=j}})\times (1+D_{C^y_{z=L_z}})\,.
\end{align}
The above equation should be thought of as relation between gauge fixed (temporal gauge) Ising gauge theory partition function with the defects inserted. 

Like in the 2D Ising model example \eqref{eq:TFID^2}, the sum over vertical spin flip defects ($D_{C^z_{y=j}}$) can be incorporated into the transfer matrix by introducing an ancilla at the plaquette centers in quantum Hamiltonian. The horizontal spin flip defect can be thought of as imposing a non-local Gauss's law. 

Let us denote the partition function with two duality defects  inserted by $Z_{\mathcal{D}^2}$.  Let $\{s_{\rm e}\}_z$ and $\{s_{\rm e}\}_{z+1}$ be two spin configurations on spatial ($xy$ plane) at $z$ and $z+1$ and the Hilbert space formed by linear span of states $|\{s_e\}_z\rangle$ as $\bar{\mathcal{H}}$. Let $v=(x,y)$ be a vertex in the 2D spatial plane and $C^y=\{(x+\frac{1}{2},y)|y\in\mathbb{Z}_{L_y}\}$ be collection of edges that form a line in the dual lattice (in 2D plane) along the $y$-direction. We define the Gauss's law operator $G_v=\prod_{e\in\partial^*v}X_e$ and $G_{C^y}=\prod_{e\in C^y}X_e$. Then
\begin{align}
    Z_{\mathcal{D}^2}=\frac{\sqrt{2}}{2^{L_yL_z}}\text{Tr}_{\bar{\mathcal{H}}}\left((1+G_{C^y})\prod_v(1+G_v)\hat{\rm T}^{L_z}\right)\,.
\end{align}
Now let us impose the Gauss's law on the states $\ket{\Psi}\in\bar{\mathcal{H}}$
\begin{align}
    G_v\ket{\Psi}=\ket{\Psi}\,,\qquad G_{C^y}\ket{\Psi}=\ket{\Psi}\,.
\end{align}
We restrict our states to Gauss's law invariant state\footnote{Any state that is not invariant under the Gauss's law will not contribute to the partition function.} and define this restricted Hilbert space to be $\mathcal{H}$. Then, we have
\begin{align}
    Z_{\mathcal{D}^2}=\frac{\sqrt{2}}{2^{L_yL_z}}2^{L_xL_y}\text{Tr}_{\mathcal{H}}\left(\hat{\rm T}^{L_z}\right)\,.
\end{align}
The factors that appear in front of the trace in the above equation would change the Hamiltonian by a constant, and hence it is safe to ignore this. In the $\tau$-continuum limit, we can write the transfer matrix as
\begin{align}
    \hat{\rm T}=e^{-\tau\hat{\rm H}_{\text{IG-Cond}}}\approx 1-\tau \hat{\rm H}_{\text{IG-Cond}}\,.
\end{align}
The $2+1$D quantum Hamiltonian with the condensation defect is
\begin{align}
    \hat{\rm H}_{\text{IG-Cond}}=-\lambda \sum_{\substack{v=(x,y)\\
    x\neq i_0}}\raisebox{-10pt}{\begin{tikzpicture}
    \draw[-,black!30,line width=1.0]
(0,0)--(1,0)--(1.5,0.5)--(0.5,0.5)--(0,0);
        \node at (0.5,0) {$Z$};
        \node at (1.25,0.25) {$Z$};
        \node at (1,0.5) {$Z$};
        \node at (0.25,0.25) {$Z$};
        \node at (0,0) {$v$};
    \end{tikzpicture}}-\lambda \sum_{v=(i_0,y)}\raisebox{-10pt}{\begin{tikzpicture}
    \draw[-,black!30,line width=1.0]
(0,0)--(1,0)--(1.5,0.5)--(0.5,0.5)--(0,0);
        \node at (0.5,0) {$Z$};
        \node at (1.25,0.25) {$Z$};
        \node at (1,0.5) {$Z$};
        \node at (0.25,0.25) {$Z$};
        \node at (0.75,0.25) {$Z$};
        \node at (0.75,0.25)[circle,fill,inner sep=1.5pt]{};
        \node at (0,0) {$v$};
    \end{tikzpicture}}-\sum_{v=(x,y)}\left(\raisebox{-5pt}{\begin{tikzpicture}
    \draw[-,black!30,line width=1.0]
(0,0)--(1,0);
        \node at (0.5,0) {$X$};
        \node at (1,0) {$v$};
    \end{tikzpicture}}+\raisebox{-8pt}{\begin{tikzpicture}
    \draw[-,black!30,line width=1.0]
(0,0)--(0,1);
        \node at (0,0.5) {$X$};
        \node at (0,1) {$v$};
    \end{tikzpicture}}\right)\,,
\end{align}
with the Gauss's law
\begin{align}
\raisebox{-10pt}{
    \begin{tikzpicture}
    \draw[-,black!30,line width=1.0]
(0,0)--(1,0)--(2,0);
\draw[-,black!30,line width=1.0](1.5,0.5)--(1,0)--(0.5,-0.5);
        \node at (0.5,0) {$X$};
        \node at (1.25,0.25) {$X$};
        \node at (1.5,0) {$X$};
        \node at (0.75,-0.25) {$X$};
        \node at (1,0) {$v$};
    \end{tikzpicture}}=1\,, \qquad \raisebox{-30pt}{
    \begin{tikzpicture}
    \draw[-,black!30,line width=1.0]
(0,0)--(1,0);
\draw[-,black!30,line width=1.0]
(-0.5,-0.5)--(0.5,-0.5);
\draw[-,black!30,line width=1.0]
(0.5,0.5)--(1.5,0.5);
\draw[-,black!30,line width=1.0](1,1)--(2,1);
        \node at (0.5,0) {$X$};
        \node at (0,-0.5) {$X$};
        \node at (1,0.5) {$X$};
        \node at (1.5,1) {$X$};
        \node at (1.75,1.25)[circle,fill,inner sep=0.5pt]{};
        \node at (2,1.5)[circle,fill,inner sep=0.5pt]{};
        \node at (2.25,1.75)[circle,fill,inner sep=0.5pt]{};
        \node at (-0.25,-0.75)[circle,fill,inner sep=0.5pt]{};
        \node at (-0.5,-1)[circle,fill,inner sep=0.5pt]{};
        \node at (-0.75,-1.25)[circle,fill,inner sep=0.5pt]{};
    \end{tikzpicture}}=1\,.
\end{align}
This shows that in the presence of condensation defect, the defect Hilbert space is stabilized under the $1$-form symmetry operator from which the condensation defect is made up of.
\section{Kramers-Wannier defect between generalized Ising models}\label{sec:KWdefectgenIsingmodel}
\subsection{Kramers-Wannier duality in generalized Ising models}
Let us extend the preceding construction to encompass generalized Ising models, employing the chain complex formalism. Hence, we shall examine the chain complex relevant to CSS codes herein. A comprehensive examination of this formalism is available in \cite{Okuda:2024azp}. We denote abstract symbols $\sigma_i$ to the qubits $i$ presented above, and their collective set is denoted by $\Delta_{\text{q}} = \{ \sigma_i\}_{i=1,...,n} $. Similarly, let $\Delta_{X} = \{ \sigma_\alpha\}_{\alpha=1,...,|\mathcal{S}_X|} $ ($\Delta_{Z} = \{ \sigma_\beta\}_{\beta=1,...,|\mathcal{S}_Z|} $) represent the set of abstract symbols apportioned to the stabilizers $A_\alpha$ ($B_\beta$). In the instance where the CSS code constitutes the toric code as elaborated in ~\eqref{eq:dualtoriccodeham}, $\sigma_\alpha\in \Delta_X$ represents a vertex, $\sigma_i\in\Delta_{\text{q}}$ corresponds to an edge, and $\sigma_\beta\in\Delta_Z$ signifies a face, all situated within a square lattice. We use $C_k$ ($k=\text{q},X,Z$) to denote the group of chains $c_k$ with $\mathbb{Z}_2$ coefficients, that is, the formal linear combinations.
\begin{align}
c_k = \sum_{\sigma \in \Delta_k} a(c_k ; \sigma) \sigma \, ,
\label{eq:chaindef}
\end{align}
with $a(c_k ; \sigma) = \{ 0,1 \text{ mod }2\}$. As a useful notation, for a chain $c_k \in C_k$, we write
\begin{align} \label{eq:def-pauli-chain}
P(c_k) = \prod_{\sigma \in \Delta_k} P(\sigma)^{a(c_k;\sigma)} \, ,
\end{align}
with $P(\sigma)$ a single-qubit Pauli or Hadamard operator acting on the qubit at $\sigma \in \Delta_k$. Similarly, we write
\begin{align}
    s(c_k)=\prod_{\sigma \in \Delta_k}s_{\sigma}^{a(c_k;\sigma)}\,,
\end{align}
where $s_{\sigma}=\pm 1$ is the spin at cell $\sigma$.
Let us define the intersection of a chain $c_k=\sum_{\sigma\in\Delta_k} a(c;\sigma)\sigma$ with a set $\Delta_l$ as
\begin{align}
    c_k\sqcap \Delta_l=\left(\sum_{\sigma\in\Delta_k} a(c;\sigma)\sigma\right)\sqcap \Delta_l=\sum_{\sigma\in\Delta_k\cap\Delta_l}a(c;\sigma)\sigma\,.
\end{align}
For the chain $c_k$, we define it's set $\mathsf{S}(c_k)$ as
\begin{align}
    \mathsf{S}(c_k)=\{\sigma\in\Delta_k|a(c;\sigma)\neq 0\}\,.
\end{align}

A CSS code can be specified by the following chain complex~\cite{Kitaev_2003,freedman2001projective,2007JMP....48e2105B}(see also~\cite{2021PRXQ....2d0101B,Kubica:2018lhn}):
\begin{align}
0 \overset{\delta}{\longrightarrow} C_Z\overset{\delta_Z}{\longrightarrow} C_\text{q} \overset{\delta_X}{\longrightarrow} C_X \overset{\delta}{\longrightarrow} 0 \, .
\label{eq:CSS-complex}
\end{align}
The nilpotency condition $\delta_X \circ \delta_Z = 0$ is equivalent to the commutativity of the stabilizers.

Let us consider the chain complex in \eqref{eq:CSS-complex} and write down a cluster state given below 
\begin{subequations}
    \begin{align}
    &|\Psi_{\mathcal{C}}\rangle=\mathcal{U}_{CZ}\ket{+}^{\Delta_{\rm q}\cup\Delta_X}\,, \quad\\ &\mathcal{U}_{CZ}=\prod_{\substack{\sigma_{\rm q}\in\Delta_{\rm q}\,,\\
    \sigma_X\in\Delta_X}}CZ_{\sigma_{\rm q},\sigma_X}^{a(\delta\sigma_{\rm q};\sigma_X)}\,.
\end{align}
\end{subequations}
The cluster state is a short range entangled state where nearby qubits in the cell $\sigma_{\rm q}$ is entangled with nearby qubits in the cell $\sigma_{X}$. Now, consider the product state defined below:
\begin{align}
    \langle\Omega(\{J(\sigma_{\rm q})\})|=\bigotimes_{\sigma_X\in\Delta_X}\bra{+}_{\sigma_X}\bigotimes_{\sigma_{\rm q}\in\Delta_{\rm q}}\bra{0}_{\sigma_{\rm q}}e^{J(\sigma_{\rm q})X(\sigma_{\rm q})}\,.
\end{align}
Taking an overlap of the cluster state with the product state  produces the partition function of a an Ising model defined by the boundary map $\delta_{X}$
\begin{align}
    \bra{\Omega(\{J(\sigma_{\rm q})\})}\Psi_{\mathcal{C}}\rangle=2^{-|\Delta_X|}2^{-|\Delta_{\rm q}|/2}\mathcal{Z}_X(\{J(\sigma_{\rm q})\})\,,
\end{align}
where
\begin{align}
    \mathcal{Z}_X(\{J(\sigma_{\rm q})\})=\sum_{\{s_{\sigma_X}=\pm 1\}_{\sigma_X\in \Delta_{X}}}\exp\left(\sum_{\sigma_{\rm q}\in \Delta_{\rm q}}J(\sigma_{\rm q})s(\delta_X \sigma_{\rm q})\right)\,.
\end{align}
The above partition function is Kramers-Wannier dual to an Ising model model defined by the dual boundary map $\delta_Z^*$
\begin{align}
    \mathcal{Z}_{Z}(\{J(\sigma_{\rm q})\})=\sum_{\{s_{\sigma_Z}=\pm 1\}_{\sigma_Z\in\Delta_Z}}\exp\left(\sum_{\sigma_{\rm q}\in\Delta_{\rm q}}J(\sigma_{\rm q})s(\delta^*_Z\sigma_{\rm q})\right)\,,
\end{align}
and can be obtained by taking the overlap
\begin{align}
    \bra{\tilde{\Omega}(\{J(\sigma_{\rm q})\})}\Psi_{\mathcal{C}}^*\rangle=2^{-|\Delta_Z|}2^{-|\Delta_{\rm q}|/2}\mathcal{Z}_X(\{J(\sigma_{\rm q})\})\,,
\end{align}
where
\begin{subequations}
\begin{align}
    \bra{\tilde{\Omega}(\{J(\sigma_{\rm q})\})}&=\bigotimes_{\sigma_Z\in\Delta_Z}\bra{+}_{\sigma_Z}\bigotimes_{\sigma_{\rm q}\in\Delta_{\rm q}}\bra{0}_{\sigma_{\rm q}}e^{J(\sigma_{\rm q})X(\sigma_{\rm q})}\,,\\
    |\Psi_{\mathcal{C}}^*\rangle&=\prod_{\substack{\sigma_{\rm q}\in\Delta_{\rm q}\,,\\
    \sigma_Z\in\Delta_Z}}CZ_{\sigma_{\rm q},\sigma_Z}^{a(\delta\sigma_{\rm q};\sigma_Z)}\ket{+}^{\Delta_{\rm q}\cup\Delta_Z}\,.
\end{align}
\end{subequations}
The duality explicitly is given by
\begin{align}
    \mathcal{Z}_{X}(\{J(\sigma_{\rm q})\})=\frac{2^{\frac{|\Delta_X|}{2}}(\prod\limits_{\sigma_{\rm q}\in\Delta_{\rm q}}\sinh (2J(\sigma_{\rm q}))^{\frac{1}{2}})(|\text{Ker }\delta_X^*|)^{\frac{1}{2}}}{2^{\frac{|\Delta_Z|}{2}}(|\text{Ker }\delta_Z|)^{\frac{1}{2}}\sqrt{|\mathcal{L}|}}\sum_{[z_{\rm q}]\in \mathcal{L}}\mathcal{Z}^{\text{twisted}}_Z(\{J^*(\sigma_{\rm q})\};z_{\rm q})\,,
    \label{eq:ZXpartitionduality}
\end{align}
where
\begin{align}
   \mathcal{Z}^{\text{twisted}}_Z(\{J^*(\sigma_{\rm q})\};z_{\rm q})=\sum_{\{s_{\sigma_Z}=\pm 1\}_{\sigma_Z\in\Delta_Z}}\exp\left(\sum_{\sigma_{\rm q}\in\Delta_{\rm q}}(-1)^{\#(z_{\rm q}\cap \sigma_{\rm q})}J^*(\sigma_{\rm q})s(\delta^*_Z\sigma_{\rm q})\right)\,. 
\end{align}
\subsection{Duality defect via strange correlator}
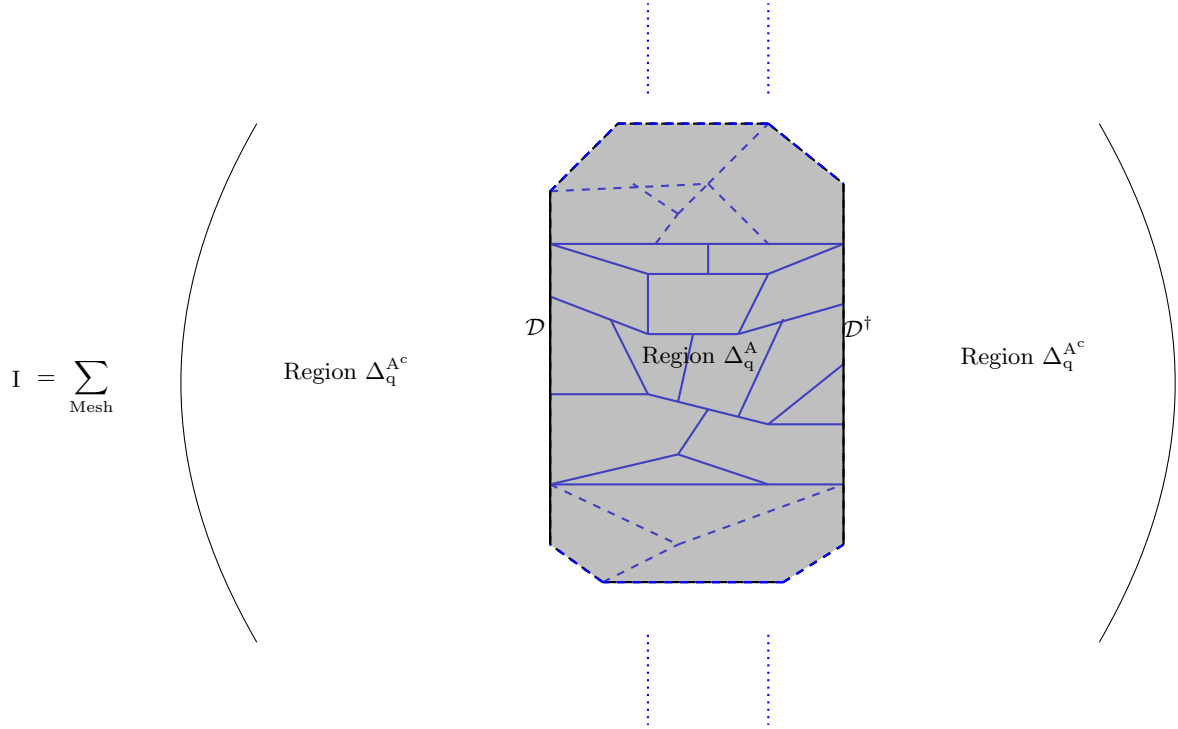
\begin{figure}
     \centering
     \begin{tikzpicture}[scale=0.8]
	\begin{pgfonlayer}{nodelayer}
		\node [style=none] (0) at (-5.25, 4) {};
		\node [style=none] (1) at (-5.25, -4) {};
		\node [style=none] (2) at (4.5, -4) {};
		\node [style=none] (3) at (4.5, 4) {};
		\node [style=none] (4) at (-2, 3) {};
		\node [style=none] (5) at (1, 1) {};
		\node [style=none] (6) at (-2, -1) {};
		\node [style=none] (7) at (-2, 1) {};
		\node [style=none] (8) at (2, 3) {};
		\node [style=none] (9) at (2, -2) {};
		\node [style=none] (10) at (-1, -3) {};
		\node [style=none] (11) at (-5.25, 2.25) {};
		\node [style=none] (12) at (-5.25, -1) {};
		\node [style=none] (13) at (4.5, 0) {};
		\node [style=none] (14) at (4.5, 2) {};
		\node [style=none] (15) at (4.5, -2) {};
		\node [style=none] (16) at (-0.5, 1) {};
		\node [style=none] (17) at (0, -1.5) {};
		\node [style=none] (18) at (-3.25, 1.5) {};
		\node [style=none] (19) at (2.5, 1.5) {};
		\node [style=none] (20) at (2, -4) {};
		\node [style=none] (21) at (1, -1.75) {};
		\node [style=none] (22) at (0, 4) {};
		\node [style=none] (23) at (0, 3) {};
		\node [style=none] (24) at (-1, -1.25) {};
		\node [style=none] (25) at (-5.25, 5.75) {};
		\node [style=none] (26) at (4.5, 6) {};
		\node [style=none] (27) at (4.5, -6) {};
		\node [style=none] (28) at (-5.25, -6) {};
		\node [style=none] (29) at (-1, -6) {};
		\node [style=none] (30) at (0, 6) {};
		\node [style=none] (31) at (-1.75, 4) {};
		\node [style=none] (32) at (2, 4) {};
		\node [style=none] (33) at (-1, 5) {};
		\node [style=none] (34) at (-2.5, 6) {};
		\node [style=none] (35) at (2, 8) {};
		\node [style=none] (36) at (-3, 8) {};
		\node [style=none] (37) at (-3.5, -7.25) {};
		\node [style=none] (38) at (2.5, -7.25) {};
		\node [style=none] (39) at (-0.25, 0.25) {Region  $\Delta_{\rm q}^{\rm A}$};
		\node [style=none] (40) at (0, -3.25) {};
		\node [style=none] (41) at (-5.75, 1.25) {$\mathcal{D}$};
		\node [style=none] (42) at (5, 1.25) {$\mathcal{D}^{\dagger}$};
		\node [style=none] (43) at (-12, -0.25) {Region $\Delta_{\rm q}^{\rm A^c}$};
		\node [style=none] (44) at (10.5, 0.25) {Region $\Delta_{\rm q}^{\rm A^c}$};
		\node [style=none] (45) at (-2, 9) {};
		\node [style=none] (46) at (-2, 12) {};
		\node [style=none] (47) at (2, 12) {};
		\node [style=none] (48) at (2, 9) {};
		\node [style=none] (49) at (-2, -9) {};
		\node [style=none] (50) at (-2, -12) {};
		\node [style=none] (51) at (2, -9) {};
		\node [style=none] (52) at (2, -12) {};
		\node [style=none] (53) at (-20.5, -0.75) {$\mathlarger{\sum}\limits_{\text{Mesh}}$};
		\node [style=none] (54) at (-23, -0.5) {$\rm I$};
		\node [style=none] (55) at (-22, -0.5) {$=$};
		\node [style=none] (56) at (-15, 8) {};
		\node [style=none] (57) at (-15, -9.25) {};
		\node [style=none] (58) at (13, 8) {};
		\node [style=none] (59) at (13, -9.25) {};
	\end{pgfonlayer}
	\begin{pgfonlayer}{edgelayer}
		\draw [style=grayfillline] (0.center) to (1.center);
		\draw [style=grayfilldashedline] (27.center) to (38.center);
		\draw [style=grayfilldashedline] (38.center) to (37.center);
		\draw [style=grayfilldashedline] (37.center) to (28.center);
		\draw [style=grayfilldashedline] (28.center) to (25.center);
		\draw [style=grayfilldashedline] (25.center) to (36.center);
		\draw [style=grayfilldashedline] (36.center) to (35.center);
		\draw [style=grayfilldashedline] (35.center) to (26.center);
		\draw [style=blueline] (0.center) to (4.center);
		\draw [style=blueline] (4.center) to (8.center);
		\draw [style=blueline] (8.center) to (3.center);
		\draw [style=blueline] (8.center) to (5.center);
		\draw [style=blueline] (5.center) to (14.center);
		\draw [style=blueline] (5.center) to (7.center);
		\draw [style=blueline] (11.center) to (7.center);
		\draw [style=blueline] (7.center) to (4.center);
		\draw [style=blueline] (12.center) to (6.center);
		\draw [style=blueline] (6.center) to (9.center);
		\draw [style=blueline] (9.center) to (13.center);
		\draw [style=blueline] (9.center) to (15.center);
		\draw [style=blueline] (10.center) to (1.center);
		\draw [style=blueline] (10.center) to (20.center);
		\draw [style=blueline] (20.center) to (2.center);
		\draw [style=blueline] (10.center) to (17.center);
		\draw [style=blueline] (6.center) to (18.center);
		\draw [style=blueline] (19.center) to (21.center);
		\draw [style=blueline] (0.center) to (22.center);
		\draw [style=blueline] (22.center) to (3.center);
		\draw [style=blueline] (22.center) to (23.center);
		\draw [style=blueline] (1.center) to (20.center);
		\draw [style=blueline] (24.center) to (16.center);
		\draw [style=dashedline] (0.center) to (25.center);
		\draw [style=dashedline] (1.center) to (28.center);
		\draw [style=bluedashedline] (25.center) to (30.center);
		\draw [style=bluedashedline] (31.center) to (33.center);
		\draw [style=bluedashedline] (33.center) to (30.center);
		\draw [style=bluedashedline] (30.center) to (32.center);
		\draw [style=bluedashedline] (33.center) to (34.center);
		\draw [style=bluedashedline] (30.center) to (35.center);
		\draw [style=bluedashedline] (1.center) to (29.center);
		\draw [style=bluedashedline] (29.center) to (2.center);
		\draw [style=bluedashedline] (37.center) to (29.center);
		\draw [blue,dotted,thick] (49.center) to (50.center);
		\draw [blue,dotted,thick] (51.center) to (52.center);
		\draw [blue,dotted,thick] (48.center) to (47.center);
		\draw [blue,dotted,thick] (45.center) to (46.center);
		\draw [in=120, out=-120] (56.center) to (57.center);
		\draw [bend left] (58.center) to (59.center);
		\draw [style=bluedashedline] (38.center) to (37.center);
		\draw [style=bluedashedline] (37.center) to (28.center);
		\draw [style=bluedashedline] (27.center) to (38.center);
		\draw [style=bluedashedline] (36.center) to (35.center);
		\draw [style=bluedashedline] (25.center) to (36.center);
		\draw [style=bluedashedline] (35.center) to (26.center);
		\draw [style=grayfilldashedline] (27.center)
			 to (38.center)
			 to (37.center)
			 to (28.center)
			 to (25.center)
			 to (36.center)
			 to (35.center)
			 to (26.center);
		\draw (26.center) to (27.center);
		\draw [style=grayfilldashedline] (26.center) to (27.center);
		\draw [style=bluedashedline] (27.center) to (38.center);
		\draw [style=bluedashedline] (38.center) to (37.center);
		\draw [style=bluedashedline] (37.center) to (28.center);
		\draw [style=bluedashedline] (25.center) to (36.center);
		\draw [style=bluedashedline] (36.center) to (35.center);
		\draw [style=bluedashedline] (35.center) to (26.center);
		\draw (26.center) to (27.center);
		\draw [style=normalline] (26.center) to (27.center);
		\draw [style=normalline] (25.center) to (28.center);
	\end{pgfonlayer}
\end{tikzpicture}
     \caption{The figure illustrate the identity \eqref{eq:ZXpartitiondualitydefect}. The sum is over all possible symmetry defect networks inside the region defined by $\Delta_{\rm q}^{\rm A}$ and shaded grey. This summation over symmetry defects is equivalent to gauging the symmetries inside the shaded region and produces Kramers-Wannier defects on the boundary of the shaded region. When the gauged model is equivalent to the original model, the Kramers-Wannier defect is a duality defect. }
     \label{fig:dualitydefect_identity}
 \end{figure}

Here we generalize the discussion of duality defect in Section~\ref{sec:dualitydefectIsing} for generalized Ising model. Now let us discuss the duality defect in this generalized Ising model. Let us define $\Delta_{\rm q}^{\rm A}$ and $\Delta_{\rm q}^{\rm A^c}$ to be non-empty subset of $\Delta_{\rm q}$. The duality defect is obtained by gauging the global symmetry of the generalized Ising model on region $\Delta_{\rm q}^{\rm A}$. As in Section~\ref{sec:dualitydefectIsing}, to construct the duality defect, let us consider $\ket{\Psi_{\mathcal{C}}}$ on $\Delta_{\rm q}^{\rm A^c}$ and $\ket{\Psi_{\mathcal{C}}^*}$ on $\Delta_{\rm q}^{\rm A}$ and stitch them together to obtain $\ket{\Psi_{\mathcal{C}}^{\rm A|A^c}}$. Now we define the following sets:
\begin{subequations}
\begin{align}
    \Delta_Z^{\rm A}\equiv\{\sigma_Z\in\Delta_Z|\delta\sigma_Z\sqcap \Delta_{\rm q}^{\rm A}\neq 0\}\subset\Delta_Z\,,&\quad \Delta_Z^{\rm A^c}\equiv(\Delta_{Z}^{\rm A})^{\rm c}\,,\\
     \Delta_X^{\rm A^c}\equiv\{\sigma_{X}\in\Delta_X|\delta^*\sigma_X\sqcap \Delta_{\rm q}^{\rm A^c}\neq 0\}\subset\Delta_X\,,&\quad \Delta_X^{\rm A}\equiv(\Delta_X^{\rm A^c})^{\rm c}\,,\\
    \Delta_{Z}^{\rm A|A^c}\equiv\{\sigma_Z\in\Delta_{Z}|\delta_Z\sigma_Z\neq \delta_Z\sigma_Z\sqcap\Delta_{\rm q}^{\rm A}\}\subset\Delta_{\rm Z}^{\rm A}\,,&\quad \Delta_{X}^{\rm A|A^c}\equiv\{\sigma_X\in\Delta_{X}|\delta_X^*\sigma_X\neq \delta_X^*\sigma_X\sqcap\Delta_{\rm q}^{\rm A}\}\subset\Delta_{\rm X}^{\rm A}\,.
\end{align}
\label{eq:setAandA^cdef}
\end{subequations}
Given these sets of cells, we can define the associated chain groups using the same definition in \eqref{eq:chaindef} but with the above underlying sets. Let us consider the chain groups $C_{Z}^{\rm A}$, $C_{\rm q}^{\rm A}$, and $C_{X}^{\rm A}$ formed out of the cells in $\Delta_{Z}^{\rm A}$, $\Delta_{\rm q}^{\rm A}$, and $\Delta_{X}^{\rm A}$. Now, we consider the chain complex
\begin{align}
   0 \overset{\delta}{\longrightarrow} C_Z^{\rm A}\overset{\delta_Z^{\rm A}}{\longrightarrow} C_\text{q}^{\rm A} \overset{\delta_X^{\rm A}}{\longrightarrow} C_X^{\rm A} \overset{\delta}{\longrightarrow} 0\,,
   \label{eq:CSS-chain-complex-A}
\end{align}
with the boundary maps defined as  $\delta_Z^{\rm A} \sigma_Z:=\delta_Z\sigma_Z\sqcap \Delta_{\rm q}^{\rm A}$ and $\delta_X^{\rm A} \sigma_{\rm q}:=\delta_X\sigma_{\rm q}\sqcap \Delta_{X}^{\rm A}$. Due to the definition of $\Delta_X^{\rm A}$, $\delta_X^{\rm A}\circ \delta_Z^{\rm A}=0$\footnote{It is easier to prove $(\delta_{Z}^{\rm A})^*\circ(\delta_{X}^{\rm A})^*=0$ where $(\delta_{X}^{\rm A})^*$ and $(\delta_{Z}^{\rm A})^*$ are dual boundary maps.}. Similarly, we consider the chain groups $C_{Z}^{\rm A^c}$, $C_{\rm q}^{\rm A^c}$, and $C_{X}^{\rm A^c}$ formed out of the cells in $\Delta_{Z}^{\rm A^c}$, $\Delta_{\rm q}^{\rm A^c}$, and $\Delta_{X}^{\rm A^c}$. These chain groups give the following chain complex
\begin{align}
     0 \overset{\delta}{\longrightarrow} C_Z^{\rm A^c}\overset{\delta_Z^{\rm A^c}}{\longrightarrow} C_\text{q}^{\rm A^c} \overset{\delta_X^{\rm A^c}}{\longrightarrow} C_X^{\rm A^c} \overset{\delta}{\longrightarrow} 0\,,
   \label{eq:CSS-chain-complex-A^c}
\end{align}
 with the boundary maps defined as  $\delta_Z^{\rm A^c} \sigma_Z:=\delta_Z\sigma_Z\sqcap \Delta_{\rm q}^{\rm A^c}$ and $\delta_X^{\rm A^c} \sigma_{\rm q}:=\delta_X\sigma_{\rm q}\sqcap \Delta_{X}^{\rm A^c}$. Due to the definition of $\Delta_Z^{\rm A^c}$, $\delta_Z^{\rm A^c} \sigma_Z=\delta_Z\sigma_Z$, and therefore $\delta_X^{\rm A^c}\circ\delta_Z^{\rm A^c}=0$. We define
 \begin{align}
    \mathcal{L}^{\rm A^c}:&=\frac{|\text{Ker}\,\delta_{X}^{\rm A^c}|}{|\text{Im}\,\delta_{Z}^{\rm A^c}|}\,.
    \label{eq:LA^c}
 \end{align}
As a matter of fact, we note that the  quotient chain complexes can be identified as follows
\begin{subequations}
 \begin{align}
     0 \overset{\delta}{\longrightarrow} C_Z/C_{Z}^{\rm A^c}\overset{\delta_Z}{\longrightarrow} C_\text{q}/C_{\rm q}^{\rm A^c} \overset{\delta_X}{\longrightarrow} C_X/C_X^{\rm A^c} \overset{\delta}{\longrightarrow} 0\,\equiv 0 \overset{\delta}{\longrightarrow} C_Z^{\rm A}\overset{\delta_Z^{\rm A}}{\longrightarrow} C_\text{q}^{\rm A} \overset{\delta_X^{\rm A}}{\longrightarrow} C_X^{\rm A} \overset{\delta}{\longrightarrow} 0\,,\\
     0 \overset{\delta}{\longrightarrow} C_Z/C_{Z}^{\rm A}\overset{\delta_Z}{\longrightarrow} C_\text{q}/C_{\rm q}^{\rm A} \overset{\delta_X}{\longrightarrow} C_X/C_X^{\rm A} \overset{\delta}{\longrightarrow} 0\,\equiv 0 \overset{\delta}{\longrightarrow} C_Z^{\rm A^c}\overset{\delta_Z^{\rm A}}{\longrightarrow} C_\text{q}^{\rm A^c} \overset{\delta_X^{\rm A}}{\longrightarrow} C_X^{\rm A^c} \overset{\delta}{\longrightarrow} 0\,.
 \end{align}
 \end{subequations}
This follows from the definition of the sets in \eqref{eq:setAandA^cdef} and the chain complexes \eqref{eq:CSS-chain-complex-A} and \eqref{eq:CSS-chain-complex-A^c}.
Now we define the cluster state
\begin{align}
    |\Psi_{\mathcal{C}}^{\rm A|A^c}\rangle&=\prod_{\sigma_X\in\Delta_{X}^{\rm A^c}}\prod_{\sigma_{\rm q}\in(\delta_X^{\rm A^c})^*\sigma_X}CZ_{\sigma_X,\sigma_{\rm q}}\prod_{\sigma_Z\in\Delta_{Z}^{\rm A}}\,\prod_{\sigma_{\rm q}\in\delta_Z^{\rm A}\sigma_Z}CZ_{\sigma_Z,\sigma_{\rm q}}\nonumber\\
    &\hspace{2cm}\prod_{\sigma_Z\in\Delta_Z^{\rm A|A^c}}\,\prod_{\sigma_X\in\mathsf{S}(\delta_{X}(\delta_Z\sigma_Z\sqcap\Delta_{\rm q}^{\rm A}))}CZ_{\sigma_Z,\sigma_X}\ket{+}^{\Delta_{\rm q}}\ket{+}^{\Delta_{Z}^{\rm A}}\ket{+}^{\Delta_X^{\rm A^c}}\,.
\end{align}
The above state is stabilized by
\begin{subequations}
\begin{align}
    X(\sigma_{\rm q})Z(\delta^*_Z\sigma_{\rm q})\,\quad \sigma_{\rm q}\in\Delta_{\rm q}^{\rm A}\,,&\qquad X(\sigma_{\rm q})Z(\delta_X\sigma_{\rm q})\, \sigma_{\rm q}\in \Delta_{\rm q}^{\rm A^c}\,,\\
 X(\sigma_Z)Z(\delta_Z\sigma_Z)\,\quad \sigma_Z\in\Delta_Z^{\rm A}\setminus\Delta_Z^{\rm A|A^c}\,,&\qquad X(\sigma_X)Z(\delta^*\sigma_X)\,\quad \sigma_X\in \Delta_X^{\rm A^c}\setminus \Delta_X^{\rm A|A^c}\,,\\
   X(\sigma_Z)Z(\delta_Z^{\rm A}\sigma_Z)Z(\delta_X(\delta_Z\sigma_Z\sqcap \Delta_{\rm q}^{\rm A}))\,\quad \sigma_Z\in\Delta_Z^{\rm A|A^c}\,,&\qquad X(\sigma_X)Z((\delta_X^{\rm A^c})^*\sigma_X)Z(\delta_Z^*(\delta_X^*\sigma_X\sqcap \Delta_{\rm q}^{\rm A^c}))\,\quad \sigma_X\in\Delta_X^{\rm A|A^c}\,.
\end{align}
\end{subequations}
After taking the overlap with $\ket{+}^{\Delta_Z^{\rm A}}$ and $\ket{+}^{\Delta_X^{\rm A^c}}$, we get a topologically ordered ground state
\begin{align}
    \ket{\Phi^{*\rm A}}=\bra{+}^{\Delta_{Z}^{\rm A}}\bra{+}^{\Delta_{X}^{\rm A^c}}\ket{\Psi_{\mathcal{C}}^{\rm A|A^c}}\,.
\end{align}
This has the same set of local stabilizers as that of $\rm H^A\ket{\Phi}$ where $\mathrm{H}^{\rm A}=\bigotimes_{\sigma_{\rm q}\in\Delta_{\rm q}^{\rm A}}H(\sigma_q)$ and $\ket{\Phi}=\bra{+}^{\Delta_X}\ket{\Psi_{\mathcal{C}}}$. They are stabilized by different logical operators. $\mathrm{H}^{\rm A}\ket{\Phi}$ is stabilized by $\mathrm{H}^{\rm A}X(z_{\rm q})\mathrm{H}^{\rm A}$ for non-trivial cycle $z_{\rm q}$ satisfying $\delta_X z_{\rm q}=0$. Hence, summing over the action of logical operators of $\mathrm{H}^{\rm A}\ket{\Phi}$ on $\ket{\Phi^{*\rm A}}$, produces the equality
\begin{align}
    \frac{1}{\mathcal{N}_{\Phi}}\mathrm{H}^{\rm A}\ket{\Phi}=\frac{1}{\mathcal{N}_{\Phi^{*\rm A}}}\frac{1}{|\mathcal{L}|}\sum_{[z_{\rm q}]\in\mathcal{L}}X(z_{\rm q}\sqcap\Delta_{\rm q}^{\rm A^c})Z(z_{\rm q}\sqcap \Delta_{\rm q}^{\rm A})\ket{\Phi^{*\rm A}}\,,
\end{align}
where (see Appendix~\ref{app:topologicallyorderedstatenorm} for calculations of the norms)
\begin{subequations}
\begin{align}
    &\mathcal{N}_{\Phi}=\sqrt{\langle \Phi|\Phi\rangle}=\sqrt{\frac{|\text{Ker}\,\delta_X^*|}{2^{|\Delta_X|}}}\,,\quad \\
    &\mathcal{N}_{\Phi^{*\rm A}}=\sqrt{\bra{\Phi^{*\rm A}}\frac{1}{|\mathcal{L}|}\sum_{[z_{\rm q}]\in\mathcal{L}}X(z_{\rm q}\sqcap\Delta_{\rm q}^{\rm A^c})Z(z_{\rm q}\sqcap \Delta_{\rm q}^{\rm A})\ket{\Phi^{*\rm A}}}=\sqrt{\frac{|\mathcal{L}^{\rm A^c}|}{|\mathcal{L}|}\frac{|\text{Ker}\,(\delta_{X}^{\rm A^c})^*||\text{Ker}\,\delta_{Z}^{\rm A}|}{2^{|\Delta_{Z}^{\rm A}|+|\Delta_{X}^{\rm A^c}|}}}\,.
\end{align}
\end{subequations}
Now consider the overlap with the product state
\begin{align}
    \bigotimes_{\sigma_{\rm q}\in\Delta_{\rm q}^{\rm A^c}}\bra{0}_{\sigma_{\rm q}}e^{J(\sigma_{\rm q})X(\sigma_{\rm q})} \bigotimes_{\sigma_{\rm q}\in\Delta_{\rm q}^{\rm A}}\bra{0}_{\sigma_{\rm q}}e^{J^*(\sigma_{\rm q})X(\sigma_{\rm q})}\,.
\end{align}
This yield the duality
\begin{align}
    \mathcal{Z}_X(\{J(\sigma_{\rm q})\})&=\frac{2^{-\frac{|\Delta_Z^{\rm A}|}{2}}2^{-\frac{|\Delta_{X}^{\rm A^c}|}{2}}2^{\frac{|\Delta_{X}|}{2}}\mathcal{N}}{|\mathcal{L}|\prod\limits_{\sigma_{\rm q}\in\Delta_{\rm q}^{\rm A}}(\sinh(2J^*(\sigma_{\rm q})))^{\frac{1}{2}}}\sum_{[z_{\rm q}]\in \mathcal{L}}\sum_{\bigg\{\substack{s_{\sigma_X}=\pm 1\\
    (-1)^{h_{\sigma_X}}=s_{\sigma_X}}\bigg\}_{\sigma_X\in\Delta_X^{\rm A^c}}}\sum_{\bigg\{\substack{s_{\sigma_Z}=\pm 1\\
    (-1)^{h_{\sigma_Z}}=s_{\sigma_Z}}\bigg\}_{\sigma_Z\in\Delta_Z^{\rm A}}}\nonumber\\
    &\qquad\exp\left(\sum_{\sigma_{\rm q}\in\Delta_{\rm q}^{\rm A^c}}J(\sigma_{\rm q})s(\delta_X\sigma_{\rm q})+\sum_{\sigma_{\rm q}\in\Delta_{\rm q}^{\rm A}}(-1)^{\#(z_{\rm q}\cap \sigma_{\rm q})}J^*(\sigma_{\rm q})s(\delta_Z^*\sigma_{\rm q})\right)\nonumber\\
    &\hspace{3cm}\prod_{\substack{\sigma_Z\in\Delta_{Z}^{\rm A|A^c}\\
    \sigma_X\in\mathsf{S}(\delta_X(\delta_Z\sigma_Z\sqcap\Delta_{\rm q}^{\rm A}))}}(-1)^{h_{\sigma_X}h_{\sigma_Z}}\times s(\delta_X(z_{\rm q}\sqcap \Delta_{\rm q}^{\rm A^c}))\,,
    \label{eq:ZXpartitiondualitydefect}
\end{align}
where
\begin{align}
    \mathcal{N}&=\sqrt{\frac{|\mathcal{L}||\text{Ker}\,\delta^*_X|}{|\mathcal{L}^{\rm A^c}||\text{Ker}\,(\delta_{X}^{\rm A^c})^*||\text{Ker}\,\delta_{Z}^{\rm A}|}}\,.
\end{align}
 The relation \eqref{eq:ZXpartitiondualitydefect} is relating the partition function and it's gauged version in the region defined by $\Delta_{\rm q}^{\rm A}$. Pictorially this is illustrated in Figure~\ref{fig:dualitydefect_identity}. The sum over the homological cycles in \eqref{eq:ZXpartitiondualitydefect} is equivalent to the sum over the defect networks in the Figure~\ref{fig:dualitydefect_identity}.
  From the above discussions, we infer the partition function with duality defect to be the term that correspond to the trivial homology class in the sum over homology classes. Explicitly it is given by 
\begin{align}
    \mathcal{Z}^{\mathcal{D}}_X(J(\sigma_{\rm q}))&=\frac{2^{-\frac{|\Delta_Z^{\rm A}|}{2}}2^{-\frac{|\Delta_{X}^{\rm A^c}|}{2}}2^{\frac{|\Delta_{X}|}{2}}\mathcal{N}}{\prod\limits_{\sigma_{\rm q}\in\Delta_{\rm q}^{\rm A}}(\sinh(2J^*(\sigma_{\rm q})))^{\frac{1}{2}}}\sum_{\bigg\{\substack{s_{\sigma_X}=\pm 1\\
    (-1)^{h_{\sigma_X}}=s_{\sigma_X}}\bigg\}_{\sigma_X\in\Delta_X^{\rm A^c}}}\sum_{\bigg\{\substack{s_{\sigma_Z}=\pm 1\\
    (-1)^{h_{\sigma_Z}}=s_{\sigma_Z}}\bigg\}_{\sigma_Z\in\Delta_Z^{\rm A}}}\nonumber\\
    &\hspace{3cm}\exp\left(\sum_{\sigma_{\rm q}\in\Delta_{\rm q}^{\rm A^c}}J(\sigma_{\rm q})s(\delta_X\sigma_{\rm q})+\sum_{\sigma_{\rm q}\in\Delta_{\rm q}^{\rm A}}J^*(\sigma_{\rm q})s(\delta_Z^*\sigma_{\rm q})\right)\nonumber\\
    &\hspace{7cm}\times \prod_{\substack{\sigma_Z\in\Delta_{Z}^{\rm A|A^c}\\
    \sigma_X\in\mathsf{S}(\delta_X(\delta_Z\sigma_Z\sqcap\Delta_{\rm q}^{\rm A}))}}(-1)^{h_{\sigma_X}h_{\sigma_Z}} \,.
    \label{eq:ZXpartitiondualitydefect2}
\end{align}
The phase factors $(-1)^{h_{\sigma_X}h_{\sigma_Z}}$ indicate that duality defect is inserted between generalized Ising model in $\Delta_{\rm q}^{\rm A^c}$ and it's gauged model in $\Delta_{\rm q}^{\rm A}$.
\subsection{Moving and fusing defects}
 \subsubsection{Movement of duality defects}
Now let us consider the two regions $\Delta_{\rm q}^{\rm A}$ and $\Delta_{\rm q}^{\rm A^c}$ where the two theories are defined.  
We consider the identity associated with deformation of the region $\Delta_{\rm q}^{\rm A}$. Now suppose that we deform the region by decreasing the size of $\Delta_{\rm q}^{\rm A}$, i.e., $ |\Delta_{\rm q}^{\rm A^c}|\longrightarrow |\Delta_{\rm q}^{\rm A^c}|+1$. Let us consider a $\sigma_{\rm q}\in\Delta_{\rm q}^{\rm A}$. Now we ungauge the dual lattice interaction to original lattice interaction via the  following identity 
\begin{align}
    &\frac{\exp\left(J^*(\sigma_{\rm q})\prod\limits_{\sigma_{Z}\in\mathsf{S}(\delta^*_Z\sigma_{\rm q})}s_{\sigma_{Z}}\right)}{\sqrt{2}\left(\sinh(2J^*(\sigma_{\rm q}))\right)^{\frac{1}{2}}}(-1)^{\sum\limits_{\sigma_{Z}\in\mathsf{S}(\delta^*_Z\sigma_{\rm q})}h_{ \sigma_{Z}}\left(\sum\limits_{\sigma_X\in \mathsf{S}(\delta_X^{\rm A^c}\sigma_{\rm q} )}h_{ \sigma_{X}}\right)}\nonumber\\
    &\hspace{2cm}=\frac{1}{2^{|\mathsf{S}(\delta_X\sigma_{\rm q}\sqcap\Delta_{X}^{\rm A})|}}\sum\limits_{\substack{s_{\sigma_X}=\pm 1\\
    \sigma_X\in\mathsf{S}(\delta_X^{\rm A}\sigma_{\rm q})}}\exp\left(J(\sigma_{\rm q})\prod_{\sigma_X\in\mathsf{S}(\delta_X\sigma_{\rm q})}s_{\sigma_X}\right)(-1)^{\sum\limits_{\sigma_{Z}\in\mathsf{S}(\delta^*_Z\sigma_{\rm q})}h_{\sigma_{Z}}\left(\sum\limits_{\sigma_X\in\mathsf{S}(\delta_X^{\rm A}\sigma_{\rm q}) }h_{\sigma_X}\right)}\,.
    \label{eq:genIsingmovement1}
\end{align}
The configuration $\Delta_{\rm q}^{\rm A}$ and the associated $\Delta_{X}^{\rm A}$ and $\Delta_{X}^{\rm A^c}$ are that of the L.H.S. of the above equation. The variables $h_{\sigma_{Z}}$ and $h_{\sigma_{X}}$ take values in 0 and 1. Explicitly $s_{\sigma_{X}}=(-1)^{h_{\sigma_{X}}}$ and $s_{\sigma_{Z}}=(-1)^{h_{\sigma_{Z}}}$. This is a generalization of the relation \eqref{eq:Isingdefectmovement} in 2D Ising model with duality defect. Another similar identity is 
\begin{align}
    &\exp\left(J(\sigma_{\rm q})\prod_{\sigma_{X}\in\mathsf{S}(\delta_X\sigma_{\rm q})}s_{\sigma_X}\right)(-1)^{\sum\limits_{\sigma_{Z}\in\mathsf{S}(\delta^*_Z\sigma_{\rm q}\sqcap\Delta_{Z}^{\rm A})}h_{\sigma_{Z}}\left(\sum\limits_{\sigma_{X}\in\mathsf{S}(\delta_X\sigma_{\rm q}) }h_{\sigma_X}\right)}\nonumber\\
    &\hspace{2cm}=\frac{\sqrt{2}}{2^{|\mathsf{S}(\delta^*_Z\sigma_{\rm q}\sqcap\Delta_{Z}^{\rm A^c})|}}\sum_{\substack{s_{\sigma_{Z}}=\pm 1\\
    \sigma_{Z}\in\mathsf{S}(\delta^*_{Z}\sigma_{\rm q}\sqcap\Delta_{Z}^{\rm A^c})}}\frac{\exp\left(J^*(\sigma_{\rm q})\prod\limits_{\sigma_{Z}\in\mathsf{S}(\delta^*_Z\sigma_{\rm q})}s_{\sigma_{Z}}\right)}{(\sinh(2J^*(\sigma_{\rm q})))^{\frac{1}{2}}}(-1)^{\sum\limits_{\sigma_{Z}\in\mathsf{S}(\delta^*_Z\sigma_{\rm q}\sqcap\Delta_{Z}^{\rm A^c})
    }h_{ \sigma_{Z}}\left(\sum\limits_{\sigma_X\in \mathsf{S}(\delta_{X}\sigma_{\rm q}) }h_{ \sigma_{X}}\right)}\,.
    \label{eq:genIsingmovement2}
\end{align}
This is a generalization of the identity \eqref{eq:Isingdefectmovement2} in the 2D Ising model with a duality defect. 
\subsubsection{Fusion of duality defects}
Here let us restrict to the situation where the duality defects is topological in any directions, i.e., using the movement relations \eqref{eq:genIsingmovement1} and \eqref{eq:genIsingmovement2} or a slight modifications of them, the defects can be moved\footnote{In terms of chain complex, if the dimension of the cells in $\Delta_Z$, $\Delta_{\rm q}$ and $\Delta_X$ decrease by $1$ successively, then the duality defect is expected to be topological. This is the case with the example of 2D Ising model and 3D Ising gauge theory/3D Ising model.}.
Now let us consider a configuration of two duality defects placed far away and wrap around all but one cycle of the underlying geometry. 
Now we apply the movement relations to bring them close and then fuse. Let us denote the hyperplane on which we do the fusion to be $\mathcal{P}$\footnote{In $2$D Ising model this was a line while in $3$D anisotropic plaquette Ising model and $3$D Ising gauge theory, this was a two dimensional plane.}. Formally, this can be specified by the subsets $\Delta_{\rm q}^{\mathcal{P}}\subset \Delta_{\rm q}$ and $\Delta_{X}^{\mathcal{P}}\subset\Delta_{X}$. Let the configuration of two parallel defects just before fusion be specified by $\Delta_{\rm q}^{\rm A}$. Upon fusion, we encounter a sum of the form
\begin{align}
    &\sum_{\substack{\{h_{\sigma_{Z}}\}\\
    \sigma_Z\in\Delta_{Z}^{\rm A}}}(-1)^{\sum\limits_{\sigma_{Z}}h_{ \sigma_{Z}}\left(\sum\limits_{\sigma_X\in \mathsf{S}(\delta_X(\delta_Z \sigma_Z\sqcap\Delta_{X}^{\rm A}))\cap \Delta_{\rm q}^{\mathcal{P}}}h_{ \sigma_X}+k_{\sigma_X}\right)}=2^{|\Delta_{Z}^{\rm A}|}\left(\sum\limits_{\substack{c\in  C_{X}^{\mathcal{P}}\\
    \delta^*_Z(\delta^*_X c\sqcap \Delta_{\rm q}^{\rm A})=0}}\delta_{\{h_{\sigma_X}\},\{k_{\sigma_X}+\mathcal{I}^c_{\sigma_X}\}}\right)\,,\nonumber\\
    &\qquad\text{ where}\quad \mathcal{I}^c_{\sigma_X}=\begin{cases}
        1\quad \text{ if } \#(\sigma_X\cap c)\neq 0\,,\\
        0\quad \text{ otherwise }
    \end{cases}\,,
\end{align}
and $C_{X}^{\mathcal{P}}$ is the chain group formed from the set $\Delta_{X}^{\mathcal{P}}$.
Now suppose that we denote the duality defect by $\mathcal{D}$ and the spin flip defect along the chain $c\in C_{X}^{\mathcal{P}}$ by $D_{c}$, then
\begin{align}
    \mathcal{D}\times\mathcal{D}\propto \sum\limits_{\substack{c\in  C_{X}^{\mathcal{P}}\\
    \delta^*_Z(\delta^*_X c\sqcap \Delta_{\rm q}^{\rm A})=0}}D_{c}\,.
    \label{eq:condensationdefect}
\end{align}
\subsection{Quantum Hamiltonian from partition function}
In this section, we look at certain classical statistical mechanical models for which we can define a quantum Hamiltonian via the transfer matrix. To construct such statistical mechanical models, we look at another chain complex obtained by foliating the chain complex in \eqref{eq:CSS-complex}.
By employing the given chain complex for CSS code, a higher-dimensional symmetry-protected topological (SPT) state, known as the cluster state, can be obtained through a procedure referred to as foliation~\cite{2016PhRvL.117g0501B}. This cluster state, termed the foliated cluster state, serves as a crucial resource for measurement-based quantum computation. The representation of the foliated cluster state can be articulated using the chain complex formalism. To further elaborate, let us consider the one-dimensional chain complex
\begin{align}\label{eq:1dcomplex}
    0\overset{\delta}{\longrightarrow} C_w \overset{\delta}{\longrightarrow} C_0\overset{\delta}{\longrightarrow} 0\,.
\end{align}
 where $C_w$ represents the group of chains composed of 1-cells, and $C_0$ corresponds to the group of chains constituted by 0-cells with $\mathbb{Z}_2$ coefficients. We denote the collection of 1-cells and 0-cells from which the chain groups $C_w$ and $C_0$ are formed by $\Delta_{w}$ and $\Delta_0$ respectively. A 1-cell is an interval and is denoted by $\sigma_{[w,w+1]}$. The boundary of $\sigma_{[w,w+1]}$ is 
\begin{align}
    \delta\sigma_{[w,w+1]}=\sigma_{w}+\sigma_{w+1}\,.
\end{align}
The construction of the chain complex for the foliated cluster state involves forming the tensor product of the chain complex~\eqref{eq:CSS-complex} with the one-dimensional chain complex~\eqref{eq:1dcomplex}. We take the tensor product of the cells as cross product of the underlying sets for each cell
\begin{subequations}
\begin{align}
\bm\Delta_{Z,w}&=\{\sigma_Z\otimes\rho_{[w,w+1]}=\sigma_Z\times[w,w+1]|\,\sigma_Z\in\Delta_{Z}\,,\rho_{[w,w+1]}\in\Delta_{w}\}\\
    \bm\Delta_{Z}&=\{\sigma_{Z}\otimes \rho_{w}=\sigma_Z\times \{w\}|\,\sigma_Z\in \Delta_Z\,,\rho_{w}\in\Delta_{0}\}\,,\\
    \quad\bm\Delta_{\mathrm{q},w}&=\{\sigma_{\rm q}\otimes \rho_{[w,w+1]}=\sigma_{\rm q}\times [w,w+1]|\,\sigma_{\rm q}\in\Delta_{\rm q}\,, \rho_{[w,w+1]}\in \Delta_w \}\,,\\
    \bm\Delta_{\rm q}&=\{\sigma_{\rm q}\otimes \rho_{w}=\sigma_{\rm q}\times \{w\}|\,\sigma_{\rm q}\in \Delta_{\rm q}\,,\rho_{w}\in\Delta_{0}\}\,,\\
    \quad\bm\Delta_{X,w}&=\{\sigma_{X}\otimes \rho_{[w,w+1]}=\sigma_X\times[w,w+1]|\,\sigma_X\in\Delta_{X}\,, \rho_{[w,w+1]}\in \Delta_w \}\,,\\
    \bm\Delta_{\rm X}&=\{\sigma_X\otimes\rho_{w}=\sigma_X\times \{w\}|\,\sigma\in\Delta_X\,,\rho_w\in\Delta_{0}\}\,.
\end{align}
\end{subequations}
We denote $\bm\Delta_{Q_1} = \bm \Delta_{Z} \cup \bm\Delta_{\text{q},w} $ and $\bm \Delta_{Q_2} = \bm\Delta_{\text{q}} \cup \bm\Delta_{X,w}$. In the foliated cluster state, qubits are positioned at $\bm\Delta_{Q_1} \cup \bm\Delta_{Q_2}$. We denote $\bm C_{Q_1} = \bm C_{Z} \oplus \bm C_{\text{q},w}$ and $\bm C_{Q_2} = \bm C_{\text{q}} \oplus \bm C_{X,w}$. Consequently, the foliated chain complex can be expressed as
\begin{align}
    0 \overset{\bm \delta}{\longrightarrow} \bm C_{Z,w} \overset{\bm \delta}{\longrightarrow} \bm C_{Q_1} \overset{\bm \delta}{\longrightarrow} \bm C_{Q_2} 
    \overset{\bm \delta}{\longrightarrow} \bm C_X \overset{\bm \delta}{\longrightarrow} 0 \, .
    \label{eq:chain-complex-foliated-rewritten}
\end{align}
The boundary maps are specified by how it acts on the elementary cells and then extended linearly. On the elementary cells, the action is 
\begin{itemize}
    \item For $\bm\sigma_{Z,w}\in\bm\Delta_{Z,w}$ and $\bm\sigma_{Z,w}=\sigma_Z\otimes\rho_{[w,w+1]}$
    \begin{align}
    \bm\delta  (\sigma_Z\otimes\rho_{[w,w+1]})&=\delta_Z\sigma_Z\otimes \rho_{[w,w+1]} +\sigma_Z\otimes \delta\rho_{[w,w+1]}=\delta_Z\sigma_Z\times [w,w+1] +\sigma_Z\times \{w\}+\sigma_Z\times \{w+1\}\,.
\end{align}
\item For $\bm\sigma_{Z}\in\bm\Delta_{Z}$ and $\bm\sigma_{Z}=\sigma_Z\otimes\rho_{w}$
    \begin{align}
    \bm\delta  (\sigma_Z\otimes\rho_{w})&=\delta_Z\sigma_Z\otimes \rho_{w} =\delta_Z\sigma_Z\times \{w\} \,.
\end{align}
\item For $\bm\sigma_{\mathrm{q},w}\in\bm\Delta_{\mathrm{q},w}$ and $\bm\sigma_{\mathrm{q},w}=\sigma_{\rm q}\otimes\rho_{[w,w+1]}$
    \begin{align}
    \bm\delta  (\sigma_{\mathrm{q}}\otimes\rho_{[w,w+1]})&=\delta_X\sigma_{\mathrm{q}}\otimes \rho_{[w,w+1]} +\sigma_{\mathrm{q}}\otimes \delta\rho_{[w,w+1]}=\delta_{X}\sigma_{\mathrm{q}}\times [w,w+1] +\sigma_{\mathrm{q}}\times \{w\}+\sigma_{\mathrm{q}}\times \{w+1\}\,.
\end{align}
\item For $\bm\sigma_{\mathrm{q}}\in\bm\Delta_{\mathrm{q}}$ and $\bm\sigma_{\mathrm{q}}=\sigma_{\mathrm{q}}\otimes\rho_{w}$
    \begin{align}
    \bm\delta  (\sigma_{\mathrm{q}}\otimes\rho_{w})&=\delta_X\sigma_{\mathrm{q}}\otimes \rho_{w} =\delta_X\sigma_{\mathrm{q}}\times \{w\} \,.
\end{align}
\item For $\bm\sigma_{X,w}\in\bm\Delta_{X,w}$ and $\bm\sigma_{X,w}=\sigma_X\otimes\rho_{[w,w+1]}$
    \begin{align}
    \bm\delta  (\sigma_X\otimes\rho_{[w,w+1]})&=\sigma_X\otimes \delta\rho_{[w,w+1]}=\sigma_X\times \{w\}+\sigma_X\times \{w+1\}\,.
\end{align}
\item For $\bm\sigma_{X}\in\bm\Delta_{X}$ and $\bm\sigma_{X}=\sigma_X\otimes\rho_{w}$
    \begin{align}
    \bm\delta  (\sigma_X\otimes\rho_{w})&=0 \,.
\end{align}
\end{itemize}
Now let us consider the first three chain groups in the complex \eqref{eq:chain-complex-foliated-rewritten}. We can equate 
\begin{align}
    0 \overset{}{\longrightarrow} \bm C_{Z,w} \overset{\bm \delta}{\longrightarrow} \bm C_{Q_1} \overset{\bm \delta}{\longrightarrow} \bm C_{Q_2} 
    \overset{}{\longrightarrow}0\equiv 0 \overset{}{\longrightarrow} C_Z\overset{\delta_Z}{\longrightarrow} C_\text{q} \overset{\delta_X}{\longrightarrow} C_X \overset{}{\longrightarrow} 0\,.
\end{align}
Setting $J(\sigma_{\rm q})=J$ for $\sigma_{\rm q}=\bm\sigma_{Z}$ and $J(\sigma_{\rm q})=K$ for $\sigma_{\rm q}=\bm\sigma_{\mathrm{q},w}$, we can rewite the duality~\eqref{eq:ZXpartitionduality} as
\begin{align}
    \mathcal{Z}_{\bm Q_2}(J,K) = \frac{ 2^{\frac{|\bm \Delta_{Q_2}|}{2}} (\sinh 2J)^{|\bm \Delta_Z|/2} (\sinh 2K)^{|\bm \Delta_{\text{q},w}|/2}(|\text{Ker}\,(\bm C_{Q_1} \overset{\bm \delta^*}{\longleftarrow} \bm C_{Q_2})|)^{\frac{1}{2}}  }{2^{\frac{|\bm \Delta_{Z,w}|}{2}} (|\text{Ker}\,(\bm C_{Z,w} \overset{\bm \delta}{\longrightarrow} \bm C_{Q_1})|)^{\frac{1}{2}}|\bm{ \mathcal{L}}|}
\sum_{ [\bm z_{Q_1}] \in \bm{\mathcal{L} } } \mathcal{Z}^{\text{twisted}}_{Z,w} (J^*,K^*;\bm z_{Q_1}) \, ,
\end{align}
where we introduced a twisted partition function
\begin{align}
\label{eq:twisted-CSS-pf}
&\mathcal{Z}^{\text{twisted}}_{Z,w} (J^*,K^*;\bm z_{Q_1}) \nonumber \\
&\quad =
\sum_{ \{ s_{\bm\sigma} = \pm 1 \}_{\bm\sigma \in \bm \Delta_{Z,w}} } 
\exp \Big( J^{*} \sum_{\bm \sigma_Z \in \bm \Delta_Z } (-1)^{\#(\bm z_{Q_1} \cap \bm \sigma_Z)}   s(\bm \delta^* \bm \sigma_Z)
+ K^{*} \sum_{\bm \sigma_{\mathrm{q},w} \in \bm \Delta_{\text{q},w} } (-1)^{\#(\bm z_{Q_1} \cap \bm \sigma_{\mathrm{q},w})} s(\bm \delta^* \bm \sigma_{\mathrm{q},w}) 
\Big) \, .
\end{align}

Similarly, we can rewrite the duality relation \eqref{eq:ZXpartitiondualitydefect} in the presence of duality defect as
\small{
\begin{align}
\begin{split}
    \mathcal{Z}_{\bm Q_2}(J,K)&=\frac{2^{-\frac{|\bm\Delta_{Z,w}^{\rm A}|}{2}}2^{-\frac{|\bm\Delta_{Q_2}^{\rm A^c}|}{2}}2^{\frac{|\bm\Delta_{Q_2}|}{2}}\mathcal{N}}{|\bm{\mathcal{L}}|(\sinh 2J^*)^{|\bm\Delta_z\cap \bm\Delta_{Q_1}^{\rm A}|/2}(\sinh 2K^*)^{|\bm\Delta_{q,w}\cap\bm\Delta_{Q_1}^{\rm A}|/2}}\sum_{[\bm z_{Q_1}]\in\bm{\mathcal{L}}}\,\, \sum_{\bigg\{\substack{s_{\bm\sigma}=\pm 1\\
    (-1)^{h_{\bm\sigma}}=s_{\bm\sigma}}\bigg\}_{\bm\sigma\in\bm\Delta_{Q_2}^{\rm A^c}}}\, \sum_{\bigg\{\substack{s_{\bm\sigma}=\pm 1\\
    (-1)^{h_{\bm\sigma}}=s_{\bm\sigma}}\bigg\}_{\bm\sigma\in\bm\Delta_{Z,w}^{\rm A}}}\\
    &\exp \Big( J \sum_{\bm \sigma_Z \in \bm \Delta_Z\cap \bm\Delta_{Q_1}^{\rm A^c} }   s(\bm \delta \bm \sigma_Z)
+ K \sum_{\bm \sigma_{\mathrm{q},w} \in \bm \Delta_{\text{q},w}\cap \bm\Delta_{Q_1}^{\rm A^c} } s(\bm \delta \bm \sigma_{\mathrm{q},w}) 
+ J^{*} \sum_{\bm \sigma_Z \in \bm \Delta_Z\cap \bm\Delta_{Q_1}^{\rm A} } (-1)^{\#(\bm z_{Q_1} \cap \bm \sigma_Z)}   s(\bm \delta^* \bm \sigma_Z)
\\
&\hspace{2cm}+ K^{*} \sum_{\bm \sigma_{\mathrm{q},w}\in \bm \Delta_{\text{q},w}\cap \bm\Delta_{Q_1}^{\rm A} } (-1)^{\#(\bm z_{Q_1} \cap \bm \sigma_{\mathrm{q},w})} s(\bm \delta^* \bm \sigma_{\mathrm{q},w}) 
\Big)\prod_{\substack{\bm \rho\in\bm\Delta_{Z,w}^{\rm A|A^c}\\
\bm\tau\in\mathsf{S}(\bm\delta(\bm\delta\bm\rho\sqcap\bm\Delta_{Q_1}^{\rm A}))}}(-1)^{h_{\bm\rho}h_{\bm\tau}}\times s(\bm\delta(\bm z_{Q_1}\sqcap \bm\Delta_{Q_1}^{\rm A^c}))\,,
\end{split}
\end{align}
where
\begin{subequations}
\begin{align}
    &\mathcal{N}=\sqrt{\frac{|\mathcal{L}||\text{Ker}\left(\bm\delta^*: \bm C_{Q_2} \longrightarrow \bm C_{Q_1}\right)|}{|\mathcal{L}^{A^c}||\text{Ker}\left((\bm\delta^{\rm A^c})^c:\bm C_{Q_2}/\bm C_{Q_2}^{\rm A}\longrightarrow \bm C_{Q_1}/\bm C_{Q_1}^{\rm A}\right)||\text{Ker}\left(\bm\delta^A:\bm C_{Z,w}/\bm C_{Z,w}^{\rm A^c}\longrightarrow \bm C_{Q_1}/\bm C_{Q_1}^{\rm A^c}|\right)}}\,,\\
    & \mathcal{L}^{\rm A^c}:=\frac{\text{Ker}\left(\bm\delta^{\rm A^c}:\bm C_{ Q_1}/\bm C_{Q_1}^{\rm A}\longrightarrow\bm C_{ Q_2}/\bm C_{Q_2}^{\rm A}\right)}{\text{Im}\left(\bm\delta^{\rm A^c}:\bm C_{Z,w}/\bm C_{Z,w}^A\longrightarrow \bm C_{ Q_1}/\bm C_{Q_1}^{\rm A}\right)}\,.
\end{align}
\end{subequations}
 The phase factors $(-1)^{h_{\bm\rho}h_{\bm\tau}}$ indicate that duality defect is inserted between the generalized Ising model in $\bm\Delta_{Q_1}^{\rm A^c}$ and it's gauged model in $\bm\Delta_{Q_1}^{\rm A}$. The factor $s(\bm\delta(\bm z_{Q_1}\sqcap \bm\Delta_{Q_1}^{\rm A^c}))$ comes from the junction between the duality defect and $\mathbb{Z}_2$ the invertible defect.
\subsubsection{Quantum Hamiltonian for the two generalized Ising models}
First, let us derive the quantum Hamiltonian for the classical statistical mechanical models. The partition function of one of the classical statistical mechanical model is 
\begin{align}
    \mathcal{Z}^{\text{CSS-dual}}(J^*,K^*)=\mathcal{Z}_{Z,w}(J^*,K^*)=\sum_{ \{ s_{\bm\sigma} = \pm 1 \}_{\bm\sigma \in \bm \Delta_{Z,w}} } 
\exp \Big( J^{*} \sum_{\bm \sigma_Z \in \bm \Delta_Z }   s(\bm \delta^* \bm \sigma_Z)
+ K^{*} \sum_{\bm \sigma_{\mathrm{q},w} \in \bm \Delta_{\text{q},w} }  s(\bm \delta^* \bm \sigma_{\mathrm{q},w}) 
\Big)\,.
\end{align}
The quantum Hamiltonian for this model can be easily obtained using the transfer matrix. Let us define the Hilbert space of the quantum Hamiltonian as the configuration of spins $\ket{s_{\bm \sigma_{Z,w}}}$ that live on cells $\bm\sigma_{Z,w}$ in the $w=0$ plane. We relabel these spin configurations as $\ket{\{s_{\sigma_Z}\}}$ by identifying the cells $\bm\sigma_{Z,w}$ and $\bm\sigma_{\mathrm{q},w}$ with the cells  $\sigma_Z$ and $\sigma_{\rm q}$ respectively on the $w=0$ plane. Writing the partition function as $\text{Tr}(\hat{\rm T}^{L_w})$ with $\hat{\rm T}=\exp(-\tau \hat{\rm H}_{\text{CSS-dual}})\sim 1-\tau \hat{\rm H}_{\text{CSS-dual}}$ and setting $\exp(-2J^*)=\lambda\tau$ and $K^*=\tau$, we get the quantum Hamiltonian
\begin{align}
    \hat{\rm H}_{\text{CSS-dual}}=-\sum_{\sigma_{\rm q}\in \Delta_{\mathrm{q}}}Z(\delta^*\sigma_{\rm q})-\lambda\sum_{\sigma_Z\in\Delta_Z}X(\sigma_Z)\,.
\end{align} 

Now let us consider the partition function of the other classical statistical mechanical model
\begin{align}
     \mathcal{Z}^{\text{CSS}}(J,K)=\mathcal{Z}_{\bm Q_2}(J,K)=\sum_{ \{ s_{\bm\sigma} = \pm 1 \}_{\bm\sigma \in \bm \Delta_{\bm Q_2}} } 
\exp \Big( J \sum_{\bm \sigma_Z \in \bm \Delta_Z }   s(\bm \delta \bm \sigma_Z)
+ K \sum_{\bm \sigma_{\mathrm{q},w} \in \bm \Delta_{\text{q},w} }  s(\bm \delta \bm \sigma_{\mathrm{q},w}) 
\Big)\,.
\end{align}
This partition function has a local symmetry generated by spin flip on $\bm\delta^*\bm\sigma_X$ when $\bm\Delta_X$ is non-empty. Like in the 3D classical Ising gauge theory, one can gauge fix all the spins $s_{\bm \sigma_{X,w}}$ that appear on the temporal cells $\bm \sigma_{X,w}$ to $1$ except the last temporal cell $\bm\sigma_{X,w=L_w}$. The spin value $s_{\sigma_{X,w=L_w}}$ in the cell $\bm\sigma_{X,w=L_w}$ is the holonomy value. When the holonomy value is $1$ in all $\bm\sigma_{X,w=L_w}$, we essentially reduce the interactions of the form $K \sum\limits_{\bm \sigma_{\mathrm{q},w} \in \bm \Delta_{\text{q},w} }  s(\bm \delta \bm \sigma_{\mathrm{q},w}) $ to $K \sum\limits_{\bm \sigma_{\mathrm{q},w} \in \bm \Delta_{\text{q},w} }  s_{\bm \sigma_{\mathrm{q}}\times\{w+1\}}s_{\bm \sigma_{\mathrm{q}}\times\{w\}}$. However, when the value of the holonomy is $-1$ for the cell $\bm\sigma_{X,w=L_w}$, we have the interaction $-K\sum\limits_{\bm \sigma_{\mathrm{q},w} \in \bm \delta^*\bm\sigma_{X,w}}s_{\bm \sigma_{\mathrm{q}}\times\{w+1\}}s_{\bm \sigma_{\mathrm{q}}\times\{w\}}$ where the coupling is flipped to $-K$. To define the Hilbert space, we consider the spin configurations in the $w=0$ time slice. We use the notation $\Delta_Z$, $\Delta_{\rm q}$, and $\Delta_X$ to denote the collection of cells $\bm\Delta_Z$, $\bm\Delta_{\rm q}$, and $\bm\Delta_X$ in the $w=0$ plane. Defining the Hilbert space as formed by all possible spin configurations $\ket{\{s_{\sigma_{\rm q}}\}}$ where $\sigma_{\rm q}\in\Delta_{\rm q}$, we can write the partition function up to an irrelevant overall factor
\begin{align}
    \mathcal{Z}^{\text{CSS}}(J,K)=\text{Tr}\left(\prod\limits_{\sigma_X\in\Delta_X}(1+G_{\sigma_X})\hat{\rm T}^{L_w}\right)\,,
\end{align}
where $G_{\sigma_{X}}=\prod\limits_{\sigma_{\rm q}\in\delta^*\sigma_{X}}X(\sigma_{\rm q})$ is the Gauss's law. The product $\prod\limits_{\sigma_X\in\Delta_X}(1+G_{\sigma_X})$ essentially sum over all possible holonomy configurations. This is a generalization of what we had in the 3D Ising gauge theory. Equating $\hat{\rm T}=\exp(-\tau \hat{\rm H}_{\text{CSS}})\sim 1-\tau \hat{\rm H}_{\text{CSS}}$ and setting $\exp(-2K)=\tau$ and $J=\lambda\tau$, we get the quantum Hamiltonian expressed in terms of the boundary map in \eqref{eq:CSS-complex}
\begin{align}
    \hat{\rm H}_{\text{CSS}}=-\lambda\sum_{\sigma_{Z}\in \Delta_{Z}}Z(\delta\sigma_{Z})-\sum_{\sigma_Z\in\Delta_Z}X(\sigma_Z)\,,
    \label{eq:HCSS}
\end{align}
with the Gauss's law $G_{\sigma_X}\ket{\Psi}=\ket{\Psi}$ on the physical states $\ket{\Psi}$. In the absence of local symmetry, we have \eqref{eq:HCSS} without the Gauss's law.
\subsubsection{Quantum Hamiltonian for the duality defect}
Now, let us consider the generalized Ising model with a single duality defect inserted. We assume that the duality defect is translationally invariant in the time direction ($w$ direction) so that we can derive the corresponding Hamiltonian with defect. The defect location is defined by the definition of region $\bm\Delta_{Q_1}^{\rm A}$. 
\begin{align}
    \begin{split}
        \mathcal{Z}^{\text{(CSS)}}_{\mathcal{D}}(J,K)&= \sum_{\bigg\{\substack{s_{\bm\sigma}=\pm 1\\
    (-1)^{h_{\bm\sigma}}=s_{\bm\sigma}}\bigg\}_{\bm\sigma\in\bm\Delta_{Q_2}^{\rm A^c}}}\, \sum_{\bigg\{\substack{s_{\bm\sigma}=\pm 1\\
    (-1)^{h_{\bm\sigma}}=s_{\bm\sigma}}\bigg\}_{\bm\sigma\in\bm\Delta_{Z,w}^{\rm A}}}\\
    &\exp \Big( J \sum_{\bm \sigma_Z \in \bm \Delta_Z\cap \bm\Delta_{Q_1}^{\rm A^c} }   s(\bm \delta \bm \sigma_Z)
+ K \sum_{\bm \sigma_{\mathrm{q},w} \in \bm \Delta_{\text{q},w}\cap \bm\Delta_{Q_1}^{\rm A^c} } s(\bm \delta \bm \sigma_{\mathrm{q},w}) 
+ J^{*} \sum_{\bm \sigma_Z \in \bm \Delta_Z\cap \bm\Delta_{Q_1}^{\rm A} }    s(\bm \delta^* \bm \sigma_Z)
+ K^{*} \sum_{\bm \sigma_{\mathrm{q},w} \in \bm \Delta_{\text{q},w}\cap \bm\Delta_{Q_1}^{\rm A} }  s(\bm \delta^* \bm \sigma_{\mathrm{q},w}) 
\Big)\\
&\prod_{\substack{\bm \rho\in\bm\Delta_{Z,w}^{\rm A|A^c}\\
\bm\tau\in\mathsf{S}(\bm\delta(\bm\delta\bm\rho\sqcap\bm\Delta_{Q_1}^{\rm A}))}}(-1)^{s_{\bm\rho}s_{\bm\tau}}\,.
    \end{split}
\end{align}
Let us make the following definitions
\begin{subequations}
    \begin{align}
    \bm\Delta_{\rm q}^{\mathrm{A}}&=\{\bm\sigma\in\bm\Delta_{\text{q}}|\bm\sigma\in\bm\delta\bm\rho\text{ for some }\bm\rho\in  \bm\Delta_{Q_1}^{\rm A}\}\,,\\
    \bm\Delta^{\mathrm{A^c}}_{\rm q}&=\{\bm\sigma\in\bm\Delta_{\text{q}}|\bm\sigma\in\bm\delta\bm\rho\text{ for some }\bm\rho\in  \bm\Delta_{Q_1}^{\rm A^c}\}\,,\\
    \bm\Delta_X^{\rm A^c}&=\{\bm\sigma_X\in\bm\Delta_X|\bm\delta^*\bm\sigma_X\sqcap \bm\Delta_{\rm q}^{\rm A^c}\neq 0\}\,,\\
    \bm \Delta_Z^{\rm A|A^c}&=\{\bm\sigma\in\bm\Delta_Z\cap\bm\Delta_{Q_1}^{\mathrm{A}}| \bm\delta\bm\sigma_Z\sqcap\bm\Delta^{\mathrm{A^c}}_{\rm q}\neq 0\} \, .
\end{align}
\label{eq:Delta_qdef}
\end{subequations}
We note that if $\bm \Delta_{X}$ is non-empty, we have the full chain complex \eqref{eq:chain-complex-foliated-rewritten}, then, flipping the spins on $\bm \delta^*\bm\sigma_{X}\sqcap \bm\Delta_{Q_2}^{\rm A^c}$ for some $\bm\sigma_X=\sigma_X\times\{w\}\in\bm\Delta_X$ is a local symmetry of the generalized Ising model defined on $\bm\Delta_{Q_1}^{\rm A^c}$\footnote{The Ising interactions for generalized Ising model defined on $\bm\Delta_{Q_1}^{\rm A^c}$ are of the type $s(\bm\delta\bm\sigma_Z)$ and $s(\bm\delta\bm\sigma_{\mathrm{q},w})$.}. We denote this flip by $g_{\sigma_X\times\{w\}}$. Using this redundancy, we can gauge fix all the $\bm\sigma_{X,w}$ to be 1 except at the last cell $\bm\sigma_{X,w=L_w}$. The value of the spin at $\bm\sigma_{X,w=L_w}$ is the value of the holonomy. When the value of the holonomy is $1$, we have
\begin{align}
   \begin{split}
        \mathcal{Z}^{\text{(CSS)}}_{\mathcal{D}}(J,K)&= \sum_{\bigg\{\substack{s_{\bm\sigma}=\pm 1\\
    (-1)^{h_{\bm\sigma}}=s_{\bm\sigma}}\bigg\}_{\bm\sigma\in\bm\Delta_{Q_2}^{\rm A^c}}}\, \sum_{\bigg\{\substack{s_{\bm\sigma}=\pm 1\\
    (-1)^{h_{\bm\sigma}}=s_{\bm\sigma}}\bigg\}_{\bm\sigma\in\bm\Delta_{Z,w}^{\rm A}}}\\
    &\exp \Big( J \sum_{\bm \sigma_Z \in \bm \Delta_Z\cap \bm\Delta_{Q_1}^{\rm A^c} }   s(\bm \delta \bm \sigma_Z)
+ K \sum_{\bm \sigma_{\mathrm{q},w} \in \bm \Delta_{\text{q},w}\cap \bm\Delta_{Q_1}^{\rm A^c} } s(\bm \delta \bm \sigma_{\mathrm{q},w}\cap \bm\Delta^{\mathrm{A^c}}_{\rm q} )
+ J^{*} \sum_{\bm \sigma_Z \in \bm \Delta_Z\cap \bm\Delta_{Q_1}^{\rm A} }    s(\bm \delta^* \bm \sigma_Z)
+ K^{*} \sum_{\bm \sigma_{\mathrm{q},w} \in \bm \Delta_{\text{q},w}\cap \bm\Delta_{Q_1}^{\rm A} }  s(\bm \delta^* \bm \sigma_{\mathrm{q},w}) 
\Big)\\
&\prod_{\substack{\bm \rho\in\bm\Delta_{Z,w}^{\rm A|A^c}\\
\bm\tau\in\mathsf{S}(\bm\delta(\bm\delta\bm\rho\sqcap\bm\Delta_{Q_1}^{\rm A})\sqcap\bm\Delta_{\text{q}})}}(-1)^{h_{\bm\rho}h_{\bm\tau}}\,.
    \end{split} 
\end{align}
To construct the corresponding quantum model, we can construct the transfer matrix of this model using the same strategy as before. In the presence of non-trivial holonomy, the coupling $K$ in front of an Ising interaction of the form $s(\bm\delta\bm\sigma_{\mathrm{q},w}\sqcap \bm\Delta_{\rm q}^{\rm A^c})$ where $\bm\sigma_{X,w}\in\bm\delta\bm\sigma_{\mathrm{q},w}$ is flipped to $-K$. This can be incorporated into the transfer matrix by inserting the Gauss's law $G_{\sigma_X}\equiv \prod\limits_{\substack{\sigma_{\rm q}\in \delta^*\sigma_X\sqcap \Delta_{\rm q}^{\rm A^c}\\
\sigma_X\in\Delta_X^{\rm A^c}}}X(\sigma_{\rm q})$ where $\Delta_X^{\rm A^c}$ and $\Delta_{\rm q}^{\rm A^c}$ are the restriction of the set $\bm \Delta_X^{\rm A^c}$ and $\bm \Delta_{\rm q}^{\rm A^c}$ to the $w=0$ plane.
\begin{align}
    \mathcal{Z}^{\text{(CSS)}}_{\mathcal{D}}(J,K)\equiv \text{Tr}\left(\prod_{\substack{\sigma_X\in\Delta_X^{\rm A^c}}}(1+G_{\sigma_X})\hat{\rm T}^{L_w}\right)\,,
\end{align}
where $L_w$ is the number of foliated layers in the $w$ direction.
We note that, the above partition function still has $\tilde{g}_{\sigma_X}=\prod_w g_{\sigma_X\times\{w\}}$ as a symmetry. Spin configurations that are related by $\tilde{g}_{\sigma_X}$ has the same Boltzmann weight in the partition function and the same value in any non-zero correlation function. This in fact make the Gauss's law to commute with the transfer matrix $\hat{\rm T}$ as in the 3D Ising gauge theory. 
We need to take $L_w\rightarrow\infty$ limit to obtain the quantum Hamiltonian. Generalizing the arguments in the previous sections
\begin{align}
    \hat{\rm T}=e^{-\tau\hat{\rm H}_{\text{CSS}|\text{CSS-dual}}}\sim 1-\tau\hat{\rm H}_{\text{CSS}|\text{CSS-dual}}\, .
\end{align}
We have
\begin{align}
    e^{-2K}=\tau\,, \quad J=\lambda\tau\,,\quad e^{-2J^*}=\lambda\tau\,, \quad K^*=\tau\, .
\end{align}
Now let us consider the following collection of cells in the $w=0$ plane: $\Delta_{Z}$ and $\Delta_{\mathrm{q}}$. We also consider the collection $\Delta_{\rm q}^{\rm A}$, $\Delta_{\rm q}^{\rm A^c}$, and $\Delta_{Z}^{\rm A|A^c}$ defined to be the restriction of the sets in \eqref{eq:Delta_qdef} to the $w=0$ plane.
The quantum Hamiltonian is
\begin{align}
    \hat{\rm H}_{\text{CSS}|\text{CSS-dual}}&=-\lambda\sum_{\sigma_Z\in\Delta_Z\cap \bm\Delta_{Q_1}^{\rm A^c}}Z(\delta\sigma_Z)-\sum_{\sigma_{\mathrm{q}}\in\Delta^{\mathrm{A^c}}_{\rm q}\setminus (\Delta_{\rm q}^{\mathrm{A}}\cap\Delta^{\mathrm{A^c}}_{\rm q})}X(\sigma_{\mathrm{q}})-\lambda\sum_{\sigma_Z\in\Delta_Z\cap\bm\Delta_{Q_1}^{\rm A}}X(\sigma_Z)-\sum_{ \sigma_{\mathrm{q}}\in  \Delta^{\mathrm{A}}_{\rm q}\setminus (\Delta_{\rm q}^{\mathrm{A}}\cap\Delta^{\mathrm{A^c}}_{\rm q})}Z(\delta^*\sigma_{\rm q}\sqcap\Delta_Z)\nonumber\\
    &\qquad-\sum_{ \sigma_{\rm q}\in \Delta_{\rm q}^{\mathrm{A}}\cap\Delta^{\mathrm{A^c}}_{\rm q}}X( \sigma_{\rm q})Z( \delta^* \sigma_{\rm q}\sqcap\Delta_Z^{\rm A|A^c})\,,
    \label{eq:HCSSdualitydefect}
\end{align}
 with the Gauss's law
 \begin{align}
     G_{\sigma_X}\ket{\Psi}=\ket{\Psi}\,\qquad\text{ for }\qquad\sigma_X\in \Delta_X^{\rm A^c}\,,
 \end{align}
 on the physical states $\ket{\Psi}$. In the absence of local symmetry, we have \eqref{eq:HCSSdualitydefect} without the Gauss's law.
\subsubsection{Quantum Hamiltonian with condensation defect}
Here, we restrict our discussion to the case when the duality defect in the corresponding statistical model is fully topological. After fusing two duality defects, we get the condensation defect as obtained in \eqref{eq:condensationdefect}. Here, we first gauge fix the temporal cells $\bm\sigma_{X,w}$ to be 1 except the one at the cell $\bm\sigma_{X,w=L_w}$. Let us embed the chain complex cells in an underlying manifold. Let $\mathcal{P}$ be a codimension-1 hyperplane that include the $w$ direction to be the hyperplane along which the two duality defects fuse. This hyperplane could be defined in terms of a collection $\bm\Delta_{\mathrm{q},w}^{\mathcal{P}}\subset\bm\Delta_{\mathrm{q},w}$ and $\bm\Delta_{Q_2}^{\mathcal{P}}\subset \bm\Delta_{Q_2}$. Consider the configuration $\bm\Delta_{Z,w}^{\rm A}$ just before the fusion of the two duality defects. Using the movement relations, we can fuse the duality defects along the plane $\mathcal{P}$. While doing this, we encounter the sum
\begin{align}
    \sum_{\substack{\{h_{\bm\sigma_{Z,w}}\}\\
    \bm\sigma_Z\in\bm\Delta_{Z,w}^{\rm A}}}(-1)^{\sum\limits_{\substack{\bm\sigma_{Z,w}\\
    w\neq L_w}}h_{\bm\sigma_{Z,w}}\left(\sum\limits_{\substack{\bm\sigma_{\rm q}\in \bm\delta\bm\sigma_Z\sqcap\bm\Delta_{Q_2}^{\mathcal{P}}\\
    \bm\sigma_Z\in\bm\delta\bm\sigma_{Z,w}}}h_{\bm\sigma_{\rm q}}+k_{\sigma_{\rm q}}\right)}(-1)^{\sum\limits_{\substack{\bm\sigma_{Z,w}\\
    w=L_w}}h_{\bm\sigma_{Z,w}}\left(\sum\limits_{\substack{\bm\sigma_{\rm q}\in \bm\delta\bm\sigma_Z\sqcap\bm\Delta_{Q_2}^{\mathcal{P}}\\
    \bm\sigma_Z\in\bm\delta\bm\sigma_{Z,w}}}h_{\bm\sigma_{\rm q}}+k_{\sigma_{\rm q}}+\sum\limits_{\substack{\bm\sigma_{X,w}\in \bm\delta\bm\sigma_{\mathrm{q},w}\sqcap\bm\Delta_{Q_2}^{\mathcal{P}}\\
    \bm\sigma_{\mathrm{q},w}\in\bm\delta\bm\sigma_{Z,w}}}h_{\bm\sigma_{X,w}}+k_{\sigma_{X,w}}\right)}\,.
\end{align}
For the above sum to be non-zero, we need the sum inside the paranthesis to be zero mode 2. The solutions can be written in terms of a chain $c_{ Q_2}\sqcap \bm\Delta_{Q_2}^{\mathcal{P}}$ with cells in the underlying set $\bm\Delta_{Q_2}^{\mathcal{P}}$. The solution is given by $h_{\bm\sigma_{Q_2}}=k_{\bm\sigma_{Q_2}}+\mathcal{I}^{c_{\rm Q_2}}_{\bm\sigma_{Q_2}}$ (with $\bm\sigma_{Q_2}$ being $\sigma_{X,w}$ or $\sigma_{\rm q}$) where
\begin{align}
    \mathcal{I}_{\bm\sigma_{Q_2}}^{c_{Q_2}}=\begin{cases}
        1\qquad \text{if }\#(\bm\sigma_{Q_2}\cap c_{Q_2})\neq 0\,,\\
        0\qquad\text{otherwise}\,.
    \end{cases}
\end{align}
Then essentially, we have
\begin{align}
    \mathcal{D}\times\mathcal{D}\propto \sum_{\substack{c_{Q_2}\in \bm C_{Q_2}^{\mathcal{P}}\\
    \bm\delta^*c_{Q_2}\sqcap \bm\Delta_{\mathrm{q},w}^{\mathcal{P}}=0}}D_{c_{Q_2}}\,,
\end{align}
where $\bm C_{Q_2}^{\mathcal{P}}$ is the chain group formed from the underlying set $\bm\Delta_{Q_2}^{\mathcal{P}}$.
The quantum Hamiltonian with the condensation defect can be derived following the same line of argument as in $3D$-Ising gauge theory. Let us denote the partition function with condensation defect inserted to be $\mathcal{Z}_{\mathcal{D}^2}^{\text{CSS}}$. We can construct the quantum Hamiltonian by taking the $\tau$-continuum limit of the transfer matrix. We use again the notation $\Delta_Z$, $\Delta_{\rm q}$, and $\Delta_X$ to denote the collection of cells $\bm\Delta_Z$, $\bm\Delta_{\rm q}$, and $\bm\Delta_X$ in the $w=0$ plane and $\delta$ to denote the restriction of the boundary map $\bm\delta$ to the $w=0$ plane. We define the Hilbert space by the basis states formed by all spin configurations $\ket{\{s_{\sigma_{\rm q}}\},\{s_{\sigma_{\rm q}^a}\}}$ where $\sigma_{\rm q}\in\Delta_{\rm q}$ and ancilla d.o.f on cells $\sigma_{\rm q}^a\in\Delta_{\rm q}^a$ where $\Delta_{\rm q}^a\equiv \{\sigma_{\rm q}^a|\sigma_{\rm q}\times \{w=0\}\in\bm\delta\bm\sigma_{\mathrm{q},w=0}\,,\quad \bm\sigma_{\mathrm{q},w=0}\in \bm\Delta_{\mathrm{q},w}^{\mathcal{P}}\}$. The ancilla d.o.f is introduced to capture the sum over condensation defects. With this definition of Hilbert space, the partition function takes the form
\begin{align}
    \mathcal{Z}^{\text{CSS}}\propto\text{Tr}\left(\prod_{\sigma_{X}\in\Delta_X}(1+G_{\sigma_X})(1+G_{z_{\rm q}^*}^{\mathcal{P}})\mathrm{\hat{T}}^{L_w}\right)\,,
\end{align}
where $G_{\sigma_X}=\prod\limits_{\sigma_{\rm q}\in\delta^*\sigma_X}X(\sigma_{\rm q})$  is the Gauss's law operator and $G_{z_{\rm q}^*}^{\mathcal{P}}=\prod\limits_{\sigma_{\rm q}\in z_{\rm q}^*,\delta^*z_{\rm q}^*=0}X(\sigma_{\rm q})$ is the Gauss's law operator parallel to the plane $\mathcal{P}'=\mathcal{P}\cap \{w=0\}$ the intersection of hyperplane $\mathcal{P}$ and $\{w=0\}$ hyperplane. The product $\prod\limits_{\sigma_{X}\in\Delta_X}(1+G_{\sigma_X})$ essentially sum over all possible holonomy configurations. The term $(1+G_{z_{\rm q}^*}^{\mathcal{P}})$ imposes a non-local Gauss's law to account for the condensation defect. Due to this Gauss's law, any symmetry of the form $X(z_{\rm q}^*)$ where $z_{\rm q}^*$ is parallel to the hyperplane $\mathcal{P}'$ can be absorbed to the condensation defect. Let us consider a normal vector to the hyperplane $\mathcal{P}'$. This normal vector define a notion of positive and negative region whose boundary is $\mathcal{P}'$. Let us denote it by $\mathcal{V}^+$ and $\mathcal{V}^-$\footnote{The positive and negative regions might be connected if the underlying manifold $w=0$ hyperplane is is not infinite or doesn't have a boundary. However, we are only interested in the cells near to the hyperplane $\mathcal{P}'$ and $\mathcal{V}^+$ and $\mathcal{V}^-$ can be thought of as region that is homeomorphic to $\mathcal{P}'\times [0,\epsilon]$ and $\mathcal{P}'\times [-\epsilon,0]$ respectively.}. Now let us define the following set: $\Delta_{Z}^{a,-}=\{\sigma_{Z}|\sigma_Z\in\delta^*\sigma_{\rm q}^a\,,\quad \sigma_Z\in \mathcal{V}^-\}$. To get the quantum Hamiltonian we take $L_w\rightarrow\infty$ and write $\hat{\rm T}=e^{-\tau\hat{\rm H}_{\text{CSS-cond}}}\approx 1-\tau \hat{\rm H}_{\text{CSS-cond}}$. The $2+1$D quantum Hamiltonian with condensation defect is 
\begin{align}
    \hat{\rm H}_{\text{CSS-cond}}=-\lambda\sum_{\sigma_{Z}\in\Delta_Z\setminus\Delta_Z^{a,-}}Z(\delta\sigma_Z)-\lambda\sum_{\sigma_{Z}\in\Delta_Z^{a,-}}Z(\delta\sigma_Z)Z(\delta\sigma_{Z}\sqcap\Delta_{\rm q}^a)-\sum_{\sigma_{\rm q}\in\Delta_{\rm q}}X(\sigma_{\rm q})\,,
    \label{eq:HCSS-cond}
\end{align}
with Gauss's law
\begin{align}
    G_{\sigma_X}\ket{\Psi}=\ket{\Psi}\qquad \text{ for }\qquad\sigma_X\in\Delta_X^{\rm A^c}\,, \text{ and }\qquad G_{z_{\rm q}^*}\ket{\Psi}=\ket{\Psi}\qquad \text{for}\qquad \delta^*z_{\rm q}^*=0\,,
\end{align}
on the physical states $\ket{\Psi}$. In the absence of local symmetry, we have \eqref{eq:HCSS-cond} without the Gauss's law.
\section{Conclusion}
In this work, we have constructed duality defects (or KW defects) in a variety of lattice models by employing the strange correlator formalism. Our approach was demonstrated through three illustrative examples. First, we rederived the well-known duality defect of the two-dimensional Ising model~\cite{Aasen:2016dop} using our method. We then extended the construction to the three-dimensional anisotropic plaquette Ising model, and finally to the interface between the three-dimensional Ising gauge theory and the three-dimensional Ising model. In each case, we analyzed the manifestation of the duality defect both in the classical models and in their corresponding quantum Hamiltonian realizations, related through the standard classical-to-quantum correspondence (i.e., a $d$-dimensional classical model corresponds to a $(d-1)+1$-dimensional quantum Hamiltonian model). Beyond these concrete examples, we presented a generalization of the construction to chain-complex–based Ising models, thereby obtaining duality defects for both the classical models and their associated quantum Hamiltonians.

Several natural directions for future work emerge from our results. One promising avenue is to extend the present formalism from $\mathbb{Z}_2$ classical spins to $\mathbb{Z}_N$ systems. The chain-complex framework is expected to generalize straightforwardly to $\mathbb{Z}_N$ qudits, at least when $N$ is prime\footnote{For prime $N$, the chain groups form vector spaces.}. It would also be of significant interest to explore degrees of freedom associated with more general finite groups beyond cyclic ones, and to investigate the corresponding duality defects in their partition function or path-integral formulations. A generalization to lattice models with fusion higher-category symmetry in higher dimensions is also worth exploring. 

In our analysis, we encounter junctions between the duality defect and the $\mathbb{Z}_2$ symmetry defect. However, we have not yet examined the corresponding F-symbols that relate different fusion channels of these defects. It would therefore be a valuable exercise to derive the F-symbols for the 2D models within this framework.

Another interesting direction would be to explore possible applications of partition functions with duality defects in dimensions $d>2$. In Refs.~\cite{Aasen:2016dop,Aasen:2020jwb}, duality defect partition functions were utilized in the presence of a Dehn twist to extract the conformal spins of operators in the corresponding continuum CFT. In addition, Ref.~\cite{Aasen:2020jwb} employed duality defects to determine universal g-factor ratios of conformal boundary conditions directly on the lattice and analyzed degeneracies in certain gapped spin chains via duality operators. It would be worthwhile to seek analogous applications of duality defects in higher dimensions through the strange correlator construction.
\pagebreak\\
\textbf{Acknowledgements:} We thank Anirudh Deb for collaboration during the early stages of this project. A.P.M. gratefully acknowledges Takuya Okuda and Hiroki Sukeno for collaborations on related works~\cite{Okuda:2024azp,Okuda:2024jzh} that inspired and informed this study. A.P.M, would also like to thank Fei Yan and Tzu-Chieh Wei for  discussions on the duality defect in the plaquette Ising model. Y.S. thanks Shu-Heng Shao and Yunqin Zheng for valuable discussions. We thank Shu-Heng Shao, Yunqin Zheng and Paul Fendley for comments on the draft. APM was mainly supported by the U.S. National Science Foundation under Award No. PHY 2310614. YS was supported by the Simons Collaboration on Ultra-Quantum Matter, which is a grant from the Simons Foundation (651444, SHS).
\\
\appendix

\section{Strange Correlator in 2D Ising Model}\label{app:2DIsingSC}
We start with the overlap of $\bra{0}e^{JX_e}$ with $\ket{{\rm TC}_{\rm GS}}$ defined in \eqref{eq:2dIsingClusterState} and \eqref{eq:2dIsingTCGS} as
\begin{align}\label{eq:2dIsingSC}
  (\langle 0|e^{J X_{\rm e}})^{\rm \Delta_e}  |\rm TC_{GS}\rangle&=(\langle 0|e^{J X_{\rm e}})^{\rm \Delta_e} \langle +|^{\rm \Delta_v}\prod_{\rm v\in\Delta_v}\prod_{\rm e\in\partial^*v}CZ_{\rm v,e}\ket{+}^{\rm\Delta_v}\ket{+}^{\rm\Delta_e}\nonumber\\
  &=\bra{0}^{\rm\Delta_e}\langle +|^{\rm \Delta_v}\prod_{\rm e\in\partial^*v}CZ_{\rm v,e}\prod_{\rm e}e^{J X_{\rm e}\prod_{\rm v\in \partial e}Z_{\rm v}}\ket{+}^{\rm\Delta_v}\ket{+}^{\rm\Delta_e}\nonumber\\
  &=\bra{0}^{\rm\Delta_e}\langle +|^{\rm \Delta_v}\prod_{\rm e}e^{J\prod_{\rm v\in \partial e}Z_{\rm v}}\ket{+}^{\rm\Delta_v}\ket{+}^{\rm\Delta_e}\nonumber\\
  &=\frac{1}{2^{|\Delta_{\rm e}|/2}}\frac{1}{2^{|\Delta_{\rm v}|}}\sum_{\{s_{\rm v}\}}e^{J\sum_{\langle {\rm v,v'} \rangle}s_{\rm v} s_{{\rm v}'}}\nonumber\\
  &=\frac{1}{2^{|\Delta_{\rm e}|/2}}\frac{1}{2^{|\Delta_{\rm v}|}}Z_{\rm Ising}\,.
\end{align}
where in second equality, when used $X_e CZ_{\rm v,e} = CZ_{\rm v,e}X_{\rm e} Z_{\rm v}$. In the third equality, $CZ_{\rm v,e}$ becomes $1$ when it acts on $\bra{0}^{\Delta_{\rm e}}$ and $X_e$ also becomes $1$ when it acts on $\ket{+}^{\Delta_{\rm e}}$. In the fourth equality, we expanded the product $\ket{+}^{\Delta_{\rm v}} = \otimes_{\rm v} \frac{1}{\sqrt{2}} \left( \ket{0}_{\rm v} + \ket{1}_{\rm v}\right)$ and $Z_{\rm v}$ in the exponential became its eigenvalue $s_{\rm v} \in \{-1,1\}$.

Now let's rewrite \eqref{eq:TCGSandTCGS*relation} here
\begin{align}\label{eq:appTCGSandTCGS*relation}
   \frac{1}{\mathcal{N}_{\text{TC}}}\mathsf{H}|\mathrm{TC}_{GS}\rangle=\frac{1}{\mathcal{N}^*_{\text{TC}}} \frac{1}{H_1(M,\mathbb{Z}_2)}\sum_{z_1\in H_1(M,\mathbb{Z}_2)}Z(z_1)|\rm TC^*_{GS}\rangle\,.
\end{align}
We want to take an overlap of both sides above with $ \bra{0}^{\Delta_{\rm e}}e^{J \sum_{\rm e}X_{\rm e}} \mathsf{H}$. On the left side, the Hadamard operator $H_e$ cancels out and we evaluate it using \eqref{eq:2dIsingSC}. To evaluate the right side, first we use $\bra{0}_{\rm e}e^{J X_{\rm e}} H_e=(\sinh (2J))^{1/2}\bra{0}_{\rm e}e^{J^*X_{\rm e}}$ with $J^* = -\frac{1}{2}\ln(\tanh(J))$. Then we focus on the following overlap which is evaluated as
\begin{align}
    \bra{0}^{\Delta_{\rm e}}e^{J \sum_{\rm e}X_{\rm e}} \mathsf{H}|{\rm TC^*_{GS}}\rangle\ &=  \bra{0}^{\Delta_{\rm e}}e^{\sum_{\rm e}J^* X_{\rm e}}Z(z_1)|\rm TC^*_{GS}\rangle\ \\
    &=  \bra{0}^{\Delta_{\rm e}}e^{\sum_{\rm e} J^* (-1)^{\# (z_1,{\rm e})} X_{\rm e}}|\rm TC^*_{GS}\rangle\ \\
    &= \bra{0}^{\Delta_{\rm e}}e^{\sum_{\rm e}J^* (-1)^{\#(z_1,{\rm e})}X_{\rm e}}
   \bra{+}^{\Delta_{\rm p}} \prod_{\rm p\in\Delta_p}\prod_{\rm e\in\partial p}CZ_{\rm p,e}\ket{+}^{\rm\Delta_p}\ket{+}^{\rm\Delta_e} \\
   &= \bra{0}^{\Delta_{\rm e}} \bra{+}^{\Delta_{\rm p}} 
    \prod_{\rm p\in\Delta_p}\prod_{\rm e\in\partial p}CZ_{\rm p,e} e^{\sum_{\rm e} J^* (-1)^{\#( z_1, {\rm e} )}X_{\rm e} \prod_{{\rm p}\in \partial^*{\rm e}} Z_{\rm p}} \ket{+}^{\rm\Delta_p}\ket{+}^{\rm\Delta_e} \\
    &= \bra{0}^{\Delta_{\rm e}} \bra{+}^{\Delta_{\rm p}} 
    \prod_{\rm p\in\Delta_p} e^{\sum_{\rm e} J^* (-1)^{\#( z_1,{\rm e} )} \prod_{{\rm p}\in \partial^*{\rm e}} Z_{\rm p}} \ket{+}^{\rm\Delta_p}\ket{+}^{\rm\Delta_e}\\
    &= \frac{1}{2^{|\Delta_{\rm p}|}}\frac{1}{2^{|\Delta_{\rm e}|/2}} \sum_{\{s_{\rm p}\}} e^{\sum_{\langle {\rm p,p'}\rangle} J^{*} (-1)^{\#(z_1, {\rm p,p'})} s_{\rm p} s_{\rm p'} }
\end{align} where in the second equality, we passed $e^{J^*X_e}$ through $Z(z_1)$ which gave a $(-1)^{\#(z_1,{\rm e})}$ factor in the exponential, where ${\#( {\rm e} , z_1} )$ equals $0$ if ${\rm e} \in z_1$ and $1$ otherwise. Then we absorbed $Z(z_1)$ on to $\bra{0}^{\Delta_e}$. In the third equality, we re-expressed the Toric code $\ket{\rm TC^*_{GS}}$ state using its corresponding Cluster State. In the fourth equality, we used $X_{\rm e} CZ_{\rm p ,e} =  CZ_{\rm p ,e} X_{\rm e} Z_{\rm p}$. In the fifth equality, we absorbed $CZ_{\rm p,e}$ on to $\bra{0}^{\Delta_{\rm e}}$ and absorbed $X_{\rm e}$ on to $\ket{+}^{\Delta_{\rm e}}$. Finally, in the last equality, we expand $\ket{+} = \frac{1}{\sqrt{2}} \left(\ket{0} + \ket{1}\right)$ from which $Z_p$ in the exponential picks the eigenvalue $s_{\rm p}$ and rewrite $e$ as $\langle \rm p,p'\rangle$ making $(-1)^{\#(z_1, {\rm \langle p,p'\rangle})}$ the intersection number. The sum in the last equality is the Ising partition function on the dual lattice with a $\mathbb{Z}_2$ defect inserted along $z_1$ cycle. This means that taking overlap of 
$ \bra{0}^{\Delta_{\rm e}}e^{J \sum_{\rm e}X_{\rm e}} \mathsf{H}$ on both sides of \eqref{eq:appTCGSandTCGS*relation}, we get
\begin{align}
    \frac{1}{\mathcal{N}_{\rm TC}} \frac{1}{2^{|\Delta_{\rm e}|/2}}\frac{1}{2^{|\Delta_{\rm v}|}}Z_{\rm Ising} = \frac{1}{\mathcal{N}^*_{\text{TC}}} \frac{1}{H_1(M,\mathbb{Z}_2)}\sum_{z_1\in H_1(M,\mathbb{Z}_2)}\frac{1}{2^{|\Delta_{\rm p}|}}\frac{1}{2^{|\Delta_{\rm e}|/2}} Z_{\rm Ising}(J^*, z_1)\,.
\end{align}
Now, we need to find the normalization constants  $\mathcal{N}_{\text{TC}}$ and $\mathcal{N}^*_{\text{TC}}$. Let's start with the definition of $\mathcal{N}_{\text{TC}}$
\begin{align}
    \mathcal{N}^2_{\text{TC}} &= \braket{\rm TC_{GS} | TC_{GS}} \\
    &= \bra{+}^{\rm\Delta_v}\bra{+}^{\rm\Delta_e}\prod_{\rm v\in\Delta_v}\prod_{\rm e\in\partial^*v}CZ_{\rm v,e}\ket{+}^{\Delta_{\rm v}} \bra{+}^{\Delta_{\rm v}}\prod_{\rm v\in\Delta_v}\prod_{\rm e\in\partial^*v}CZ_{\rm v,e}\ket{+}^{\rm\Delta_v}\ket{+}^{\rm\Delta_e}\,.
\end{align}
Expand all $\ket{+}$ and $\bra{+}$ in the $Z$ eigenstates, and use $CZ_{\rm v,e}\ket{h_{\rm v}}_{\rm v}\ket{h_{\rm e}}_{\rm e} = (-1)^{h_{\rm v} h_{\rm e}}\ket{h_{\rm v}}_{\rm v}\ket{h_{\rm e}}_{\rm e}$ for $h_{\rm v}, h_{\rm e} \in \{0,1\}$ to get
\begin{align}
    \mathcal{N}^2_{\text{TC}} = \frac{1}{2^{|\Delta_{\rm v}|}}\frac{1}{2^{2|\Delta_{\rm e}|}}\sum_{\{h_{\rm e}\},\{h_{\rm v}\},\{k_{\rm v}\}} \prod_{e=\langle {\rm v,v'}\rangle} (-1)^{h_{\rm e}(h_{\rm v} + h_{\rm v'} + k_{\rm v} + k_{\rm v'})} 
\end{align}
where $\partial \rm e = \{v,v'\}$. Performing the sum over $\{h_{\rm e}\}$, we will get non-zero value when either $h_{\rm v}=k_{\rm v}$ and $h_{\rm v'} = k_{v'}$ or $h_{\rm v}=k_{\rm v}+1$ and $h_{\rm v'} = k_{v'}+1$ which gives us
\begin{align}
    \mathcal{N}^2_{\text{TC}} = \frac{1}{2^{|\Delta_{\rm v}|}}\frac{1}{2^{2|\Delta_{\rm e}|}}\sum_{\{h_{\rm v}\},\{k_{\rm v}\}} 2^{|\Delta_{\rm e}|}\prod_{\langle \rm v,v'\rangle }\left(\delta_{h_{\rm v}, k_{\rm v}}\delta_{h_{\rm v'},k_{\rm v'}} + \delta_{h_{\rm v}, k_{\rm v}+1} \delta_{h_{\rm v'}, k_{\rm v'}+1} \right)
\end{align}
Now we expand the product and use $\delta_{h_{\rm v}, k_{\rm v}}\delta_{h_{\rm v}, k_{\rm v +1}} = 0$, 
\begin{align}
    \mathcal{N}^2_{\text{TC}} &= \frac{1}{2^{|\Delta_{\rm v}|}}\frac{1}{2^{|\Delta_{\rm e}|}}\sum_{\{h_{\rm v}\},\{k_{\rm v}\}} \left(\delta_{\{h_{\rm v}\}, \{k_{\rm v}\}} + \delta_{\{h_{\rm v}\}, \{k_{\rm v} + 1\}} \right) \\
    &= \frac{1}{2^{|\Delta_{\rm v}|}}\frac{1}{2^{|\Delta_{\rm e}|}} 2^{|\Delta_{\rm v}| + 1}
\end{align}
where $\delta_{\{h_{\rm v}\}, \{k_{\rm v}\}} = \prod_{\rm v} \delta_{h_{\rm v},k_{\rm v}}$. The normalization constant is
\begin{align}
    \mathcal{N}_{\text{TC}} = \frac{1}{2^{\frac{|\Delta_{\rm e}|-1}{2}}}\,.
\end{align}
Let's now find $\mathcal{N}^*_{\rm TC}$.
\begin{subequations}
\begin{align}
    {\mathcal{N}^*_{\text{TC}}}^2 &= \bra{\rm TC_{GS}^*} \frac{1}{H_1(M,\mathbb{Z}_2)}\sum_{z_1 \in H_1(M, \mathbb{Z}_2)}Z(z_1)\ket{\rm TC_{GS}^*} \\
    &= \bra{+}^{\rm\Delta_p}\bra{+}^{\rm\Delta_e}  \prod_{\rm p\in\Delta_p}\prod_{\rm e\in\partial p}CZ_{\rm p,e}\ket{+}^{\Delta_{\rm p}}\frac{1}{H_1(M,\mathbb{Z}_2)}\sum_{z_1 \in H_1(M, \mathbb{Z}_2)}Z(z_1)\bra{+}^{\Delta_{\rm p}} \prod_{\rm p\in\Delta_p}\prod_{\rm e\in\partial p}CZ_{\rm p,e}\ket{+}^{\rm\Delta_p}\ket{+}^{\rm\Delta_e}
\end{align}
\end{subequations}
Expand all $\ket{+}$ and $\bra{+}$ in the $Z$ eigenstates, and use $CZ_{\rm p,e}\ket{h_{\rm v}}_{\rm p}\ket{h_{\rm e}\}_{\rm e} = (-1)^{h_{\rm v}} h_{\rm e}}\ket{h_{\rm p}}_{\rm p}\ket{h_{\rm e}}_{\rm e}$ for $h_{\rm p}, h_{\rm e} \in \{0,1\}$ and use $Z\ket{h_{\rm e}}_{\rm e} = (-1)^{h_{\rm e}}\ket{h_{\rm e}}_{\rm e}$ to get
\begin{align}
    {\mathcal{N}^*_{\text{TC}}}^2 = \frac{1}{H_1(M,\mathbb{Z}_2)} \frac{1}{2^{|\Delta_{\rm p}|}}\frac{1}{2^{2|\Delta_{\rm e}|}}\sum_{z_1 \in H_1(M, \mathbb{Z}_2)}\sum_{\{h_{\rm e}\},\{h_{\rm p}\},\{k_{\rm p}\}} \prod_{{\rm e}|\partial^*{\rm e}=\{\rm p,p'\}} (-1)^{h_{\rm e}(h_{\rm p} + h_{\rm p'} + k_{\rm p} + k_{\rm p'}+ \#(z_1,\rm e))}\,. 
\end{align}
Now, we perform the sum over $\{h_{\rm e}\}$ to get
\begin{align}
    {\mathcal{N}^*_{\text{TC}}}^2 = \frac{1}{H_1(M,\mathbb{Z}_2)} \frac{1}{2^{|\Delta_{\rm p}|}}\frac{1}{2^{2|\Delta_{\rm e}|}}\sum_{z_1 \in H_1(M, \mathbb{Z}_2)}\sum_{\{h_{\rm p}\},\{k_{\rm p}\}}2^{|\Delta_{\rm e}|} \prod_{{\rm e}|\partial^*{\rm e}=\{\rm p,p'\}}\left(\delta_{h_{\rm p}, k_{\rm p}}\delta_{h_{\rm p'},k_{\rm p'}+\#(z_1,\rm e)} + \delta_{h_{\rm p}, k_{\rm p}+1} \delta_{h_{\rm p'}, k_{\rm p'}+1+\#(z_1,\rm e)} \right)\,.
\end{align}
If we expand the product over $\rm e$, then in sum of $z_1$ over ${H_1(M,\mathbb{Z}_2)}$ only the trivial cycle survives. After this, all steps are similar to the case when we found $\mathcal{N}_{\rm TC}$. We finally get
\begin{align}
\mathcal{N}^*_{\text{TC}} =\frac{1}{\sqrt{H_1(M,\mathbb{Z}_2)}}\frac{\sqrt{2}}{2^{\frac{|\Delta_{\rm p}|}{2}}}\,.
\end{align}
\section{Norm of topologically ordered states}\label{app:topologicallyorderedstatenorm}
We compute the norm of various topologically ordered states that appear in Section~\ref{sec:KWdefectgenIsingmodel}. 
\begin{enumerate}
    \item 
    \begin{subequations}
    \begin{align}
        \langle \Phi|\Phi\rangle&=\langle \Psi_{\mathcal{C}}|+\rangle^{\Delta_X}\langle +|^{\Delta_X}|\Psi_{\mathcal{C}}\rangle\\
        &=\frac{1}{2^{2|\Delta_{X}|+|\Delta_{\rm q}|}}\sum_{\substack{\{h_{ \sigma_{\rm q}}\},\sigma_{\rm q}\in\Delta_{\rm q}\\
        \{h_{\sigma_{X}}\},\{k_{\sigma_X}\},\sigma_X\in\Delta_{X}}}(-1)^{\sum_{\sigma_{\rm q}\in\Delta_{\rm q}}h_{\sigma_{\rm q}}\left(\sum_{\sigma_X\in\delta \sigma_{\rm q}}h_{\sigma_X}+k_{\sigma_X}\right)}\\
        &=\frac{1}{2^{2|\Delta_{X}|+|\Delta_{\rm q}|}}2^{|\Delta_{\rm q}|+|\Delta_{X}|}\sum_{ z_X\in C_{X},\delta^*_X z_X=0}\delta_{\{h_{\sigma_X}\},\{k_{ \sigma_X}+\mathcal{I}^{ z_X}_{\sigma_X}\}}\,\qquad \text{ where}\quad \mathcal{I}^{ z_X}_{\sigma_X}=\begin{cases}
        1\quad \text{ if } \#(\sigma_X\cap  z_X)\neq 0\,,\\
        0\quad \text{ otherwise }
    \end{cases}\\
    &=\frac{|\text{Ker}\,\delta^*_X|}{2^{|\Delta_{X}|}}\,.
    \end{align}
    \end{subequations}
    \item 
    \begin{subequations}
    \begin{align}
        \langle \Phi^*|\Phi^*\rangle&=\langle \Psi^*_{\mathcal{C}}|+\rangle^{\Delta_{Z}}\langle +|^{\Delta_{Z}}|\Psi^*_{\mathcal{C}}\rangle\\
        &=\frac{1}{2^{2|\Delta_{Z}|+|\Delta_{\rm q}|}}\sum_{\substack{\{h_{ \sigma_{\rm q}}\},\sigma_{\rm q}\in\Delta_{\rm q}\\
        \{h_{\sigma_Z}\},\{k_{\sigma_Z}\},\sigma_Z\in\Delta_{Z}}}(-1)^{\sum_{\sigma_{\rm q}\in\bm\Delta_{\rm q}}h_{\sigma_{\rm q}}\left(\sum_{\sigma_Z\in\delta^*\sigma_{\rm q}  }h_{\sigma_Z}+k_{\sigma_Z}\right)}\\
        &=\frac{1}{2^{2|\Delta_{Z}|+|\Delta_{\rm q}|}}2^{|\Delta_{\rm q}|+|\Delta_{Z}|}\sum_{z_{Z}\in C_{Z},\delta_Z z_Z=0}\delta_{\{h_{\sigma_Z}\},\{k_{\sigma_Z}+\mathcal{I}^{ z_Z}_{\sigma_Z}\}}\,\qquad \text{ where}\quad \mathcal{I}^{z_Z}_{\sigma_Z}=\begin{cases}
        1\quad \text{ if } \#(\sigma_Z\cap  z_Z)\neq 0\,,\\
        0\quad \text{ otherwise }
    \end{cases}\\
&=\frac{|\text{Ker}\,\delta_Z|}{2^{|\Delta_{Z}|}}\,.
    \end{align}
    \end{subequations}
    \item
    \begin{subequations}
    \begin{align}
        &\langle \Phi^*|\frac{1}{|\mathcal{L}|}\sum_{[ z_{\rm q}]\in \mathcal{L}}Z(z_{\rm q})|\Phi^*\rangle=\langle \Psi^* _{\mathcal{C}}|+\rangle^{\Delta_{Z}}\frac{1}{|\mathcal{L}|}\sum_{ [z_{\rm q}]\in \mathcal{L}}Z(z_{\rm q})\langle +|^{\Delta_{Z}}|\Psi^*_{\mathcal{C}}\rangle\\
       & =\frac{1}{|\mathcal{L}|}\frac{1}{2^{2|\Delta_{Z}|+|\Delta_{\rm q}|}}\sum_{[ z_{\rm q}]\in\mathcal{L}}\sum_{\substack{\{h_{ \sigma_{\rm q}}\},\sigma_{\rm q}\in\Delta_{\rm q}\\
        \{h_{\sigma_Z}\},\{k_{\sigma_Z}\},\sigma_Z\in\Delta_{Z}}}(-1)^{\sum\limits_{\substack{\sigma_{\rm q}\in\Delta_{\rm q}\\
        \#(\sigma_{\rm q}\cap z_{\rm q})\neq 0}}h_{\sigma_{\rm q}}\left(\sum\limits_{\sigma_Z\in\delta^*_Z\sigma_{\rm q}
        }(h_{\sigma_Z}+k_{\sigma_Z})+1\right)}\nonumber\\
        &\hspace{8cm}\times(-1)^{\sum\limits_{\substack{\sigma_{\rm q}\in\Delta_{\rm q}\\
        \#(\sigma_{\rm q}\cap  z_{\rm q})= 0}}h_{\sigma_{\rm q}}\left(\sum\limits_{\sigma_Z\in\delta^*_Z\sigma_{\rm q}
        }(h_{\sigma_Z}+k_{\sigma_Z})\right)}\\
        &=\frac{1}{|\mathcal{L}|}\frac{2^{|\Delta_{\rm q}|}}{2^{2|\Delta_{Z}|+|\Delta_{\rm q}|}}\sum_{\substack{
        \{h_{\sigma_Z}\},\{k_{\sigma_Z}\}\\
        \sigma_Z\in\Delta_{Z}}}\sum_{z_Z\in C_{Z},\delta_Z z_Z=0}\delta_{\{h_{\sigma_Z}\},\{k_{\sigma_Z}+\mathcal{I}^{z_Z}_{\sigma_Z}\}}\,\qquad \text{ where}\quad \mathcal{I}^{ z_Z}_{\sigma_Z}=\begin{cases}
        1\quad \text{ if } \#(\sigma_Z\cap  z_Z)\neq 0\,,\\
        0\quad \text{ otherwise }
    \end{cases}\\
&=\frac{1}{|\mathcal
L|}\frac{|\text{Ker}\,\delta_Z|}{2^{|\Delta_{Z}|}}\,.
        \end{align}
        \end{subequations}
        Going from the second equality to the third, we made the observation that in the sum over nontrivial homologies, only the trivial homology gives a nonzero value.
        \item 
        \begin{subequations}
        \begin{align}
            \langle \Phi^{*\rm A}|\Phi^{*\rm A}\rangle&=\langle \Psi^{\rm A|A^c}_{\mathcal{C}}|+\rangle^{\Delta_{Z}^{\rm A}}|+\rangle^{\Delta_{X}^{\rm A^c}}\langle +|^{\Delta_{X}^{\rm A^c}}\langle +|^{\Delta_{Z}^{\rm A}}|\Psi^{ \rm A|A^c}_{\mathcal{C}}\rangle\\
            &=\frac{1}{2^{2(|\Delta_{Z}^{\rm A}|+|\Delta_{X}^{\rm A^c}|)+|\Delta_{\rm q}|}}\sum_{\substack{\{h_{\sigma_{\rm q}}\},\sigma_{\rm q}\in\Delta_{\rm q}\\
            \{h_{\sigma_X}\},\{k_{\sigma_X}\},\sigma_X\in\Delta_{X}^{\rm A^c}\\
            \{h_{\sigma_Z}\},\{k_{\sigma_Z}\},\sigma_Z\in\Delta_{Z}^{\rm A}}}(-1)^{\sum\limits_{\sigma_{\rm q}\in\Delta_{\rm q}^{\mathrm{A}^c}}h_{\sigma_{\rm q}}\left(\sum\limits_{\sigma_{X}\in\delta_X\sigma_{\rm q}\sqcap\Delta_{X}^{\mathrm{A}^c}}(k_{\sigma_X}+h_{\sigma_X})\right)}\nonumber\\
            &\hspace{3cm}\times(-1)^{\sum\limits_{\sigma_{\rm q}\in\Delta_{\rm q}^{\mathrm{A}}}h_{\sigma_{\rm q}}\left(\sum\limits_{\sigma_Z\in\delta^*_Z\sigma_{\rm q}\sqcap\Delta_{Z}^{\rm A}}(k_{\sigma_Z}+s_{\sigma_Z})\right)}(-1)^{\sum_{\sigma_Z\in\Delta_{Z}^{\rm A|A^c}}h_{\sigma_Z}\left(\sum_{\sigma_X\in\delta_X(\delta_Z\sigma_Z\sqcap \Delta_{\rm q}^{\mathrm{A}})}h_{\sigma_X}\right)}\nonumber\\&\hspace{6cm}\times(-1)^{\sum_{\sigma_Z\in\Delta_{Z}^{\rm A|A^c}}k_{\sigma_Z}\left(\sum_{\sigma_X\in\delta_X(\delta_Z\sigma_Z\sqcap \Delta_{\rm q}^{\mathrm{A}})}k_{\sigma_X}\right)}\\
            &=\frac{2^{|\Delta_{\rm q}|}}{2^{2(|\Delta_{Z}^{\rm A}|+|\Delta_{X}^{\rm A^c}|)+|\Delta_{\rm q}|}}\sum_{\substack{
            \{h_{\sigma_X}\},\{k_{\sigma_X}\},\sigma_X\in\Delta_{X}^{\rm A^c}\\
            \{h_{\sigma_Z}\},\{k_{\sigma_Z}\},\sigma_Z\in\Delta_{Z}^{\rm A}}}\left(\sum_{ z_X\in C_{X}^{\rm A^c},(\delta_{X}^{\rm A^c})^*( z_X)=0}\delta_{\{h_{\sigma_X}\},\{k_{\sigma_X}+\mathcal{I}^{ z_X}_{\sigma_X}\}}\right)\nonumber\\
            &\hspace{8cm}\times\left(\sum_{ z_Z\in C_{Z}^{\rm A},\delta_{Z}^{\rm A}( z_Z)=0}\delta_{\{h_{\sigma_Z}\},\{k_{\sigma_Z}+\mathcal{I}^{ z_Z}_{\sigma_Z}\}}\right)\\
            &=\frac{|\text{Ker}\,(\delta_{X}^{\rm A^c})^*||\text{Ker}\,\delta_{Z}^{\rm A}|}{2^{|\Delta_{Z}^{\rm A}|+|\Delta_{X}^{\rm A^c}|}}\,.
        \end{align}
        \end{subequations}
        \item A very similar calculations as in 3 \& 4 leads to 
        \begin{align}
            \langle \Phi^{*\mathrm{A}}|\frac{1}{|\mathcal{L}|}\sum_{[ z_{\rm q}]\in\mathcal{L}}X( z_{\rm q}\cap \mathrm{A}^c)|\Phi^{*\rm A}\rangle=\frac{|\mathcal{L}^{\rm A^c}|}{|\mathcal{L}|}\frac{|\text{Ker}\,(\delta_{X}^{\rm A^c})^*||\text{Ker}\,\delta_{Z}^{\rm A}|}{2^{|\Delta_{Z}^{\rm A}|+|\Delta_{X}^{\rm A^c}|}}\,,
        \end{align}
        where $\mathcal{L}^{\rm A^c}$ is defined in \eqref{eq:LA^c}.
\end{enumerate}
\section{Triangular lattice Ising model}\label{app:triangularlatticeIsing}
In this section, we work out the KW defect in 2D Ising model on a triangular lattice. 
\subsection{2D traingular lattice Ising model and Kramers-Wannier defect}
\subsubsection{Hamiltonian}
We consider 2D triangular lattice on a torus with spins defined on the vertices. The Hamiltonian of the Ising model is 
\begin{align}
    \mathrm{H}_{\text{Ising}}^{\triangle}=-J\sum_{\langle \rm v,v'\rangle}s_{\rm v}s_{\rm v'}\,.
    \label{eq:triangleIsing}
\end{align}
The partition function is given by
\begin{align}
    Z_{\text{Ising}}^{\triangle}(J)=\sum_{\{s_{\rm v}\}}\exp(J\sum_{\langle\rm v ,v'\rangle}s_{\rm v}s_{\rm v'})\,.
    \label{eq:2DtriangIsingmodelpartitionfn}
\end{align}
The dual lattice to a traingular lattice is a hexagonal lattice. Hence, the Ising model that is KW dual to the triangular lattice Ising model is the hexagonal lattice Ising model. The Hamiltonian is given by
\begin{align}
    \mathrm{H}_{\text{Ising}}^{\hexagon}=-J\sum_{\langle \rm v,v'\rangle}s_{\rm v}s_{\rm v'}\,.
\end{align}
The partition function takes the same form as in \eqref{eq:2DtriangIsingmodelpartitionfn} but on the hexagonal lattice
\begin{align}
    Z_{\text{Ising}}^{\hexagon}(J)=\sum_{\{s_{\rm v}\}}\exp(J\sum_{\langle\rm v ,v'\rangle}s_{\rm v}s_{\rm v'})\,.
    \label{eq:2DhexagonIsingmodelpartitionfn}
\end{align}
The two classical models are obtained from the following chain complex
\begin{align}
    0\longrightarrow C_{2}\overset{\partial}{\longrightarrow}C_1\overset{\partial}{\longrightarrow} C_0\overset{}{\longrightarrow}0\,,
    \label{eq:chaincomplextriangularlattice}
\end{align}
with the triangular lattice Ising model originating from the boundary map $C_1\overset{\partial}{\longrightarrow} C_0$ and the hexagonal lattice Ising model originating from $C_{1}\overset{\partial^*}{\longrightarrow}C_2$\footnote{Comparing with the general chain complex formalism in Section~\ref{sec:KWdefectgenIsingmodel}, $C_Z\equiv C_2$, $C_{\rm q}\equiv C_1$ and $C_X\equiv C_0$.}. Note that $C_2$ is the chain group constituting chains of plaquettes, $C_1$ is the chain group constituting chains of edges and $C_0$ is the chain group constituting chains of vertices.
\subsubsection{$\mathbb{Z}_2$ symmetry defect}
The operation given by flipping $s_{\rm v}\longrightarrow-s_{\rm v}$  $\forall\,\rm v$ is a $\mathbb{Z}_2$ symmetry of the Hamiltonian \eqref{eq:triangleIsing}. A defect of this symmetry can be created by applying the spin flip operation on the spins inside a  region $S$. This will essentially flip the coupling $J$ on the links that intersect with a loop $C$ in the dual lattice. A general $\mathbb{Z}_2$ symmetry defect is created along $C$ (a loop in the dual lattice) for any cycle $C$ in $H_1(M,\mathbb{Z}_2)$.
\subsubsection{Kramers-Wannier duality}
The Kramers-Wannier duality relates the two partition functions on the triangular and hexagonal lattices
\begin{align}
    Z_{\text{Ising}}^{\triangle}(J)=\frac{2^{\frac{|\Delta_{\rm v}|}{2}}(\sinh(2J))^{|\Delta_{\rm e}|/2}}{2^{\frac{|\Delta_{\rm p}|}{2}}\sqrt{|H_1(M,\mathbb{Z}_2)|}}\sum_{z_1\in H_1(M,\mathbb{Z}_2)}Z_{\text{Ising}}^{\hexagon}(J^*,z_1)\,,
\end{align}
where $\Delta_{\rm v}$,$\Delta_{\rm e}$, and $\Delta_{\rm p}$ denote the set of vertices, edges and plaquettes of the triangular lattice. $H_1(M,\mathbb{Z}_2)$ denote the set of edges that form a line and have no boundary on the traingular lattice. The twisted partition function $Z_{\text{Ising}}^{\hexagon}(J^*,z_1)$ is same as \eqref{eq:twistedIsingpartitionfn} with the interactions between plaquette centers in the triangular lattice.
\subsection{Duality defect via strange correlator}
Here, we apply the construction of duality defect through strange correlator to the chain complex~\eqref{eq:chaincomplextriangularlattice}.  The relation \eqref{eq:ZXpartitiondualitydefect} takes the form
\begin{align}
    Z_{\text{Ising}}^{\triangle}(J)&=\frac{2^{-\frac{|\Delta_{\rm p}^{\rm A}|}{2}}2^{-\frac{|\Delta_{\rm v}^{A^c}|}{2}}2^{\frac{|\Delta_{\rm v}|}{2}}}{|H_1(M,\mathbb{Z}_2)|(\sinh(2J^*))^{\frac{|\Delta_{\rm e}^{\rm A}|}{2}}}\sum_{z_{1}\in H_1(M,\mathbb{Z}_2)}\sum_{\Big\{\substack{s_{\sigma_{\rm v}}=\pm 1\\
    (-1)^{h_{\sigma_{\rm v}}}=s_{\sigma_{\rm v}}}\Big\}\rvert_{\sigma_{\rm v}\in\Delta_{\rm v}^{\rm A^c}}}\sum_{\Big\{\substack{s_{\sigma_{\rm p}}=\pm 1\\
    (-1)^{h_{\sigma_{\rm p}}}=s_{\sigma_{\rm p}}}\Big\}\rvert_{\sigma_{\rm p}\in\Delta_{\rm p}^{\rm A}}}\nonumber\\
    &\hspace{2cm}\exp\left(\sum_{\sigma_{\rm e\in\Delta_{\rm e}^{\rm A^c}}}Js(\partial\sigma_{\rm e})+\sum_{\sigma_{\rm e\in\Delta_{\rm e}^{\rm A}}}(-1)^{\#(z_1\cap \sigma_{\rm e})}J^*s(\partial^*\sigma_{\rm e})\right)\prod_{\substack{\sigma_{\rm p}\in\Delta_{\rm p}^{\rm A|A^c}\\
    \sigma_{\rm v}\in\partial(\partial\sigma_{\rm p}\sqcap \Delta_{\rm e}^{\rm A})}}(-1)^{h_{\sigma_{\rm v}}h_{\sigma_{\rm p}}}\times s(\partial(z_1\sqcap\Delta_{\rm e}^{\rm A^c}))\,.
\end{align}
The above is the identity that relates the partition function of triangular lattice Ising model and gauged version of the triangular lattice Ising model in a region defined by $\Delta_{\rm e}^{\rm A}$. The partition function in the presence of two duality defects are given by
\begin{align}
    Z_{\text{Ising}}^{\triangle, \mathcal{D}}&=\frac{2^{-\frac{|\Delta_{\rm p}^{\rm A}|}{2}}2^{-\frac{|\Delta_{\rm v}^{A^c}|}{2}}2^{\frac{|\Delta_{\rm v}|}{2}}}{(\sinh(2J^*))^{\frac{|\Delta_{\rm e}^{\rm A}|}{2}}}\sum_{\Big\{\substack{s_{\sigma_{\rm v}}=\pm 1\\
    (-1)^{h_{\sigma_{\rm v}}}=s_{\sigma_{\rm v}}}\Big\}\rvert_{\sigma_{\rm v}\in\Delta_{\rm v}^{\rm A^c}}}\sum_{\Big\{\substack{s_{\sigma_{\rm p}}=\pm 1\\
    (-1)^{h_{\sigma_{\rm p}}}=s_{\sigma_{\rm p}}}\Big\}\rvert_{\sigma_{\rm p}\in\Delta_{\rm p}^{\rm A}}}\nonumber\\
    &\hspace{2cm}\exp\left(\sum_{\sigma_{\rm e\in\Delta_{\rm e}^{\rm A^c}}}Js(\partial\sigma_{\rm e})+\sum_{\sigma_{\rm e\in\Delta_{\rm e}^{\rm A}}}J^*s(\partial^*\sigma_{\rm e})\right)\prod_{\substack{\sigma_{\rm p}\in\Delta_{\rm p}^{\rm A|A^c}\\
    \sigma_{\rm v}\in\partial(\partial\sigma_{\rm p}\sqcap \Delta_{\rm e}^{\rm A})}}(-1)^{h_{\sigma_{\rm v}}h_{\sigma_{\rm p}}}\,.
\end{align}
\subsection{Moving and fusing defect}
Let us look at the movement relations that move the defects. Th movement equations takes exactly the same form in the 2D Ising model on a square lattice \eqref{eq:Isingdefectmovement} and \eqref{eq:Isingdefectmovement2}. Let us look at the fusion of two duality defects in Figure~\ref{fig:triangulardualitydefect}. After fusing two duality defects, we get a sum of the form 
\begin{align}
    (-1)^{\sum\limits_{\substack{\alpha_{i},\,
    i\text{ even }}}h_{\alpha_i}\left(h_{b_{i}}+h_{b'_{i+1}}+h_{c_i}+h_{c_{i+1}}\right)+h_{\alpha'_i}\left(h_{b'_{i+1}}+h_{b_{i+1}}\right)+h_{\alpha_{i+1}}\left(h_{b_{i+1}}+h_{b''_{i+1}}\right)+h_{\alpha'_{i+1}}\left(h_{b''_{i+1}}+h_{b_{i+2}}+h_{c_{i+1}}+h_{c_{i+2}}\right)}\,.
\end{align}
This sum imposes the constraint $h_{b_{i}}=h_{b'_i}=h_{b''_i}$ $\forall\, i$ and $h_{b_i}=h_{c_i}$ $\forall\, i$ or $h_{b_i}=h_{c_i}+1$ $\forall\, i$. 
\begin{figure}[h!]
    \centering
    \input{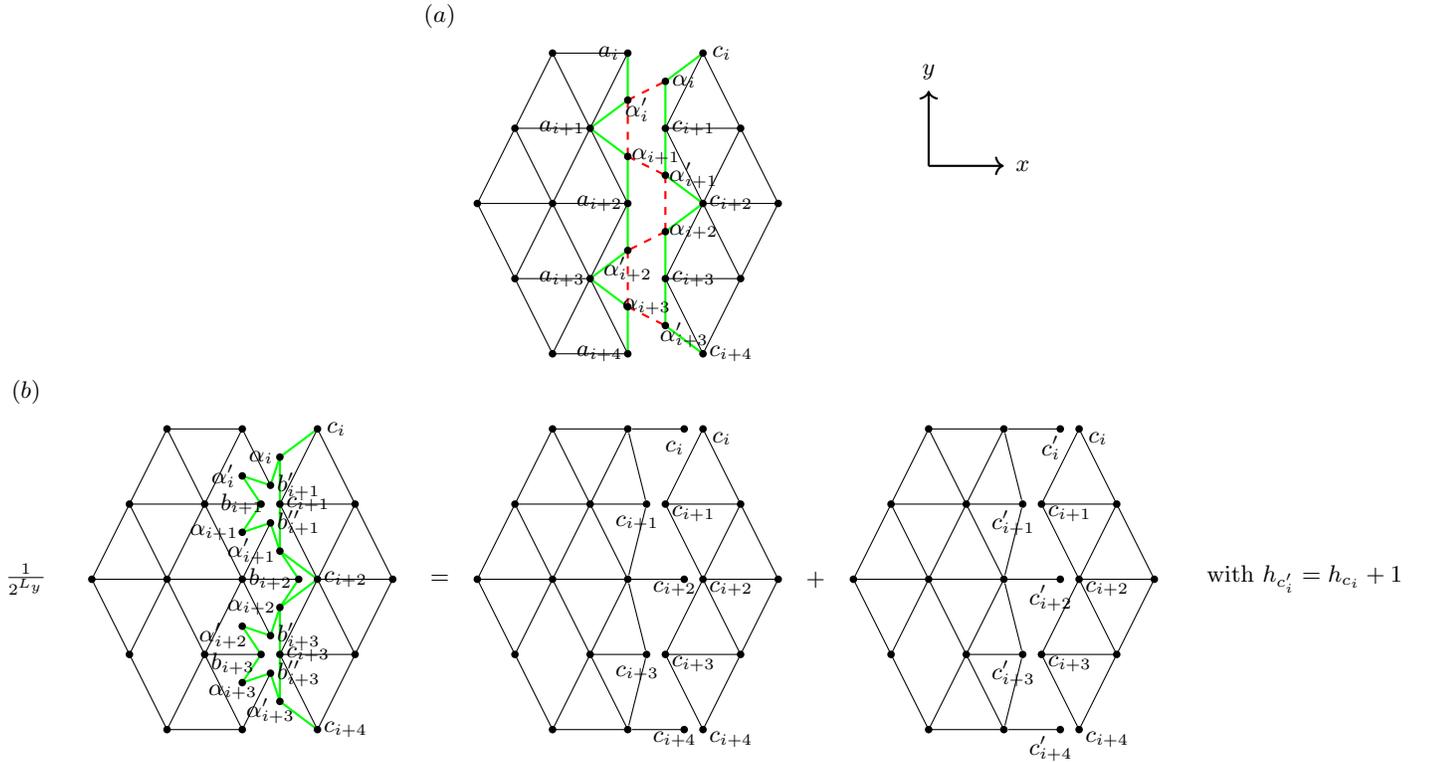}
    \caption{The figure illustrate the fusion of two duality defects. (a) Two duality defects are placed near to each other just before fusion. The dashed red edges are the dual Ising interaction. (b) Fusion of two duality defects produces identity defect and spin flip defect. $L_y$ is the number of vertices along the red edges that wind around $y$ direction on the dual lattice.}
    \label{fig:triangulardualitydefect}
\end{figure}
Suppose $\mathcal{D}$ denote the duality defect and $D_{\eta}$ denote the spin flip defect and the absence of defect insertion as $\text{I}$, then
\begin{align}
    \mathcal{D}^2=\text{I}+D_{\eta}\,.
\end{align}
\pagebreak
\bibliography{Ref}
\end{document}